\documentclass[11pt,twoside,openany]{book}
 
\usepackage[top = 3cm, bottom = 3cm, left = 2.5cm, right = 2.5cm]{geometry}

\usepackage{amsmath,amssymb}
\usepackage{graphicx}
\usepackage{cancel}

\usepackage{color}
\definecolor{indigo}{RGB}{0,0,120}
\usepackage[colorlinks=true, linkcolor=indigo, citecolor=blue, urlcolor=indigo]{hyperref}

\def\imply{\Rightarrow}
\newcommand{\pt}{\noindent {$\bullet$~}}

\newcommand{\tl}[1]{\tilde{#1}}
\newcommand{\dd}[2]{\frac {\partial #1}{\partial #2}}
\newcommand{\deldel}[2]{\frac {\delta #1}{\delta #2}}
\newcommand{\pdr}{\partial}
\newcommand{\DD}[2]{\frac {d #1}{d #2}}
\newcommand{\grad}{{\bf \nabla}}
\newcommand{\diver}[1]{{\bf \nabla \cdot {#1}}}

\newcommand{\beq}{\begin{equation}}
\newcommand{\eeq}{\end{equation}}
\newcommand{\beqs}{\begin{eqnarray}}
\newcommand{\eeqs}{\end{eqnarray}}

\newcommand{\half}{\frac{1}{2}}
\newcommand{\ov}[1]{\frac{1}{#1}}

\def\al{\alpha} 		
\def\del{\delta}	
\def\eps{\epsilon} 
\def\la{\lambda}		
\def\sig{\sigma}
\def\tht{\theta}
\def\om{\omega}		
\def\Om{\Omega}
\def\g{\gamma}
\newcommand{\G}{{\Gamma}}

\newcommand{\bfS}{{\bf S}}
\newcommand{\bfZ}{{\bf Z}}
\newcommand{\bfP}{{\bf P}}
\newcommand{\bfQ}{{\bf Q}}
\newcommand{\bfa}{{\bf a}}
\newcommand{\bfb}{{\bf b}}
\newcommand{\bff}{{\bf f}}
\newcommand{\bfg}{{\bf g}}
\newcommand{\bfh}{{\bf h}}
\newcommand{\bfu}{{\bf u}}
\newcommand{\bfv}{{\bf v}}
\newcommand{\bfw}{{\bf w}}
\newcommand{\bfx}{{\bf x}}
\newcommand{\bfy}{{\bf y}}

\newcommand{\bfj}{{\bf j}}
\newcommand{\bfM}{{\bf M}}
\newcommand{\bfk}{{\bf k}}
\newcommand{\bfr}{{\bf r}}
\newcommand{\bfs}{{\bf s}}

\newcommand{\bfA}{{\bf A}}
\newcommand{\bfE}{{\bf E}}
\newcommand{\bfB}{{\bf B}}
\newcommand{\bfC}{{\bf C}}
\newcommand{\bfD}{{\bf D}}
\newcommand{\bfl}{{\bf l}}

\newcommand{\bfL}{{\bf L}}
\newcommand{\bfT}{{\bf T}}
\frontmatter

\newcount\colveccount  
\newcommand*\colvec[1]{\global\colveccount#1  \begin{pmatrix} \colvecnext} \def\colvecnext#1{#1 \global\advance\colveccount-1
        \ifnum\colveccount>0 \\ \expandafter\colvecnext
        \else \end{pmatrix} \fi}
\usepackage{subcaption}

\begin{document}

\thispagestyle{empty}

\frontmatter

\begin{center}

\vspace*{5mm}

{\Large \bf Conservative regularization of neutral fluids and plasmas}
\vspace{1cm}
\\{\Large by}\\ 
\vspace{1cm}
{\Large \textbf{Sonakshi Sachdev}} 

\vspace{2cm}
\large 

{\it A thesis submitted in partial fulfilment of the requirements for \\ the degree of Doctor of Philosophy in Physics} \\
\vspace*{0.8cm}
to  
\vspace*{0.8cm} 

Chennai Mathematical Institute

\begin{figure}[h]
\begin{center}
 \includegraphics[width = 5cm]{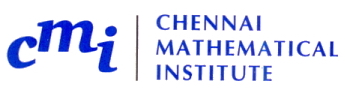}
 \end{center}
\end{figure}

\vspace{0.1cm}

Plot H1, SIPCOT IT Park, Siruseri,\\
Kelambakkam, Tamil Nadu 603103, \\
India

\vspace{2cm}

Submitted March 2020,

\vspace{1cm}

Defended 13th July, 2020

\end{center}
\normalsize


\newpage
\thispagestyle{empty}
\large
\begin{flushleft}
Advisor:

Dr. Govind S Krishnaswami, \emph{Chennai Mathematical Institute (CMI).}\\ 
\vspace*{1cm}
Doctoral Committee Members: 
\begin{enumerate}
\item Dr. Alok Laddha, \emph{Chennai Mathematical Institute (CMI).}\\
\item Dr. R Shankar, \emph{Institute of Mathematical Sciences (IMSc).}
\end{enumerate}
\end{flushleft}



\newpage

\chapter*{\center Declaration}

\large
This thesis titled \emph{Conservative regularization of neutral fluids and plasmas} is a presentation of my original research work, carried out with my collaborators and under the guidance of Prof. Govind S Krishnaswami at Chennai Mathematical Institute. This work has not formed the basis for the award of any degree, diploma, associateship, fellowship or other titles in Chennai Mathematical Institute or any other university or institution of higher education.

\vspace*{15mm}
\begin{flushright}
Sonakshi Sachdev,\\
Chennai Mathematical Institute\\
March, 2020.
\end{flushright}

\vspace*{20mm}

In my capacity as the supervisor of the candidate's thesis, I certify that the above statements are true to the best of my knowledge.
\vspace*{15mm}
\begin{flushright}
Govind S Krishnaswami\\
Thesis Supervisor, \\
Chennai Mathematical Institute\\
March, 2020.
\end{flushright}



\chapter*{Acknowledgments}
\addcontentsline{toc}{chapter}{Acknowledgments}

First and foremost, I would like to thank my advisor, Prof. Govind S. Krishnaswami for his guidance, encouragement and patience. I am grateful for the time he has devoted to our discussions. His insights on the subject have been immensely helpful in bringing this thesis to fruition. He has been a source of inspiration and motivation at every stage of the process. I am most grateful to him for his unwavering support, persistent belief in my abilities, helpful advice and criticism when needed. His dedication to the project cannot be overstated. This thesis  would not have been possible without his nurturing.

I also take this opportunity to thank Prof. A Thyagaraja for his mentorship. This work has greatly benefitted from his expertise, encouragement and practical suggestions. With his annual trips to CMI from the UK and the prompt responses via email, I never felt that he was not around. Discussions with him were immensely fun and educative.

Collaboration with Sachin Phatak on this project is also gratefully acknowledged. 

I extend my thanks to the doctoral committee members Prof. R. Shankar and Prof. Alok Laddha for regular discussions and overseeing my thesis. I also thank Prof. K G Arun, my Faculty Advisor during MSc. I am extremely grateful to Prof. Abhijit Sen and Prof. Mark Birkinshaw for their helpful comments and questions on the thesis.

I would like to pay my special regards to Shikha Basu, Dr. Pranab Basu and Anirban Biswas among other inspiring teachers for instilling in me a love for physics and mathematics from a young age.

I would also like to express my deepest gratitude to the faculty members and my wonderful colleagues at Chennai Mathematical Institute for their continued support. I have  enjoyed discussing various aspects of physics, as well as life, with them. I would like to specially thank Keerthan Ravi, Arjun Arul, Himalaya Senapati, Vishnu T R, Ritam Basu, Chinmay Kalaghatgi, Kedar Kolekar and  Debangshu Mukherjee for their invaluable support during the course of this project. 

Special thanks go to the CMI staff, especially Rajeshwari Nair, for being extremely helpful throughout my time as a student here. 

I thank the Infosys Foundation, J N Tata Trust and Science and Engineering Research Board (SERB), Government of India for funding. I wish to acknowledge the funding received from the International Travel Support Scheme of SERB for attending the Les Houches plasma physics school, 2019. I also wish to thank both SERB and CMI for funds to attend the SERB Nonlinear Dynamics School at Pune University (2018) as well as the school on mathematical aspects of fluid flows at Kacov, Czech Republic in 2019, the CNSD conferences in 2018 and 2019 and academic visits to other institutes.

My oldest friends, Rohit Ghosh, Kingshuk Mishra and Sayantan Chaudhuri have always been a huge help in all matters of life, big and small. For that, I shall always be grateful to them.

I convey my warmest regards to Monica Leekha Oberoi (aunt),  Vasudha Sachdev (aunt), Sonit Singh Sachdev (uncle), my grandparents and every other person in my supportive family network for nudging me forward in life. 

I am indebted to my parents, Veenu Sachdev (mother) and Nonit Singh Sachdev (father) and sister Neelakshi Sachdev from whom I have received unconditional love and support. I owe them everything. I dedicate this thesis to them.



\newpage

\chapter*{\center Abstract}
\addcontentsline{toc}{chapter}{Abstract}

\normalsize

Ideal systems of equations such as those of Euler and MHD may develop singular structures like shocks, vortex sheets and current sheets. Among these, vortical singularities arise due to vortex stretching which can lead to unbounded growth of enstrophy. Viscosity and resistivity provide dissipative regularizations of these singularities. 

In analogy with the dispersive KdV regularization of the 1D inviscid Burgers' equation, we propose a local conservative regularization of ideal 3D compressible flows, MHD and two-fluid plasmas (with potential applications to high vorticity flows with low dissipation). The regularization involves introducing a vortical `twirl' term $\lambda^2 \bfw \times (\grad \times \bfw)$ in the velocity equation. The cut-off length $\lambda $ must be $\propto 1/\sqrt{\rho}$ to ensure the conservation of a `swirl' energy. The latter  includes positive kinetic, compressional, magnetic and vortical contributions, thus leading to a priori bounds on enstrophy. The extension to two-fluid plasmas involves additionally magnetic `twirl' terms $\lambda_l^2 (q_l/m_l) \bfB \times (\grad \times \bfw_l)$ in the ion and electron velocity equations ($l = i,e$) and a solenoidal addition to the current in Amp\`ere's law. A Hamiltonian-Poisson bracket formulation is developed using the swirl energy as Hamiltonian.  We also establish a minimality property of the twirl regularization. A swirl velocity field $\bfv_*$ is shown to transport vortex and magnetic flux tubes (with conserved flow/magnetic helicity) as well as $\bfw/\rho$ and $\bfB/\rho$, thus generalizing the Kelvin-Helmholtz and Alfv\'en theorems.

The steady regularized equations are used to model a rotating vortex, MHD pinch and vortex sheet. Our regularization could facilitate numerical simulations of neutral and charged fluids and a statistical treatment of vortex and current filaments in 3D. 

Finally, we briefly describe a conservative regularization of shock-like singularities in compressible flow generalizing both the KdV and nonlinear Schr\"odinger equations to the adiabatic dynamics of a gas in 3D.




\chapter*{List of papers}
\addcontentsline{toc}{chapter}{List of papers}

\large
This thesis is based on the following papers:

\begin{itemize}

	\item Krishnaswami G S, Sachdev S and Thyagaraja A, {\it Local conservative regularizations of compressible MHD and neutral flows}, \href{https://aip.scitation.org/doi/10.1063/1.4942621}{Phys. Plasmas {\bf 23}, 022308 (2016)}, \\ \href{https://arxiv.org/abs/arXiv:1602.04323}{[arXiv:1602.04323]}.
	
	\item Krishnaswami G S, Sachdev S and Thyagaraja A, {\it Conservative regularization of compressible flow and ideal magnetohydrodynamics}, \href{https://arxiv.org/abs/arXiv:1510.01606}{[arXiv:1510.01606]}.
	
	\item Krishnaswami G S, Sachdev S and Thyagaraja A, {\it Conservative regularization of compressible dissipationless two-fluid plasmas}, \href{https://aip.scitation.org/doi/10.1063/1.5016088}{Phys. Plasmas {\bf 25}, 022306 (2018)}, \\ \href{https://arxiv.org/abs/arXiv:1711.05236}{[arXiv:1711.05236]}.
	
	\item Krishnaswami G S, Phatak S, Sachdev S and Thyagaraja A, {\it Nonlinear dispersive regularization of inviscid gas dynamics}, \href{https://doi.org/10.1063/1.5133720}{AIP Advances, 10, 025303 (2020)}, \href{https://arxiv.org/abs/1910.07836}{[arXiv:1910.07836]}.
		
\end{itemize}

Though not directly related to this thesis topic, the following article was also published during the course of the PhD:

\begin{itemize}

\item Krishnaswami G S and Sachdev S, {\it Algebra and geometry of Hamilton's quaternions}, \href{https://www.ias.ac.in/article/fulltext/reso/021/06/0529-0544}{ Resonance, 21, 6, June 2016}, \href{https://arxiv.org/abs/1606.03315}{[arXiv:1606.03315]}.

\end{itemize}


\small

\tableofcontents

\mainmatter

\chapter{Introduction}
\label{s:intro}

\normalsize

Mathematical models for various physical systems require regularization. Quantum mechanics regularizes the ultraviolet (UV) divergence in the energy radiated by a blackbody, with $\hbar$ playing the role of a regulator. UV regularization and renormalization are necessary to extract finite physical quantities from divergent renormalizable quantum field theories in particle physics. Somewhat analogously, ideal flows and plasmas can develop singular structures such as shocks, vortex/current sheets associated, for instance, with discontinuities in density, velocity/magnetic field and require regularization.

In particular, three-dimensional (3D) fluid dynamics fundamentally involves vortex stretching, a process which in the standard Euler equations leads (as indicated in the classic work of Taylor and Green \cite{taylor-green} on Navier-Stokes (NS) with very low viscosity, see also \cite{Frisch,KRSreenivasan-onsager,agafontsev-kuznetsov-mailybaev}) to unbounded growth of the fluid enstrophy [enstrophy density is the square of local vorticity $\bfw = \grad \times \bfv$ where $\bfv$ is the velocity field]. Vorticity may also diverge in the presence of singular structures such as vortex sheets, with discontinuous tangential velocity. This is analogous to the loss of single-valuedness of $u$ and development of singularities in derivatives of $u$ in the 1D Hopf or ``kinematic wave'' equation (KWE) ($u_t + u u_x = 0$) which is used to model wave-breaking. This equation admits a {\it dissipative} regularization in the well-known viscous Burgers equation ($u_t + u u_x = \nu u_{xx}$), and thereby provides an excellent, exactly soluble [via the Cole-Hopf transformation] model of random arrays of 1D shocks, traffic flows etc. On the other hand, the Hopf equation also admits a {\it dispersive} regularization via the KdV equation ($u_t -6 u u_x + u_{xxx} = 0$). The KdV equation has been  extensively discussed \cite{Miura,davidson} as the paradigmatic, conservatively regularized extension of the KWE with applications in many fields (E.g. solitons and integrable systems, shallow water waves, ion acoustic waves, long internal ocean waves and blood pressure waves). It is this latter example that provides the motivation for the results presented in this thesis and the papers on which it is based \cite{thyagaraja,govind-sonakshi-thyagaraja-pop,govind-sonakshi-thyagaraja-two-fluid,govind-sachin-sonakshi-thyagaraja-r-gas-dynamics}. Other well-known examples of 1D conservative systems such as the nonlinear Schr\"{o}dinger equation (NLS) also show that effective analysis and computation are greatly facilitated when the dynamics imply bounded motions rather than the development of singularities which prevent a proper understanding of the system dynamics and statistical mechanics.

In 3D, the analogue of the Hopf equation is the Euler equation of inviscid fluid dynamics. The latter has a standard dissipative regularization in the Navier-Stokes equations, on which almost all of modern fluid dynamics rests. We seek a consistent and well-motivated 3D analogue of a KdV-like dissipationless regularization of the Euler equation. The physical principles guiding our choice of regularization terms are 
\begin{enumerate}

\item  minimality in nonlinearity and derivatives (local)\footnote{There are other interesting conservative regularizations of the 3D Euler equations, motivated partly by numerical schemes or involving averaging procedures, such as the Euler-$\alpha$ and vortex blob regularizations \cite{chorin,holm-marsden-ratiu}. The incompressible Euler-$\alpha$ equations are the geodesic equations for the H1 metric on the group of volume preserving diffeomorphisms of the flow domain. They correspond to the energy functional $\rho \int \left(\half \bfv^2 + \half \al^2 (\pdr_i v_j)^2 \right) d\bfr = \rho \int \left(\half \bfv^2 + \half \al^2 \bfw^2 \right) d\bfr$ for $\grad \cdot \bfv = 0$, with $\al$ a regularizing length. However, the resulting Euler-$\alpha$ equation of motion for $\bfv$ is highly non-local as it involves the advecting velocity $(1 - \al^2 \grad^2)^{-1} \bfv$.} 

\item preservation of symmetries (rotations, translations, parity, time reversal) 

\item validity for general initial data

\item existence of a conserved energy and bounded enstrophy

\item presence of a short distance cutoff 

\item retention of the continuity equation for density $\rho_t + \grad \cdot (\rho \bfv) = 0$  

\item absence of entropy production.

\end{enumerate}

Based on these principles, in Chapter \ref{s:r-euler-r-mhd}, we propose and study  what we call the regularized Euler (R-Euler) equations for compressible barotropic (pressure a function of $\rho$ but not specific entropy $s$, $p  = p(\rho, \cancel s))$ flow given by
	\beq
	\dd{\bfv}{t} + (\bfv \cdot \grad) \bfv = -\frac{\grad p}{\rho} - \underline{\la^2 \bfw \times (\grad \times \bfw)} \quad \text{and} \quad \rho_t + \grad \cdot (\rho \bfv) = 0.
	\label{e:R-Euler-compress}
	\eeq
Unlike in 1D, there is no KdV-like regularizer linear in velocity that preserves Eulerian symmetries (say, for instance, time-reversal for $\grad \times \grad \times \grad \times \bfv$). The `twirl' term $- \la^2 \bfw \times (\grad \times \bfw)$ is quadratic in velocities and should be significant in flows with large vorticity or its curl. The short-distance regulator $\la$ has dimensions of length. We show that if $\la$ satisfies the constitutive relation $\la^2 \rho = $ constant then (\ref{e:R-Euler-compress}) admits the conserved energy (\ref{e:cons-enrgy-r-euler}). For incompressible flow with constant $\rho$, $\la$ is a constant. The twirl force density $- \la^2 \rho \, \bfw \times (\grad \times \bfw)$ can be thought of as a vortical counterpart of the magnetic Lorentz force density $\bfj \times \bfB = - \bfB \times (\grad \times \bfB)/\mu_0$ familiar from non-relativistic MHD, with $\la^2 \rho$ replacing the constant $1/\mu_0$. The constitutive relation $\la^2 \rho = $ constant implies that $\la$ is like a position-dependent mean free path: smaller in denser regions and vice-versa. The twirl term preserves the Galilean (rotation, translation and boost) symmetries of ideal compressible flow. The above-mentioned conserved energy for the system is given by
	\beq
	E_{\rm R-Euler} = \int {\cal E}_{\rm R-Euler}\: d\bfr = \int \left[ \frac{\rho \bfv^2}{2} + U(\rho) + \frac{\la^2 \rho}{2} \bfw^2 \right] d\bfr.
	\label{e:cons-enrgy-r-euler}
	\eeq
Here, $U(\rho)$ is the compressional energy density. For a polytropic gas with specific heat ratio $\g$ ($p \propto \rho^{\g}$), $U(\rho) = p/(\g - 1)$. All terms in $E_{\rm R-Euler}$ are positive, so that enstrophy is a priori bounded above
	\beq
	\int \bfw^2 d\bfr \leq \frac{2 E_{\rm R-Euler}}{\la^2 \rho} 
	\label{e:enstrophy-bound}
	\eeq
Although introduced as a formal regularizer, it is conceivable that such a twirl term could arise in a Chapman-Enskog-like expansion of kinetic equations in the Knudsen number. 

The twirl term $-\la^2 \bfw \times (\grad \times \bfw)$ in (\ref{e:R-Euler-compress}) is a minimal (in the sense of effective local field theory, see Appendix \ref{a:minimality}) nonlinear dispersive regularization of the ideal equations leading to bounded enstrophy and conservation laws. Indeed, as with the NS regularization of Euler, the twirl term in (\ref{e:R-Euler-compress}) increases the spatial order by unity. On the other hand, Ladyzhenskaya's `hyperviscosity' regularization \cite{ladyzhenskaya-1} of the Euler [and Navier-Stokes] equation involves the fourth order term $\eps (\grad^2)^2 \bfv$ with $\eps$ constant. In \cite{ladyzhenskaya-2}, she also considers a nonlinear regularization term $\nu_3 \grad^2 \bfv$ where the viscosity coefficient $\nu_3$ depends on the sum of squares of the components of the rate of strain tensor. Both these dissipative regularizations serve to balance, in principle, the nonlinear vortex-stretching mechanism of 3D inviscid flow. Our conservative nonlinear twirl term is similarly responsible for controlling the growth of vorticity at short distances of order $\la$.

The above ideas on conservative regularization are also applicable to charged fluids that occur in plasma physics. Plasma physics finds extensive applications in astrophysics, physics of fusion devices like tokamaks, stellarators and in inertial confinement and in technological applications \cite{Hazeltine-Meiss, Wesson, Lifshitz-Pitaevski, Rosenbluth-Sagdeev, Kulsrud, choudhuri, Michel}. Plasmas have extremely complex dynamics when they interact with self-generated and externally applied electric and magnetic fields. The dynamics of such systems are governed both by Maxwell's equations and either a kinetic or fluid model representing the co-evolution of the plasma variables. In kinetic descriptions appropriate distribution functions are introduced for the ions and electrons of the plasma. They are evolved according to equations such as the Boltzmann-Fokker-Planck system. The charge and current densities derived from the distribution functions are then used to evolve the fields. In fluid models only the first few ``principal moments'' like the number densities, velocities, temperatures, stresses and heat fluxes appear. It is often the case that the fluid description provides a relatively tractable system which can be used to describe a variety of phenomena actually observed in experiments and in the cosmos. Among fluid models, the simplest ones are generalizations of the well-known dissipationless Euler equations of neutral fluid dynamics to include the effects of electromagnetic body forces. A typical example is provided by the classic model known as Ideal (one-fluid) Magnetohydrodynamics (MHD) (see, for example, \cite{Goedbloed-Poedts, JBT}) which has found very wide application in both fusion plasma theory and in astrophysical theories. This theory was used by Alfv\'{e}n to describe plasma waves in a magnetized fluid (see the classic text by Stix \cite{Stix}) and to show that in the absence of dissipation [resistivity and viscosity and possibly thermal diffusivity] the magnetic field is ``frozen'' into the flow. This result has wide application to both solar physics and to important classes of instabilities known to occur in tokamak plasmas (``ideal ballooning and kink modes'', see \cite{Hazeltine-Meiss,Wesson,Goedbloed-Poedts, JBT}). It is generally the case that even the simplest ideal  descriptions of fluid and plasma equations involves rather complicated nonlinear partial differential equations. One does not have useful exact, analytically derived solutions valid for experimentally relevant situations. The only generally applicable methods are numerical methods. The dissipationless two-fluid (ion and electron) equations are similar in their qualitative properties to the Euler equations of inviscid fluid dynamics and ideal MHD. They possess several conservation laws but involve energy transfer mechanisms which can lead to short-wavelength singularities like vortex and current sheets, shocks and finite-time unbounded behaviour of enstrophy and current density. It is usually the case that ultraviolet singularities of these types are resolved by viscosity, thermal conductivity and electrical resistivity. All these are entropy-producing effects and are not consistent with the conservation properties of the dissipationless models.

Thus, in Chapter \ref{s:r-euler-r-mhd} of this thesis, we propose a dissipationless regularization of ideal MHD that we refer to as R-MHD. Our compressible barotropic R-MHD equations are 
	\beqs
	\dd{\rho}{t} + \grad \cdot (\rho \bfv) &=& 0, \quad \dd{\bfv}{t} + (\bfv \cdot \grad) \bfv = -\frac{\grad p}{\rho} + \frac{\bfj \times \bfB}{\rho} \underline{- \la^2 \bfw \times (\grad \times \bfw)} \quad \text{and} \cr
	\dd{\bfB}{t} &=& - \grad \times \bfE = \grad \times(\bfv \times \bfB \underline{- \la^2 \bfB \times (\grad \times \bfw)}) \quad \text{and} \quad \grad \cdot \bfB = 0.
	\eeqs
These R-MHD equations are seen to include both vortical $\la^2 \bfw \times (\grad \times \bfw)$ and magnetic $\la^2 \bfB \times (\grad \times \bfw)$ twirl regularizers in the velocity and induction equations. As before, the regularizing length $\la$ must satisfy the constitutive relation $\la^2 \rho = $ constant. We note that in plasma physics there are natural length-scales which are inversely proportional to the square-root of the number density $n$. For example, in S.I.units, the electron collisionless skin-depth $\del_e = c/\om_{pe} \propto 1/\sqrt{n_e}$ where $\om_{pe} = \sqrt{\frac{e^2 n_e}{m_e \eps_0}}$ is the electron plasma frequency and $n_e$ is the electron number density. Thus if $\la \approx \del_e$ then $\la^2 \rho$ will indeed be a constant. In any event, it is well-known that ideal MHD is not valid at length scales of order $\del_e$. Another example is provided by the electron Debye length $\la_D = \sqrt{k_B T_e \eps_0 /n_e e^2}$ in an isothermal plasma with electron temperature $T_e$. Thus, having a cut-off of this kind will not affect any major consequence of ideal MHD on meso- and macro-scales and yet provide a finite upper bound to the enstrophy of the system depending on the regulator $\la$. This constitutive law implies a conserved energy in R-MHD. In addition to kinetic, compressional and vortical contributions,  this conserved energy also has a magnetic contribution:
	\beq
	E_{\rm R-MHD} = \int {\cal E}_{\rm  R-MHD} \: d\bfr = \int \left( \frac{\rho \bfv^2}{2}+ U(\rho) + \frac{\la^2 \rho \bfw^2}{2} + \frac{\bfB^2}{2 \mu_0} \right) d\bfr.
	\eeq

A Hamiltonian formulation for R-Euler and R-MHD is made possible by taking the total conserved energy (referred to as `swirl' energy) of the system as the Hamiltonian and using the elegant Poisson structures \cite{arnold} for compressible flow due to Morrison and Greene \cite{morrison-greene,morrison-review,morrison-aip} anticipated in Landau's \cite{landau} paper on quantum theory of superfluids (cf. Equations (1.7,1.8)) and developed by London \cite{london}. This formalism shows that the extended systems formally share the Hamiltonian, non-canonical Poisson structures of the original, singular conservative dynamics. The existence of a positive definite Hamiltonian and bounded enstrophy should facilitate the formulation of a valid statistical mechanics of 3D vortex tubes, extending the work of Onsager on 2D line vortices. The same remark also applies to the 2D statistical mechanics of line current filaments developed by Edwards and Taylor \cite{edwards-taylor} and many other authors in ideal MHD theory. 

In Chapter \ref{s:two-fluid}, the ideas used in regularizing ideal Eulerian flows and MHD are generalized  to two-fluid plasmas. In two-fluid plasmas, each species (e.g. ions, electrons) is treated as a fluid with dynamical density and velocity producing changing electric charge and current densities. Maxwell's equations govern the evolution of the electric and magnetic fields which in turn affect the motion of the charged fluids via the Lorentz force. Thus, the dynamical variables of a two-fluid plasma are: $\bfE$, $\bfB$, ion and electron velocities $\bfv_{i,e}$, number densities $n_{i,e}$ and partial pressures $p_{i,e}$. Of particular interest are  quasineutral plasmas where $n_e \approx n_i$ with $\bfE$ nevertheless nontrivial and determined by Ohm's law (see Section \ref{s:quasineutral-two-fluid}). Another important limiting case is Hall-magnetohydrodynamics (Hall-MHD) which is a good approximation for phenomena on length scales less than the ion inertial  length $(\del_i)$ but greater than the electron inertial length $(\del_e)$ so that the magnetic field is frozen into the electron fluid. On length scales much greater than the Debye lengths $(\la_D)$, gyro-radii $(\sqrt{k_B T_{i,e} m_{i,e}}/e B)$ and skin-depths $(\del_{i,e})$ and frequencies less than cyclotron $(eB/m_i)$ and plasma frequencies $\om_{pi}$, one obtains the previously introduced one-fluid ideal MHD in which the magnetic field is frozen into the center of mass velocity of ions and electrons.   

We regularize the two-fluid model  by introducing vortical and magnetic `twirl' terms $\lambda_l^2 ({\bf w}_l + \frac{q_l}{m_l} {\bf B}) \times (\nabla \times {\bf w}_l)$ \footnote{Replacing $\bfw \to ({\bf w}_l + \frac{q_l}{m_l} {\bf B})$ is motivated by the fact that in the presence of an electromagnetic field the momentum ${m \bfv}$ of a charged particle of mass $m$ is replaced by the canonical momentum  ${\bf P} = m\bfv + e \bfA$.} in the ion/electron velocity equations ($l = i,e$) where $q_{i,e} = \pm e$ and ${\bf w}_l = \nabla \times {\bf v}_l$ are vorticities:	
	\beqs
	\pdr_t \bfv_l &+& \bfv_l \cdot \grad \bfv_l = - \ov{n_l m_l} \grad p_l + \frac{q_l}{m_l} (\bfE + \bfv_l \times \bfB) - \la_l^2 \bfw_l \times (\grad \times \bfw_l) - \frac{\la_l^2 q_l}{m_l} \bfB \times (\grad \times \bfw_l), \cr
	\pdr_t n_{l} &+& \grad \cdot (n_{l} \bfv_{l}) = 0, \quad \grad \times \bfB = \mu_0 \bfj_* + \mu_0 \eps_0 \frac{\pdr \bfE}{\pdr t}, \quad \frac{\pdr \bfB}{\pdr t} = - \grad \times \bfE,  \cr
	 \grad \cdot \bfE &=& \frac{e(n_i - n_e)}{\eps_0} \quad \text{and} \quad \grad \cdot \bfB = 0. 
	\eeqs
As before, the cut-off lengths $\lambda_l$ must be inversely proportional to the square-roots of the number densities $(\lambda_l^2 n_l = C_l)$ and may be taken proportional to Debye lengths or skin-depths. A novel feature is that the `flow' current $\bfj_{\rm flow} = \sum_l q_l n_l {\bf v}_l$ in Amp\`ere's law  is augmented by a solenoidal `twirl' current $\bfj_{\rm twirl} = \sum_l \nabla \times (\nabla \times \lambda_l^2 {\bf j}_{{\rm flow},l})$ so that $\bfj_* = \bfj_{\rm flow} + \bfj_{\rm twirl}$. The resulting equations imply conserved linear and angular momenta and a positive definite swirl energy density ${\cal E}^*$ which includes an enstrophic contribution $\sum_l (\lambda_l^2 \rho_l {\bf w}_l^2)/2$. Furthermore, our full two-fluid equations follow from Poisson brackets (PB) proposed by Spencer-Kaufman \cite{spencer-kaufman} and Holm-Kuperschmidt \cite{holm-kuperschmidt}  with the Hamiltonian 
	\beq 
	H = \int {\cal E}^* d\bfr = \int \sum_l \left[ \half \rho_l \left(\bfv_l^2 + \la_l^2 \bfw_l^2 \right)    + \frac{p_l}{\g - 1} + \frac{\bfB^2}{2 \mu_0} + \frac{\eps_0 \bfE^2}{2} \right] d\bfr.
	\eeq
Moreover, we obtain regularized quasineutral, Hall and one fluid MHD models by taking the successive limits (i) $\eps_0 \to 0$, (ii) $m_e/m_i \ll 1$ and (iii) $ e \to \infty$ along with $\la_e/\la_i \to 1$. However, we have not identified PBs for the quasineutral two-fluid or Hall MHD models. 

Finally, in Chapter \ref{s:r-gas-dynamics}, in analogy with the dispersive KdV regularization of the inviscid Burgers equation, we briefly touch upon a  conservative regularization of {\it shock-like} discontinuities in ideal gas dynamics which leads to an elegant 3D generalization of both the KdV and nonlinear Schr\"odinger equations (for a more complete treatment of this, see \cite{sachin-thesis}).

Inclusion of these twirl regularizations should lead to more controlled numerical simulations of Euler, NS and MHD equations without finite time blowups of enstrophy. It is important to distinguish between purely numerical instabilities which have nothing to do with physical properties of the system and real physical instabilities. In particular, these regularized models are capable of handling three-dimensional tangled vortex line and sheet interactions in engineering and geophysical fluid flows, as well as corresponding current filament and sheet dynamics which occur in astrophysics (E.g. as in solar prominences and coronal mass ejections, pulsar accretion disks and associated turbulent jets, and on a galactic scale, jets driven by active galactic nuclei) and in strongly nonlinear phenomena such as edge localised modes in tokamaks. There is no known way of studying many of these phenomena at very low collisionality  [i.e. at very high, experimentally relevant Reynolds, Mach and Lundquist numbers] with unregularized continuum models. Thus, we  note that recent theories \cite{henneberg-cowley-wilson,chandra-thyagaraja,lashmore-mccarthy-thyagaraja,thyagaraja-valovic-knight} of the nonlinear evolution of ideal and visco-resistive plasma turbulence in a variety of fusion-relevant devices (and many geophysical situations) can be numerically investigated in a practical way using our regularization.

Thus, the development of regularized compressible flow and MHD presented in this thesis (minimal extension of ideal equations with Hamiltonian-PB structure, conservation laws, bounded enstrophy, identification of appropriate boundary conditions, applications etc.) brings these 3D models a step closer to what KdV achieves for 1D flows.

{\noindent \bf Organization of this thesis}

We begin in Section \ref{s:formulation-r-euler-r-mhd} by giving the equations of twirl regularized compressible flow and their extension to compressible MHD. Criteria for the choice of regularization term and its physical interpretation are provided. Local conservation laws for `swirl' energy, helicity, linear and angular momenta are derived in Section \ref{s:cons-laws} followed by boundary conditions for the R-Euler equations in Section \ref{s:BC}. The corresponding results for R-MHD may be found in Section \ref{s:R-MHD-cons-laws-alfven}. Regularized versions of the Kelvin-Helmholtz and Alfv\'en theorems on freezing-in of vorticity and magnetic field into the swirl velocity ($\bfv_* =\bfv + \la^2 \grad \times \bfw$) are derived in Section \ref{s:Kelvin-circulation-Kelvin-Helmholtz-swirl-vel}. Integral invariants associated with closed curves, surfaces and volumes moving with the swirl velocity field are discussed in Section \ref{s:integral-inv-v-star}. Poisson brackets for compressible and incompressible R-Euler and R-MHD are introduced in Section \ref{s:pb-for-fluid} and Section \ref{s:PB-for-R-MHD}. The regularized equations are shown to be Hamilton's equations for the swirl energy. The Poisson algebra of conserved quantities is obtained paying special attention to boundary conditions. The Poisson bracket formulation is used in Section \ref{s:other-const-laws-and-regs} to identify new regularization terms (involving new constitutive relations) that guarantee bounded higher moments of vorticity and its curl while retaining the symmetries of the ideal equations. Section \ref{s:examples} contains several applications to steady flows. The regularized equations are used to model a rotating columnar vortex and MHD pinch, channel flow, plane flow, a plane vortex sheet and propagating spherical and cylindrical vortices. These examples elucidate many interesting physical consequences. They show that our conservatively regularized flows are indeed more regular than the corresponding Eulerian solutions. 

In Chapter \ref{s:two-fluid}, we extend our local conservative regularization of compressible ideal MHD to two-fluid (ion-electron) plasmas. The equations for regularized two-fluid plasmas are introduced in Section \ref{s:reg-eqns-two-fluid-compress}. Here, we also discuss the local conservation laws for linear and angular momenta and energy (along with the boundary conditions). A scheme for constructing a hierarchy of regularized plasma models (quasineutral, Hall and ideal MHD), starting with the two-fluid model and taking the successive limits $\eps_0 \to 0$, $m_e/m_i \to 0$ and $e \to \infty$ (along with $\la_e/\la_i \to 1$) is elaborated upon in Section \ref{s:heirarchy-of-models}. In Section \ref{s:PB-two-fluid}, the Poisson bracket  formalism for regularized compressible two-fluid models is discussed. In Section \ref{s:reg-field-curl-PBs-Hamiltonian}, we exploit the PB formulation to propose a way of regularizing magnetic field gradients in compressible one- and two-fluid plasma models. We also briefly compare our regularization with XMHD \cite{kimura-morrison,abdelhamid-kawazura-yoshida} which an alternate way of regularizing magnetic though not vortical singularities within a one-fluid setup. 

Chapter \ref{s:r-gas-dynamics} contains a brief summary of our work on a conservative regularization of shock-like singularities in ideal gas dynamics.

In Chapter \ref{s:discussion}, we conclude by placing our conservative regularization of ideal Euler flow, MHD and two-fluid plasma models in a wider physical context and discuss several open questions. 

Some properties of the Poisson brackets for compressible  flow and a novel proof of the Jacobi identity are given in Appendix \ref{a:PB-properties}. A Lagrangian formulation of our twirl-regularized compressible fluid equations using Clebsch variables is given in Appendix \ref{a:Lagrangian}. In Appendix \ref{a:minimality}, we prove that among symmetry-preserving conservative regularization terms (involving $\bfv$ and its derivatives) that can be added to the Euler equation while retaining the usual continuity equation and standard Hamiltonian formulation, the twirl term is minimal and unique. Finally, in Appendix \ref{a:time-averaged-inequality} we prove an interesting inequality in R-MHD involving the time average of a quantity which includes the twirl term $\bfw \times (\grad \times \bfw)$. This is   unlike our a priori bounds on kinetic energy and enstrophy (\ref{e:enstrophy-bound}) which do not involve derivatives of vorticity. This inequality  could be useful in checking the accuracy of numerical schemes.

\chapter{Conservative regularization of Euler and  ideal MHD}
\label{s:r-euler-r-mhd}

\section{Formulation of regularized compressible flow and MHD}
\label{s:formulation-r-euler-r-mhd}

A detailed introduction to this chapter was given in Chapter \ref{s:intro}. This chapter is based on \cite{govind-sonakshi-thyagaraja-pop} and \cite{govind-sonakshi-thyagaraja-r-euler-arxiv}. For compressible, barotropic flow with mass density $\rho$ and velocity field $\bfv$, the continuity and Euler equations are
	\beq
	\dd{\rho}{t} + \grad \cdot (\rho {\bf v}) = 0 \quad \text{and} \quad
	\dd{\bf v}{t} + \left( {\bf v} \cdot \grad \right) {\bf v} = - \frac{\grad  p}{\rho}.
	\eeq
The pressure $p$ is related to $\rho$ through a constitutive relation in barotropic flow. Let us introduce the stagnation pressure $\sigma$ and specific enthalpy $h$ for adiabatic flow of an ideal gas (or specific Gibbs free energy for isothermal flow) through the equation
	\beq
	\sigma = \left( \frac{\gamma}{\gamma -1} \right) \frac{p}{\rho} + \half {\bf v}^2 \equiv h + \half \bfv^2 \quad \text{where} \quad
	\frac{p}{\rho^\gamma} = \text{constant} \quad \text{with} \;\; \gamma = C_p/C_v.
	\eeq
Then using the identity $\half \grad {\bf v}^2 = {\bf v} \times \left( \grad \times {\bf v} \right) + \left({\bf v} \cdot \grad \right) {\bf v}$, the Euler equation may be written in terms of vorticity $\bfw = \grad \times \bfv$;
	\beq
	\dd{\bf v}{t} + {\bf w} \times {\bf v} = - \grad \sigma.
	\label{e:Euler-eqn}
	\eeq
In \cite{thyagaraja} a `twirl' regularization term $-\la^2 \bfT$ was introduced into the incompressible $(\grad \cdot \bfv = 0)$ Euler equations
	\beq
	\frac{D \bfv}{Dt} \equiv \dd{\bf v}{t} + \left( {\bf v} \cdot \grad \right) {\bf v} = - \frac{\grad p}{\rho} - \la^2 {\bf w} \times (\grad \times {\bf w}) \quad \text{with} \quad \bfT = \bfw \times (\grad \times \bfw).
	\label{e:R-Euler-eqn}
	\eeq
Here t$D/Dt$ is the material derivative. The twirl term is a singular\footnote{Spatial order of the equation is increased just as in going from the Euler equation to the Navier Stokes equation.} perturbation, making R-Euler $2^{\rm nd}$ order in space derivatives of $\bfv$ while remaining $1^{\rm st}$ order in time. The regularizing vector may be written ${\bf T} = {\bf w} \times \grad (\grad \cdot {\bf v}) - {\bf w} \times \grad^2 {\bf v}$. For incompressible flow it becomes ${\bf T} = - {\bf w} \times \grad^2 {\bf v}$. The parameter $\la$ with dimensions of length is a constant for incompressible flow. We will see that $\la$ acts as a short-distance regulator that prevents the enstrophy $\int \bfw^2 \: d\bfr$ from diverging. Unlike a lattice or other cut-off R-Euler ensures bounded enstrophy while retaining locality and all the space-time symmetries and conservation laws of the Euler equation. The sign of $\bfT$ ensures that the conserved energy $E^*$ obtained below (\ref{e:cons-egy-incompress}) is positive definite. The twirl acceleration is clearly absent in irrotational or constant vorticity flows. Since $\bfT$ involves derivatives of $\bfw$, it kicks in when vorticity develops large gradients and thereby prevents unbounded growth of enstrophy. As discussed below, $\bfT$ is chosen to have as few spatial derivatives and nonlinearities as possible. A linear term in $\bfv$ (as in KdV) preserving the symmetries of the Euler equation does not exist. The twirl term $- \la^2 \bfT$ is a conservative analogue of the viscous dissipation term $\nu \grad^2 \bfv$ in the incompressible NS equations
	\beq
	\dd{\bfv}{t} + (\bfv \cdot \grad) \bfv = - \frac{\grad p}{\rho} + \nu \grad^2 \bfv, \quad \grad \cdot \bfv = 0.
	\eeq
Kinematic viscosity $\nu$ and the regulator $\la$ play similar roles. The momentum diffusive time scale in NS is set by $\nu k^2$ where $k$ is the wave number of a mode. On the other hand in the nonlinear twirl term of R-Euler, the dispersion time-scale of momentum is set by $\la^2 k^2 |\bfw|$. So for high vorticity and short wavelength modes, the twirl effect would be more efficient in controlling enstrophy than pure viscous diffusion. 

It is instructive to compare incompressible Euler, R-Euler and NS under rescaling of coordinates and velocities ($\bfr = L \bfr'$, $\bfv = U \bfv'$ so that $t = (L/U) t'$). The incompressible Euler equations for vorticity
	\beq
	\dd{\bfw}{t} + \grad \times (\bfw \times \bfv) = 0 \quad \text{and} \quad \grad \cdot \bfv = 0,
	\eeq
are invariant under such rescalings. The NS equation is {\em not} invariant under independent rescalings of $\bfr$ and $\bfv$ unless $LU = 1$:
	\beq
	\dd{\bfw'}{t'} + \grad' \times (\bfw' \times \bfv') = \left( \frac{\nu}{L U} \right) \grad'^2 \bfw'.
	\eeq
As is well-known, flows with the same Reynolds number ${\cal R} = LU/\nu$ are similar. Interestingly, the R-Euler equation $\pdr \bfw/\pdr t + \grad \times (\bfw \times \bfv) = - \la^2 \grad \times (\bfw \times (\grad \times \bfw))$ is invariant under rescaling of time alone: $\bfr = \bfr', t = t'/U, \bfv = U \bfv'$ but not under independent rescalings of time and space. With both viscous and twirl regularizations present, under the rescaling $\bfr =  L \bfr', \bfv = U \bfv'$, we get
	\beq
	\dd{\bfw'}{t'} + \grad' \times (\bfw' \times \bfv') = \frac{\nu}{LU} \grad'^2 \bfw' - \frac{\la^2}{L^2} \grad' \times (\bfw' \times (\grad' \times \bfw')).
	\eeq
We may also compare the relative sizes of the dissipative viscous and conservative twirl stresses in vorticity equations. Under the usual rescaling $\bfr = L \bfr', \bfv = U \bfv'$ ($t = (L/U) t'$, $\bfw = (U/L) \bfw'$) and $|\grad'| = k$, $F_{visc} \sim (\nu/L^2) k^2 \om$ whereas $F_{twirl} \sim (\la^2 U/L^3) k^2 \om^2$ where $\om$ is the magnitude of the non-dimensional vorticity. Then $F_{twirl}/F_{visc} \sim {\cal R} \om (\la/L)^2$. This shows that at any given Reynolds number ${\cal R} = LU/\nu$ and however small $\la/L$ is taken, at sufficiently large vorticity the twirl force will always be larger than the viscous force.

Since $\bfT$ is quadratic in $\bfw$ (or $\bfv$), it should be important in high-vorticity or high-speed flows. Thus it is natural to seek a generalization of the twirl regularization to compressible flows. Consider adiabatic flow of an ideal compressible fluid whose pressure and density are related by $(p/p_0) = (\rho/\rho_0)^\gamma$. The {\it compressible} R-Euler equations are
	\beq
	\dd{\rho}{t} + \grad \cdot (\rho \bfv) = 0 \quad \text{and} \quad
	\dd{\bf v}{t} + \left( {\bf v} \cdot \grad \right) {\bf v} = - \frac{\gamma}{\gamma - 1} \grad \left(\frac{p}{\rho} \right) - \la^2 {\bf w} \times (\grad \times {\bf w}).
	\label{e:continuity-R-Euler}
	\eeq
For compressible flows we find that $\la(\bfr,t)$ and $\rho(\bfr,t)$ must satisfy a constitutive relation taking the form,
	\beq
	\la^2 \rho = \text{constant} = \la_0^2  \rho_0,
	\label{e:constitutive-relation}
	\eeq
to ensure that a positive-definite conserved energy exists for an arbitrary flow [more general constitutive relations are possible, see Section \ref{s:other-const-laws-and-regs}]. The constant $\la_0^2 \rho_0$ depends on the fluid and not the specific flow. We also note that the introduction of the twirl force entails a modification of the stress tensor $S_{ij} = p \del_{ij}$ appearing in the ideal Euler equation $\rho (Dv_i/Dt) = - \pdr_j S_{ij}$. The regularized stress tensor is $S_{ij} = p \del_{ij} + \la^2 \rho \left( \frac{w^2}{2} \del_{ij} - w_i w_j \right)$. 

As before, we write the R-Euler equation as
	\beq
	\dd{\bf v}{t} + {\bf w} \times {\bf v} = - \grad \sigma - \la^2 {\bf w} \times (\grad \times {\bf w}).
	\label{e:reg-Euler-3D}
	\eeq
Here ${\bf w} \times {\bf v}$ is the `vorticity acceleration' and $- \la^2 \bfw \times (\grad \times \bfw)$ is the twirl acceleration while $\grad \sigma$ includes acceleration due to pressure gradients. The regularization term increases the spatial order of the Euler equation by one (since $\bfw=\grad \times \bfv$), just as $\nu \grad^2 \bfv$ does in going from Euler to NS. However the boundary conditions required by the above conservative regularization involve the first spatial derivatives of $\bfv$, unlike the no-slip condition of NS. Furthermore, the regularizing viscous stress in NS is linear in $\bfv$ as opposed to the quadratically nonlinear twirl stress. The twirl term involves three derivatives and should be important at high wave numbers, as is the dispersive $u_{xxx}$ term in KdV. The R-Euler equation is invariant under parity (all terms reverse sign) and under time-reversal (all terms retain their signs). It is well-known that NS is not invariant under time-reversal, since it includes viscous dissipation. Moreover, we shall see that R-Euler possesses local conservation laws for energy, flow helicity, linear and angular momenta, in common with the Euler system.

The R-Euler equation takes a compact form in terms of the `swirl' velocity field $\bfv_* = \bfv + \la^2 \grad \times \bfw$:
	\beq
	\dd{\bfv}{t} + \bfw \times \bfv_* = - \grad \sigma.
	\label{e:R-Euler-v*}
	\eeq
Here $\bfw \times \bfv_*$ is a regularized version of the Eulerian vorticity acceleration $\bfw \times \bfv$. The swirl velocity $\bfv_*$ plays an important role in the regularized theory, as will be demonstrated. In fact, the continuity equation can be written with $\bfv_*$ replacing $\bfv$:
	\beq
	\dd{\rho}{t} + \grad \cdot (\rho \bfv_*) = 0.
	\label{e:v*-continuity-eqn}
	\eeq
This is a consequence of the constitutive relation  (\ref{e:constitutive-relation}) which implies $\grad \cdot (\rho \bfv_*) =  \grad \cdot (\rho \bfv + \la^2 \rho (\grad \times \bfw)) = \grad \cdot (\rho \bfv)$. Taking the curl of the R-Euler momentum balance equation we get the R-vorticity equation:
	\beq
	{\bf w}_t + \grad \times ( {\bf w} \times {\bf v}) = - \grad \times \left( \la^2 {\bf w} \times (\grad \times {\bf w}) \right) 
	\qquad 
	\text{or} \qquad 
	\bfw_t + \grad \times (\bfw \times \bfv_*) = 0.
	\label{e:R-vorticity-eqn-compressible}
	\eeq
The incompressible regularized evolution equations possess a positive definite integral invariant [with suitable boundary data]:
	\beq
	\DD{E^*}{t} = \DD{}{t} \left( \int_V \left[ \half \rho {\bf v}^2 + \half \la^2 \rho {\bf w}^2 \right] \: d\bfr \right) = 0.
	\label{e:cons-egy-incompress}
	\eeq
For compressible flow, $E^*$ is {\em not} conserved if $\la$ is a constant length. On the other hand, we do find a conserved energy if we include compressional potential energy and also let the field $\la(\bfr,t)$ be a dynamical length governed by the constitutive relation $\la^2 \rho = \la_0^2 \rho_0 = $ constant (\ref{e:constitutive-relation}). As a consequence, $\la$ is not an independent propagating field like $\bfv$ or $\rho$, its evolution is determined by that of $\rho$. Here $\la_0$ is some constant short-distance cut-off (e.g. a mean-free path at mean density) and $\rho_0$ is a constant mass density (e.g. the mean density). $\la$ is smaller where the fluid is denser and larger where it is rarer. This is reasonable if we think of $\la$ as a position-dependent mean-free-path. However, it is only the combination $\la_0^2 \:\rho_0$ that appears in the equations. So compressible R-Euler involves only one new dimensional parameter, say $\la_0$. A dimensionless measure of the cutoff $n \la^3 = \la_0^3 \, n_0^{3/2} \,n^{-1/2}$ may be obtained by introducing the number density $n = \rho/m$ where $m$ is the molecular mass. It is clearly smaller in denser regions and larger in rarified regions. As noted in the introduction, if we take $(\la/L)^2 \propto a^3/L^3$ where $a \propto n^{-1/3}$ and $L$ are inter-particle spacing and macroscopic system size, then $\la^2 \rho$ would be a constant. The conservation of $E^*$ implies an a priori bound on enstrophy; no such bound is available for Eulerian flows, where enstrophy could diverge due to vortex stretching \cite{Frisch, KRSreenivasan-onsager}. Note that boundedness of enstrophy under R-Euler evolution may still permit $\bfw$ to develop discontinuities or mild divergences for certain initial conditions.

The KdV and R-Euler equations are conservative regularizations in one and three dimensions. The dimensional reduction of R-Euler provides a possible regularization of ideal flows in $2$ dimensions. However, for incompressible 2D flow, the twirl term becomes a gradient and does not affect the evolution of vorticity (see Section \ref{s:plane-flow}). This is to be expected as {\it incompressible} 2D Euler flows do not require regularization: there is no vortex stretching, enstrophy and all moments of $\bfw^2$ are conserved. On the other hand, the twirl term leads to a new and non-trivial regularization of compressible flow in 2D (see Section \ref{s:plane-flow}).


It is possible to show that the twirl term is unique among regularization terms that are at most quadratic in $\bfv$ with at most $3$ spatial derivatives subject to the following physical requirements (1) it must preserve Eulerian symmetries and (2) admit a Hamiltonian formulation with the standard Landau Poisson brackets and continuity equation. A proof of this uniqueness result will be given in Appendix \ref{a:minimality}.


In the light of possible astrophysical applications, we briefly note two important generalisations of the R-Euler system. Suppose a conservative body force ${\bf F}= - \rho \grad V$ is operative, where the potential $V$ arises for instance from gravity. Then (\ref{e:reg-Euler-3D}) has the additional term $- \grad V$, signifying acceleration due to the body force. Evidently, we may now set,
	\beq
	\sigma = \left( \frac{\gamma}{\gamma -1} \right) \frac{p}{\rho} + \frac{{\bf v}^2}{2} + V \equiv h + \frac{\bfv^2}{2} 
	\eeq
where the new enthalpy includes a contribution from potential energy. The conservation laws of the next Section generalize upon including the potential energy of the body force.

A much less trivial extension will also be briefly indicated: in compressible ideal MHD the body force is the magnetic Lorentz force ${\bf j \times B}$, which has to be related to the fluid motion through Maxwell's equations for a quasineutral, compressible, ideal fluid. The governing equations for mass density $\rho$, magnetic field $\bfB$ and velocity $\bfv$ take the following forms:
	\beq
	\dd{\rho}{t} + \grad \cdot (\rho \bfv) = 0, \quad
		\dd{\bfv}{t} + (\bfv \cdot \grad ) \bfv = - \ov{\rho} \grad p + \frac{\bfj \times \bfB}{\rho}
	\quad \text{and} \quad
	\dd{\bfB}{t} = \grad \times (\bfv \times \bfB)\label{e:unreg-ideal-MHD}.
	\eeq
The electric body force cancels out when one adds the momentum equations for electrons and ions. Thus one  arrives at the above momentum equation for the center of mass velocity $\bfv$ of the electrons and ions in the quasineutral plasma treated as a single fluid. In non-relativistic plasmas, the displacement current term in Ampere's law can be neglected, allowing us to express the electric current as the curl of the magnetic field: $\mu_0 \bfj = \grad \times \bfB$. In particular, $\bfj$ is not an independent dynamical variable, its evolution is determined by that of $\bfB$. So the magnetic body force may be written as $ (\grad \times \bfB) \times \bfB/\rho \mu_0$. In MHD, the constitutive equation relating the electric and magnetic fields to the fluid motion is the ideal Ohm's law: $\bfE + (\bfv \times \bfB)=0$, which leads to the above expression for Faraday's law.

The regularized compressible MHD (R-MHD) equations follow from arguments similar to those presented for neutral compressible flows. The continuity equation, $\rho_t + \grad \cdot (\rho \bfv) = 0$ is unchanged. As noted, it may be written in terms of swirl velocity: $\rho_t + \grad \cdot (\rho \bfv_*) = 0$. As in regularized fluid theory, we introduce the twirl acceleration on the RHS of the momentum equation, where $\la$ is again subject to (\ref{e:constitutive-relation}):
	\beq
	\dd{\bfv}{t} + (\bfv \cdot \grad) \bfv = -\frac{\grad p}{\rho} + \frac{\bfj \times \bfB}{\rho} - \la^2 \bfw \times (\grad \times \bfw) = -\frac{\grad p}{\rho} - \frac{\bfB \times(\grad \times\bfB)}{\mu_0\rho} - \la^2 \bfw \times (\grad \times \bfw).
	\label{e:R-MHD-Euler-v}
	\eeq
The twirl regularization term is the vortical analogue of the magnetic Lorentz force term with $1/\mu_0$ replaced with $\la^2 \rho$. This is also evident in the R-MHD stress tensor $S_{ij} = p \del_{ij} + \la^2 \rho \left( \half w^2 \del_{ij} - w_i w_j \right) +  \left( \half B^2 \del_{ij} - B_i B_j \right)/\mu_0$ appearing in the momentum equation $\rho (Dv_i/Dt) = - \pdr_j S_{ij}$. Equation (\ref{e:R-MHD-Euler-v}) can be obtained from the unregularized equation (\ref{e:Euler-eqn}) by replacing $\bfv$ with $\bfv_*$ in the vortex acceleration term:
	\beq
	\dd{\bfv}{t} + \bfw \times \bfv_* = -\ov{\rho}\grad p - \half \grad \bfv^2 + \frac{\bfj \times \bfB}{\rho}.
	\label{e:R-MHD-Euler-v*}
	\eeq
Similarly, the regularized Faraday law in R-MHD is obtained by replacing $\bfv$ by $\bfv_*$ in (\ref{e:unreg-ideal-MHD}) i.e.,
	\beq
	\dd{\bfB}{t} = \grad \times (\bfv_* \times \bfB) = \grad \times \left(\bfv \times \bfB - \la^2 \bfB \times (\grad \times \bfw) \right).
	\label{e:R-MHD-Faraday}
	\eeq
The regularization term in Faraday's law is the curl of the `magnetic' twirl $- \la^2 \bfB \times (\grad \times \bfw)$ term in analogy with the `vortical' twirl term $-\la^2 \bfw \times (\grad \times \bfw)$. The regularized Faraday equation is $3^{\rm rd}$ order in space derivatives of $\bfv$ and first order in $\bfB$. From (\ref{e:R-MHD-Faraday}), we deduce that the potentials ($\bfA, \phi$) in any gauge must satisfy
	\beq
	\pdr_t \bfA = \bfv_* \times \bfB - \grad \phi.
	\label{e:A-evolution-R-MHD}
	\eeq
It turns out that compressible R-MHD possesses conservation laws similar to those deduced in \cite{thyagaraja} for incompressible R-MHD, see Section \ref{s:R-MHD-cons-laws-alfven}. One can readily include conservative body forces like gravity into R-MHD. The inclusion of regularization terms arising from electron inertia and Hall effect \cite{thyagaraja} and extension to the two-fluid plasma system will be presented in Chapter 3.

\section{Conservation laws for  regularized compressible flow and MHD}

\subsection{Conservation laws for regularized compressible fluid flow}
\label{s:cons-laws}

{\noindent \bf Swirl Energy Conservation:} Under compressible R-Euler evolution, the ``swirl'' energy density and flux vector
	\beq
	{\cal E}^* = \left[ \frac{\rho \bfv^2}{2}+ U(\rho) + \frac{\la^2 \rho \bfw^2}{2} \right] 
	\quad \text{and} \quad
		\bff = \rho \sigma \bfv + \la^2 \rho (\bfw \times \bfv_*) \times \bfw.
	\label{e:swirl-egy-density-current}
	\eeq
satisfy the local conservation law $\dd{{\cal E}^*}{t} + \grad \cdot \bff = 0$. Here $U(\rho) = p/(\gamma - 1)$ is the compressional potential energy for adiabatic flow. Given suitable boundary conditions [BCs, discussed below], the system obeys a global energy conservation law:
	\beq
	\frac{d E^*}{dt} = 0 \quad \text{where} \quad E^* = \int \left[ \frac{\rho \bfv^2}{2}+ U(\rho) + \frac{\la^2 \rho \bfw^2}{2} \right] \: d\bfr.
	\label{e:swirl-energy-R-Euler}
	\eeq

{\noindent \bf Flow Helicity Conservation:} The R-Euler equations possess a local conservation law for helicity density $\bfv \cdot \bfw$ and its flux $\bff_{\cal K}$:
	\beq
	\pdr_t (\bfv \cdot \bfw)
	+ \grad \cdot \left( \sig \bfw + (\bfw \times \bfv_*) \times \bfv \right)  = 0.
	\label{e:helicity-current-conservation}
	\eeq
This local conservation law implies global conservation of helicity ${\cal K} = \int \bfv \cdot \bfw \, d\bfr$, provided $\bff_{\cal K} \cdot \hat n = 0$ on the boundary $\pdr V$ of the flow domain $V$. Here $\hat n$ is the unit outward-pointing normal vector on the surface $\pdr V$.

{\noindent \bf Momentum Conservation:} Flow momentum is ${\bf P} = \int \rho \bfv \; d\bfr$. Momentum density ${\cal P}_i = \rho v_i$ and the stress tensor $\Pi_{ij}$ satisfy
	\beq
	\dd{{\cal P}_i}{t} + \pdr_j \Pi_{ij} = 0 \quad \text{where} \quad
	\Pi_{ij} = \Pi_{ji} = \rho v_i v_j + p \del_{ij} + \rho \la^2 \left(\half \bfw^2 \del_{ij} - w_i w_j \right).
	\label{e:momentum-current-tensor}
	\eeq
For $\bfP$ to be globally conserved, we expect to need a translation-invariant flow domain $V$. In $V = \mathbb{R}^3$,  $\bfv$ must decay to zero and  $\rho$ to a constant sufficiently fast as $\bfr \to \infty$ to ensure $d{\bf P}/dt = 0$. Periodic BCs in a cuboid also ensure global conservation of $\bf P$.

{\noindent \bf Angular Momentum Conservation:} For regularized compressible flow, we define the angular momentum density as $\vec {\cal L} = \rho {\bf r} \times \bfv$. We find that the angular momentum satisfies the local conservation law:
	\beq
	\dd{{\cal L}_i}{t} + \pdr_l \Lambda_{il} = 0 \quad \text{where} \quad
	\Lambda_{il} = \eps_{ijk} r_j \Pi_{kl}.
	\label{e:ang-mom-current-conservation}
	\eeq
$\Lambda_{il}$ is the angular momentum flux tensor. For $\bfL = \int \vec {\cal L} \: d\bfr$ to be globally conserved, the system must be rotationally invariant. For instance, decaying BC in an infinite domain would guarantee conservation of $\vec {\cal L}$. We also note that in symmetric domains [axisymmetric torus or circular cylinder] corresponding components of angular momentum or linear momentum associated with the symmetry may also be conserved. The situation here is similar to typical Eulerian systems.

\subsection{Boundary conditions}
\label{s:BC}

In the flow domain $\mathbb{R}^3$, it is natural to impose decaying BCs ($\bfv \to 0$ and $\rho \to$ constant as $|\bfr| \to \infty$) to ensure that total energy $E^*$ is finite and conserved. For flow in a cuboid, periodic BCs ensure finiteness and conservation of energy. For flow in a bounded domain $V$, demanding global conservation of energy leads to another natural set of BCs. Now $d E^*/dt = - \int_{\pdr V} \bff \cdot \hat n \: dS$ where $\bff$ is the energy current (\ref{e:swirl-egy-density-current}) and $\pdr V$ the boundary surface. $\bff \cdot \hat n = 0$ if the following conditions hold:
	\beq
	\bfv \cdot \hat n = 0 \quad
	\text{and} \quad
	\bfw \times \hat n = 0.
	\label{e:BCs-for-energy-cons}
	\eeq
These BCs are, for instance, satisfied at the top and bottom of a bucket of rigidly rotating fluid. The BC $\bfv \cdot \hat n = 0$ also ensures global conservation of mass as $\DD{}{t} \int \rho d\bfr = - \int \rho \bfv \cdot \hat n \: dS$. Since the R-Euler equation is $2^{\rm nd}$ order in spatial derivatives of $\bfv$, it is consistent to impose conditions on both $\bfv$ and its $1^{\rm st}$ derivatives. These boundary conditions imply that the twirl acceleration is tangential to the boundary surface $\bfT \cdot \hat n = (\bfw \times \la^2 (\grad \times \bfw)) \cdot \hat n = (\hat n \times \bfw) \cdot (\la^2 \grad \times \bfw) = 0$. It is interesting to note that the BCs ensuring helicity conservation (see Section \ref{s:direct-proofs}) are `orthogonal' to those for energy conservation
	\beq
	\bfv \times \hat n = 0 \quad 
	\text{and} \quad
	\bfw \cdot \hat n = 0 \quad \imply \quad \bff_{\cal K} \cdot \hat n = 0.
	\eeq
So helicity and energy cannot both be globally conserved simultaneously with these BCs [in bounded domains]. However, periodic or decaying BC would ensure simultaneous conservation of both. Similarly, neither angular momentum nor linear momentum is conserved in a finite flow domain with the BCs that ensure energy conservation. However, with sufficiently rapidly decaying BCs, energy, momentum, angular momentum and helicity can all be conserved simultaneously.

\subsection{Direct proofs of the conservation laws}
\label{s:direct-proofs}

We derive the stated conservation relations for R-Euler flows from the equations of motion (\ref{e:continuity-R-Euler},\ref{e:reg-Euler-3D},\ref{e:R-vorticity-eqn-compressible}) and the imposed BC's. Later these conservation laws will also be obtained using Poisson brackets.

{\noindent \bf Swirl energy conservation:} To prove the local conservation law for ${\cal E}^*$ (\ref{e:swirl-egy-density-current}) we begin by computing the time derivative of each term in the energy density.
\beqs
	\dd{}{t} \left(\half \rho {\bf v}^2 \right) &=& \half {\bf v}^2 \dd{\rho}{t} + \rho {\bf v} \cdot \dd{\bf v}{t}
	= - \half {\bf v}^2 \grad \cdot (\rho {\bf v}) - \rho {\bf v} \cdot \grad \sigma - \la^2 \rho {\bf v} \cdot {\bf T},
	\cr
	\dd{}{t} \left( \frac{p}{\gamma - 1} \right) &=& \frac{p_o}{\gamma - 1} \dd{}{t} \left(\frac{\rho}{\rho_o} \right)^\gamma
		= -\frac{\gamma}{\gamma -1} \left( \frac{p}{\rho} \right) \grad \cdot (\rho {\bf v}),
	\cr
	\dd{}{t} \left( \half \la^2 \rho {\bf w}^2 \right) &=& \la_0^2 \rho_0 {\bf w} \cdot \dd{\bf w}{t} = \la^2 \rho {\bf w} \cdot
	\left[ \grad \times ({\bf v} \times {\bf w}) - \grad \times (\la^2 {\bf T}) \right].
	\eeqs
It follows that:
	\beq
	\dd{{\cal E}^*}{t} = - \sig \grad \cdot (\rho {\bf v}) - \rho {\bf v} \cdot \grad \sigma
	- \la^2 \rho \left[ {\bf v} \cdot {\bf T} - {\bf w} \cdot
	\grad \times ({\bf v} \times {\bf w}) \right]
	- \la^2 \rho {\bf w} \cdot \grad \times (\la^2 {\bf T}).
	\eeq
Since $\la$ is a free parameter, the coefficient of each power of $\la$ must be shown to be a divergence. It follows from straightforward but somewhat lengthy algebra [which we omit for brevity] that this is indeed the case, leading to a local conservation equation $\pdr{{\cal E}^*}/\pdr t +  \grad \cdot \bff = 0$ with the energy flux vector $\bff$ given in (\ref{e:swirl-egy-density-current}). It should be noted that this local conservation law crucially depends on the constitutive relation (\ref{e:constitutive-relation}). The conservation of $E^* = \int {\cal E}^* \: d\bfr$ follows from Gauss' divergence theorem and our choice of boundary conditions ($\bfv \cdot \hat n = 0$ and $\bfw \times \hat n = 0$), which follow from writing
	\beq
	\bff \cdot \hat n =	\rho \sigma \bfv \cdot \hat n + \la^2 \rho (\bfw \times \bfv_*) \cdot (\bfw \times \hat n).
	\eeq
 
{\noindent \bf Flow helicity conservation:} To obtain the local conservation law for $\bfv \cdot \bfw$, we use the regularized equations (\ref{e:reg-Euler-3D},\ref{e:R-vorticity-eqn-compressible}) to write
	\beq
	\bfw \cdot \bfv_t = - \bfw \cdot ( \grad \sigma - \la^2 \bfT)  \quad
	\text{and} \quad
	\bfv \cdot \bfw_t = \bfv \cdot (\grad \times (\bfv \times \bfw) - \grad \times (\la^2 \bfT)).
	\eeq
Now $\bfv \cdot (\grad \times (\bfv \times \bfw)) = - \grad \cdot (\bfv \times (\bfv \times \bfw))$ since $(\bfv \times \bfw) \cdot \bfw = 0$. Similarly, $\bfv \cdot (\grad \times (\la^2 \bfT)) = \grad \cdot (\la^2 \bfT \times \bfv)$ since $\bfT \cdot \bfw = 0$. Combining these two, the time derivative of flow helicity density is a divergence $\grad \cdot \bff_{\cal K}$,
	\beq
	\pdr_t (\bfv \cdot \bfw) = \bfw \cdot \bfv_t + \bfv \cdot \bfw_t
	= - \bfw \cdot \grad \sigma
	- \grad \cdot ( \bfv \times (\bfv \times \bfw) ) - \grad \cdot (\la^2 \bfT \times \bfv)
	= \grad \cdot \left(\sig \bfw + (\bfw \times \bfv_*) \times \bfv \right),
	\eeq
as $\bfw$ is solenoidal. Writing
	\beq
	\bff_{\cal K} \cdot \hat n
	= \sig \bfw \cdot \hat n + (\bfw \times \bfv_*) \cdot (\bfv \times \hat n),
	\eeq
we infer BCs $\bfw \cdot \hat n = 0$ and
$\bfv \times \hat n = 0$ that ensure global helicity conservation [decaying BCs would of course also work].

{\noindent \bf Linear and angular momentum conservation:} The proof of local conservation of momentum density $\rho \bfv$ uses the continuity and R-Euler equations:
	\beq
	\rho \dd{v_i}{t} = - \rho v_j \pdr_j v_i - \pdr_i p - \rho \la^2 T_i
	\quad \text{and} \quad
	v_i \dd{\rho}{t} = - v_i \pdr_j (\rho v_j).
	\eeq
By the constitutive relation, $\la^2 \rho$ is a constant, so
	\beq
	\dd{{\cal P}_i}{t} = - \pdr_j (\rho v_i v_j) - \pdr_i p - \rho \la^2 T_i = - \pdr_j \left[ \rho v_i v_j + p \del_{ij} + \rho \la^2 \left(\half \bfw^2 \del_{ij} - w_i w_j \right) \right] \equiv - \pdr_j \Pi_{ij}.
	\eeq
Thus, we have local conservation of momentum $\pdr {{\cal P}_i}/\pdr t + \pdr_j \Pi_{ij} = 0$. The time derivative of angular momentum density $\vec {\cal L} = \rho {\bf r} \times \bfv$ is calculated using the local conservation law for momentum density and the symmetry of $\Pi_{ij}$:
	\beq
	\dd{{\cal L}_i}{t} = \eps_{ijk} r_j \dd{(\rho v_k)}{t} 
	= - \eps_{ijk} r_j \pdr_l \Pi_{kl} = - \pdr_l \left( \eps_{ijk} r_j \Pi_{kl} \right) = - \pdr_l \Lambda_{il}.
	\eeq
So angular momentum satisfies $\pdr {\cal L}_i/ \pdr t + \pdr_l \Lambda_{il} = 0$ where $\Lambda_{il}$ is the angular momentum flux tensor  (\ref{e:ang-mom-current-conservation}).

\subsection{Conservation laws for R-MHD and boundary conditions}
\label{s:R-MHD-cons-laws-alfven}

{\noindent \bf Swirl energy conservation}: In R-MHD, we obtain the following local energy conservation law:
		\beqs
	     && \dd{{\cal E}_{\rm mhd}^*}{t} + \grad \cdot \bff_{\rm mhd} = 0 \;\; \rm{where} \quad
	      {\cal E}_{\rm mhd}^*=\left( \frac{\rho \bfv^2}{2}+ U(\rho) + \frac{\la^2 \rho \bfw^2}{2} + \frac{\bfB^2}{2 \mu_0} \right)  \quad \text{and} \cr  && \bff_{\rm mhd} = \left( \rho \sigma \bfv + \la^2 \rho (\bfw \times \bfv_*) \times \bfw \right) 
		     + \ov{\mu_0} \left[ \bfB \times  (\bfv_* \times \bfB) + \la^2 \left( \bfw \times ((\grad \times \bfB) \times \bfB) \right)  \right]
	\eeqs
is the energy flux vector and $E^*_{\rm mhd} = \int_V{\cal E}^*_{\rm mhd}\: d\bfr$ is the the total `swirl' energy of barotropic compressible R-MHD. 

{\noindent \sc Proof:} The time derivative of the swirl energy density is calculated using the evolution equations (\ref{e:R-MHD-Euler-v},\ref{e:R-MHD-Faraday}) for $\bfv, \bfw, \bfB$ and $\rho$: 
\beqs
	\dd{}{t} \left(\half \rho {\bf v}^2 \right) &=& \half {\bf v}^2 \dd{\rho}{t} + \rho {\bf v} \cdot \dd{\bf v}{t}
	= - \half {\bf v}^2 \grad \cdot (\rho {\bf v}) - \rho {\bf v} \cdot \grad \sigma - \la^2 \rho {\bf v} \cdot {\bf T} + \bfv \cdot (\bfj \times \bfB),
	\cr
	\dd{}{t} \left( \frac{p}{\gamma-1} \right) &=& \frac{p_o}{\gamma-1} \dd{}{t} \left(\frac{\rho}{\rho_o} \right)^\gamma
		= -\frac{\gamma}{\gamma -1} \left( \frac{p}{\rho} \right) \grad \cdot (\rho {\bf v}),
	\cr
	\dd{}{t} \left( \half \la^2 \rho {\bf w}^2 \right) &=&  \la_0^2 \rho_0 {\bf w} \cdot \dd{\bf w}{t} = \la^2 \rho {\bf w} \cdot
	\left( \grad \times ({\bf v} \times {\bf w}) - \grad \times (\la^2 {\bf T}) + \ov{\rho} \grad \times (\bfj \times \bfB) \right) \cr
	\dd{}{t} \left( \frac{\bfB^2}{2\mu_0} \right) &=& \ov{\mu_0} \bfB \cdot \dd{\bfB}{t} = \ov{\mu_0} \bfB \cdot \left( \grad \times (\bfv_* \times \bfB) \right).
	\eeqs
Therefore the time derivative of energy density is :
	\beqs
	\dd{{\cal E}_{\rm mhd}^*}{t} &=& - \sig \grad \cdot (\rho {\bf v}) - \rho {\bf v} \cdot \grad \sigma
	- \la^2 \rho \left[ {\bf v} \cdot {\bf T} - {\bf w} \cdot
	\grad \times ({\bf v} \times {\bf w}) \right]
	- \la^2 \rho {\bf w} \cdot \grad \times (\la^2 {\bf T})
	\cr 
	&& + \: {\mu_0}^{-1}   \Bigg( \Bigg.\bfv \cdot ((\grad \times \bfB) \times \bfB) + \la^2 \bfw \cdot \grad \times ((\grad \times \bfB) \times \bfB) \cr
	&& + \: \bfB \cdot \left(\grad \times ((\bfv + \la^2 \grad \times \bfw) \times \bfB ) \right) \Bigg. \Bigg) .
	\eeqs
The first line containing terms independent of $\bfB$ has already been expressed as the divergence of the R-Euler fluid energy current $\bff = \rho \sigma \bfv + \la^2 \rho (\bfw \times \bfv_*) \times \bfw$. Now we split the terms containing $\bfB$ into those of order $\la^0$ and those quadratic in $\la$ and express each as a divergence using the vector identity $\grad \cdot (\bfA \times \bfB) = \bfB \cdot \grad \times \bfA - \bfA \cdot \grad \times \bfB$:
	\beqs
\la^0: &&	\hspace{-.6cm}	\bfB \cdot ( \grad \times (\bfv \times \bfB)) + \bfv \cdot ((\grad \times \bfB) \times \bfB) = \grad \cdot \left[ (\bfv \times \bfB) \times \bfB \right] \cr
\la^2: &&	\hspace{-.6cm}
\bfw \cdot \grad \times \left( (\grad \times \bfB) \times \bfB \right) + \bfB \cdot \grad \times \left( (\grad \times \bfw) \times \bfB \right) \cr
&=& - \grad \cdot \left[ \bfw \times ((\grad \times \bfB) \times \bfB) + \bfB \times ((\grad \times \bfw) \times \bfB) \right].
	\eeqs
Thus we obtain the abovementioned conserved energy current density for regularized compressible MHD. Boundary conditions on the surface $\pdr V$ of the flow domain $V$ that ensure  global conservation of $E^*_{\rm mhd}$ are \beq
	\bfv \cdot \hat n = 0, \quad
	\bfw \times \hat n = 0, \quad 
	(\grad \times \bfw) \cdot \hat n = 0 	\quad \text{and} \quad
	\bfB \cdot \hat n = 0.
	\label{e:R-MHD-energy-BCs}
	\eeq
The R-MHD equations of motion (\ref{e:R-MHD-Euler-v},\ref{e:R-MHD-Faraday}) are $3^{\rm rd}$ order in $\bfv$ and $1^{\rm st}$ order in $\bfB$. So we must impose BCs on $\bfB$, $\bfv$, the $1^{\rm st}$ and $2^{\rm nd}$ derivatives of $\bfv$. It also follows from (\ref{e:R-MHD-energy-BCs}) that $\bfB \cdot \bfw = 0$ and $\bfv_* \cdot \hat n = 0$ on the boundary. These BCs follow from writing
	\beqs
	&& \bff_{\rm mhd} \cdot \hat n = \rho \sigma \bfv \cdot \hat n + \la^2 \rho \:(\bfw \times \bfv_*) \cdot (\bfw \times \hat n) \cr
	&& + \ov{\mu_0} \left[\bfB^2 (\bfv_* \cdot \hat n) - (\bfv_* \cdot \bfB) \bfB \cdot \hat n
	+ \la^2 \left\{  (\bfw \cdot \bfB) (\grad \times \bfB \cdot \hat n) - (\grad \times \bfB \cdot \bfw)  (\bfB \cdot \hat n)	\right\} \right].
	\eeqs
{\noindent \bf Magnetic helicity conservation}: We define magnetic helicity as ${\cal K}_B = \int_V \bfA \cdot \bfB \:d\bfr$. This is the magnetic analogue of flow helicity ${\cal K} = \int_V \bfv \cdot \bfw d\bfr$ where we make the replacements $\bfv \to \bfA, \bfw \to \bfB$. Despite appearances, ${\cal K}_B$ is gauge-invariant for decaying boundary conditions or if $\bfB$ is tangential to $\pdr V$. For, under a gauge transformation $\bfA \to \bfA + \grad \tht$,
	\beq
	{\cal K}_B \to {\cal K}_B + \int_V \bfB \cdot \grad \tht d\bfr = K + \int_V \grad \cdot (\tht \bfB) \: d\bfr = K + \int_{\pdr V} \tht \bfB \cdot \hat n \: dS.
	\eeq
Magnetic helicity density is locally conserved in any gauge with potentials $(\bfA,\phi)$
	\beq
	\dd{(\bfA \cdot \bfB)}{t} + \grad \cdot (\bfA \times (\bfv_* \times \bfB) + \bfB \phi) = 0.
	\eeq
{\noindent \sc Proof}: Using (\ref{e:R-MHD-Faraday}, \ref{e:A-evolution-R-MHD}) the time derivative of $\bfA \cdot \bfB$ is
	\beq
	\dd{(\bfA \cdot \bfB)}{t} = \bfA \cdot \dd{\bfB}{t} + \bfB \cdot \dd{\bfA}{t} = \bfA \cdot \grad \times (\bfv_* \times \bfB) + \bfB \cdot (\bfv_* \times \bfB - \grad\phi).
	\eeq
The second term is zero. Using the vector identity $\grad \cdot (\bfA \times \bfD) = \bfD \cdot \grad \times \bfA - \bfA \cdot \grad \times \bfD$ and $\grad \cdot \bfB$ = 0 we may write
	\beqs
	\dd{(\bfA \cdot \bfB)}{t} &=& \bfA \cdot \grad \times (\bfv_* \times \bfB) - \bfB \cdot \grad \phi \cr
	&=& - \grad \cdot (\bfA \times (\bfv_* \times \bfB)+\bfB\phi) + (\bfv_* \times \bfB) \cdot (\grad \times \bfA) \cr
	&=& - \grad \cdot (\bfA \times (\bfv_* \times \bfB) + \bfB\phi) .
		\eeqs
Thus we get the local conservation law for magnetic helicity density as stated above.
	$\bfA \times (\bfv_* \times \bfB) + \bfB \phi$ is the flux of magnetic helicity\footnote{In the laboratory gauge used in the Poisson brackets of Section \ref{s:PB-for-R-MHD}, $\phi = \bfv_* \cdot \bfA$ so the magnetic helicity current is $(\bfA \cdot \bfB) \bfv_*$ in this gauge.}. Global conservation of ${\cal K}_B$ requires the flux of magnetic helicity across the boundary surface to be zero. This is guaranteed by the conditions $\bfB \cdot \hat n = 0$, $\bfv \cdot \hat n = 0$ and $(\grad \times \bfw) \cdot \hat n = 0$. This is because 
	\beqs
	\left( \bf A \times (\bfv_* \times \bfB) \right) \cdot \hat n &=& (\bfv_* \cdot \hat n) (\bfA \cdot \bfB) - (\bfv_* \cdot \bfA) (\bfB \cdot \hat n)
	 = (\bfA \cdot \bfB) \left( \bfv \cdot \hat n + \la^2 (\grad \times \bfw) \cdot \hat n \right) \cr
	&&- (\bfv_* \cdot \bfA) (\bfB \cdot \hat n).
	\eeqs
Note that for conservation of ${\cal K}_B$ it suffices that both $\bfB$ and $\bfv_*$ be tangential to $\pdr V$. The BC $\bfB \cdot \hat n = 0$ also guarantees gauge-invariance of ${\cal K}_B$. Moreover, unlike for flow helicity, the BCs that guarantee $E^*$ conservation also ensure conservation of ${\cal K}_B$ (though not vice versa). In an infinite domain energy and magnetic helicity are conserved if $\bfv , \bfB \to 0$ and $\rho \to $ constant as $\bfr \to \infty$. For a finite flow domain, we may also impose periodic BC for energy and magnetic helicity conservation. 


{\noindent \bf Cross helicity conservation:} Cross helicity $\int \bfv \cdot \bfB \: d\bfr$ measuring the degree of linkage of vortex and magnetic field lines is locally conserved in R-MHD:
	\beq
	\pdr_t (\bfv \cdot \bfB) + \grad \cdot (\sigma\bfB + \bfv \times (\bfv_* \times \bfB)) = 0.
	\eeq
The cross helicity current may be obtained from the magnetic helicity current by replacing $\phi \to \sigma$ and $\bfA \to \bfv$. To see this, we express $\pdr_t (\bfv \cdot \bfB)$ as a divergence 
	\beqs
	\pdr_t (\bfv \cdot \bfB) &=& \bfB \cdot \bfv_t + \bfv \cdot \bfB_t = \bfB \cdot ( - \grad \sigma + \bfv_* \times \bfw) + \bfv \cdot (\grad \times (\bfv_* \times \bfB)) \cr
	&=& - \bfB \cdot \grad \sigma + \bfB \cdot \bfv_* \times \bfw + \bfv_* \times \bfB \cdot \bfw +\grad \cdot ((\bfv_* \times \bfB) \times \bfv) \cr
	&=& -\grad \cdot (\sigma\bfB + \bfv \times (\bfv_* \times \bfB)). 
	\eeqs
Boundary conditions that lead to global cross helicity conservation are $\bfv_* \cdot \hat n =0$ and $\bfB \cdot \hat n = 0$. 


{\noindent \bf Locally conserved linear and angular momenta:} The momentum density ${\cal P}_i = \rho v_i$ and stress tensor $\Pi_{ij}$ satisfy a local conservation law
	\beq
	\dd{{\cal P}_i}{t} + \pdr_j \Pi_{ij} = 0, \quad \text{where} \quad \Pi_{ij} = \rho v_i v_j + p \del_{ij} + \la^2 \rho \left( \half w^2 \del_{ij} - w_i w_j \right) + \ov{\mu_0} \left( \half B^2 \del_{ij} - B_i B_j \right).
	\label{e:mom-cons-r-mhd}
	\eeq
$\bfB$ and $\bfw$ enter $\Pi_{ij}$ in the same manner since the twirl force ($-\la^2 \rho \bfw \times (\grad \times \bfw)$) and magnetic Lorentz force ($- (\bfB \times (\grad \times \bfB))/\mu_0$) are of the same form. The proof is as follows
	\beq
	\dd{{\cal P}_i}{t} = v_i \dd{\rho}{t} + \rho \dd{v_i}{t} = - \pdr_j \left( \rho v_i v_j + p \del_{ij} + \la^2 \rho \left( \half w^2 \del_{ij} - w_i w_j \right) \right) + \ov{\mu_0} ((\grad \times \bfB) \times \bfB)_i.
	\eeq
The first term is known from the conservation of momentum in R-Euler flow and the second comes from the magnetic force. The magnetic force term can be expressed as a divergence leading to the above-mentioned result:
	\beq
	((\grad \times \bfB) \times \bfB)_i = - \half \pdr_i B^2 + B_j \pdr_j B_i = - \pdr_j \left( \half B^2 \del_{ij} - B_i B_j \right).
	\eeq
We define angular momentum density in R-MHD as $\vec {\cal L} = \rho \bfr \times \bfv$\footnote{While the angular momentum density depends on the choice of origin, the total angular momentum does not.}. Using the local conservation of $\rho \bfv$ we find that $\vec {\cal L}$ too is locally conserved in R-MHD:
	\beq
	\dd{{\cal L}_i}{t} = \eps_{ijk} r_j \dd{\rho v_k}{t} = - \pdr_l \left( \eps_{ijk} r_j \Pi_{kl} \right) = - \pdr_l \Lambda_{il}.
	\label{e:ang-mom-cons-r-mhd}
	\eeq
Linear momentum $\int {\cal P}_i \, d\bfr$ and angular momentum $\int {\cal L}_i \, d\bfr$ are globally conserved for appropriate boundary conditions (e.g. decaying BC in an infinite domain or periodic BC in a cuboid for linear momentum).

\subsection{Regularized Kelvin-Helmholtz and Alfv\'{e}n freezing-in theorems and swirl velocity}
\label{s:Kelvin-circulation-Kelvin-Helmholtz-swirl-vel}

{\noindent \bf Regularized Kelvin-Helmholtz freezing-in theorem}: For incompressible ideal flow, it is well known that vorticity is frozen into the velocity field: $\bfw_t + \bfv \cdot \grad \bfw - \bfw \cdot \grad \bfv = 0$ or $\bfw_t + {\cal L}_\bfv \bfw = 0$. Here ${\cal L}_\bfv \bfw$ is the Lie derivative of $\bfw$ along $\bfv$, which is also the commutator of vector fields $[\bfv,\bfw]$. Kelvin's and Helmholtz's theorems on vorticity follow from the freezing of $\bfw$ into $\bfv$. This result has an extension to the compressible,  regularized theory. We show that $\bfw/\rho$ is frozen into the swirl velocity $\bfv_* = \bfv + \la^2 \grad \times \bfw$ (\ref{e:v*-continuity-eqn}). The R-vorticity equation (\ref{e:R-vorticity-eqn-compressible}) can be written as
	\beqs
	\dd{\bfw}{t} + \grad \times (\bfw \times \bfv_*) = 0
	\quad \imply \quad  \dd{\left\{(\bfw/\rho) \rho\right\}}{t} + \grad \times \left(\rho \frac{\bfw}{\rho} \times  \bfv_* \right) = 0 \imply \cr
	\frac{\rho \pdr}{\pdr t} \frac{\bfw}{\rho} + \frac{\bfw}{\rho} \dd{\rho}{t} + \bfw (\grad \cdot \bfv_*) - \bfv_* (\grad \cdot \bfw) + (\bfv_* \cdot \grad) \left[\rho \left(\frac{\bfw}{\rho} \right)\right]
	- \left[\rho \left(\frac{\bfw}{\rho} \right)\cdot \grad\right] \bfv_* = 0.
	\eeqs
We use the continuity equation (\ref{e:v*-continuity-eqn}) to write $\rho_t = - \rho \grad \cdot \bfv_* - \bfv_* \cdot \grad \rho$. The last term is one that appears in the Lie derivative ${\cal L}_{\bfv_*}(\bfw/\rho)$ and the penultimate term also contributes to ${\cal L}_{\bfv_*}(\bfw/\rho)$ upon using the Leibnitz rule. Thus
	\beq
	\rho \dd{(\bfw / \rho)}{t} - \frac{\bfw}{\rho} (\bfv_* \cdot \grad) \rho - \bfw \grad \cdot \bfv_* + \bfw \grad \cdot \bfv_* + \rho \bfv_* \cdot \grad \left(\frac{\bfw}{\rho} \right) + \frac{\bfw}{\rho} (\bfv_* \cdot \grad) \rho - \left[\rho \left(\frac{\bfw}{\rho} \right) \cdot \grad \right] \bfv_* = 0.
	\eeq
So dividing by $\rho$ we obtain the freezing-in of $\bfw/\rho$ into $\bfv_*$:
	\beq
	\dd{(\bfw/\rho)}{t} + (\bfv_* \cdot \grad) (\bfw/\rho)  - ((\bfw/\rho) \cdot \grad) \bfv_* = 0 \quad
	\text{or} \quad
	\dd{(\bfw/\rho)}{t} + {\cal L}_{\bfv_*} (\bfw/\rho) = 0.
	\label{e:freezing-in-w-by-rho-into-v*}
	\eeq
Indeed, it is well-known in Eulerian compressible, barotropic flow [$\la \rightarrow 0$] that $\bfw/\rho$ is frozen into $\bfv$.

{\noindent \bf Regularized Alfv\'{e}n's Theorem:} $\bfB/\rho$ is frozen into the swirl velocity $\bfv_*$ (\ref{e:v*-continuity-eqn}), i.e., it is Lie dragged along $(\pdr_t, \bfv_*)$:
	\beq
	\dd{}{t}\left(\frac{\bfB}{\rho}\right) + {\cal L}_{\bfv_*} \frac{\bfB}{\rho} = 
	\dd{}{t}\left(\frac{\bfB}{\rho}\right) + (\bfv_* \cdot \grad) \frac{\bfB}{\rho} - \left(\frac{\bfB}{\rho} \cdot \grad \right) \bfv_* = 0.
	\label{e:Bbyrho-frozen-in-v*}
	\eeq
{\sc Proof:} Multiplying and dividing by $\rho$ in the regularized Faraday's law (\ref{e:R-MHD-Faraday}) and using Leibnitz rule we get:
	\beq
	\pdr_t \bfB =  \pdr_t \left(\rho \frac{\bfB}{\rho} \right) 
	= \frac{\bfB}{\rho} \dd{\rho}{t} + \rho \pdr_t \frac{\bfB}{\rho} = \grad \times \left(\rho \bfv_* \times \frac{\bfB}{\rho} \right) \;\; \imply \;\; 
	 \rho\dd{}{t}\left(\frac{\bfB}{\rho}\right)
	= \grad \times \left(\rho \bfv_* \times \frac{\bfB}{\rho} \right) - \frac{\bfB}{\rho} \dd{\rho}{t}.
	\eeq
Using the continuity equation expressed in terms of $\bfv_*$ (\ref{e:v*-continuity-eqn}), this simplifies to
	\beqs
	\rho \dd{}{t}\left(\frac{\bfB}{\rho}\right)
	&=& \grad \times \left(\rho \bfv_* \times \frac{\bfB}{\rho} \right) + \frac{\bfB}{\rho} \grad \cdot (\rho \bfv_*) \cr
	&=& \rho \bfv_*\grad \cdot \left(\frac{\bfB}{\rho}\right) - \frac{\bfB}{\rho} \grad \cdot (\rho \bfv_*) + \left(\frac{\bfB}{\rho}\right) \cdot \grad (\rho \bfv_*) - \rho \bfv_* \cdot \grad  \left(\frac{\bfB}{\rho}\right)
	+ \frac{\bfB}{\rho} \grad \cdot (\rho \bfv_*)\cr
	&=& \rho \bfv_* \left(\bfB \cdot \grad \ov{\rho}\right) + \bfv_* (\grad \cdot \bfB) 
	+ \rho \left(\frac{\bfB}{\rho}\right) \cdot \grad \bfv_* + \bfv_* \left(\frac{\bfB}{\rho}\right) \cdot \grad \rho
	- (\rho \bfv_* \cdot \grad ) \left(\frac{\bfB}{\rho}\right) \cr
	&=& \bfB \cdot \grad \bfv_*
	- (\rho \bfv_* \cdot \grad ) \left(\frac{\bfB}{\rho}\right).
	\eeqs
where we used the Leibnitz rule and  $\grad \cdot \bfB = 0$. Thus we get the above-mentioned result.


{\noindent \bf Swirl energy in terms of swirl velocity:} It is useful to note that the conserved swirl energy $E^*$ (in both R-Euler and R-MHD) can be expressed compactly in terms of $\bfv_*$ (for appropriate BC):
	\beq
	E^* = \int_V \left[ \frac{\rho \bfv^2}{2}+ U(\rho) + \frac{\la^2 \rho \bfw^2}{2} + \frac{\bfB^2}{2 \mu_0} \right] \: d\bfr  \; = \; 	\int_V \left(\ov{2}\rho \bfv_*\cdot \bfv + U(\rho) + \frac{\bfB^2}{2 \mu_0} \right) \: d\bfr
	\equiv E_{\bfv_*}^*.
	\eeq
So up to a boundary term, $\bfv \cdot \bfv_*$ accounts for both kinetic and enstrophic energies. To see this, we begin by substituting for  $\bfv_* = \bfv + \la^2 \grad \times \bfw $ in $E_{\bfv_*}^*$ and use the divergence of a cross product to get
	\beqs
	E_{\bfv_*}^* &=& \int_V \left( \frac{\rho \bfv^2}{2} + \frac{\la^2 \rho}{2} (\grad \times \bfw)\cdot \bfv + U(\rho) + \frac{\bfB^2}{2 \mu_0} \right) \: d\bfr \cr
	&=& \; \int_V \left( \frac{\rho \bfv^2}{2} + \frac{\la^2 \rho}{2} \bfw^2 + U(\rho) + \frac{\bfB^2}{2 \mu_0} + \frac{\la^2 \rho}{2} \grad \cdot(\bfw \times \bfv)\right) \: d\bfr \cr
	&=& \int_V \left( \ov{2}\rho \bfv^2 + \ov{2}\la^2 \rho \bfw^2 + U(\rho) + \frac{\bfB^2}{2 \mu_0}\right)\: d\bfr + \half \la^2 \rho \int_{\pdr V} (\bfw \times \bfv) \cdot \hat n \: dS.
	\eeqs
The boundary term vanishes if $\bfv \times \hat n = 0$ or $\bfw \times \hat n = 0$. In both R-Euler and R-MHD, the BCs for $E^*$ conservation include $\bfw \times \hat n = 0$. So it is possible to express $E^*$ in terms of $\bfv_*$ with the same BCs that lead to $E^*$ conservation. Moreover, in R-Euler the BCs that guarantee conservation of flow helicity include $\bfv \times \hat n = 0$. So in R-Euler it is possible to express $E^*$ in terms of $\bfv_*$ with the BCs that lead to either $E^*$ or flow helicity conservation.


{\noindent \bf Time evolution of $\bfv_*$:} In compressible R-Euler flow, the evolution equation for $\bfv_*$ is  
	\beq
	\bfv_{*t} + \bfw \times \bfv_* + \grad \sigma = \frac{\la^2}{\rho} \grad \cdot (\rho \bfv_*) \grad \times \bfw - \la^2 \grad \times \left( \grad \times (\bfw \times \bfv_*) \right).
	\eeq
\normalsize
Here $\sigma = h + \half (\bfv_* - \la^2 \grad \times \bfw)^2$ and $\bfw$ satisfies (\ref{e:R-vorticity-eqn-compressible}). This is a {\em local} formulation of R-Euler in terms of $\bfv_*$, $\rho$ and $\bfw$. In R-MHD, for $\sigma$ as above, the evolution equation for $\bfv_*$ becomes
	\beq
	\bfv_{*t} + \bfw \times \bfv_* + \grad \sigma = \frac{\la^2}{\rho} \grad \cdot (\rho \bfv_*) \grad \times \bfw - \la^2 \grad \times \left( \grad \times (\bfw \times \bfv_*) \right) + \frac{\bfj \times \bfB}{\rho} + \la^2 \grad \times \left(\grad \times \left(\frac{\bfj \times \bfB}{\rho}\right)\right).
	\eeq

\section{Integral invariants associated to swirl velocity}
\label{s:integral-inv-v-star}
\subsection{Swirl Kelvin theorem: Circulation around a contour moving with $\bfv_*$ is conserved}

We show here that the circulation $\G$ of $\bfv$ around a closed contour $C^*_t$ (that moves with $\bfv_*$) is independent of time. This is a regularized version of the Kelvin circulation theorem.
	\beq
		\frac{d\G}{dt} = \frac{d}{dt} \oint_{C^*_t} \bfv \cdot d\bfl = \frac{d}{dt} \int_{S^*_t} \bfw \cdot d\bfS = 0.
	\label{e:swirl-kelvin-theorem}
	\eeq
Here $S^*_t$ is any surface moving with $\bfv_*$ spanning $C^*_t$. Note that the circulation is that of $\bfv$ while the advecting velocity is $\bfv_*$.
{\flushleft \sc Proof:} When the time derivative is taken inside the integral sign to act on Eulerian quantities transported by $\bfv_*$, we introduce the operator $D_t^* \equiv \frac{D^*}{Dt} = \pdr_t + \bfv_* \cdot \grad$:
	\beq
	\DD{}{t} \oint_{C^*_t} \bfv \cdot d\bfl 
	= \oint_{C^*_t} \frac{D^* \bfv}{Dt} \cdot d\bfl + \oint_{C^*_t} \bfv \cdot \frac{D^* d\bfl}{Dt}.
	\eeq
Since $d \bfl$ is a line element that moves with $\bfv_*$, $\frac{D^* d\bfl}{Dt} = d \frac{D^* \bfl}{Dt} = d \bfv_*$. To see this we make use of the flow map from the fixed initial coordinates $\bfx_0$ to the coordinates $\bfx$ at time $t$.
	\beq
      dx_i = \frac{\partial x_i}{\partial x_{0 j}}dx_{0 j} \quad \imply \quad
      \frac{d^{*}}{dt}(dx_i)=\frac{\partial}{\partial x_{0 j}}\left(\frac{d^{*}x_i}{d t}\right)dx_{0 j}=\frac{\partial v_{*i}}{\partial x_{0 j}}dx_{0 j}=\frac{\partial v_{* i}}{\partial x_k} dx_k = dv_{* i}.
      \eeq
Thus
	\beq
	\frac{d\G}{dt} = \oint_{C^*_t} \left( \dd{\bfv}{t} + \bfv_* \cdot \grad \bfv \right) \cdot d\bfl + \oint_{C^*_t} \bfv \cdot d\bfv_*.
	\eeq
Using the R-Euler equation $\bfv_t = - \bfw \times \bfv_* - \grad \sigma$ and the vector identity $\bfv_* \cdot \grad \bfv =   \grad \bfv \cdot \bfv_*- \bfv_* \times (\grad \times \bfv)$ where $(\grad \bfv \cdot \bfv_*)_i = v_{*j} \pdr_i v_j$ we get
	\beq
	\frac{d\G}{dt} = \oint_{C^*_t} \grad \bfv \cdot \bfv_* \cdot d\bfl + \oint_{C^*_t} \bfv \cdot d \bfv_* - \oint_{C^*_t} \grad \sig \cdot d\bfl.
	\eeq
$\grad \sig$ integrates to zero around a closed contour. Finally, using $\bfv \cdot d \bfv_* = v_j \pdr_i v_{*j} dl^i$ and $\grad \bfv \cdot \bfv_* \cdot d\bfl = \bfv_{*j} \pdr_i v_j dl^i$ we get
	\beq
	\DD{\G}{t} = \oint_{C^*_t} \pdr_i (\bfv_* \cdot \bfv) dl^i = \oint_{C^*_t} d (\bfv_* \cdot \bfv) = 0.
	\eeq
The final equality of (\ref{e:swirl-kelvin-theorem}) follows from Stokes' theorem $\G = \int_{S^*_t} (\grad \times \bfv) \cdot d\bfS$.

\subsection{Swirl Alfv\'en theorem on conservation of magnetic flux}

We show that the line integral $\Phi = \oint_{C_t^*} \bfA \cdot d\bfl$ over a closed contour $C^*_t$ moving with $\bfv_*$ is a constant of the motion.

{\flushleft \sc Proof:} Using the equation of motion for $\bfA$: $\dd{\bfA}{t} = \bfv_* \times \bfB - \grad \phi$ we can write 
	\beqs
	\DD{}{t}\oint_{C^*_t} \bfA \cdot d\bfl &=& \oint_{C^*_t} \frac{D^*\bfA}{Dt}  \cdot d\bfl  + \oint_{C^*_t} \bfA \cdot d\bfv_* = \oint_{C^*_t} \left( \bfv_* \times \bfB - \grad \phi + \bfv_* \cdot \grad \bfA \right) \cdot d\bfl  \cr
	+ \oint_{C^*_t} \bfA \cdot d\bfv_* &=& \oint_{C^*_t} \left( v_{*j} \pdr_i A_j dl^i + A_i\pdr_j v_{*i}dl^j \right) = \oint_{C^*_t} \grad(\bfv_* \cdot \bfA) \cdot d\bfl = 0.
	\eeqs
We used the identity 
$(\bfv_* \times \bfB + \bfv_* \cdot \grad \bfA)_i = v_{*j} \; \pdr_i A_j$ and wrote $(d\bfv_*)_i = \pdr_j v_{*i} dl^j$ as in our proof of the swirl Kelvin theorem. Now if $S^*$ is any surface spanning the contour $C^*$ and $\bfB = \grad \times \bfA$ is the magnetic field, from Stokes' theorem we see that $\Phi = \int_{S^*} \bfB \cdot d\bfS$ is a constant of the motion. This is the regularized version of Alfv\'en's frozen-in flux theorem.

\subsection{Surfaces of vortex and magnetic flux tubes move with $\bfv_*$}

Given any smooth function $S(\bfr,t)$ we may consider its level surfaces at a given instant of time. We define an evolution of such a surface through an equation for $S({\bf r},t)$:
	\beq
	\frac{\partial S}{\partial t}+{\bf v}_{*}.\nabla S = D^{*}_{t}S = 0 \quad \text{where the operator} \quad D^{*}_{t}\equiv \frac{\partial }{\partial t}+{\bf v}_{*}.\grad.
	\label{e:material-surface-adv-v_*}
	\eeq
It follows that level surfaces of $S$ are advected by ${\bf v}_{*}$. Suppose the equation $({\bf w}/\rho) \cdot \grad S = 0$ holds at $t=0$, it implies that $\bfw$ is tangential to the level surfaces of $S$ at $t = 0$. For $\bfw$ to remain tangential to the level surfaces of $S$ at all times, $D^*_t(\frac{\bfw}{\rho} \cdot \grad S)$ must vanish. This is indeed so as a consequence of the freezing of $\bfw/\rho$ into $\bfv_*$ (\ref{e:freezing-in-w-by-rho-into-v*}) and the advection of $S$ by $\bfv_*$:
    \beqs
	D^{*}_{t}\left[\left(\frac{\bfw}{\rho}\right)\cdot\nabla S\right] &=& \left(\frac{\bfw}{\rho}\right).\nabla {\bf v}_{*} \cdot\nabla S + \left(\frac{\bfw}{\rho}\right)\cdot D^{*}_{t} \nabla S 
	    = \left(\frac{\bfw}{\rho}\right).\nabla {\bf v}_{*}\cdot \nabla S \cr
	    && + \left(\frac{\bfw}{\rho}\right)\cdot \left[-\nabla \left({\bf v_{*}\cdot \nabla S}\right)+{\bf v_{*}}\cdot \nabla \nabla S\right] = 0.
	\eeqs
In particular, the surface of a vortex tube is advected by $\bfv_*$ (and {\em not} by $\bfv$). As in the case of vorticity, $\bfB/\rho$ is frozen into $\bfv_*$ by virtue of (\ref{e:Bbyrho-frozen-in-v*}). Thus magnetic flux tubes, like vortex tubes, are transported by $\bfv_*$.

\subsection{Curves advected by $\bfv_*$}
 Consider the level surfaces of two functions, $\alpha({\bf x},t)$ and $\beta({\bf x},t)$, advected 
by ${\bf v}_{*}$:
	\beq
	\al_t + \bfv_*\cdot \grad \alpha = 0 \quad \text{and} \quad \beta_t + \bfv_* \cdot \grad \beta = 0.
	\label{e:alpha-beta-advected-by-v*}
	\eeq
If $\alpha$ and $\beta$ are not functions of each other, the curve defined by the [solenoidal] direction vector, ${\bf Z}=\grad \alpha \times \grad \beta$ is a space curve, varying with time. We show that this space curve moves with $\bfv_*$, i.e. that ${\bf Z}/\rho$ is `frozen' into $\bfv_*$:
    \beq
    \bfZ_t = \grad \alpha_t \times \grad \beta + \grad \alpha \times \grad \beta_t .
    \eeq
From (\ref{e:alpha-beta-advected-by-v*}) and the identity $\grad a \times \grad b = \grad\times (a\grad b)$, we get:
	\beq
	\bfZ_t = - \grad(\bfv_* \cdot \grad \alpha) \times \grad \beta + \grad(\bfv_* \cdot \grad \beta)\times \grad \alpha
	= \grad \times \left[ (\bfv_* \cdot \grad \beta) \: \grad \al - (\bfv_* \cdot \grad \alpha) \: \grad \beta \right]
	= \grad \times (\bfv_* \times \bfZ).
	\eeq  
A solenoidal field satisfying $\bfZ_t = \grad \times (\bfv_* \times \bfZ)$ is termed a `Helmholtz' field associated to $\bfv_*$ \cite{thyagaraja-IITM}. Combining this with the continuity equation, we find that $\bfZ/\rho$ is frozen into $\bfv_*$:
	\beq
    \frac{\partial}{\partial t}\left(\frac{{\bf Z}}{\rho}\right)+\bfv_* \cdot \grad \left(\frac{{\bf Z}}{\rho}\right) = \frac{D^*}{Dt}\left(\frac{{\bf Z}}{\rho}\right) = \left(\frac{{\bf Z}}{\rho}\right)\cdot\grad \bfv_*.
    \eeq
Not every Helmholtz field is expressible as $\bfZ = \grad \al \times \grad \beta$ for a pair of functions advected by $\bfv_*$. We will show in Section \ref{s:helmholtz-field-g-helicity-g-tube} that such a Helmholtz field has zero `$\bfZ$-helicity', unlike Helmholtz fields like vorticity and magnetic field which lead to generally non-trivial flow and magnetic helicity.

\subsection{Analogue of Reynolds' transport theorem for volumes advected by $\bfv_*$}

There is useful version of Reynolds' transport theorem for volumes advected by the swirl velocity $\bfv_*$. Suppose $f(\bfx,t)$ is a scalar function associated with a volume $V^*$ moving with $\bfv_*$, then
	\beq
  \frac{d}{dt}\int_{V^*_t} f d\bfx =   \int_{V^*_t} D^{*}_{t}\left(\frac{f}{\rho}\right)\rho d \bfx.
  \label{e:reynolds-transport-theorem}
	\eeq
It is useful to develop briefly the ``Lagrangian'' theory underlying Reynolds' transport theorem. Let ${\bf x}(t)$ be the location of a ``fluid particle'' being transported by the swirl velocity ${\bf v}_{*}({\bf x},t)$. By definition $\partial^0 \bfx/\partial t = {\bf v}_{*}({\bf x},t)$ where the `Lagrangian' time derivative is taken holding the initial position $\bfx_0$ fixed unlike the `local' Eulerian time derivative. If ${\bf v}_{*}({\bf x},t)$ is known, integration gives, ${\bf x}={\bf x}({\bf x}_{0},t)$, so that at any instant the fluid position is a function of $t$ and initial location ${\bf x}_{0}$. The Jacobian, $J=\frac{\partial (x,y,z)}{\partial (x_{0},y_{0},z_{0})}$ relates the volume elements in the two coordinates $\bfx_0$ and $\bfx$ : $Jd\bfx_{0} = d\bfx$. It is a standard result \cite{chorin-marsden} that:
	\beq
 	\ov{J} \dd{^0 J}{t} = \grad \cdot {\bf v_*}
  \eeq
where ${\bf v}_{*}$ is the advecting velocity and the RHS is the standard Eulerian divergence taken at ${\bf x}$ at the instant $t$. Using the continuity equation :$D^*_t \rho = -\rho \grad \cdot \bfv_*$ we get
	\beq
	\grad \cdot \bfv_* = -\ov{\rho} D^*_t \rho = \ov{J} \dd{^0 J}{t} \quad \imply \quad D^*_t (\rho J) =0.
	\eeq
In fact, $\rho J = \rho_0$ where $\rho_0 = \rho(\bfx, t=0)$ as $J(t=0) = 1$. Now if $f({\bf x},t)$ is a scalar function associated with a volume $V$ moving with $\bfv_*$ we have
	\beq
  \frac{d}{dt}\int_{V^*_t} f d\bfx 
  = \frac{d}{dt}\int_{V^*_{0}}fJ \, d\bfx_0 
  = \int_{V^*_{0}} D^*_t\left(\frac{f}{\rho} \rho J\right)d\bfx_{0} 
  = \int_{V^*_{0}} D^*_t \left(\frac{f}{\rho} \right)\rho J d\bfx_{0} 
  = \int_{V^*_t} D^{*}_{t}\left(\frac {f}{\rho}\right)\rho \,d\bfx.
	\eeq
We have used $D^*_t (\rho J) = 0$, $D^*_t d\bfx_0 = 0$ and $Jd\bfx_0 = d\bfx$.

\subsection{Conservation of mass in a volume moving with $\bfv_*$}

Suppose a volume $V_t^*$ moves with $\bfv_*$. The mass of fluid within such a volume is independent of time. From (\ref{e:reynolds-transport-theorem}),
	\beq
	\frac{d}{dt} \int_{V_t^*} \rho \: d\bfx 
	= \int_{V^*_t} \rho D^*_t\left(\frac{\rho}{\rho} \right) \: d\bfx =	0
	\eeq

\subsection{Conservation of flow helicity in a closed vortex tube}

As we have noted, vortex tubes move with $\bfv_*$. Here we show that the flow helicity ${\cal K}$ associated with such a tube enclosing a volume $V^*_t$ is independent of time:
	\beq
	\frac{d{\cal K}}{dt} = \frac{d}{dt}\int_{V^*_t} \bfw \cdot \bfv \: d\bfx = 0.
	\eeq
{\flushleft \sc Proof} : Applying (\ref{e:reynolds-transport-theorem}) to ${\cal K} $ and using the freezing in condition $D_t^*(\bfw/\rho) = (\bfw/\rho) \cdot \grad \bfv_*$ and equation of motion (\ref{e:R-Euler-v*}) we get 
	\beqs
	\dot {\cal K} &=& \int_{V_t^*} D^{*}_{t}
	\left(\frac{\bfw}{\rho} \cdot {\bfv} \right) \, \rho d\bfx
	= \int_{V^*_t} \left[D^{*}_{t}\left(\frac{\bf w}{\rho}\right)\cdot{\bf v}+\left(\frac{\bf w}{\rho}\right)\cdot D^{*}_{t} ({\bf v})\right] \rho d\bfx  \cr               
	&&= \int_{V^*_t} {\bf w}\cdot \left[ \grad\bfv_* \cdot \bfv + {\bf v}_{*}\cdot {\bf \grad v}+{\bf v}_{*}\times {\bf w} - \grad \sigma \right] d\bfx.
  \eeqs
The middle two terms combine (${\bf v}_{*}\cdot {\bf \grad v}+{\bf v}_{*}\times {\bf w} = \grad \bfv \cdot \bfv_*$) to give
	\beq
	\frac{d {\cal K}}{dt} = \int_{V^*_t} {\bf w}\cdot[\grad\bfv_* \cdot \bfv + \grad\bfv \cdot \bfv_* - \grad \sigma ]d\bfx
		= \int_{V^*_t} \bfw \cdot \grad [\bfv \cdot \bfv_* - \sig] \,d\bfx
		= \int_{\pdr V^*_t} (\bfv \cdot \bfv_* - \sigma) \bfw \cdot \hat n \, dS = 0.
	\eeq
Here we used $\grad \cdot \bfw = 0$ and the fact that $\bfw$ is tangential to the surface (vortex tube) bounding the volume $V^*_t$.

\subsection{Conservation of magnetic helicity in a magnetic flux tube}
In R-MHD, the magnetic helicity ${\cal K}_B$ (but {\it not} flow helicity) associated with a volume $V^*_t$ bounded by a closed magnetic flux tube is independent of time:
	\beq
	\frac{d {\cal K}_B}{dt}= \frac{d}{dt}\int_{V^*_t} \bfB \cdot \bfA \,d\bfx = 0.
   	\eeq
This is a consequence of the fact that $\bfB$ is tangential to the boundary of such a volume by the freezing of $\bfB/\rho$ into $\bfv_*$.

{\flushleft \sc Proof} : As before, we apply (\ref{e:reynolds-transport-theorem}) to $d{\cal K}_B/dt$ and use the freezing-in condition $D^*_t(\bfB/\rho) = (\bfB/\rho) \cdot \grad \bfv_*$ and equation for the evolution of the vector potential (\ref{e:A-evolution-R-MHD}) to get\footnote{$\phi$ is arbitrary, it depends on the choice of gauge. In the PB formulation $\phi = \bfv_* \cdot \bfA$}
	\beqs
	\frac{d{\cal K}_B}{dt}&=& \int_{V_t^*} D^{*}_{t}\left[\left(\frac{\bfB}{\rho}\right)  {\bf \cdot A} \right] \rho \, d\bfx =  \int_{V^*_t} \left[D^{*}_{t}\left(\frac{\bfB}{\rho}\right)\cdot{\bfA}+\left(\frac{\bfB}{\rho}\right)\cdot D^{*}_{t} (\bfA)\right] \rho \,d\bfx \cr
    &=& \int_{V^*_t} {\bf B}\cdot[{\bf \grad v}_{*}\cdot{\bf A} +\bfv_* \times \bfB - \grad \phi +\bfv_* \cdot \grad \bfA]\,d\bfx   
	= \int_{V^*_t} {\bf B}\cdot \grad \left[\bfv_* \cdot \bfA - \phi \right] \, d\bfx \cr
	&=& \int_{\pdr V^*_t} \left[\bfv_* \cdot \bfA - \phi \right] {\bf B} \cdot \hat n \, d\bfx = 0.
      \eeqs
The last equality follows as $\grad \cdot \bfB = 0$ and since $\bfB$ is tangential to a surface that moves with $\bfv_*$ ($V_t^*$ is a magnetic flux tube).

\subsection{Helmholtz fields $\bfg$ and their conserved helicities in $\bfg$-tubes}
\label{s:helmholtz-field-g-helicity-g-tube}

The conservation of flow and magnetic helicity in vortex and magnetic flux tubes are special cases of a more general result. Recall that a Helmholtz field \cite{thyagaraja-IITM} is a solenoidal vector field $\bfg$ that evolves according to $\bfg_t + \grad \times (\bfg \times \bfv_*) = 0$. If $\bfg$ is a Helmholtz field, then $\bfg/\rho$ is frozen into $\bfv_*$, i.e., $D^*_t (\bfg/\rho) = (\bfg/\rho) \cdot \grad \bfv_*$. A Helmholtz field in a simply-connected region (one where every closed curve can be continuously shrunk to a point while remaining in the region) is expressible in terms of a `vector potential' ${\bf u}$:
	\beq
	\bfg = \grad \times {\bf u} \quad \text{with} \quad {\bf u}_t + \bfg \times \bfv_* + \grad \tht  = 0
	\eeq
for some scalar function $\tht(\bfx,t)$. Examples of Helmholtz fields in R-Euler and R-MHD include $\bfw$ and $\bfB$. The corresponding vector potentials are $\bfv$ and $\bfA$, with $\tht$ corresponding to the stagnation enthalpy $\sigma$ and electrostatic potential $\phi$ respectively. 

If $\bfg$ is a Helmholtz field then its flux through a surface $S^*_t$ spanning a closed contour $C^*_t$ moving with $\bfv_*$ is conserved, generalizing the Kelvin and Alfv\'en theorems:
	\beq
	\DD{}{t} \oint_{C^*_t} \bfu \cdot d\bfl = \DD{}{t} \int_{S^*_t} \bfg \cdot d\bfS = 0.
	\eeq
Given a Helmholtz field, a closed surface everywhere tangent to $\bfg$ is called a $\bfg$-tube, generalizing vortex tubes and magnetic flux tubes. The freezing of $\bfg/\rho$ into $\bfv_*$ then implies that a $\bfg$-tube moves with $\bfv_*$. Associated to a Helmholtz field $\bfg$ and its vector potential $\bf u$ is a $\bfg$-helicity density, $\bfg \cdot {\bf u}$. It follows from the transport theorem and the above equations of motion that the $\bfg$- helicity in a $\bfg$-tube is independent of time:
	\beq
	\DD{}{t} \int_{V^*_t} \bfg \cdot {\bf u} \: d\bfx = \int_{V^*_t} D^*_t \left(\frac{\bfg}{\rho} \cdot {\bf u}\right) \rho \: d\bfx = 0
	\eeq
{\flushleft \bf Note:} If $\bfZ = \grad \al \times \grad \beta$ is a Helmholtz field defined by two independent scalar functions advected by $\bfv_*$, then its vector potential is of the form $\bfu = \al \grad \beta + \grad \gamma$ where $\gamma$ is a scalar function. The corresponding $\bfZ$-helicity in a moving volume $V^*_t$, $\int_{V^*_t} \bfZ \cdot \grad \gamma \: d\bfx = \int_{\pdr V^*_t} \gamma \bfZ \cdot d\bfS - \int_{V^*_t} \gamma \grad \cdot \bfZ \: d\bfx$ is identically zero since $\bfZ$ is solenoidal and tangential to the boundary $\pdr V^*_t$.

\section{Poisson brackets for the R-Euler equations}
\label{s:pb-for-fluid}

Commutation relations among `quantized' fluid variables were proposed by Landau \cite{landau} in an attempt at a quantum theory of superfluid He-II. As a byproduct, one obtains Poisson brackets (PB) among {\it classical} fluid variables allowing a Hamiltonian formulation for compressible flow. Suppose $F$ and $G$ are two functionals of $\rho$ and $\bfv$, then their equal-time PB (see \cite{morrison-greene,morrison-review}) is
	\beqs
	\{ F, G \} &=& \int \left[ \frac{\bfw}{\rho} \cdot \left( \deldel{F}{\bfv} \times \deldel{G}{\bfv} \right) - \deldel{F}{\bfv} \cdot \grad G_{\rho} + \deldel{G}{\bfv} \cdot \grad F_{\rho} \right] d\bfr \cr
	&=& \int \left[ \frac{\bfw}{\rho} \cdot \left( \deldel{F}{\bfv} \times \deldel{G}{\bfv} \right) +\grad \cdot \left( \deldel{F}{\bfv} \right) G_{\rho} - \grad \cdot \left(\deldel{G}{\bfv} \right) F_{\rho} \right] d\bfr.
	\label{e:pb-between-functionals-of-rho-v}
	\eeqs
The two formulae are related by integration by parts. If $\rho$ and mass current ${\bf M} = \rho \bfv$ are taken as the basic variables, then
	\beq
	\{ F , G \} = - \int \left[ \rho \left( \deldel{F}{\bf M} \cdot \grad G_{\rho} - \deldel{G}{\bf M} \cdot \grad F_{\rho} \right) + M_i \left( \deldel{F}{\bf M} \cdot \grad \deldel{G}{M_i} - \deldel{G}{\bf M} \cdot \grad \deldel{F}{M_i} \right) \right] \: d\bfr.
	\eeq
We will show that this PB, along with our conserved swirl energy hamiltonian $H$ leads to the R-Euler equations. The PB is manifestly anti-symmetric and the dimension of $\{ F, G \}$ is that of $FG/\hbar$. The PB of $F[\rho,\bfv]$ with a constant (independent of $\rho$ and $\bfv$) is zero. The Leibnitz rule $\{ FG, H \} = F \{ G, H\} + \{F, H \} G$ for three functionals follows from the (\ref{e:pb-between-functionals-of-rho-v}) upon using the Leibnitz rule for functional derivatives. In other words, the PB $\{ F, G \}$ is a derivation in each entry holding the other fixed.

From (\ref{e:pb-between-functionals-of-rho-v}) we deduce the PB among basic dynamical variables subject to the constitutive relation $\la^2 \rho =$ constant:
	\beqs
	&& 
	\{ \rho(\bfx), \bfv(\bfy) \} = - \grad_\bfx \del(\bfx-\bfy) =  \frac{(\grad_\bfy - \grad_\bfx)}{2} \del(\bfx-\bfy), \quad \{ v_i(\bfx), v_j(\bfy) \} = \frac{\om_{ij}}{\rho} \: \del(\bfx-\bfy),
	\cr 
	&& \{ \rho(\bfx), \rho(\bfy) \} = 0, \quad \{ \rho(\bfx), \la(\bfy) \} = 0, \quad \{ \la^2(\bfx), \bfv(\bfy) \} = - \frac{\la^2(\bfx)}{\rho(\bfx)} \{ \rho(\bfx), \bfv(\bfy) \}.
	\label{e:PB-among-basic-var}
	\eeqs
Here $\om_{ij} = \pdr_i v_j - \pdr_j v_i$ is the dual of vorticity, $w_i = \eps_{ijk} \omega_{jk}/2$ or $\omega_{ij} = \eps_{ijk} w_k$. (\ref{e:PB-among-basic-var}) generalises Gardner's PB $\{ u(\bfx), u(\bfy) \} = \half (\pdr_\bfy - \pdr_\bfx) \del(\bfx-\bfy)$ for KdV \cite{gardner}. The $\{ v_i , v_j \}$ is akin to the PB between canonical momenta of a charged particle in a $\bfB$ field
	\beq
	\left\{ p_i - ({e}/{c}) A_i(\bfx), p_j - ({e}/{c}) A_j(\bfx) \right\} = ({e}/{c}) F_{ij}(\bfx) \quad \text{where} \quad F_{ij} = \eps_{ijk} B_k.
	\eeq
$\bfB$ is analogous to $\bfw$ and $F_{ij}$ to $\omega_{ij}$. The Morrison-Greene PBs among functionals (\ref{e:pb-between-functionals-of-rho-v}) follow from the basic PBs (\ref{e:PB-among-basic-var}) by postulating that the PB is a derivation in either entry. For instance, denoting functional derivatives by subscripts we have:
	\beqs
	\{F[\rho], G[\bfv]\} &=& \int \frac{\del F}{\del \rho(x)}\frac{\del G}{\del v_i(y)}\{\rho(x), v_i(y)\} \:d\bfx\: d\bfy 
	= \int \frac{\del F}{\del \rho(x)}\frac{\del G}{\del v_i(y)}\pdr_{y^i} \del(x -y) \:d\bfx\: d\bfy \cr
	&=& -\int  F_{\rho} \grad \cdot G_\bfv \: d\bfx.\cr
 \{F[\bfv], G[\bfv]\} &=& \int \frac{\del F}{\del v_i(x)}\frac{\del G}{\del v_j(y)}\{v_i(x), v_j(y)\} d\bfx \; d\bfy \cr
    &=& \int \frac{\del F}{\del v_i(x)}\frac{\del G}{\del v_j(y)}\frac{\eps_{ijk} w_k(x)}{\rho(x)}\del(x-y) d\bfx \;d\bfy \cr
    &=& \int \frac{\bfw}{\rho} \cdot (F_{\bfv} \times G_{\bfv}) d\bfx.
	\eeqs
Some useful PBs follow from (\ref{e:PB-among-basic-var}). For instance $\rho$ commutes with vorticity:
	\beqs
	(a) && \{ \rho(\bfx), \bfw(\bfy) \} = 0 = \{ \la(\bfx), \bfw(\bfy) \}, \cr
	(b) && \{ v_i(\bfx), w_j(\bfy) \} = \eps_{jkl} \pdr_{\bfy^k} \left( \rho^{-1} \, \om_{il}(\bfy) \del(\bfx-\bfy) \right) = (\del_{jk} \pdr_{\bfy^i} - \del_{ij} \pdr_{\bfy^k}) (\rho^{-1} w_k(\bfy) \del(\bfx-\bfy)), \cr
	(c) && \{ w_i(\bfx), w_j(\bfy) \} = \eps_{ikl} \eps_{jmn} \pdr_{\bfx^k} \pdr_{\bfy^m} \left( \rho^{-1} {\om_{ln}(\bfx \; \text{or} \; \bfy)} \, \del(\bfx-\bfy) \right), \cr
	(d) && \{ v_k(\bfx), \om_{ij}(\bfy) \} = \pdr_{\bfy^i} \left( \rho^{-1}\om_{k j}(\bfy) \, \del(\bfx-\bfy) \right) - (i \leftrightarrow j), \cr
	(e) && \{ (\bfv \cdot \bfw)(\bfx), \rho(y)\bfy \} = - (\bfw(\bfx) \cdot \grad_\bfx) \del(\bfx-\bfy), \cr
	(f) && \{ (\grad \cdot \bfv)(\bfx) , \rho(\bfy) \} = - \grad^2_\bfx \del(\bfx-\bfy).
	\label{e:list-of-useful-PB}
	\eeqs
Some PBs of ${\bf M} = \rho \bfv$ and $\bfv_*$ are collected in Appendix \ref{a:pb-mass-curr-and-v*}. Properties of PBs among linear functionals are discussed in Appendix \ref{a:solenoidal-irrot-pb}. The basic PBs may also be written in Fourier space, which should be useful for numerics in a periodic domain:
	\beqs
	\{ \tl \rho(\bfk), \tl \rho(\bfk') \} = 0, &&
	\{ \tl \rho(\bfk), v_j(\bfk') \} = -i k_j (2\pi)^3 \del(\bfk + \bfk'), \quad
	\{ \tl v_i(\bfk) , \tl v_j(\bfk') \} = \widetilde{\left( \frac{\om_{ij}}{\rho} \right)}(\bfk + \bfk'),
	\cr
	\text{where} \quad
	\tl \rho(\bfk) &=& \int \rho(\bfx) e^{- i \bfk \cdot \bfx} \: d\bfx, \quad
	v_i(x) = \int \tl v_i(k) e^{i \bfk \cdot \bfx} \: \frac{d \bfk}{(2\pi)^3}, \quad \text{etc.}	
	\eeqs
The Jacobi identity is $\{ \{ F[\rho,\bfv], G[\rho,\bfv] \}, H[\rho, \bfv] \} + {\rm cyclic} = 0$. Using the PB among $\rho$ and $\bfv$, it is straightforward to check the Jacobi identity in some special cases, e.g., for coordinate functionals $F = \rho(x), G = \rho(y)$ and $H = \bfv(z)$ or for two $\bfv$'s and a $\rho$. It is not so straightforward to check the Jacobi condition in general, see the discussion in \cite{morrison-aip}. In Appendix \ref{a:jacobi} we give an elementary proof of the Jacobi identity for three linear functionals of $\rho$ and $\bfv$. It involves a remarkable integral identity. In \ref{s:Jacobi-general-proof} we extend the proof to exponentials of linear functionals and use a functional Fourier transform to establish the identity for a much wider class of nonlinear functionals. The Jacobi identity should also follow by interpreting these PBs as among functions on the dual of a Lie algebra, see \cite{holm-kupershmidt}. Furthermore, one formally expects the Jacobi identity to hold if we regard these PB as the semi-classical limit of commutators in Landau's quantized superfluid model.


\subsection{Equations of motion from Hamiltonian and Poisson brackets}

We show in this section that the continuity and R-Euler equations 
	\beq
	\dd{\rho}{t} + \grad \cdot (\rho \bfv) = 0, \quad
	{\rm and} \quad
	\dd{\bfv}{t} + ({\bfv} \cdot \grad) \bfv = - \grad U'(\rho) - \la^2 \bfw \times \grad \times \bfw
	\eeq
follow from Hamilton's equations $\pdr \rho/\pdr t = \{ \rho, H \}$ and $\pdr \bfv/\pdr t = \{ \bfv, H \}$ for the swirl hamiltonian
	\beq
	H = \int \left[ \frac{\rho \bfv^2}{2}+ U(\rho) + \frac{\la^2 \rho \bfw^2}{2} \right] \: d\bfr.
	\eeq
We call the $3$ terms kinetic (KE), potential (PE) and enstrophic (EE) energies. By the constitutive relation $\la^2 \rho$ is a constant. Here $U'(\rho) = h(\rho)$, e.g., for adiabatic flow $U(\rho) = p/(\gamma - 1)$ so that $U'(\rho) = h(\rho) = \gamma/(\gamma - 1) (p/\rho)$ and $\grad U'(\rho) = \grad h = \grad p/\rho$. For the continuity equation, we note that only KE contributes to $\{ H, \rho \}$ since $\{ \rho, \rho \} = \{ \bfw , \rho \} = 0$:
	\beqs \nonumber
	\{ H, \rho(\bfy) \} 
	&=& - \int_V \rho(\bfx) v_i(\bfx) \pdr_{\bfx^i} \del(\bfx-\bfy) \: d\bfx
	= \int_V \pdr_i [\rho(\bfx) v_i(\bfx)] \: \del(\bfx-\bfy) \: d\bfx \cr
	&& - \int_{\pdr V} \rho(\bfx) v_i(\bfx) n_i \, \del(\bfx-\bfy) dS
	= \grad \cdot (\rho \bfv).
	\eeqs 
The boundary term vanishes as $y$ is in the interior and $x$ on the boundary ($\bfv \cdot \hat n = 0$ also ensures this). To get the R-Euler equation, we evaluate $\{ H, \bfv \}$. The individual PBs are
	\beqs
	&& \{ KE, v_i \} = (\bfv \cdot \grad) v_i - \int_{\pdr V} \bfv^2 n_i \del(\bfx-\bfy) dS,
	\{ PE, v_i \} = \pdr_{i} U'(\rho) - \int_{\pdr V} U'(\rho) n_i \del(\bfx-\bfy) dS \cr
	&& {\rm and} \quad \{ EE, v_i \} = \la^2 (\bfw \times (\grad \times \bfw))_i - \int_{\pdr V} \la^2 ((\bfw \times \hat n) \times \bfw)_i \del(\bfx-\bfy) \: dS.
	\eeqs
The boundary terms vanish as before. The equation of motion for $\bfv$ then follows:
	\beq
	\{\bfv, H\} = \frac{\pdr \bfv}{\pdr t} = - (\bfv \cdot \grad) \bfv - \grad U'(\rho) -\la^2 \bfw \times (\grad \times \bfw).
	\eeq
For this to agree with the Euler equation $U'(\rho)$ must be chosen to be the enthalpy $h(\rho)$.

\subsection{Poisson brackets among locally conserved quantities and symmetry generators}
\label{s:PB-cons-qty}

We work out the PBs among locally conserved quantities of regularized compressible flow. As one might expect, linear and angular momenta and helicity Poisson commute with the {\it swirl} hamiltonian
	\beq
	\{ P_i, H \} = \{ L_i, H \} = \{ {\cal K}, H \} = 0.
	\eeq
BC are important: we would not expect linear or angular momenta to be conserved in a finite container that breaks translation or rotation invariance. Decaying BC ($\bfv \to 0, \rho \to$ constant) in an infinite domain would guarantee the above PB. More generally, we show below that the above PB may be expressed in terms of the conserved (regularized) currents of momentum, angular momentum and helicity. So these PB vanish provided the corresponding currents have zero flux across the boundary. 
	
$\{ P_i, H \}$ can be expressed as the divergence of the momentum current $\Pi_{ij}$ using $\rho \grad U' = \grad p$ and the constitutive relation:
	\beq
	\{ P_i, H \} = - \int_V \left( \pdr_i p +\pdr_j(\rho v_i v_j) + \la^2 \rho \left(\ov{2} \pdr_i \bfw^2 - \pdr_j (w_j w_i) \right) \right) d\bfr = - \int_{\pdr V} \Pi_{ij} n_j dS.
	\eeq
This vanishes if the momentum current (\ref{e:momentum-current-tensor}) has zero flux across the boundary. Similarly, $\{L_i , H \}$ can be expressed as a boundary term after dropping some terms using antisymmetry of $\eps$:
	\beq
	\{L_i , H \} = \eps_{ijk} \int_{\pdr V} x_j \left[ -\rho v_l v_k - p \delta_{kl} - \la^2 \rho \left( \half \bfw^2 \del_{lk} - w_l w_k \right) \right] n_l \: dS = - \int_{\pdr V} \Lambda_{il} n_l \: dS.
	\eeq
This vanishes if the regularized angular momentum current (\ref{e:ang-mom-current-conservation}) has zero flux across the boundary. The PB of the $H$ with flow helicity can be expressed in terms of the regularized helicity current. Let us first consider the unregularized $H$, for which $ \{KE + PE,{\cal K}\}$ gives
	 \beqs \nonumber
	 && \int_V \left[ - v_j(\bfx) w_i(\bfy) \om_{ij}(\bfx) \del(\bfx-\bfy) + \rho(\bfx) v_j(\bfx) v_i(\bfy) \eps_{ilk} \pdr_{\bfy^l} \left( \frac{\om_{jk}(\bfy)}{\rho(\bfy)} \del(\bfx-\bfy) \right) \right] \: d\bfx \: d\bfy 
	\cr && - \int_V \left( \half v^2(\bfx) + U'(\rho(\bfx)) \right) \left[w_i(\bfy) \pdr_{\bfx_i} \del(\bfx-\bfy)\right ]\: d\bfx \: d\bfy 
	= \int_{\pdr V} [\bfv \times (\bfv \times \bfw) + \sigma \bfw ] \cdot \hat n \: dS.
	\eeqs
$\bfv \times (\bfv \times \bfw) + \sigma \bfw$ is the unregularized ($\la \to 0$) helicity current. Using (\ref{e:constitutive-relation}) and repeated integration by parts we get
	\beqs
	\{ EE, {\cal K} \} &=& \iint_V \left\{ \half \la^2 \rho \bfw^2 , \bfv \cdot \bfw \right\} \, d\bfx \, d\bfy \cr 	&=& -\la^2 \rho \iint_V \left( w_i(\bfx) v_j(y) \eps_{ikl} \eps_{jmn} \pdr_{\bfx^k} \pdr_{\bfy^m} \left( \frac{\om_{nl}(\bfx)}{\rho(\bfx)} \del(\bfx-\bfy) \right) \right) \, d\bfx \, d\bfy
	\cr
	 &=& \int_{\pdr V} \la^2 (\bfT \times \bfv) \cdot \hat n \, dS - \int_{\pdr V} \int_{\pdr V} \la^2 \bfw \cdot ((\bfw \times \hat n) \times (\bfv \times \hat n)) \, dS \, dS.
	\eeqs
We conclude that $\{ H, {\cal K} \} = \int_{\pdr V} \bfj_{\cal K} \cdot \hat n - \int_{\pdr V} \int_{\pdr V} \la^2 \bfw \cdot ((\bfw \times \hat n) \times (\bfv \times \hat n)) \, dS \, dS$ where $\bfj_{\cal K}$ is the conserved helicity current (\ref{e:helicity-current-conservation}). So if we use decaying or $\bfw \cdot \hat n = 0$ and $\bfv \times \hat n = 0$ BCs, then $\bfj_{\cal K}$ has zero flux across $\pdr V$ and the double boundary term also vanishes ensuring $\{ H, {\cal K} \} = 0$. Helicity also commutes with $\bfP$ and $\bfL$ with decaying or $\bfw \cdot \hat n = 0$ and $\bfv \times \hat n = 0$ BCs
	\beq
	\{ \bfP, {\cal K} \} = \int_{\pdr V} \left[ (\bfv \times \hat n) \times \bfw + (\bfw \cdot \hat n) \bfv \right] \, dS, \quad
	\{ \bfL, {\cal K} \} = \int_{\pdr V} \bfr \times \left[ (\bfv \times \hat n) \times \bfw + (\bfw \cdot \hat n) \bfv \right] \, dS.
	\eeq
Indeed it is known that helicity is a Casimir invariant of the Poisson algebra with decaying or $\bfw \cdot \hat n = 0$ and $\bfv \times \hat n = 0$ BCs. Using $\del {\cal K}/\del \bfv = 2 \bfw$ (assuming $\bfv \times \hat n = 0$ on $\pdr V$), we have for any functional $F$ of $\rho$ and $\bfv$,
	\beq
	\{ {\cal K}, F[\rho, \bfv] \} = 2 \int_V \left[\frac{\bfw}{\rho} \cdot \left( \bfw \times \deldel{F}{\bfv} \right) - \bfw \cdot \grad F_{\rho} \right] d\bfx = -2 \int_{\pdr V} (\bfw \cdot \hat n) F_{\rho} dS = 0.
	\eeq
The PBs among $\bfP$ and $\bfL$ are
	\beqs
	\{ L_i , L_j \} &=& \eps_{ijk} L_k 
	+ \int_{\pdr V} \rho(\bfr) [(\bfr \times \bfv)_i (\bfr \times \hat n)_j - (i \leftrightarrow j)] \, dS \cr
	{\rm and} \quad 
	\{ P_i , L_j \} &=& \eps_{ijk} P_k +\int_{\pdr V} \rho(\bfr) \left[ (\bfr \times \hat n)_j v_i - (\bfr \times \bfv)_j n_i \right] \, dS.
	\eeqs
So with, say decaying BCs, both $\bfP$ and $\bfL$ transform as vectors under rotations generated by $\bfL$. Finally, the generator of Galilean boosts is ${\bf G} = \int (\rho(\bfr - t \bfv))\: d\bfr$. Unlike the densities of mass, momentum or energy, the Galilei charge density ${\cal G} = \rho(\bfr - t \bfv)$ depends explicitly on time. Despite this, $G$ is conserved  (with suitable BCs) even though it does not commute with the Hamiltonian:
	\beq
	\{G_i, H \} = \int \bfx_i \{ \rho(\bfx), H \} \:d\bfx - t \{P_i, H\}= \int \bfx_i \dot \rho \: d\bfx = - \int \bfx_i \pdr_j (\rho v_j) \: d\bfx = \del_{ij} \int \rho v_j = P_i.
	\eeq
We similarly check that $\bf G$ transforms as a vector under rotations $\{ G_i, L_j \} = \eps_{ijk} G_k$ and that $\{ {\bf G}, {\cal K} \} = 0$ and $\{ G_i, G_j \} = 0$. Finally, there is a central term in $\{ G_i, P_j \} = M \del_{ij}$ where $M$ is the total mass of fluid.

\subsection{Poisson brackets for incompressible flow}

PB for incompressible flow $(\grad \cdot \bfv = 0, \rho = \text{constant})$ are given in the literature (see \S 1.5 of \cite{marsden-ratiu}). Suppose $F[\bfv], G[\bfv]$ are two functionals of $\bfv$, then the `ideal fluid bracket' is
	\beq
	\{ F, G \} = - \ov{\rho} \int \bfv \cdot \left[ \deldel{F}{\bfv} , \deldel{G}{\bfv} \right] \: d\bfr.
	\eeq
The square brackets above denote the commutator of incompressible vector fields $[\bff , \bfg] = \bff \cdot \grad \bfg - \bfg \cdot \grad \bff$. These PBs follow from the compressible PBs when we impose the conditions
	\beq
	\grad \cdot \bfv = 0, \quad 
	\grad \cdot \deldel{F}{\bfv} = 0 = \grad \cdot \deldel{G}{\bfv} \quad \text{and} \quad \rho = \text{constant}.
	\label{conditions-incompress-PB}
	\eeq
We start with the compressible PB and impose (\ref{conditions-incompress-PB}) so that the quantity in the second parentheses below vanishes, giving 	\beqs
	\{F, G \} &=& \int \frac{\bfw}{\rho} \cdot \left[\deldel{F}{\bfv} \times \deldel{G}{\bfv} \right] d\bfr 
	= \int \frac{\eps_{ijk} \eps_{ilm} \pdr_l v_m}{\rho} \deldel{F}{v_j} \deldel{G}{v_k} d\bfr
	= \int \left[\frac{\pdr_j v_k - \pdr_k v_j}{\rho} \right] \deldel{F}{v_j} \deldel{G}{v_k}  d\bfr \cr
	&=& \ov{\rho} \int \left[ v_j \deldel{G}{v_k} \pdr_k \deldel{F}{v_j} - v_k \deldel{F}{v_j} \pdr_j \deldel{G}{v_k} \right] + \left[ v_j \deldel{F}{v_j} \pdr_k \deldel{G}{v_k} - v_k \deldel{G}{v_k} \pdr_j \deldel{F}{v_j} \right] d\bfr \cr
	&=& - \ov{\rho} \int \bfv \cdot \left[ F_\bfv , G_\bfv \right] d\bfr.
	\eeqs 
\subsubsection{Incompressible R-Euler from PB}

The incompressible R-Euler equation (\ref{e:R-Euler-compress}) follows from the above PB and Hamiltonian (with $\la$ and $\rho$ constant)
	\beqs
	H &=& \rho \int \left( \half \bfv^2 + \half \la^2 \bfw^2 \right) \: d\bfx \quad \imply \quad
	\rho \dd{v_i(y)}{t} = \rho \{ v_i(y) , H \} \cr
	&=& - \int v_k(x) \left[ \deldel{v_i(y)}{v_j(x)} \pdr_j \deldel{H}{v_k(x)} - \deldel{H}{v_j(x)} \pdr_j \deldel{v_i(y)}{v_k(x)} \right] d\bfx.
	\eeqs
Here, $\del{KE}/\del{\bfv} = \rho \bfv$
and $\del{EE}/\del{\bfv} = \la^2 \rho \grad \times \bfw$ are divergence free as required, but $\del{v_i(x)}/\del{v_j(x)} = \del_{ij} \del(x-y)$ is not. Hence we will need to take care to project the equation of motion resulting from these PBs to the incompressible subspace. We will do this after calculating the PBs.
	\beqs
	\rho \{ v_i(y) , KE \} &=& - \int v_k(x) \left[ \del_{ij} \del(x-y) \pdr_j (\rho v_k(x)) - \rho v_j(x) \pdr_j \left( \del_{ik} \del(x-y) \right) \right] d\bfx\cr
	&=& - \rho \left[ v_j \pdr_i v_j + v_j \pdr_j v_i \right], \cr
	\rho \{ v_i(y), EE \} &=& - \int v_k(x) \left[ \del_{ij} \del(x-y) \pdr_j (\la^2 \rho (\grad \times \bfw)_k(x)) - \la^2 \rho (\grad \times \bfw)_j(x) \pdr_j \left( \del_{ik} \del(x-y) \right) \right] d\bfx
	\cr
	&=& - \la^2 \left[ v_j \pdr_i (\grad \times \bfw)_j + \pdr_j \left( v_i (\grad \times \bfw)_j \right) \right] \cr
	&=& - \la^2 \left[ \pdr_i \left( \bfv \cdot (\grad \times \bfw) \right) - (\grad \times \bfw)_j \pdr_i v_j + (\grad \times \bfw)_j \pdr_j v_i \right] \cr
	&=& - \la^2 \left[ \bfT_i + \pdr_i (\bfv \cdot (\grad \times \bfw)) \right].
	\eeqs
Thus the momentum equation is
	\beqs
	&& \dd{\bfv}{t} + \mathbb{P} \left( \bfv \cdot \grad \bfv + \la^2 \bfT + \grad \left(\half \bfv^2 + \la^2 \bfv \cdot \grad \times \bfw \right) \right) = 0
	\quad \text{or} \cr
	&& \dd{\bfv}{t} + \mathbb{P} \left( \bfv \cdot \grad \bfv + \la^2 \bfT + \grad \left(\bfv \cdot \bfv_* -\half \bfv^2 \right) \right) = 0
	\eeqs
where $\mathbb{P}$ is the projection to the incompressible subspace, which we can define using the Helmholtz decomposition. Given a vector field $\bfv$ we may write it as the sum of curl-free and divergence-free parts $\bfv = - \grad \phi + \grad \times \bfA$ where $\phi = {(4\pi)}^{-1} \int \frac{\grad \cdot \bfv}{|\bfr - \bfs|} d \bfs $. Then, $\mathbb{P} (\bfv) = \bfv + \grad \phi = \grad \times \bfA$. In particular, the projection of a gradient vanishes. Thus $\mathbb{P}\left(\grad \left(\bfv \cdot \bfv_* - \half \bfv^2\right) \right)= 0$ while
	\beq
	\mathbb{P}( \bfv \cdot \grad \bfv + \la^2 \bfT) =  \bfv \cdot \grad \bfv + \la^2 \bfT +\ov{\rho} \grad p \quad \text{where} \quad \frac{p(\bfr)}{\rho} = \ov{4\pi} \int \frac{\grad_s \cdot (\bfv \cdot \grad \bfv(s) + \la^2 \bfT(s))}{|\bfr - \bfs|} \: d\bfs.
	\eeq
So after projecting to the incompressible subspace we get the incompressible R-Euler equation $\bfv_t + \bfv \cdot \grad \bfv = - \grad p/\rho - \la^2 \bfT$. Note that the above definition of pressure may be written as a Poisson equation for $p$ or $\sigma$
	\beq
	\grad^2 p = - \rho \grad \cdot (\bfv \cdot \grad \bfv + \la^2 \bfT) \quad \text{or} \quad
	\grad^2 \sigma = - \grad \cdot (\bfw \times \bfv + \la^2 \bfT) = - \grad \cdot (\bfw \times \bfv_* ).
	\eeq

\section{Poisson brackets for regularized MHD}
\label{s:PB-for-R-MHD}

Poisson brackets among functionals of velocity, density and magnetic field, for ideal compressible MHD were given by Morrison and Greene in \cite{morrison-greene}. The PB of functionals $F,G$ of $\rho, \bfv, \bfB$ is
	\beqs
	&& \{ F, G \} = \int \left[\frac{\bfw}{\rho} \cdot \left( F_{\bfv} \times G_{\bfv} \right) - F_{\bfv} \cdot \grad G_{\rho} + G_{\bfv} \cdot \grad F_{\rho}\right] d\bfr \cr
	&& - \int \left[\frac{\bfB}{\rho} \cdot \left[ \left( F_{\bfv} \cdot \grad \right) G_{\bfB} - \left( G_{\bfv} \cdot \grad \right) F_{\bfB} \right]
	+ \frac{B_i}{\rho} \left(\deldel{F}{v_j} \pdr_i \deldel{G}{B_j} - \deldel{G}{v_j} \pdr_i \deldel{F}{B_j}  \right)\right] d\bfr.
	\label{e:pb-mhd-functionals-rho-v-B}
	\eeqs
There are other forms related to the above formula via integration by parts using $\diver{\bfB} = 0$ and appropriate BCs.

From these we get the PBs between $\rho, \bfv$ and $\bfB$. As before (see Section \ref{s:pb-for-fluid}) for the fluid variables $\rho,\bfv$ and $\bfw$ we have
	\beq
	\{ \rho(x), \rho(y) \} = 0, \quad
	\{ v_i(x), v_j(y) \} = \frac{\eps_{ijk}w_k(x)}{\rho(x)} \del(x-y) , \quad \text{and} \quad
	\{ v_i(x), \rho(y) \} = - \pdr_{x^i} \del(x-y),
	\eeq
Like $\bfw$, $\bfB$ Poisson commutes with $\rho$, but unlike $\bfw$ its components commute. The PB of $\bfv$ with $\bfB$  is
	\beq
	\{ v_i(x), B_j(y) \} = \ov{\rho(x)} \left[ \del_{ij} B_k(x) \pdr_{x^k} - B_j(x) \pdr_{x^i} \right] \del(x-y)
	= \ov{\rho(x)} \eps_{ilk} \eps_{jmk} B_l(x) \pdr_{x^m} \del(x-y) .
	\label{e:pb-v-B}
	\eeq
Taking the curl of (\ref{e:pb-v-B}) we get the PB of vorticity with magnetic field: 
	\beqs
	\{ w_i(x), B_j(y) \} &=& \eps_{ilm} \pdr_{x^l} \left( \ov{\rho(x)} \left[ \del_{mj} B_k(x) \pdr_{x^k} - B_j(x) \pdr_{x^m} \right] \del(x-y) \right) \quad \text{or} \cr
	\{ B_i(x), w_j(y) \} &=& - \eps_{jlm} \pdr_{y^l} \left( \ov{\rho(y)} \left[ \del_{mi} B_k(y) \pdr_{y^k} - B_i(y) \pdr_{y^m} \right] \del(x-y) \right).
	\eeqs 
MHD PBs can also be written for functionals of $\rho, \bfM = \rho \bfv$ and $\bfB$. Denoting the commutator of vector fields in the usual way,
	\beqs
	&& \{ F , G \} = - \int \left[ \rho \left( F_{\bfM} \cdot \grad G_{\rho} - G_{\bfM} \cdot \grad F_{\rho} \right) 
	+ \bfM \cdot \left[ F_{\bfM}, G_{\bfM}\right]\right]d\bfr \cr
	&& -  \int \left[\bfB \cdot \left[ \left( F_{\bfM} \cdot \grad \right) G_{\bfB} - \left( G_{\bfM} \cdot \grad \right) F_{\bfB} + \grad \left( F_{\bfM} \right) \cdot G_{\bfB} - \grad \left( G_{\bfM} \right) \cdot F_{\bfB} \right]\right] d\bfr.
	\eeqs
We use the dyadic notation in the last term e.g. $\bfB \cdot \grad(\bfC) \cdot \bfD = B_i (\pdr_i C_j) D_j$. If $\bfA$ is the magnetic vector potential $\bfB = \grad \times \bfA$, then the PBs of functionals of $\rho, \bfM$ and $\bfA$ in the laboratory gauge (to be discussed below) is given by
	\beqs
	\{ F , G \} &=& - \int \left[ \rho \left( F_{\bfM} \cdot \grad G_{\rho} - G_{\bfM} \cdot \grad F_{\rho} \right) + \bfM \cdot \left[ F_{\bfM}, G_{\bfM} \right]\right] d\bfr \cr
	&& + \int \bfA \cdot \left[ F_{\bfM} \grad \cdot G_{\bfA} - G_{\bfM} \grad \cdot F_{\bfA} - \grad \times \left( F_{\bfM} \times G_{\bfA} - G_{\bfM} \times F_{\bfA} \right) \right] \, d\bfr.
	 \label{e:PB-RMHD-A}
	\eeqs
Thus the components of $\bfA$ commute with $\rho$ and among themselves while the PB with mass current and velocity are
	\beq
	\{ M_i(x) , A_j(y) \} = (F_{ij}(x) + A_i(x) \pdr_{y^j}) \del(x-y) \; \text{and} \;
	 \{ v_i(x), A_j(y) \} = \frac{(F_{ij}(x) + A_i(x) \pdr_{y^j}) \del(x-y)}{\rho(x)}.
	\label{e:v-A-pb}
	\eeq
Here $F_{ij} = \pdr_i A_j - \pdr_j A_i = \eps_{ijk} B_k$. We check that these PBs of $\bfA$ imply the above PBs of $\bfB$. Taking the curl of $\{ \bfv(x), \bfA(y) \}$ in $y$, the second term is a curl of a gradient and vanishes and we recover (\ref{e:pb-v-B}).
The curl of (\ref{e:v-A-pb}) gives the PB between vector potential and vorticity:
	\beq
	\{ A_i(x) , w_j(y)\} = \eps_{jkl} \pdr_{y^k} \left[ \ov{\rho(y)} \left( F_{li} (y) - A_l(y) \pdr_{y^i}\right)\del(x-y) \right].
	\eeq
For {\it incompressible} ($\grad \cdot \bfv = 0$ and constant $\rho$) R-MHD, the above PBs (\ref{e:PB-RMHD-A}) in laboratory gauge reduce to the following PBs
	\beqs
	\{ F[\bfv, \bfA] , G[\bfv, \bfA] \} &=& - \ov{\rho}\int \Bigg( \Bigg. \bfv \cdot \left[ F_{\bfv}, G_{\bfv} \right]
	 + \bfA \cdot \Bigg( \Bigg. F_{\bfv} \grad \cdot G_{\bfA} - G_{\bfv} \grad \cdot F_{\bfA} \cr
	  && - \grad \times \left( F_{\bfv} \times G_{\bfA} - G_{\bfv} \times F_{\bfA} \right)  \Bigg. \Bigg) \Bigg. \Bigg) \, d\bfr  \cr
	 &=& -\ov{\rho}\int  \left[\bfv \cdot \left[ F_{\bfv}, G_{\bfv} \right]
	 + \bfA \cdot \left([F_{\bfA}, G_{\bfv}] -[ G_{\bfA}, F_{\bfv}] \right) \right]\, d\bfr.
	\eeqs
As for incompressible neutral fluids, functional derivatives with respect to $\bfv$ are assumed solenoidal: $\grad \cdot F_{\bfv} = 0$ and $\grad \cdot G_{\bfv} = 0$.

\subsection{R-MHD equations of motion from Poisson brackets}

The Hamiltonian for R-MHD is the conserved swirl energy of R-Euler with the additional magnetic energy term:
	\beq
	H = \int \left[ \frac{\rho \bfv^2 }{2}+ U(\rho) + \frac{\la^2 \rho \bfw^2}{2} + \frac{\bfB^2}{2 \mu_0} \right] \: d\bfr.
	\eeq
Since $\rho$ commutes with $\bfB$, $\{ H, \rho \}$ is the same in R-MHD as in R-Euler. So the continuity equation  $\pdr \rho/\pdr t = \{ \rho, H \} = - \grad \cdot (\rho \bfv)$ follows. On the other hand, the introduction of the magnetic field alters the evolution equation for $\bfv$. We show that our PB give the correct evolution equations for $\bfv$ and $\bfB$ in regularized compressible MHD.

\subsubsection{Evolution of $\bfA$ and $\bfB$ from Poisson brackets} 

Here we derive the evolution equation for $\bfA$ using PB :$\pdr \bfA/\pdr t = \{\bfA,H\}$. Let us evaluate $\{ \bfA ,KE + EE\}$. $\{\bfA, PE \}= \{ \bfA ,ME\} = 0$ since both $\rho$ and $\bfB$ commute with $\bfA$.  
	\beqs
     \{A_j(y), H \} &=& \int \left[\rho(x)v_i(x)\{A_j(y) ,  v_i(x)\} + \la^2 \rho w_i(x)\{A_j(y),w_i(x) \} \right]dx \cr
     &=& \int \left[v_i(x)\left( A_i(x)\pdr_{x^j} - F_{ij} \right) \del(x-y) \right] \, dx \cr
     && + \int \left[\la^2 \rho w_i(x)\eps_{ikl}\pdr_{x^k}\left(\rho(x)^{-1} \left( A_l(x) \pdr_{x^j} - F_{lj} (x) \right) \del(x-y) \right)\right] dx \cr
     &=& -\pdr_j(v_i A_i) - v_i F_{ij}+\la^2\eps_{ikl}\left[(\pdr_k w_i)F_{lj} + \rho \pdr_j \left((\pdr_k w_i)\left({A_l}/{\rho}\right) \right)\right] \cr
     &=& (\bfv \times \bfB)_j +\left(\la^2 (\grad \times \bfw)\times \bfB \right)_j -\pdr_j(\bfv \cdot\bfA) - \la^2 \rho \pdr_j \left(\grad \times \bfw \cdot \frac{\bfA}{\rho}\right)
     \eeqs
     \beq
     \imply \quad \bfA_t = \{\bfA, H\} = (\bfv_* \times \bfB) - \grad(\bfv_* \cdot \bfA)  \;\; \text{or} \;\;
     \left[ - \grad(\bfv_* \cdot \bfA) -  {\bfA}_{t} \right]  + (\bfv_* \times \bfB) = 0. 
     \label{e:pb-A-with-H}
     \eeq
In this calculation we omitted the boundary terms assuming suitable BCs (e.g. $\bfw \times \hat n = 0$ and $\bfA \times \hat n = 0$). We identify the electric field as $\bfE = - \pdr \bfA/\pdr t - \grad (\bfv_* \cdot \bfA)$. Thus in this `laboratory' gauge, the electrostatic potential $\phi = \bfv_* \cdot \bfA$. This would be the electrostatic potential in the lab frame for the case where the electrostatic potential is zero in a `plasma' frame moving at $\bfv_*$ (See eq. 24.39 of \cite{Fock}). In the lab frame, if $\bfv_* = 0$ at a point, then the electrostatic potential would be zero in this gauge at that point. This gauge is distinct from Coulomb gauge, indeed $\grad \cdot \bfA$ evolves according to
	\beq
	\pdr_t (\grad \cdot \bfA) = \grad \cdot (\bfv_* \times \bfB) - \grad^2 (\bfv_* \cdot \bfA).
	\eeq
Taking the curl of (\ref{e:pb-A-with-H}) we arrive at the regularized Faraday law governing evolution of $\bfB$
	\beq
	\pdr_t \bfB = \{ \bfB, H \}
	= \grad \times \left[ \bfv_* \times \bfB \right].
	\eeq
An ab initio calculation of $\{ \bfB, H \}$ from the PBs (\ref{e:pb-mhd-functionals-rho-v-B}) assuming the BCs $\bfv \cdot \hat n = 0$, $\bfB \cdot \hat n = 0$ and $\bfw \times \hat n= 0$ gives the same regularized Faraday's law .

\subsubsection{Evolution of velocity from Poisson brackets}

Here we show that $\pdr \bfv/\pdr t = \{ \bfv , H \}$ gives the R-Euler equation including the Lorentz force term
	\beq
	\dd{\bfv}{t} + ({\bf v} \cdot \grad) \bfv = - \grad U'(\rho) - \la^2 \bfw \times (\grad \times \bfw) + \frac{{\bf j} \times \bfB}{\rho}.
	\eeq
Recall that $H = KE + PE + EE + ME$ and the PB of $KE + PE + EE$ with velocity is the same as in R-Euler and gives rise to all but the Lorentz force term in the momentum equation. So it only remains to calculate the PB of ME with $\bfv$:
	\beqs
	\{ ME, v_i(x) \} &=& \ov{\mu_0} \int B_j(y) \{ B_j(y), v_i(x) \} dy \cr
	&=& \ov{\mu_0} \int B_j(y)\ov{\rho(x)} \left[ B_j(x) \pdr_{x^i} - \del_{ij} B_k(x) \pdr_{x^k} \right] \del(x-y) \: dy \cr
	&=& \frac{B_j}{\mu_0 \rho} \pdr_{x^i} \int B_j(y) \del(x-y) dy
	- \frac{B_k}{\mu_0 \rho} \pdr_{x^k} \int B_i(y) \del(x-y) dy \cr
	&=& - \ov{\rho \mu_0} (B_k \pdr_k B_i - B_k \pdr_i B_k) = - \ov{\rho} (\bfj \times \bfB)_i.
	\eeqs
Here $\mu_0 \bfj = \grad \times \bfB$. This gives the Lorentz force term in the momentum equation.

\subsubsection{$\grad \cdot {\bf B}$ commutes with the Hamiltonian $H$}

The Maxwell equation $\grad \cdot \bfB = 0$ is consistent with our PBs since we show below that $\grad \cdot \bfB$ commutes with $H$. So if $\grad \cdot \bfB$ is initially zero, it will remain zero under hamiltonian time evolution. Now potential energy $\int U(\rho) dx$ commutes with $\grad \cdot \bfB$ since $\{ \rho, \bfB \} =0$. Magnetic energy $\int {\bfB^2}/{2 \mu_0}$ also commutes with $\grad \cdot \bfB$ since $\{B_i, B_j\} =0$. We will show now, that $\{ KE, \grad \cdot \bfB \}$ and $\{ EE, \grad \cdot \bfB \}$ vanish separately, so that the above assertion holds:
	\beqs \nonumber
	\{ KE, \grad \cdot \bfB\} &=&\pdr_{y^j} \int \rho(x) v_i(x) \{ v_i(x), B_j(y) \} \: dx \cr
	&=& \pdr_{y^j} \int v_i(x)\left[ \del_{ij} B_k(x) \pdr_{x^k} - B_j(x) \pdr_{x^i} \right] \del(x-y) \: dx  \cr
	&=&  \pdr_i (v_j B_i) - \pdr_j \pdr_i (v_i B_j) = 0, \cr
	\{ EE,  \grad \cdot \bfB\} &=& \pdr_{y^j} \int \la^2\rho w_i(x) \{ w_i(x), B_j(y) \} \: dx \cr
	&=& \pdr_{y^j} \int \la^2 \rho w_i(x) 
	\eps_{ilm} \pdr_{x^l} \left(\ov{\rho(x)} \left[ \del_{mj} B_k(x) \pdr_{x^k} - B_j(x) \pdr_{x^m} \right] \del(x-y) \right) \: dx \cr
	&=& \pdr_{y^j} \int (\la^2 (\grad \times \bfw)_m)(x)
	  \left[ \del_{mj} B_k(x) \pdr_{x^k} - B_j(x) \pdr_{x^m} \right]\del(x-y) dx  \cr
	  &=& \pdr_j\pdr_m (\la^2 (\grad \times \bfw)_j  B_k)- \pdr_j\pdr_k(\la^2 (\grad \times \bfw)_j B_k)=0.
	\eeqs

\subsection{Poisson algebra of conserved quantities in R-MHD}
\label{s:pb-cons-qty-MHD}

Linear momentum $\bfP = \int \rho \bfv d\bfr$ commutes with itself and the R-MHD Hamiltonian $H$. To show that $\bfP$ commutes with the $H$ we need only calculate $\{ P_i , ME \}$ since it was shown to commute with $KE, PE$ and $EE$ in R-Euler with appropriate BCs:
	\beqs
	\{ P_i ,ME \} &=& \ov{\mu_0}\iint_V \rho(x) B_j(y) \left\{v_i(x) ,B_j(y)\right\}\,dx\, dy \cr
	&=& \ov{\mu_0}\iint_V B_j(y)\left[\del_{ij} B_k(x) \pdr_{x^k} - B_j(x) \pdr_{x^i}\right] \del(x - y) \,dx\, dy \cr
	&=& \ov{\mu_0}\int_V  B_j \pdr_{i} B_j \,dy +  \ov{\mu_0}\int_{\pdr V} \left[B_i\left(\bfB \cdot \hat n \right) - \bfB^2 n_i\right]\, dS \cr
	&=& - \ov{\mu_0}\int_{\pdr V}  \left(\frac{\bfB^2}{2} \del_{ij} - B_i B_j  \right) n_j \,dS.
	\eeqs
Thus $\{ P_i, H \} = - \int_{\pdr V} \Pi_{ij} n_j dS$ where $\Pi_{ij}$ is the momentum current (\ref{e:mom-cons-r-mhd}). For periodic or decaying BC this flux is zero. Angular momentum $\bfL = \int \rho \bfr \times \bfv  \: d\bfr$ also commutes with $H$. Again we only compute $\{ L_i ,ME\}$: \small
	\beqs 
	\mu_0\{ L_i ,ME \} &=&  \iint_V \eps_{ijk} x_j\rho(x) B_l(y)\left\{v_k(x), B_l(y)\right\}\,dx\, dy \cr
	&=& \iint_V \eps_{ijk} x_j B_l(y) \left[\del_{kl} B_m(x) \pdr_{x^m} - B_l(x) \pdr_{x^k}\right] \del(x - y) \,dx\, dy \cr
	&=&  \int_V \left[(\bfB \times \bfB)_i + \eps_{ijj} B^2 + \eps_{ijk}y_j \pdr_k \frac{B^2}{2} \right] dy +\int_{\pdr V} \eps_{ijk}y_j  \left [B_k \bfB \cdot \hat n- B^2 n_k \right]\, dS \cr
	&=& \int_{\pdr V} \eps_{ijk}y_j n_m \left [B_k B_m - \frac{B^2}{2} \del_{mk} \right]\, dS.
	\eeqs \normalsize
Thus $\{ L_i , H \} = - \int_{\pdr V} \Lambda_{ij} n_j dS$ where $\Lambda_{ij}$ is the angular momentum current (\ref{e:ang-mom-cons-r-mhd}). So $\{ \bfL, H \} = 0$ if this flux vanishes (as for decaying BCs). The angular momentum algebra $\{ L_i, L_j \} = \eps_{ijk} L_k$ is unaffected by the addition of $ME$. Magnetic helicity ${\cal K}_B = \int \bfA \cdot \bfB \, d\bfr$ commutes with the swirl Hamiltonian\footnote{In MHD flow helicity does not commute with $H$ due to the Lorentz force in the momentum equation.}. In fact, it is a Casimir invariant of the Poisson algebra. Since $\bfA$ commutes with $\rho$ and itself and ${\cal K}_B$ is a functional of $\bfA$ alone, by (\ref{e:PB-RMHD-A}), the PB of ${\cal K}_B$ with any functional $F[\rho,\bfM,\bfA]$ is
	\beq
	\{ {\cal K}_B, F \} = \int_V \bfA \cdot \left[ F_{\bfM} \grad \cdot {\cal K}_{B,\:\bfA}  - \grad \times \left( F_{\bfM} \times {\cal K}_{B,\:\bfA} \right) \right]\, d\bfr.	
	\eeq
To proceed, we first show that ${\cal K}_{B,\:\bfA} \equiv \del {\cal K}_B/\del \bfA = 2 \bfB$ provided $\bfA$ is normal to the boundary: 
	\beqs
	\deldel{{\cal K}_B}{A_l(y)} &=& \deldel{}{A_l(y)} \int_V A_i(x) \eps_{ijk} \pdr_j A_k(x) \: d\bfx  \cr
	&=& \int_V \eps_{ijk} \left[ \del_{il} \del(x-y) \pdr_j A_k(x) + A_i(x) \pdr_j (\del_{kl} \del(x-y)) \right] d\bfx \cr
	&=& 2B_l +	\int_{\pdr V} (\bfA \times \hat n)_l\del(x-y) \: dS.
	\eeqs
Armed with this, the PB becomes
	\beqs
	\{ {\cal K}_B, F \} &=& 2 \int_V \bfA \cdot \left[ F_{\bfM} \grad \cdot \bfB -  \grad \times \left( F_{\bfM} \times \bfB \right) \right]\, d\bfr 
	= - 2 \int_V \bfA \cdot \grad \times \left( F_{\bfM} \times \bfB \right) \:d\bfr \cr
	&=& 2\int_V \grad \cdot (\bfA \times ( F_{\bfM} \times \bfB)) \:d\bfr = 2\int_{\pdr V} \bfA \times (\bfB \times F_{\bfM}) \cdot \hat n \: d\bfr \cr
	&=& 2 \int_{\pdr V} [\bfB (F_{\bfM} \cdot \bfA) - F_{\bfM} (\bfA \cdot \bfB )] \cdot \hat n dS.
	\eeqs
Thus ${\cal K}_B$ commutes with any observable $F$ provided $\bfB \cdot \hat n = 0$, $F_{\bfM} \cdot \hat n =0$ and $\bfA \times \hat n = 0$ on the boundary $\pdr V$ of the flow domain. Taking $F = H$ and using $H_M = \ov{\rho} H_\bfv = \bfv_*$ (assuming $\bfw \times \hat n = 0$) we have
	\beqs
	\{{\cal K}_B, H \} &=& 2\int_{\pdr V} [(\bfB \cdot \hat n)(H_{\bfM} \cdot \bfA) - (\bfA \cdot \bfB )(H_{\bfM} \cdot \hat n)]dS \cr
	&=& 2\int_{\pdr V} [(\bfB \cdot \hat n)(\bfv_* \cdot \bfA) - (\bfA \cdot \bfB )(\bfv_* \cdot \hat n)]dS = 0.
 	\eeqs
Thus magnetic helicity commutes with the Hamiltonian with decaying/periodic BCs or assuming $\bfB$ and $\bfv_*$ are tangential and $\bfw$ and $\bfA$ are normal to the boundary.

In addition to magnetic helicity, cross helicity $X = \int \bfv \cdot \bfB \:d\bfr$ is also a Casimir invariant. To see this, we compute its PB with an arbitrary functional $G$ (assuming decaying BCs for simplicity) using (\ref{e:pb-mhd-functionals-rho-v-B}) and the functional derivatives $X_\bfv = \bfB$ and $X_\bfB = \bfv$: \small
	\beqs
	\{ X, G \} &=&  \int \left[ \frac{\bfw}{\rho} \cdot \left( \bfB \times G_{\bfv} \right) - \bfB \cdot \grad G_{\rho} 
	- \frac{\bfB}{\rho} \cdot \left[ \left( \bfB \cdot \grad \right) G_{\bfB} - \left( G_{\bfv} \cdot \grad \right) \bfv \right]
 	 + \frac{B_i}{\rho} \left(B_j \pdr_i G_{B_j} - G_{v_j} \pdr_i v_j  \right) \right] d\bfr \cr
	 &=& \int \left[ \frac{\grad \times \bfv}{\rho} \cdot \left( \bfB \times G_{\bfv} \right) + (\grad \cdot \bfB )G_{\rho} 
	+ B_j \pdr_j \left( \frac{B_i}{\rho} \right) G_{B_i}  - \pdr_j \left( \frac{B_i}{\rho} G_{v_j}\right)v_i - B_i \pdr_i \left( \frac{B_j}{\rho} \right) G_{B_j} \right] d\bfr \cr
	&& + \int \pdr_i \left( \frac{B_i}{\rho} G_{v_j}\right)v_j d\bfr
	\cr
	&=& \int \left[ \bfv \cdot \left( \grad \times \left(\frac{\bfB}{\rho} \times G_{\bfv}\right)\right)  - v_i \pdr_j \left( \frac{B_i}{\rho} G_{v_j}\right) + v_j \pdr_i \left( \frac{B_i}{\rho} G_{v_j}\right) \right]d\bfr = 0.
	\eeqs
\normalsize

\section{Other constitutive laws bounding higher moments of $\bfw$}
\label{s:other-const-laws-and-regs}

An interesting application of our Hamiltonian and PB formulation is to the identification of other possible conservative regularizations that preserve the symmetries of the Euler equations. An interesting class of these arise by choosing new constitutive relations. Recall that the twirl regularization term $- \la^2 \bfw \times (\grad \times \bfw)$ was selected as it is the least nonlinear term of lowest spatial order that preserves the symmetries of the Euler equation. Moreover, with the constitutive relation $\la^2 \rho =$ constant, R-Euler admits a conserved swirl energy $E^*$ (\ref{e:swirl-energy-R-Euler}) which implies bounded enstrophy. R-Euler equations are Hamilton's equations for $E^*$ and the standard PBs (\ref{e:pb-between-functionals-of-rho-v}). Retaining the same Poisson brackets as before, and choosing an unaltered form for the Hamiltonian,
	\beq
	H = \int \left[\half \rho \bfv^2 + U(\rho) + \half \la^2 \rho \bfw^2 \right] \; d\bfr,
	\label{e:hamiltonian-new-constitutive-law}
	\eeq
we will now allow for more general constitutive relations, e.g., $\la_n^2 \rho = c_n \left(\bfw^2 \right)^n$ where $c_n$ is a positive constant. The virtue of this type of constitutive law is that the $(n+1)^{\rm th}$ moment of $\bfw^2$ is bounded in the flow generated by this conserved Hamiltonian\footnote{More generally $c_n$ could depend on $\rho$ without affecting the continuity equation but resulting in additional terms in the equation of motion which ensure boundedness of $\int c_n(\rho) (\bfw^2)^{n+1} \: d\bfr$.}. From Hamilton's equation for $\rho$ we see that the continuity equation is unaltered since $\rho$ commutes with itself and $\bfw$ (in fact as long as $\la$ depends only on $\rho$ and $\bfw$, the continuity equation will remain the same). However, there is a new regularization term in the equation for $\bfv$. Indeed, from (\ref{e:list-of-useful-PB}) one finds that
	\beqs
	&& \{ \bfv , EE_n \} = \left\{ \bfv , \int \half \la_n^2 \: \rho \bfw^2 \: d\bfy \right\} 
	= c_n (n+1) \int (\bfw^2(\bfy))^n  \{ \bfv, \bfw(\bfy) \} \cdot \bfw(\bfy) \, d\bfy \cr
	&=& - \frac{(n+1) c_n}{\rho} \left[ \bfw \times \left( \grad \times (\bfw^2)^n \bfw \right) \right].
	\eeqs
Thus the equation of motion becomes
	\beqs
	&& \dd{\bfv}{t} = \{ \bfv, H \} = - \bfv \cdot \grad \bfv - \ov{\rho} \grad p - \frac{(n+1) c_n}{\rho} \left[ \bfw \times \left( \grad \times (\bfw^2)^n \bfw \right) \right] = - \grad \sigma - \bfw \times \bfv_{n*}, \cr
	&& \qquad \text{where} \quad \bfv_{n*} = \bfv + \ov{\rho} \grad \times ((n+1)c_n |\bfw|^{2n} \bfw)
	\label{e:mom-eqn-new-const-law}
	\eeqs
is a new swirl velocity. Clearly, $\grad \cdot (\rho \bfv) = \grad \cdot (\rho \bfv_{n*})$ so the continuity equation may be written as $\rho_t + \grad \cdot (\rho \bfv_{n*}) = 0$. Thus the form of the governing equations is unchanged; only the swirl velocity $\bfv_*$ is modified to $\bfv_{n*}$. When $n = 0$, this reduces to the R-Euler equation for which the first moment of $\bfw^2$ (enstrophy) is bounded. For $n > 0$ we get new regularization terms which are more nonlinear (i.e, of degree $2n+2$ in $\bfv$) than the quadratic twirl term, though the equation {\it remains} $2^{\rm nd}$ order in space derivatives. Furthermore, $P_i, L_i$ continue to be conserved as the new constitutive relation does not break translation or rotation symmetries (it only depends on the scalar $\bfw^2$). Flow helicity is also conserved being a Casimir invariant of the Poisson algebra. Finally, parity, time reversal and Galilean boost invariance are also preserved.

For R-MHD, the Hamiltonian (\ref{e:hamiltonian-new-constitutive-law}) is augmented by the magnetic energy ME $\propto \int \bfB^2 \: d\bfr$. ME does not affect the continuity equation as $\{ \rho, \bfB \} = 0$ but adds the Lorentz force term to the momentum equation (\ref{e:mom-eqn-new-const-law})
	\beq
	\pdr_t \bfv = - \bfw \times \bfv_{n *} + \frac{\bfj \times \bfB}{\rho}.
	\eeq
The R-Faraday law (\ref{e:R-MHD-Faraday}) is modified by the new constitutive relation since $\bfB$ does not commute with vorticity.  Remarkably the R-Faraday equation takes the same form as (\ref{e:R-MHD-Faraday}) with $\bfv_* \mapsto \bfv_{n*}$: $\bfB_t = \grad \times (\bfv_{n*} \times \bfB)$. Indeed,
	\beqs
	\{B_i(\bfx), EE_n \} &=& \int c_n (n+1) (\bfw)^{2n} w_j(\bfy) \{B_i (\bfx) , w_j(y)\} d\bfy \cr
	&=& \int c_n (n+1) (\bfw)^{2n} w_j(\bfy)\eps_{jlm} \pdr_{y^l}\left( \ov{\rho}(B_i \pdr_m - \del_{mi}B_k\pdr_k)\right)\del(\bfx -\bfy)\,d\bfy \cr
	&=&\int \ov{\rho}c_n(n+1) \left(\grad \times (\bfw)^{2n} \bfw\right)_m \left(B_i \pdr_m - \del_{mi}B_k\pdr_k\right)\del(\bfx -\bfy)\,d\bfy \cr
	&=& c_n(n+1) \left(\left(\grad \times (\bfw)^{2n} \bfw\right) \cdot \grad \left(\frac{B_i}{\rho}\right)  - \bfB \cdot \grad \left(\frac{(\grad \times (\bfw)^{2n} \bfw)_i }{\rho}\right)\right) \cr
	&=& \grad \times \left( \ov{\rho} \grad \times ((n+1)c_n |\bfw|^{2n} \bfw) \times \bfB \right)
	\eeqs
where we have used a vector identity for $\grad \times (\bfC \times \bfD)$ taking $\bfC = \grad \times ((n+1)c_n |\bfw|^{2n} \bfw)$ and $\bfD = \bfB /\rho$. Thus $\pdr_t \bfB = \{ \bfB , H \} = \grad \times (\bfv_{n*} \times \bfB)$. It is remarkable that the PB formalism enables us to obtain, with the help of suitable constitutive relations, regularized flows with bounded higher moments of vorticity.

\subsection{Regularizations that bound higher moments of $\grad \times \bfw$}

We use the PB formalism to derive new regularized equations for which we have an a priori bound on the $L^2$ norm of the curl of vorticity (just as we had a bound on the $L^2$ norm of vorticity earlier). This is achieved by considering the Hamiltonian
	\beq
	H = \int \left[ \half \rho \bfv^2 + U(\rho) + \frac{\bfB^2}{2 \mu_0} + \half d_1 (\grad \times \bfw)^2 \right] \: d\bfr
	\label{e:hamiltonian-with-powers-of-curl-w}
	\eeq
where $d_1$ is a positive constant. By dimensional analysis, $d_1$ may be expressed in terms of a dynamical short-distance cut off $\la(\bfr,t)$ that satisfies the constitutive relation $\la^4 \rho = d_1$. The continuity equation $\rho_t = \{ \rho, H \}= - \grad \cdot (\rho \bfv) = 0$ is unchanged from that in ideal MHD since $\{ \rho, \bfw \} = 0$. The evolution equation for $\bfv$ is of fourth order in space derivatives of $\bfv$ and turns out to be expressible in the familiar form (\ref{e:R-MHD-Euler-v*}) where $\bfv_* = \bfv + \la^4 \grad \times (\grad \times (\grad \times \bfw))$ is a new swirl velocity field. To see this we compute $\{ \bfv, H \}$. It suffices to consider only the PB with new term in $H$ (\ref{e:hamiltonian-with-powers-of-curl-w}):
	\beqs
	\left\{ v_i(x) , \int \frac{d_1}{2} (\grad \times \bfw)^2 \: dy \right\}
	&=& d_1 \int \left(\grad \times (\grad \times \bfw) \right)_m \left( \del_{km} \pdr_{y^i} - \del_{im} \pdr_{y^k} \right) \frac{w_k(y)}{\rho(y)} \del(x-y) \: d\bfy \cr
	&=& - \frac{d_1 w_k(x)}{\rho(x)} \left[ \pdr_i \left(\grad \times (\grad \times \bfw) \right)_k - \pdr_k \left(\grad \times (\grad \times \bfw) \right)_i  \right] \cr
	&=& - \la^4 \left[ \bfw \times \left(\grad \times (\grad \times (\grad \times \bfw)) \right) \right]_i.
	\eeqs
Similarly, Faraday's law of ideal MHD gets modified, but takes the same form $\bfB_t = \grad \times (\bfv_* \times \bfB)$ as in R-MHD when expressed in terms of $\bfv_*$. To see this we compute the PB with the regularization term in $H$ (\ref{e:hamiltonian-with-powers-of-curl-w}):
	\beqs
	\left\{ B_i(x) , \int \frac{d_1}{2} (\grad \times \bfw)^2 \: dy \right\}
	&=& d_1 \int (\grad \times \bfw)_j \{ B_i(x) , (\grad \times \bfw)_j \} \: d\bfy \cr
	&=& d_1 \int (\grad \times \bfw)_j \eps_{jlm} \pdr_{y^l} \{ B_i(x), w_m(y) \} \: d\bfy \cr
	&=& d_1 \int (\grad \times (\grad \times \bfw))_m \eps_{mnp} \pdr_{y^n} \Bigg( \Bigg.  \ov{\rho(y)} B_i(y) \pdr_{y^p} \: dy \cr
	 && - \:\del_{ip} B_k(y) \pdr_{y^k}  \del(x-y) \Bigg. \Bigg)  \: d\bfy
	\cr
	&=& - d_1 \: \pdr_p \left[ \rho^{-1} \left(S_p B_i - S_i B_p \right) \right] \cr
	&=& d_1 \left( \frac{\bfB}{\rho} \cdot \grad S_i + S_i \grad \cdot \frac{\bfB}{\rho} - \bfS \cdot \grad \left(\frac{B_i}{\rho} \right) \right) \cr
	\imply \quad \left\{ \bfB , \int \frac{d_1}{2} (\grad \times \bfw)^2 \: dy \right\}
&=& \grad \times \left(\la^4 \bfS \times \bfB \right) = \grad \times \left( (\bfv_* - \bfv) \times \bfB \right).
	\eeqs
Here we defined ${\bf S} = \grad \times (\grad \times (\grad \times \bfw))$. Including the usual contribution from KE, we get the regularized Faraday law $\bfB_t = \grad \times (\bfv_* \times \bfB)$. The freezing-in and integral theorems automatically generalize to this case with the above swirl velocity $\bfv_*$.

We can generalize to a model where the $(2m)^{\rm th}$ moment of $\grad \times \bfw$ is bounded by considering the Hamiltonian
	\beq
	H = \int \left[ \half \rho \bfv^2 + U(\rho) + \frac{\bfB^2}{2 \mu_0} + \ov{2} d_m (\grad \times \bfw)^{2m} \right] \: d\bfr \equiv H_{MHD} + H_m.
	\label{e:hamiltonian-higher-power-of-curl-w}	
	\eeq
The constant $d_m$ must have dimensions of $(M/L^3) L^{2m+2} T^{2m-2}$. To express it in terms of the dynamical short distance cut-off $\la$ and density $\rho$ we introduce a reference {\it constant} speed $c$: $d_m = \la^{4m} \rho c^{2 - 2m}$. The regularized equations take the same form as above when expressed in terms of an appropriate swirl velocity 
	$\bfv_{m*} = \bfv + m \la^{4m} c^{2 - 2m} \grad \times \left(\grad \times \left( \left(\grad \times \bfw \right)^{2m-2} \grad \times \bfw \right) \right).$ The new term in $H$ does not change the continuity equation. By the constitutive relation $\la^{4m} \rho c^{2-2m} = d_m$, a constant,  $\grad \cdot (\rho \bfv_{m*}) = \grad \cdot (\rho \bfv)$ which means the continuity equation can also be expressed as $\rho_t = -\grad \cdot (\rho \bfv_{m*})$. To verify the regularized Euler and Faraday laws, it suffices to compute the PBs of $\bfv$ and $\bfB$ with the regularizing term $H_m$ in (\ref{e:hamiltonian-higher-power-of-curl-w}):
	\beqs
	\left\{ v_i(x) , H_m \right\}
	&=& m d_m \int (\grad \times \bfw)^{2m-2} (\grad \times \bfw)_j \{ v_i(x) , (\grad \times \bfw)_j \} \: d\bfy \cr
	&=& m d_m \int (\grad \times \bfw)^{2m-2} (\grad \times \bfw)_j \eps_{jlm} \pdr_{y^l} \{ v_i(x), w_m(y) \} \: d\bfy \cr
	&=& m d_m \int \left(\grad \times ((\grad \times \bfw)^{2m-2} \grad \times \bfw) \right)_m \left( \del_{km} \pdr_{y^i} - \del_{im} \pdr_{y^k} \right) \frac{w_k(y)}{\rho(y)} \del(x-y) \: d\bfy \cr
	&=& - m \la^{4m} c^{2-2m}  \left[ \bfw \times \left(\grad \times (\grad \times ((\grad \times \bfw)^{2m-2} \grad \times \bfw)) \right) \right]_i.
	\eeqs
Similarly, if we define ${\bf S} = \grad \times (\grad \times ((\grad \times \bfw)^{2m-2} \grad \times \bfw))$, then:
	\beqs
	\left\{ B_i(x) , H_m \right\}
	&=& m d_m \int (\grad \times \bfw)^{2m-2} (\grad \times \bfw)_j \{ B_i(x) , (\grad \times \bfw)_j \} \: d\bfy \cr
	&=& m d_m \int (\grad \times \bfw)^{2m-2} (\grad \times \bfw)_j \eps_{jlm} \pdr_{y^l} \{ B_i(x), w_m(y) \} \: d\bfy \cr
	&=& - m d_m \: \pdr_p \left[ \rho^{-1} \left(S_p B_i - S_i B_p \right) \right] \cr
	&=& m d_m \left( \frac{\bfB}{\rho} \cdot \grad S_i + S_i \grad \cdot \frac{\bfB}{\rho} - \bfS \cdot \grad \left(\frac{B_i}{\rho} \right) \right) \cr
	\imply \quad \{\bfB, H_m\} &=& \grad \times \left(m \la^{4m} c^{2-2m} \bfS \times \bfB \right) = \grad \times \left( (\bfv_{m*} - \bfv) \times \bfB \right).
	\eeqs
Thus use of the PBs enables us to identify new regularization terms in the momentum equation that ensure bounded higher moments of $\grad \times \bfw$, without altering the continuity equation. It is remarkable that the regularized momentum and Faraday equations involve a common swirl velocity field $\bfv_{m*}$ into which both $\bfw/\rho$ and $\bfB/\rho$ are frozen.

\section{Some solutions of regularized flow equations}
\label{s:examples}

\subsection{Compressible flow model for rotating vortex}
\label{s:modeling-vortex}

\begin{figure}[h]
\begin{center}
 \includegraphics[width = 5cm]{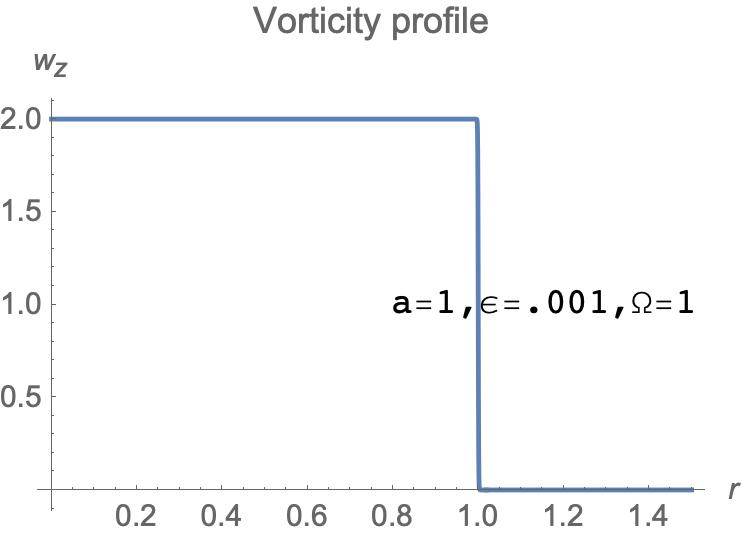}
 \hspace{0.5cm} 
 \includegraphics[width = 5cm]{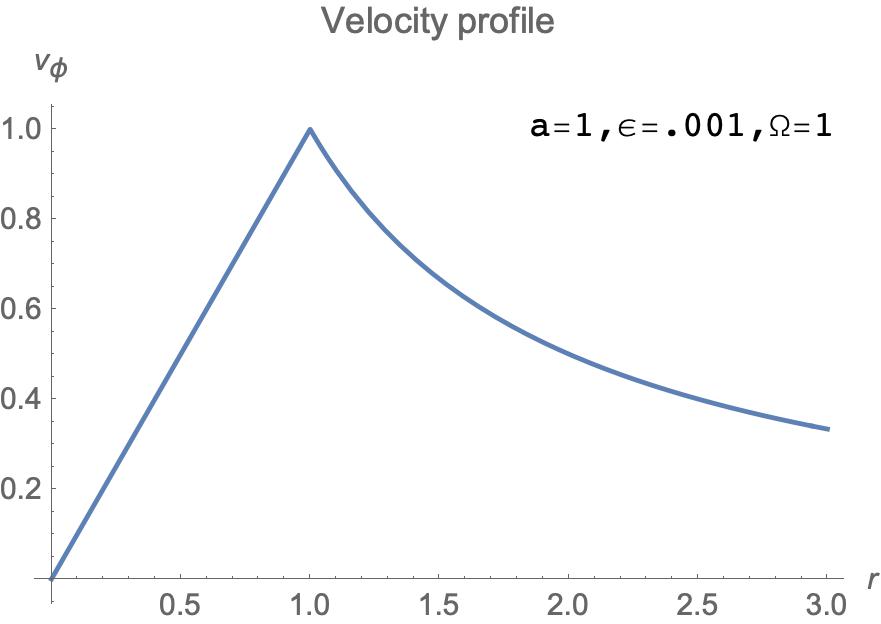}
 \hspace{0.5cm}
 \includegraphics[width = 5cm]{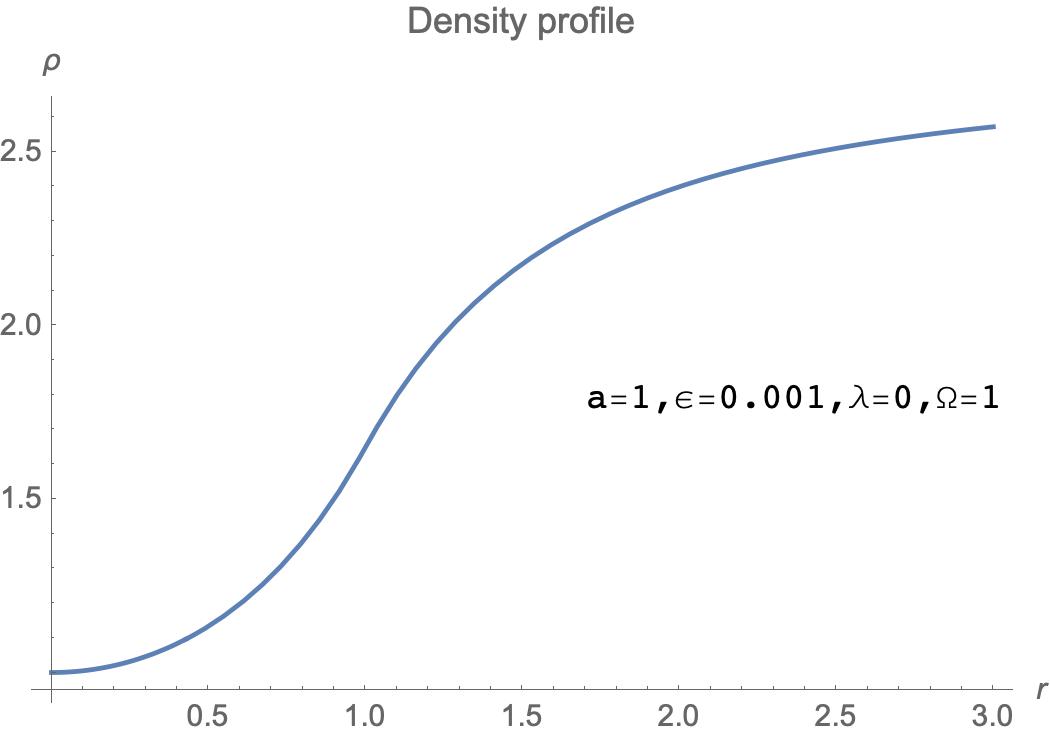}
 \caption{\footnotesize In the unregularized case, the steady Euler equations admit a solution where the vorticity profile is discontinuous while the density is not. There is no relation between the drop in vorticity and increase in density at the edge of the tornado (see Eq. (\ref{e:density-vorticity-balance-across-layer}) which says that when $\la = 0$, the change in density near the edge of the tornado is zero even when change in vorticity is nonzero). }
\label{f:tornado-figs-unreg}
 \end{center}
\end{figure}

In this section we model a {\em steady} tornado [cylindrically symmetric rotating columnar vortex with axis along $z$] using the compressible R-Euler equations. The unregularized Euler equations do not involve derivatives of vorticity, and admit solutions where the vorticity can be discontinuous or even divergent (e.g. at the edge of the tornado, see figure \ref{f:tornado-figs-unreg}). On the other hand, the R-Euler equations involve the first derivative of $\bfw$ and can be expected to smooth out large gradients in vorticity on a length scale of order $\la$ while ensuring bounded enstrophy.

Given appropriate initial profiles for $\bfv$ and $\rho$, the R-Euler equations should uniquely determine $\rho$ and $\bfv$ at later times. However, unlike the initial value problem, the {\em steady} R-Euler equations are under-determined (just like the steady Euler equations). As a consequence of this under-determinacy, the system may reach different steady states depending on the initial conditions. This is unlike dissipative systems (e.g. Navier-Stokes) which typically have a unique steady solution irrespective of initial conditions (except when there are bifurcations to multiple steady states allowed by the boundary conditions).

In our rotating vortex model, the density $\rho$ and pressure $p$ depend only on the distance from the central axis while $\bfv$ is purely azimuthal ($\bfv = v_\phi(r) \: \hat \phi$) and vorticity vertical $\bfw = w_z(r) \, \hat z$. In the steady state there is a single equation for the two unknowns $w_z$ and $\rho$, so we can determine the density profile given a suitable vorticity field. In the vortex core of radius $a$, we assume the fluid rotates at approximately constant angular velocity $\Omega$. Far from the core, $\bfw \to 0$. In a boundary layer of width $\ll a$, the $\bfw$ smoothly interpolates between its core and exterior values. As a consequence of the regularization term, we find that this decrease in vorticity is related to a corresponding increase in density (from a rare core to a denser periphery). By contrast, the unregularized Euler equations (i.e. $\la \to 0$) allow $\bfw$ to have unrestricted discontinuities across the layer while $\rho$ is continuous.

\subsubsection{Steady state regularized equations in cylindrical geometry}

Our infinitely long columnar vortex rotates about the $z$-axis and is assumed to be rotationally and translationally invariant about its axis. Hence $\bfv \cdot \hat z = 0$. We seek steady solutions of the R-Euler system. The continuity equation $\grad \cdot (\rho \bf v) = 0$ becomes $\pdr_x(\rho v_x) + \pdr_y(\rho v_y) = 0$. The incompressible 2D vector field $\rho \bfv$ can be expressed in terms of a scalar stream function $\rho \bfv = - \grad \times (\psi \hat z)$. Axisymmetry dictates that $\psi$ is a function of the cylindrical coordinate $r$ alone. It follows that $\bfv$ is purely azimuthal: $v_\phi = \psi'(r)/\rho$ (primes denote differentiation in $r$) and the continuity equation is identically satisfied. The steady state R-Euler equation is
	\beq
	\bfw \times \bfv = - \grad \sigma - \la^2 \bfw \times (\grad \times \bfw) \quad \text{where} \quad \sigma = h + \half  \bfv^2,
	\eeq
and $h$ is the specific enthalpy/Gibbs free energy for adiabatic/isothermal flow. Vorticity is vertical ($w_z = r^{-1} (r v_\phi)'$) while its curl is azimuthal $(\grad \times \bfw)_\phi = - w_z'(r)$. Thus the vorticity $\bfw \times \bfv$ and twirl accelerations both point radially:
	\beq
	(\bfw \times \bfv)_r = -w_z v_\phi \quad
	\text{and} \quad
	(\bfw \times (\grad \times \bfw))_r = w_z \dd{w_z}{r}.
	\eeq
Hence $\grad \sigma$ must also be radial and $h$ and $\rho$ functions of $r$ alone. Thus the steady R-Euler equations reduce to a single $1^{\rm st}$ order nonlinear ODE for $\rho(r)$ given $v_\phi(r)$ or $w_z(r)$. To solve it we need an equation of state relating $p$ to $\rho$.
	\beq
	w_z v_\phi = \dd{}{r} \left(h + \half v_\phi^2 \right) + \frac{\la^2}{2} \dd{w_z^2}{r} \quad \text{or} \quad
	\frac{v_\phi^2}{r} = \dd{h}{r} + \frac{\la^2}{2} \dd{w_z^2}{r}.
	\label{e:steady-state-eqn-for-vortex}
	\eeq

\subsubsection{Vortex model with rigidly rotating fluid core}

\begin{figure}[h]
\begin{center}
 \includegraphics[width = 5cm]{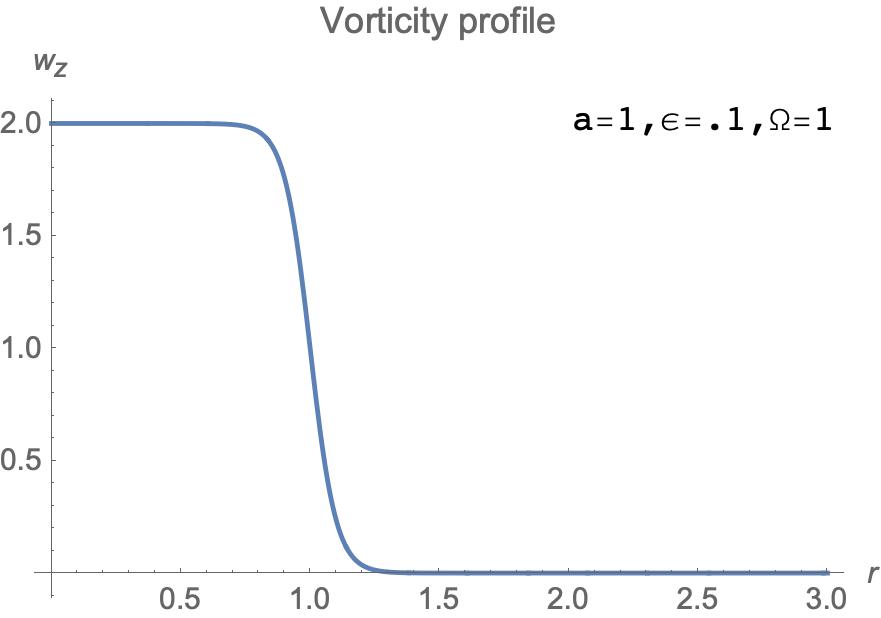}
 \hspace{.5cm} 
 \includegraphics[width = 5cm]{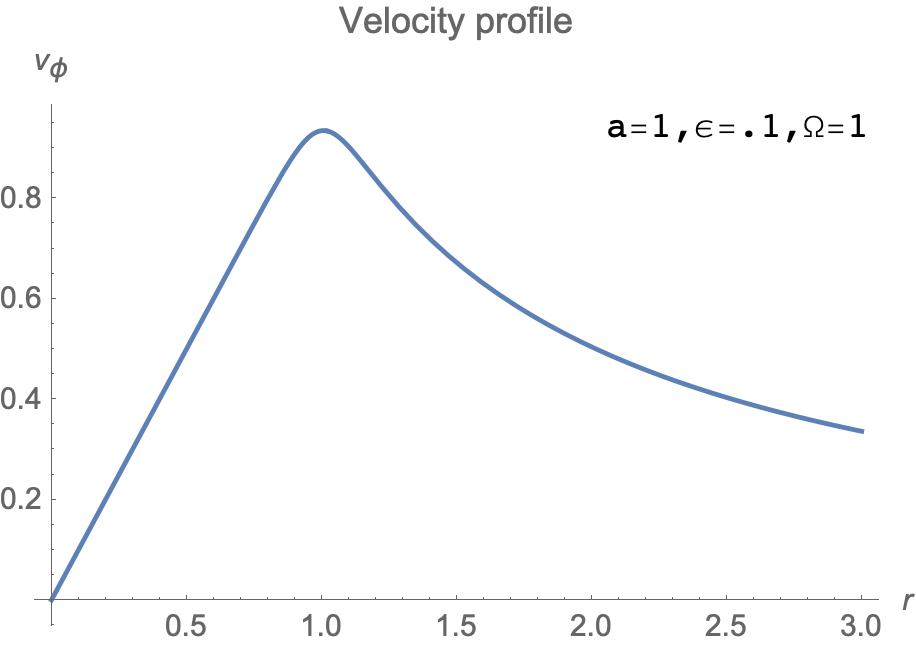}
 \hspace{.5cm} 
 \includegraphics[width = 5cm]{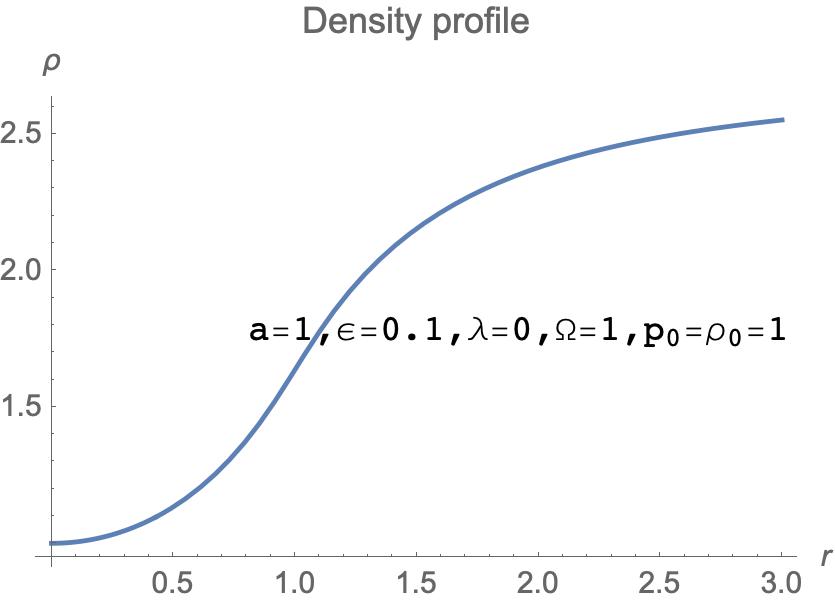}
 \hspace{.5cm}
 \includegraphics[width = 5cm]{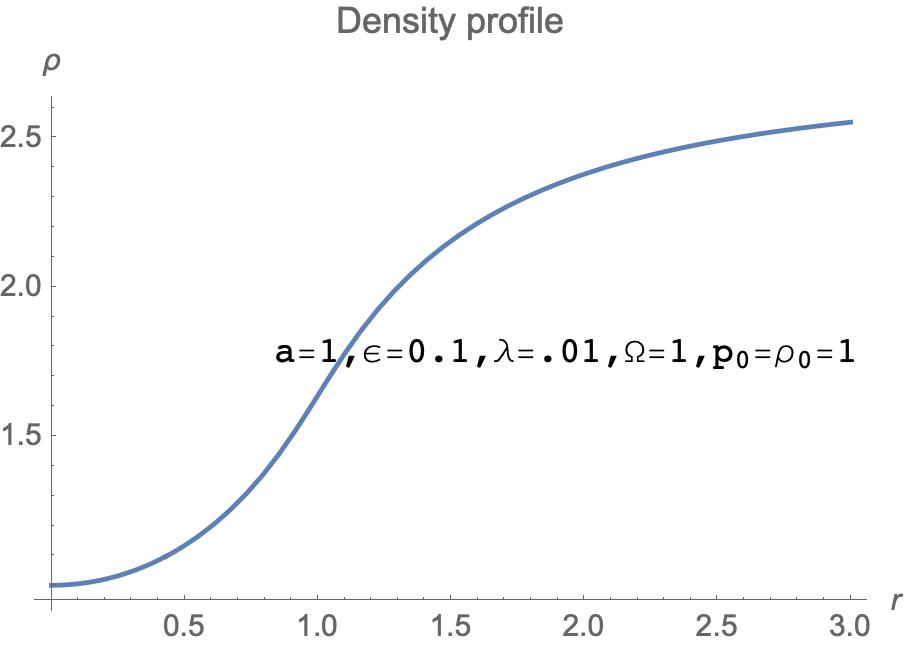}
 \hspace{.5cm}
 \includegraphics[width = 5cm]{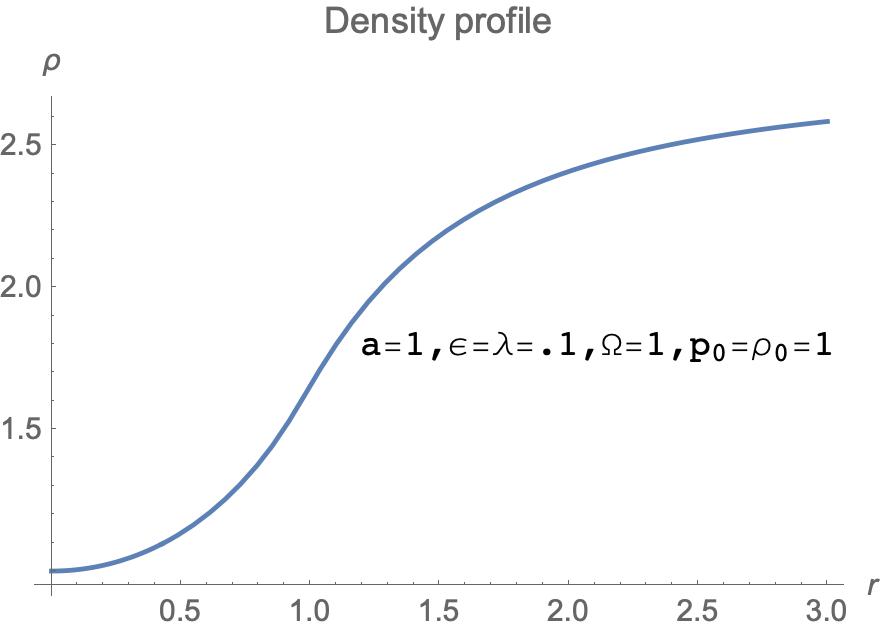}
 \hspace{.5cm}
 \includegraphics[width = 5cm]{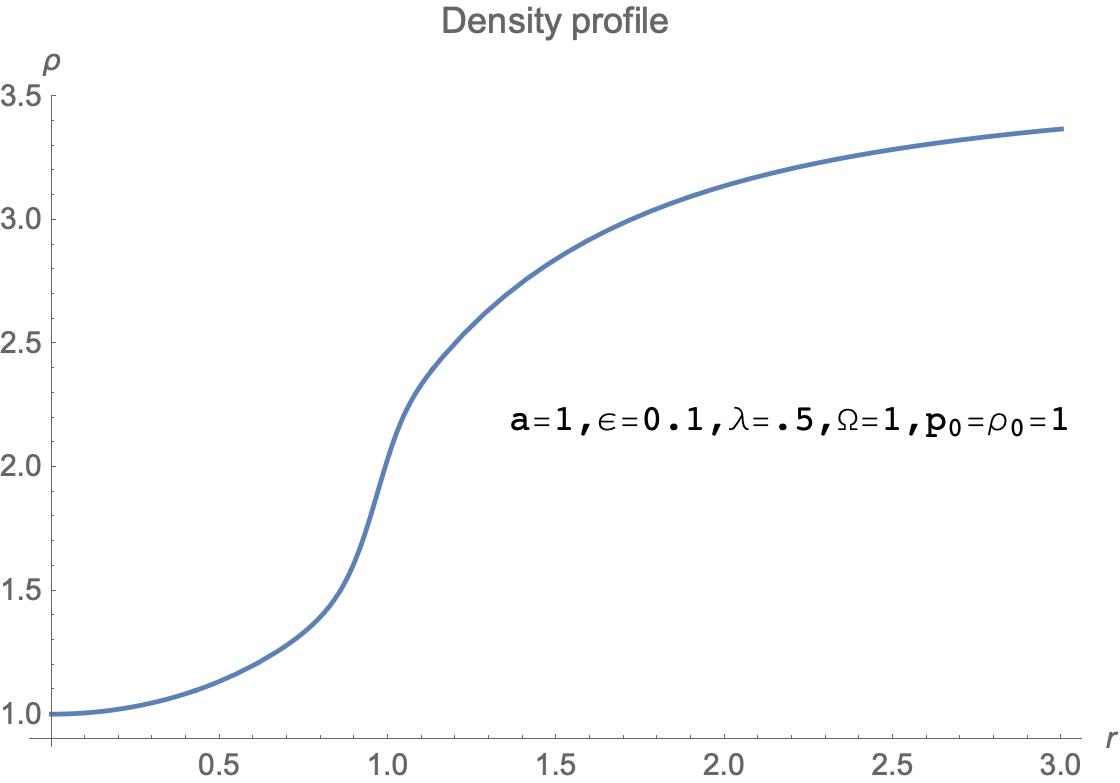} 
 \caption{\footnotesize Vorticity $w_z(r)$ and velocity $v_\phi(r)$ for rotating vortex of core radius $a$ and angular velocity $\Om$. $\rho(r)$ for isothermal flow increases outwards from core and reaches an asymptotic value. The regularization relates the drop in $w_z$ to an increase in $\rho$ in a layer of thickness $\eps$ around $r = a$. When $\eps > \lambda$, the regularization plays practically no role as seen in the third ($\la = 0$), fourth ($\la = 0.01$) and fifth ($\la = 0.1$) graphs for $\rho(r)$. This is understandable because when $\la < \eps$, the regularization scale is smaller than the length scale over which the vorticity profile varies significantly. On the other hand, when $\la > \eps$, we see (from the last graph with $\la = 0.5$ and $\eps = 0.1$) that the density increases much more rapidly over the length scale of variation of vorticity as we would expect from Eq. (\ref{e:density-vorticity-balance-across-layer}). If, however the density profile were held fixed, increasing $\lambda$ would reduce the gradient in vorticity. In all plots we have taken $a = \Om = p_0 = \rho_0 = 1$ and $\eps = 0.1$.}
\label{f:tornado-figs}
 \end{center}
\end{figure}

As a simple model for a rotating vortex of core radius $a$, we consider the vorticity distribution (see Fig. \ref{f:tornado-figs})
	\beq
	w_z(r) = 2 \Om \left[1 - \tanh \left(\frac{r-a}{\epsilon} \right) \right] \: \left[1+\tanh \left(a/\epsilon \right) \right]^{-1}.
	\label{e:vorticity-profile-tornado}
	\eeq
Over a transition layer of width $\approx 2 \eps \ll a$, the vorticity drops rapidly from $\approx 2 \Om$ to $\approx 0$. In the vortex core $r \ll a - \eps$, the flow corresponds to rigid body rotation at the constant angular velocity \\
$\Omega \hat z$, apart from higher order corrections in $\eps$. Thus in the core, the vorticity is roughly twice the \\ angular velocity and $\bfv = \Omega \hat z \times \bfr$ so that $v_\phi(r) =  \Om r $. In the exterior region, for $r \gg a + \epsilon$ the vorticity tends to zero exponentially. The velocity is obtained by integration subject to the BC $v_{\phi}(0)=0$. \scriptsize
	\beq
	v_\phi(r) = \frac{\Omega \eps^2
   \left[\text{Li}_2\left(-e^{\frac{2 (a-r)}{\eps}}\right)-\text{Li}_2\left(-e^{\frac{2 a}{\eps}}\right)\right] 
   + 2 \left[\eps (a-r) \log
   \left(e^{\frac{2(a-r)}{\eps}}+1\right) 
   + a  \left( r + \eps \log\frac{\cosh(a/\eps)}{\cosh((a-r)/\eps)}\right) 
   - a \eps \log\left(e^{\frac{2a}{\eps}} + 1 \right)\right]}
   {r \left(\tanh \frac{a}{\eps} + 1 \right)}.
	\label{e:vel-profile-tornado-exact}
	\eeq
\normalsize
The velocity profile (Fig.\ref{f:tornado-figs}) rises nearly linearly with $r/a$ in the core [rigid body motion] and drops off as $\sim 1/r$ at large distances like a typical, irrotational potential vortex. In the transition layer $a-\eps \lesssim r \lesssim a+\eps$ the radial derivative of the velocity varies rapidly.

The density can be obtained by integrating the steady R-Euler equation. We do this below in the simpler case of isothermal flow where the equation for $\rho$ is linear since $p \propto \rho$. The adiabatic case ($p/p_0 = (\rho/\rho_0)^\gamma$) is similar, but the steady state equation (\ref{e:steady-state-eqn-for-vortex}) is a nonlinear first order ODE for density: $\rho v_\phi^2/r = \gamma  \frac{p_0}{\rho_0^\gamma} \rho^{\gamma -1} \rho'(r) + \rho \la^2 (w_z^2)'/2$.

\subsubsection{Vortex with isothermal flow}

For isothermal flow the ideal gas equation of state $p V = n k_B T$ implies the pressure-density relation $p = (p_0/\rho_0) \rho$. The specific `Gibbs free energy' $h$ is obtained from the condition $\grad h = \ov{\rho} \grad p$,
	\beq
	\grad h = \frac{p_0}{\rho_0} \frac{\grad \rho}{\rho} = \frac{p_0}{\rho_0} \grad \log\frac{\rho}{\rho_0} \quad \imply \quad h = \frac{p_0}{\rho_0} \log \frac{\rho}{\rho_0}.
	\eeq
The flow is assumed purely `hydrodynamic': internal energy changes due to density variations are ignored; entropy and the internal energy equation do not play roles. The steady equation (\ref{e:steady-state-eqn-for-vortex}) in the isothermal case is
	\beq
	\frac{v_\phi^2}{r} = \frac{p_0}{\rho_0} \frac{\rho'(r)}{\rho(r)} + \frac{\la^2}{2} \dd{w_z^2}{r} \;\; \imply \;\;
	\frac{p_0}{\rho_0} \rho'(r) - \frac{v_\phi^2}{r} \rho(r) = - \frac{\la_0^2 \rho_0}{2} (w_z^2)'.
	\label{e:isothermal-steady-state-vortex-eqn}
	\eeq
This is a first order linear inhomogeneous ODE for $\rho$ with variable coefficients in the standard form\footnote{Putting $q'/q = B/A$ the equation becomes $(q \rho)' = qf/A$ whence $\rho = \ov{q} [\rho(0) q(0) + \int_0^r \frac{qf}{A} \, ds]$.}
	\beq
	A(r) \rho' + B(r) \rho = f(r), \quad \text{where} \quad
	A = \frac{p_0}{\rho_0}, \quad B = - \frac{v_\phi^2}{r} \quad \text{and} \quad f(r) = - \frac{\la_0^2 \rho_0}{2} (w_z^2)'.
	\eeq
It is convenient to take the reference values $\rho_0, \la_0, p_0$ to be at $r = 0$. The solution for $\rho(r)$ is
	\beqs
	\rho &=& \frac{\rho_0 q(0)}{q(r)} \left[ 1 + \int_0^r \frac{q(s) f(s)}{q(0) \rho(0) A} \, ds \right]
	= \frac{\rho_0 q(0)}{q(r)} \left[ 1 - \frac{\Om^2 \la_0^2 \rho_0}{2 p_0} \int_0^r \frac{q(s)}{q(0)} \left(\frac{w_z^2}{\Om^2} \right)' \, ds \right], \cr
	\text{where} \quad \frac{q(r)}{q(0)} &=& \exp \left[- \frac{\rho_0}{p_0} \int_0^r \frac{v_\phi^2}{s} \, ds \right].
	\label{e:soln-linear-inhom-1st-Order-ode}
	\eeqs
$q(r)$ is a positive monotonically decreasing function of $r$ and we can take $q(0)=1$ without loss of generality. The integrations are done numerically and the resulting density is plotted in Fig \ref{f:tornado-figs}. $\rho$ is monotonically increasing from $\rho(0)$ to an asymptotic value $\rho(\infty)$ (material has been `ejected' from the core). The above formula shows that one effect of the regularization is to increase the density relative to its Eulerian value (especially outside the core) if $w_z$ is a decreasing function of $r$ as seen in Fig. \ref{f:tornado-figs}. To get more insight into the role of the regularization we solve the steady equation approximately in the core, transition and exterior regions separately.


\pt {\bf Vortex Core $0 < r \lesssim a_- = a-\eps$}: In this region $w_z(r) \approx w_z(0) = 2\Omega$. The corresponding velocity $v_\phi(r) = r w_z(0)/2 = r\Omega$ grows linearly as for a rigidly rotating fluid. Since $\bfw$ is roughly constant, the regularization term may be ignored and (\ref{e:isothermal-steady-state-vortex-eqn}) becomes $\rho'(r) = \rho(r) \:(\rho_0/p_0) \Om^2 r$. The density grows exponentially inside the vortex core:
	\beq
	\rho(r) \approx \rho(0) \exp \left( \frac{\rho_0 \Om^2 r^2}{2 p_0} \right) 
	\approx \rho(a_-) \exp \left( \frac{\rho_0 \Om^2 (r^2 - a_-^2)}{2 p_0} \right) \quad \text{for} \;\; r \lesssim a_-.
	\eeq


\pt {\bf Outside the vortex $r \gtrsim a_+ = a+\eps$:} Here $w_z(r) \approx 0$ so the velocity decays as $v_{\phi}(r) = a_+ v_\phi(a_+)/r$. Again, ignoring the regularization term, the steady state density is determined by (\ref{e:isothermal-steady-state-vortex-eqn}):
	\beq
	\frac{\rho'(r)}{\rho(r)} = \frac{\rho_0 a_+^2 v_\phi(a_+)^2}{p_0} \ov{r^3}.
	\eeq
$\rho(r)$ monotonically increases from its value at the outer edge $\rho(a_+)$ to an asymptotic value $\rho(\infty)$ 
	\beq
	\rho(r)	= \rho(a_+) \exp \left( \frac{\rho_0 v_\phi(a_+)^2 \left(r^2 - a_+^2 \right)}{2 p_0 r^2} \right)
	= \rho(\infty) \exp \left(- \frac{\rho_0 v_\phi(a_+)^2 a_+^2}{2 p_0 r^2} \right) \quad \text{for} \;\; r \gtrsim a_+.
	\eeq
Even in this approximation, $\rho$ in the exterior depends on the regularization via $v_\phi(a_+)$.


\pt {\bf Transition layer $a_- \lesssim r \lesssim a_+$:} Here $w_z(r)$ (\ref{e:vorticity-profile-tornado}) rapidly falls from $w_z(0)$ to $0$. $v_\phi(r)$ is given by (\ref{e:vel-profile-tornado-exact}). $\rho$ is determined by 
	\beq
	\frac{\rho v_\phi^2}{r} = \frac{p_0}{\rho_0} \rho'(r) + \frac{\la^2 \rho}{2} \dd{w_z^2}{r}.
	\label{e:steady-R-Euler-for-rot-vortex}
	\eeq
To find the density we integrate this equation from $a_-$ to $r < a_+$ using the relation $\la^2 \rho = $ constant:
	\beq
	\int_{a_-}^r \frac{\rho v_\phi^2}{r} dr' = \frac{p_0}{\rho_0} \left[ \rho(r) - \rho(a_-) \right] + \frac{\la^2 \rho}{2} \left( w_z^2(r) - w_z^2(a_-) \right).	
	\eeq
Since the layer is thin ($\eps \ll a$) and $\rho$, $v_\phi$ are continuous across the layer, we may ignore the LHS. Thus the rapid decrease in $w_z$ must be compensated by a corresponding increase in $\rho$ across the layer
	\beq
	(p_0/\rho_0) \left[ \rho(r) - \rho(a_-) \right] \approx -  (\la^2 \rho/2) \left[ w_z^2(r) - w_z^2(a_-) \right].
	\label{e:density-vorticity-balance-across-layer}	
	\eeq
The increase in $\rho$ is not as rapid as the fall in $w_z$ since the latter is multiplied $\la^2$. For our vorticity profile (\ref{e:vorticity-profile-tornado}), taking $w_z(a_-) \approx w_z(0) = 2 \Om$, we get $\rho(r)$ in the transition layer
	\beq
	 \rho(r) \approx \rho(a_-) + \frac{2 (\Om \la_0)^2 \rho_0^2}{p_0} \left[1 - \frac{(1 - \tanh((r-a)/\eps))^2}{(1 + \tanh(a/\eps))^2} \right].	
	\eeq
In particular, $\rho(a_+)$ exceeds $\rho(a_-)$ by an amount determined by the regularization
	\beq
	\rho(a_+) \approx \rho(a_-) + \frac{2 (\Om \la_0)^2 \rho_0^2}{p_0} \left[1 - \frac{[1 - \tanh(1)]^2}{(1 + \tanh(a/\eps))^2} \right] \approx \rho(a_-) + 2 M^2  \rho_0 \quad \text{for} \quad \eps \ll a.
	\eeq
We see that for $\eps \ll a$ (vortex edge thin compared to core size), the twirl force causes an increase in density across the boundary layer by an amount controlled by the `twirl Mach number' $M = \la_0 \Om/c_s$ where $c_s = \sqrt{p_0/\rho_0}$ is the isothermal sound speed.

The steady R-Euler equation (\ref{e:steady-R-Euler-for-rot-vortex}) for the vortex is similar to Schr\"odinger's stationary equation for a non-relativistic quantum particle in a 1D delta potential: $E \psi(x) = - g \del(x) \psi(x) - (\hbar^2/2m)\: \psi''(x)$. $E \psi$ is like $\rho v_\phi^2/r$ on the LHS of (\ref{e:steady-R-Euler-for-rot-vortex}). The potential $- g \del(x) \psi(x)$ and kinetic $-(\hbar^2/2m) \psi''(x)$ terms mimic the pressure $(p_0/\rho_0) \rho'$ and twirl $\la^2 \rho \:(w_z^2)'/2$ terms respectively. The kinetic and twirl terms are both singular perturbations. The free particle regions $x < 0$ and $x > 0$ are like the interior and exterior of the vortex. The bound-state wave function is $\psi(x) = A \exp(- \kappa |x|)$ with $\kappa = \sqrt{-2mE}/\hbar$, so $\psi'$ has a jump discontinuity at $x=0$. The boundary layer is like the point $x=0$ where the delta potential is supported. Just as we integrated R-Euler across the transition layer, we integrate Schr\"odinger in a neighbourhood of $x=0$ to get $\psi'(\eps) - \psi'(-\eps) = - (2mg/\hbar^2)\: \psi(0)$. The discontinuity in $\psi'$ is determined by $\psi(0)$, just as the increase in $\rho$ across the layer is fixed by the corresponding drop in $w_z$ (\ref{e:density-vorticity-balance-across-layer}). Finally, $\la > 0$ regularizes Euler flow just as $\hbar > 0$  regularizes the classical theory, ensuring $E_{\rm gs} = - m g^2/2 \hbar^2$ is bounded below.

\clearpage
\subsection{A steady columnar vortex in conjunction with an MHD pinch}

A similar analysis in R-MHD involves specifying in addition to the above $w_z(r)$, a vertical (axial) current $j_z(r)$. The (solenoidal) azimuthal $B_\phi(r)$ associated with it is determined from $\mu_0 \bfj = \grad \times \bfB$, i.e. by integrating $\mu_0 j_z = {r}^{-1} (r B_\phi(r))'$. Assuming $r B_\phi(r)$ vanishes along the axis, $B_\phi(r) = {r}^{-1} \int_0^r \mu_0 s j_z(s) ds$. As in R-Euler above, the steady continuity equation $\grad \cdot (\rho(r) v_\phi(r)) \equiv 0$ is identically satisfied. The steady R-Faraday equation $\grad \times (\bfv_* \times \bfB) = 0$ is also identically satisfied since both $\bfv_* = (v_\phi - \la^2 w_z') \hat \phi$ and $\bfB$ are parallel. Thus the electric field is zero. In R-MHD, the steady momentum equation (\ref{e:R-MHD-Euler-v}) only has a non-trivial radial component. Under isothermal conditions ($p/p_0 =  \rho/\rho_0$) it becomes
	\beq
	\frac{p_0}{\rho_0} \rho' - \frac{v_\phi^2}{r} \rho = - \half \la_0^2 \rho_0 (w_z^2)' - \frac{B_\phi}{\mu_0 r} (r B_\phi)'.
	\label{e:mhd-pinch-vortex-ODE-for-rho}
	\eeq
In (\ref{e:mhd-pinch-vortex-ODE-for-rho}) the inhomogeneous term on the RHS is modified by the Lorentz force relative to (\ref{e:isothermal-steady-state-vortex-eqn}). The latter is always radially inwards (`pinching') whereas the twirl term is outwards for radially decreasing vorticity. Furthermore, the twirl term could be small for $\la_0 \ll a$. Thus the radial density variation in this magnetized columnar pinch could differ from R-Euler where there is no magnetic Lorentz force. For any given current and vorticity profiles (\ref{e:mhd-pinch-vortex-ODE-for-rho}) can be integrated to find $\rho(r)$ as we did in (\ref{e:soln-linear-inhom-1st-Order-ode}).

Another case of interest in R-MHD is a magnetized columnar vortex with an axial skin current. Thus we assume $j_z(r)$ is localized between $a - c/\om_{pe}$ and $a + c/\om_{pe}$ where $c/\om_{pe}$ is the electron collisionless skin depth and $\la \approx c/\om_{pe}$. In this case, in the interior $r < a_-$ we have the previous (tornado) interior solution with $B_\phi = 0$. In the exterior solution, $B_\phi(r) \approx \mu_0 I/2\pi r$ for $r \geq a_+$. The effect of the Lorentz force in the skin is seen from (\ref{e:mhd-pinch-vortex-ODE-for-rho}) to be opposite to that of the twirl term. The exclusion of the magnetic field within the vortex is reminiscent of the Meissner effect in superconductivity. Axial magnetic fields (screw pinch) and flows with the same symmetries (i.e., purely radial dependence) may be readily incorporated in the framework presented since the momentum equation remains purely radial and the continuity and R-Faraday laws are identically satisfied.

\subsection{Simple model for channel flow using regularized equations}
\label{s:channel-flow}

We consider flow along an infinitely long (in the $x$ direction) and infinitely wide (in the $z$ direction) channel. The channel extends from $y=0$ to a height of $y=a$. We seek a steady state solution of the regularized equations with velocity field $\bfv = (u(y),0,0)$ and density $\rho$ a function of $y$ alone. i.e., velocity and density vary with height but are translation invariant along the length and breadth of the channel. The steady state continuity equation $\grad \cdot (\rho \bfv) = 0$ is identically satisfied since $\pdr_x(\rho(y) u(y)) = 0$. For our velocity field the advection term in the momentum equation $\bfv \cdot \grad \bfv$ is identically zero and \footnote{Note that subscripts denote derivatives.}
	\beq
	\bfw = - u_y \hat z, \quad
	\bfw \times \bfv = - u u_y \hat y, \quad
	\grad \times \bfw = - u_{yy} \hat x
	\quad \text{and} \quad
	\bfT = \bfw \times (\grad \times \bfw) = u_y u_{yy} \hat y.
	\eeq
 So only the $\hat y$ component of the momentum equation survives:
	\beq
	\la^2 u_y u_{yy} = - \pdr_y{h(\rho(y))}.
	\label{e:reg-euler-channel-flow}
	\eeq
In other words, the steady state equations are underdetermined, we have a single second order nonlinear ODE for both $u(y)$ and $\rho(y)$. So given $u(y)$ and a suitable boundary value, say $\rho(0)$, we may determine the density profile. In particular, in the unregularized theory ($\la = 0$), the Euler equation  simply states that density must be a constant since $\pdr_y h(\rho) = 0$. As a consequence, the unregularized velocity $u$ can be an arbitrary function of $y$ (satisfying appropriate boundary conditions). So the regularization introduces a non-trivial dependence of $\rho(y)$ on $u(y)$.

\noindent {\em Remark on energy conservation:} For steady flow, local conservation of energy becomes $\grad \cdot \bff = 0$. $\grad \cdot \bff \equiv 0$ for channel flow since the energy current points along $\hat x$ but depends only on $y$:
	\beq
	\bff = \rho \sigma \bfv + \la^2 \rho ((\bfw \times \bfv) \times \bfw) + \la^4 \rho \bfT \times \bfw \;
	= \; \left[ \rho \left(h + \half u^2 \right) u + \la^2 \rho u u_y^2 - \la^4 \rho u_y^2 u_{yy} \right] \hat x.
	\eeq
Furthermore, the energy flux across the upper and lower walls of the channel vanish ($\bff \cdot \hat z = 0$). So energy is conserved even though our flow does {\em not} satisfy the BC $\bfw \times \hat n = 0$ that we obtained as a sufficient condition for energy conservation in Section \ref{s:cons-laws}.

\subsubsection{Isothermal channel flow}

For isothermal flow specific enthalpy is $h = (p_0/\rho_0) \log(\rho/\rho_0)$. Since $\la^2 \rho$ is a constant, the R-Euler equation (\ref{e:reg-euler-channel-flow}) becomes 
	\beq
	\frac{\la^2 \rho}{2} \dd{u_y^2}{y} = - (p_0/\rho_0) \rho_y \quad
	\text{or} \quad
	\pdr_y \left(\half \la^2 \rho u_y^2 + \frac{p_0 \rho}{\rho_0} \right) = 0.
	\label{e:R-euler-isotherm-channel-flow}
	\eeq
As $\bfw = - u_y \hat z$, this Bernoulli-like equation states constancy of the sum of enstrophic and compressional energy densities with height. The kinetic energy contribution is absent due to the assumption of a purely longitudinal velocity field that varies only with height: recall that the advection term $\bfv \cdot \grad \bfv$ is identically zero. As a consequence, this Bernoulli-like equation is very different in character from the usual one, which involves the kinetic energy of the flow and the compressional energy along streamlines. In that case, the pressure along a streamline is lower where the velocity is higher. In the present case, there is no variation of any quantity along streamlines, but only in the $y$-direction. We find that the density, and hence the pressure, is higher where the vorticity is higher! This is fundamentally a consequence of the regularizing ``twirl acceleration''.

An exact first integral of the above equation is $\half \la^2 \rho u_y^2 + (p_0/\rho_0) \rho = K$, where $K$ is an integration constant. We make use of the constitutive relation $\la^2 \rho=\la_{0}^{2}\rho_{0}$, where both $\la_0$ and $\rho_0$ are taken at the base of the channel $y=0$, and evaluate the equation there to obtain $K = \ov{2}\la_{0}^2 \rho_{0} u_y^2(0) + p_0$. For convenience, we use the reference values $p_0 = p(0)$ and $\rho_0 = \rho(0)$ to be the pressure and density at $y=0$. For instance, we consider the example of a parabolic velocity profile:
	\beq
	u(y) = 4u_{\rm max}\left[\frac{y}{a} \left(1-\frac{y}{a} \right)\right], 
	\label{e:parabolic-vel-profile}
	\eeq
where $u_{\rm max}=u(a/2)$ is the flow velocity midway up the channel, and $u(0)=0$. It follows that 
	\beq
	\dd{u}{y} = u_{y} = 4\frac{u_{\rm max}}{a}\left[1-2 \left(\frac{y}{a} \right)\right] \quad
	\text{and so} \quad u_{y}(0)=4 \frac{u_{\rm max}}{a}.
	\eeq
Thus the Bernoulli constant $K = 8\rho_0 u_{\rm max}^{2} (\la_0/a)^2 +p_{0}$. Substitution in the Bernoulli integral leads to the density profile:
      \beq
        \frac{ \rho}{\rho_0} = 1 + 32 \left(\frac{\rho_0 u_{\rm max}^{2}}{p_{0}}\right) \left(\frac{\la_0}{a} \right)^2 \left(\frac{y}{a}\right) \left(1-\frac{y}{a} \right) = \frac{p}{p_0}.
      \label{e:density-profile-isothermal-channel}
      \eeq
The resulting density profile is also parabolic. The density increases from $\rho(0)=\rho_0$ at the bottom of the channel to a maximum value half way up the channel and decreases symmetrically back to $\rho_0$ at the top. Thus, we have,
	\beq
	\frac{\rho_{\rm max}}{\rho_{0}}= 1 + 8 \left(\frac{\rho_{0} u_{\rm max}^{2}}{p_{0}} \right) \left(\frac{\la_0}{a} \right)^{2}.
	\eeq
We note that in isothermal conditions, we may write, $c_{s}^{2} = p_{0}/\rho_{0}$, where $c_s$ is the isothermal sound-speed. Since the Mach number of the flow along the centre is, $M^{2}= u_{\rm max}^2/c_s^2$, we have the relation:
	\beq
	\frac{\rho_{\rm max}}{\rho_{0}}= 1 + 8 M^{2} \left(\frac{\la_0}{a} \right)^{2}.
	\eeq
$M$ can take any value in principle. The second factor, $(\la_0/a)^2$, is by assumption a very small number. For moderate Mach numbers, the density increase is rather small. The flow superficially resembles Poiseuille flow and satisfies the same boundary conditions, but is strictly non-dissipative. It should be noted that Poiseuille flow involves a constant pressure {\em gradient along the flow} driving the latter against viscosity, whereas in the present case, there is no variation of any quantity along the flow.

\subsubsection{Adiabatic channel flow}
\label{s:channel-adibatic}

For adiabatic channel flow $(p/p_0) = (\rho/\rho_0)^\gamma$ and $h = \frac{\gamma}{\gamma-1} \frac{p}{\rho}$. We employ the same parabolic velocity profile as in the isothermal case. Since $\la^2 \rho$ is a constant, the R-Euler equation (\ref{e:reg-euler-channel-flow}) becomes a ``twirl force'' Bernoulli's equation:
	\beq
	\frac{\la^2 \rho}{2} \dd{u_y^2}{y} = - \rho \dd{h}{y} \quad
	\text{or} \quad
	\pdr_y \left(\half \la^2 \rho u_y^2 + p \right) = 0.
	\eeq
As before we obtain the exact first integral $\half \la^2 \rho u_y^2 + p = K$. Making use of the constitutive relation we evaluate the Bernoulli constant at $y =0$ by choosing $p_0 = p(0)$ and $\rho_{0}=\rho(0)$:
	\beq
	K = \half \la_{0}^2 \rho_{0} u_y(0)^2 +  p_{0}.
	\eeq	
Substitution in the exact integral to eliminate $K$, we obtain the pressure (and density) distributions
	\beq
	\frac{p}{p_0} = \left( \frac{\rho}{\rho_0} \right)^\gamma = 1 + 32 \left(\frac{\rho_{0} u_{\rm max}^2}{p_0} \right) \left(\frac{\la_0}{a} \right)^2 \left( \frac{y}{a} \right) \left(1 - \frac{y}{a} \right).
	\eeq
For adiabatic flow, $p/p_0$ varies with height in exactly the same way as $p/p_0 = \rho/\rho_0$ in the isothermal case (\ref{e:density-profile-isothermal-channel}) (though not if one writes things in terms of $M$)!

\subsection{Isothermal plane vortex sheet}
\begin{figure}[h]
\begin{center}
 \includegraphics[width = 8cm]{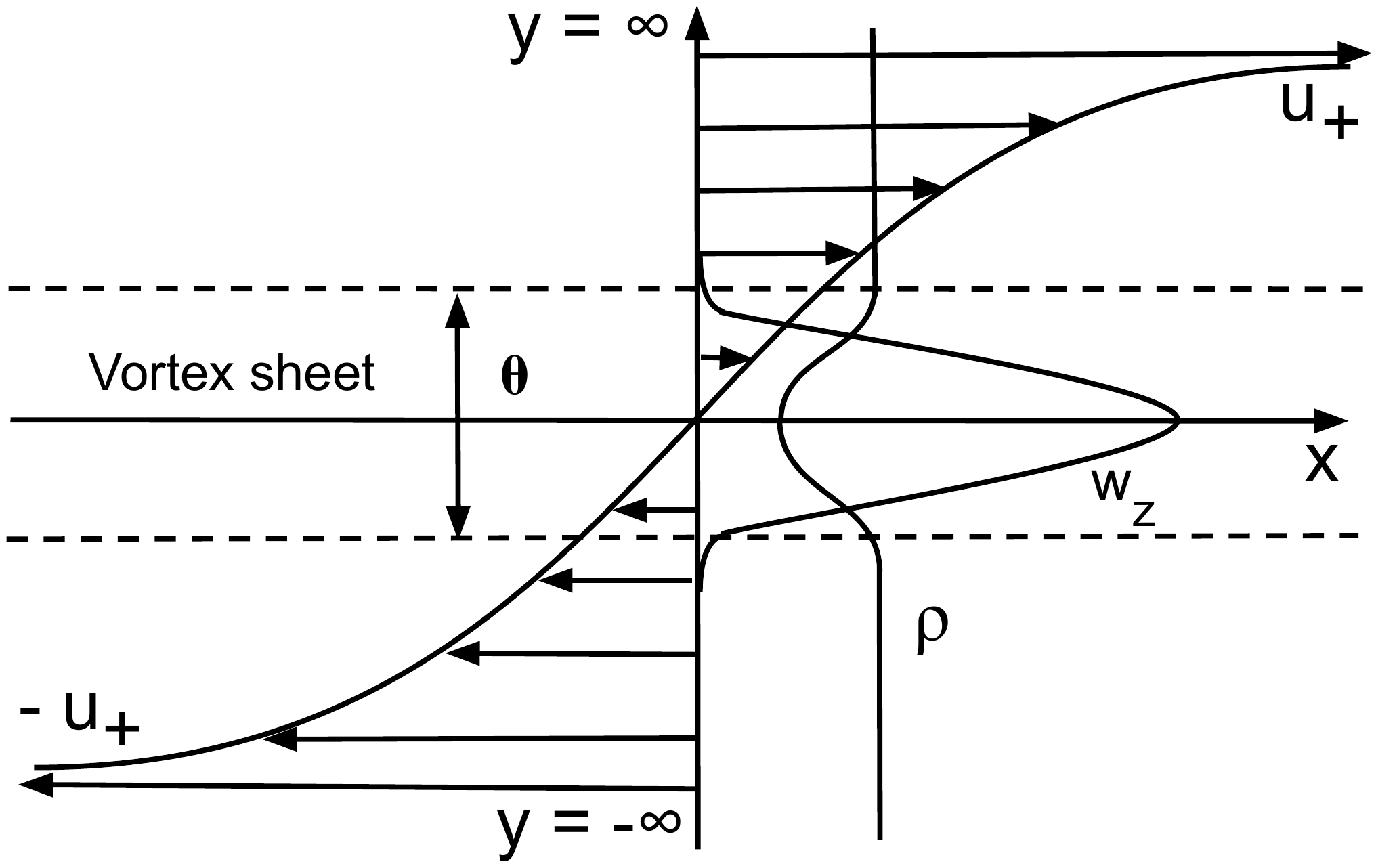}
  \caption{\small Vortex sheet configuration}
\label{f:vortex-sheet}
 \end{center}
\end{figure}

As a typical illustrative example, we consider a steady plane vortex sheet under isothermal conditions. The vortex sheet is assumed to lie in the $x$-$z$ plane and to have a thickness $\tht$ in the $y$-direction. We assume the velocity points in the $x$-direction $\bfv = (u(y),0,0)$ and approaches {\em different} asymptotic values $u_\pm$ as $y \to \pm \infty$. The density $\rho$ is also assumed to vary only with height $y$. Exactly as in channel flow, we obtain the equation for time-independent flows (\ref{e:R-euler-isotherm-channel-flow}):
	\beq
	\pdr_y \left(\half \la^2 \rho u_y^2 + p_{0} \frac{\rho}{\rho_{0}}\right) = 0.
	\eeq
The steady state is not unique and this equation can be used to find the density profile for any given vorticity profile. To model a vortex sheet of thickness $\tht$ we take the vorticity profile in $y$ to be given by
	\beq
	u_{y} = \Delta u \: \left(\frac{\theta}{\pi} \right) \left[\frac{1}{\theta^{2}+y^{2}} \right] \quad \text{where} \quad \bfw = - u_y(y) \: \hat z
	\eeq
Here $\Delta u = u_{+} - u_{-}$ and $w_{0} = -\Delta u/\pi \theta$ is the $z$-component of vorticity on the sheet. We obtain, as usual, the first integral,
	\beq
	\half \la_{0}^2 \rho_{0} u_y^2 + \frac{p_{0} \rho}{\rho_0} = K.
	\eeq
The suffix in this instance refers to quantities on the sheet ($y=0$). The Bernoulli constant $K = p_0 + \half \rho_0 (\Delta u)^2 \, \left( \frac{\la_0^2}{\pi^2 \tht^2} \right).$
We obtain the velocity profile by integration:
	\beq
	u(y) = u_- + (\Delta u) \left(\frac{\theta}{\pi} \right) \int_{-\infty}^{y} \frac{d\mu}{\theta^{2}+\mu^{2}} = u_{-} + (\Delta u) \left[\half + \frac{1}{\pi}\arctan\left(\frac{y}{\theta} \right) \right].
	\eeq
Assuming $u_+ > u_-$, the velocity monotonically increases from $u_-$ to $u_+$ with increasing height $y$. Moreover, the velocity on the sheet $u(0) = \half(u_- + u_+)$ is the average of its asymptotic values. The density profile follows from the first integral:
	\beq
	 \frac{\rho}{\rho_0} = 1 + \half \left(\frac{\la_0 }{\pi \theta} \right)^2 \left[\frac{\rho_0 (\Delta u)^{2}}{p_0} \right] \left(1- \left[\frac{\theta^{2}}{\theta^{2}+y^{2}} \right]^{2} \right).
	\eeq
In particular, the asymptotic densities are
	\beq
	\frac{\rho_{\pm \infty}}{\rho_0} = 1 + \half \left( \frac{\la_0}{\pi \theta} \right)^2 \left[\frac{\rho_0 (\Delta u)^2}{p_0} \right].
	\eeq
Thus, the density is decreased at the sheet relative to the values at $\pm \infty$. If the sheet thickness $\theta \gg \la_0/\pi$, the decrease is not significant. If the thickness is comparable to the regularizing length $\la_0$, the density decrease at the sheet can be considerable, depending upon the `relative flow Mach number' defined as, $(\Delta M)^{2} = (\rho_0/p_0)(\Delta u)^2$. Unlike velocity, the density increases from the sheet to the same asymptotic values on either side of the sheet ($y = \pm \infty$), reflecting the symmetry of the assumed vorticity profile. This is similar to the rotating vortex/tornado model (\ref{s:modeling-vortex}) where an increase in density outwards from the core of the vortex is balanced by a corresponding decrease in vorticity.

\subsection{Regularized plane flow}
\label{s:plane-flow}

It is interesting to consider the R-Euler equations for flow on the $x$-$y$ plane with $\bfv = (u(x,y),v(x,y),0)$. First consider incompressible flow $\grad \cdot \bfv = 0$ with constant $\rho$, and hence constant $\la$. The condition $u_x + v_y = 0$ is solved in terms of a stream function $u = - \psi_y$ and $v = \psi_x$ (subscripts denote partial derivatives). Vorticity points vertically $\bfw = w \hat z$ with $w = v_x - u_y = \Delta \psi$. The twirl acceleration is proportional to the gradient of $w^2$:
	\beq
	\bfw \times (\grad \times \bfw) = w \hat z \times (w_y \hat x - w_x \hat y) = w \grad w = (1/2) \grad w^2.
	\eeq
So for constant $\la$, the incompressible 2D R-Euler equation becomes 
	\beq
	\pdr_t \bfv + \bfw \times \bfv = - \grad \left( \sigma + (1/2) \la^2 w^2 \right).
	\eeq
The twirl acceleration term may be absorbed into a redefinition of stagnation enthalpy $\sigma$. In particular, the regularization drops out of the evolution equation for vorticity $\bfw_t + \grad \times (\bfw \times \bfv) = 0$, which states that $\bfw$ is frozen into the incompressible flow field $\bf v$. In other words, for incompressible plane flow, the regularization plays no role in vortical dynamics. This is to be expected: enstrophy $\int w^2 \: dx \, dy$ is bounded in incompressible 2D flows (indeed it is conserved) and there is no vortex stretching. 

By contrast, compressible flow on a plane is richer. For simplicity, consider steady flow with ${\bf v} = u \hat x + v \hat{y}$, ${\bf w} = w(x,y) \hat{z}$ and $\nabla \times {\bf w}=w_{y}\hat{x}-w_{x}\hat{y}$. The continuity equation $\grad \cdot (\rho \bfv) = 0$ is solved using a stream function: $\rho u = - \psi_{y}$, $\rho v = \psi_{x}$. The R-Euler equation becomes
	\beq
	 w u =-\sigma_y - \la^2 ww_y \quad \text{and} \quad
	 -w v = -\sigma_x -\la^2 w w_x.
	\eeq	
Using the relation, $\sigma = h + \half \bfv^2 = h + (\grad \psi)^2/2 \rho^2$, we obtain the equivalent equations:
	\beq
	 w \psi_{y} = \rho \left[ h + \frac{1}{2\rho^2} (\grad \psi)^2 \right]_y + \la^2 \rho ww_y \quad \text{and} \quad
	 w \psi_{x} = \rho \left[ h + \frac{1}{2\rho^2} (\grad \psi)^2 \right]_x + \la^2 \rho ww_x.
	\label{e:plane-flow-R-euler-stream-fn}
	\eeq
From the constitutive relation $\la^2 \rho = \la_0^2 \rho_0$ is a constant. Assuming $w$ is not zero, we get
	\beq
	\psi_y = \frac{\rho}{w} \left[ h + \frac{1}{2\rho^2}(\grad \psi)^2 \right]_y + \la_0^2 \rho_0 w_y
	\quad \text{and} \quad
	 \psi_x = \frac{\rho}{w} \left[ h + \frac{1}{2\rho^{2}}(\grad \psi)^2 \right]_x + \la_0^2 \rho_0 w_x.
	\eeq
Differentiating the first equation in $x$, the second in $y$ and subtracting, we see that, $\rho/w$ has a vanishing Jacobian with $\sig = h + (\grad \psi)^2/2\rho^2$. Thus the equations say that $\sig$ is an arbitrary function $\Sigma$ of $\rho/w$. Setting $\rho/w = \Theta$, we get
	\beq
	\psi_y = \Theta \: \Sigma'(\Theta) \: \Theta_y + \la_0^2 \rho_0 w_y \quad \text{and} \quad
	\psi_x = \Theta \: \Sigma'(\Theta) \: \Theta_x + \la_0^2 \rho_0 w_x.
	\eeq
It follows that we may integrate the equations to get $\psi = \la_0^2 \rho_0 \, w + H(w/\rho)$. Here $H$ is an arbitrary function related to $\Sigma$ through a quadrature $H = \int \Sigma'(\Theta) \Theta d \Theta$. Since $w = (\psi_x/\rho)_x + (\psi_y/\rho)_y$, a specification of $H$ reduces this to a nonlinear PDE for the two unknowns $\psi$ and $\rho$. The under-determinacy of this system is a common feature of the {\em steady} compressible R-Euler equations.

Alternatively, suppose we do not divide the R-Euler equation by $w$ [which could vanish in a region] but simply note that differentiating the first equation of (\ref{e:plane-flow-R-euler-stream-fn}) in $x$ and the second in $y$, and subtracting, we get an equation involving two Jacobians:
	\beq
\frac{\partial (w,\psi)}{\partial (x,y)}=\frac{\partial (\rho,\sig)}{\partial (x,y)}
	\eeq
We may consider the ansatz $w = J(\psi)$ where $J$ is an arbitrary function so that the LHS vanishes. For the RHS to vanish, $\sigma$ must be a function of $\rho$, say $\sigma = Z(\rho)$. Thus the `compatibility condition' on (\ref{e:plane-flow-R-euler-stream-fn}) can be satisfied by introducing two arbitrary functions $J$ and $Z$. There may be many other, much more complicated solutions of (\ref{e:plane-flow-R-euler-stream-fn}) but we do not investigate them here. Given, $J$ and $Z$ and the equation of state $p = p(\rho)$ we can eliminate $w$ and $p$ to reduce (\ref{e:plane-flow-R-euler-stream-fn}) to two nonlinear PDEs for $\psi$ and $\rho$. The simplest case could be for example, $Z(\rho) = Z_0$, a constant in which case (\ref{e:plane-flow-R-euler-stream-fn}) becomes $\psi = \la_0^2 \rho_0 J(\psi)$ upon absorbing a constant into $\psi$. Once $J$ is specified and $\psi$ determined, $\rho$ is obtained from $\sigma(\rho) = Z_0$ given an equation of state. 

Another possible solvable case occurs for subsonic flows at relatively low Mach numbers. In $0^{\rm th}$ order, we may take $w=0$. There then exists a velocity potential $\phi(x,y)$ such that $\psi$ is its conjugate function. In zeroth order, $\rho$ is constant and hence $\phi$ is clearly the standard incompressible Euler velocity potential. The pressure variations are then determined by the constancy of $\sigma$. Evidently, they must be of order the square of the Mach number. The full nonlinear equation must then be linearised about this basic irrotational flow to calculate the vorticity in the next order. We do not pursue this here.

\subsection{Incompressible 3-d axisymmetric vortex flow}

We consider the steady, incompressible R-Euler equations in an axisymmetric geometry. We have in mind applications to typical exterior flows where a spherical or cylindrical vortex capsule moves along the axis (e.g., Hill's spherical vortex). For simplicity, we consider incompressible flow $\grad \cdot \bfv = 0$ so both $\rho$ and $\la$ are a constant. We choose the axis to point along $\hat z$ and use cylindrical coordinates $(r,\phi,z)$. Axisymmetry here means $\bfv$ does not have an azimuthal component ($v_\phi = 0$) and that pressure, $v_r$ and $v_z$ are independent of $\phi$. This is to be contrasted with the rotating vortex of Section \ref{s:modeling-vortex}, where the velocity was purely azimuthal. The continuity equation $\grad \cdot \bfv = r^{-1}\pdr_r (r v_r) + \pdr_z v_z = 0$ can be solved in terms of a stream function\footnote{Beware! Subscripts on $\psi, w$ denote partial derivatives, while those on $v$ denote components.}
	\beq
	\bfv = - \grad \times \left( r^{-1} \psi(r,z) \hat \phi \right) \quad
	\text{or} \quad
	v_r =  \psi_z/r \quad \text{and} \quad
	v_z = - \psi_r/r.
	\eeq
The vorticity is purely azimuthal ($\bfw = w \hat \phi$) while the pressure gradient, vortex and twirl accelerations have no azimuthal components:
	\beqs
	&& w = \left( \pdr_z v_r - \pdr_r v_z \right) = \ov{r} \psi_{zz} + \pdr_r \left(\ov{r} \psi_r \right) = \grad^2 \left( \frac{\psi}{r} \right) - \frac{\psi}{r^3}
	\quad \text{and} \quad \bfw \times \bfv =  w v_z \hat r - w v_r \hat z  \cr
	\quad
	&& \text{and} \quad \bfT = \frac{w}{r} (r w)_r \hat r + w w_z \hat z.
	\label{e:vorticity-azimuthal-3d-vortex}
	\eeqs
Thus the steady R-Euler equations $\bfw \times \bfv = - \grad \sigma - \la^2 \bfT$ reduce to two component equations:
	\beq
	w v_z = - \sigma_r - \la^2 \frac{w}{r} (r w)_r \quad \text{and} \quad
	- w v_r = - \sigma_z - \la^2 w w_z.
	\eeq
Taking the curl of the R-Euler equation we may eliminate pressure. Expressing $\bfv$ in terms of its stream function $\psi$, we obtain
	\beq
	\dd{(w/r, \psi)}{(r,z)} = - \frac{\la^2}{r} \left( w^2 \right)_z.
	\label{e:jacobian-condition-for-R-euler}
	\eeq
This Jacobian condition can be simplified by working with $\bfv_*$ rather than $\bfv$. Recall that the steady R-Euler equation is $\bfw \times \bfv_* = - \grad \sigma$ and the R-vorticity equation [steady freezing-in of $\bfw$ into $\bfv_*$] is 
	\beq
	\grad \times (\bfw \times \bfv_*) = \grad \times (w v^*_z \hat r - w v^*_r \hat z) = \left[ (w v^*_z)_z + (w v^*_r)_r \right] \hat \phi = 0.
	\eeq
Since $\bfv_*$ is divergence-free, we may express it in terms of a stream function $\psi^*$
	\beq
	v^* = - \grad \times \left(\frac{\psi^*}{r} \hat \phi \right) \quad \text{or} \quad v^*_r = \ov{r} \psi^*_z \quad \text{and} \quad
	v^*_z = - \ov{r} \psi^*_r.
	\eeq
In terms of $\psi^*$, the R-vorticity equation reduces to a vanishing Jacobian condition:
	\beq
	\left(- \frac{w}{r} \psi^*_r \right)_z + \left( \frac{w}{r} \psi^*_z \right)_r  = 0 \quad \text{or} \quad \dd{(w/r,\psi^*)}{(r,z)} = 0.
	\label{e:jacobian-condition-psi*}
	\eeq
Thus $\psi^*$ can be an arbitrary function of $w/r$ or $w \equiv 0$. To see what this means for $\psi$ we write $\bfv = \bfv_* - \la^2 (\grad \times \bfw)$ in components and read off the relation $\psi = \psi^* + \la^2 r w$ (upto an additive constant). Thus (\ref{e:jacobian-condition-psi*}) implies a vanishing Jacobian condition on $\psi$
	\beq
\dd{(w/r, \psi - \la^2 r w)}{(r,z)} = 0.
	\label{e:vanishing-jacobian-psi}
	\eeq
One checks that this is equivalent to (\ref{e:jacobian-condition-for-R-euler}). Thus $w/r$ must be an arbitrary function of $\psi - \la^2 r w$ or $w \equiv 0$. In the latter case (irrotational incompressible flow) the regularization plays no role and $\psi$ must satisfy\footnote{In the case of irrotational flow, we could work in terms of a velocity potential which is harmonic, unlike the stream function.}
	\beq
	w = \ov{r} \psi_{zz} + \left(\ov{r} \psi_r \right)_r = 0 \quad \text{or} \quad
	\grad^2 \left( \frac{\psi}{r} \right) = \frac{\psi}{r^3}.
	\label{e:w-eq-0-in-terms-of-psi}
	\eeq
Alternatively, $w/r$ must be constant on level surfaces of $\psi - \la^2 r w$, i.e. $w/r = H(\psi - \la^2 r w)$ where $H$ is an arbitrary function. This is an exact generalisation of Lamb's Eq.(13), Art. 165, p. 245 \cite{lamb} when $\lambda=0$. The appearance of an arbitrary function is another instance of the steady underdeterminacy of the R-Euler equation. Writing $w = r^{-1} \psi_{zz} + (r^{-1} \psi_r)_r$ we get a (generally nonlinear) 2nd order PDE for $\psi$. Consider the simplest case where $H(g) = A - B g$ is a linear function ($[B] = 1/L^4$ and $[A] = 1/LT$). Then $\psi(r,z)$ must satisfy a 2nd order inhomogeneous linear PDE
	\beq
	\frac{w}{r} =  A - B \left[ \psi - \la^2 r w \right] \quad 
	\imply \quad
	\left(1 - \la^2 B r^2 \right) \left[ \ov{r^2} \psi_{zz} + \ov{r} \left( \ov{r} \psi_r \right)_r  \right] = A - B \psi.
	\eeq
The differential operator may be expressed in a more `invariant' manner in terms of the Laplacian of $\psi/r$:
	\beq
	\left[ \grad^2 - \ov{r^2} + \frac{B r^2}{1 - \la^2 B r^2} \right] \left( \frac{\psi}{r} \right) =  \frac{Ar}{1 - \la^2 B r^2}.
	\label{e:inhom-SE-for-psibyr}
	\eeq
When $A = 0$, (\ref{e:inhom-SE-for-psibyr}) becomes homogeneous and resembles the time-independent Schr\"odinger equation for a zero energy particle with wave function $f = \psi/r$ in a cylindrically symmetric non-central potential $V = r^{-2} - B r^2/(1 - \la^2 B r^2)$. If $B < 0$, then the potential is strictly positive and we would not expect any zero energy eigenstate. So when $A = 0$, we take $B > 0$.

\subsubsection{Spherical vortex}

The above equations may be used to model a spherical vortex of radius $a$ moving along the axis of symmetry in an irrotational exterior flow. An example of such irrotational flow occurs in the exterior of Hill's spherical vortex where
	\beq
	\psi = \half \: V_\infty r^2 \left[1- a^3/R^3 \right] \quad \text{for} \quad
	R^2 \equiv r^2 + z^2 > a^2.
	\eeq
This describes uniform flow far from the sphere, i.e. $v_r \to 0$ and $v_z \to - V_\infty$ as $R \to \infty$ (we go to the vortex frame and allow the fluid flow at infinity to be uniform). Furthermore, $\psi=0$ is a stream surface and hence the flow is tangential to the surface $R = a$. Within the sphere, if we choose $B=0$ in (\ref{e:inhom-SE-for-psibyr}), the regularization plays no role and we have to solve
	\beq
	\psi_{zz} + r \left(\psi_r/r \right)_r = A r^2.
	\eeq
This has a polynomial solution $\psi=\half Ar^{2} [a^{2}-r^{2}-z^{2}]$ vanishing on $R=a$. Continuity of velocity across $R =a$ implies
$A = -(3/2a^{2})V_{\infty}$. This constitutes Hill's famous ``spherical vortex'' solution. However, this makes $w_{\phi}$ {\it discontinuous} on $R=a$.
       
On the other hand, we could have chosen $A=0$ and left $B$ arbitrary in the interior. Then in spherical polar coordinates ($r = R \sin \tht, z = R \cos \tht$) (\ref{e:inhom-SE-for-psibyr}) becomes a Schrodinger equation for $f = \psi/r$
	\beq
	- \ov{R^2} \pdr_R \left(R^2 \dd{f}{R} \right) - \ov{R^2 \sin \tht} \pdr_\tht \left(\sin \tht \dd{f}{\tht} \right) + \left[ \ov{R^2 \sin^2 \tht} - \frac{B R^2 \sin^2 \tht}{1 - \la^2 B R^2 \sin^2 \tht} \right] f = 0.
	\label{e:SE-spherical-geom}
	\eeq
We must solve (\ref{e:SE-spherical-geom}) requiring $\psi=0$ on $R=a$ and regularity of $\psi$ at $R=0$. $B$ must then be chosen to match the outer solution. Note that since $w/r = -B(\psi - \la^2 r w)$, in this solution $w$ vanishes where $\psi$ does, and is therefore {\it rendered continuous} at the boundary (even for $\la = 0$), unlike in Hill's solution. We do not pursue here an explicit solution of (\ref{e:SE-spherical-geom}) for the regularized version of Hill's spherical vortex\footnote{Separation of variables does not work in (\ref{e:SE-spherical-geom}) since the `potential' $V$ depends on both $R$ and $\tht$.} but instead consider a cylindrical geometry where an explicit solution illustrating key features is easily found.

\subsubsection{Cylindrical vortex}

As the simplest special case of the above equations (\ref{e:vanishing-jacobian-psi}), we consider a cylindrical vortex (pipe-like flow). We imagine a flow with $v_\phi = 0$ as above, that is irrotational outside an infinite circular cylinder with axis along $z$ and with radius $a$. Vorticity is purely azimuthal inside the cylinder. We require the stream function, its normal derivative and $w$ to be continuous across the cylindrical surface $r = a$. The simplest irrotational flow in the region $r > a$ is a uniform flow with speed $c$ in the $-\hat z$ direction:
	\beq
	\psi = \frac{c}{2} \left(r^2 -a^2 \right) \quad \text{with} \quad v_r = v_\phi = 0 \quad \text{and} \quad v_z = - c.
	\eeq
The additive constant is chosen so that $r=a$ is a stream surface on which $\psi$ vanishes.

For $r \leq a$, $\frac{w}{r} = H(\psi - \la^2 r w)$ where $H$ is an arbitrary function. If $H = A$ is a non-zero constant, then $w(r=a^-) = a A$ cannot match the value $w = 0$ for $r > a$, so $H$ cannot be a constant. The next simplest possibility is a linear $H(g) = A - B g$. Choosing $A=0$ ensures that $w$ is continuous across the cylindrical surface $r=a$:
	\beq
	w(r,z) = - \frac{B r }{1 - \la^2 B r^2} \psi \quad \imply \quad
	w(r=a) = 0.
	\eeq
We get a zero energy Schr\"odinger eigenvalue equation for the `wave function' $f = \psi/r$ for $r \leq a$:
	\beq
	(- \grad^2 + V(r)) f = 0 \quad 
	\text{or} \quad
	- f_{zz} - \ov{r} (r f_r)_r + V(r) f(r) = 0
	\quad \text{where} \quad V(r) = \ov{r^2} - \frac{B r^2}{1 - \la^2 B r^2}.
	\eeq	
Unlike in the spherical vortex, the potential $V(r)$ is independent of both $\phi$ and $z$, so we may separate variables. $f$ could diverge at $r = 0$ in such a way that the stream function (or more importantly the velocity) is finite at $r = 0$. The BCs at $r=a$ are continuity of $\psi$ i.e. $\psi(r=a) = 0$ (which guarantees continuity of $w$) and its normal derivative $\psi_r$. The simplest interior solution is obtained by assuming that $\psi$ depends only on $r$ so that velocity is purely longitudinal $v_z = -r^{-1} \psi_r$. In this case the above Schrodinger-like equation reduces to a 2nd order linear ODE $- r^{-1} \left( r f_r \right)_r + V(r) f(r) = 0$ on the interval $0 \leq r \leq a$ with the BCs $\psi(r=a) = 0$ and $\psi'(r=a) = c a$.

\begin{figure}[h]       
    \begin{center}
    \includegraphics[width=4.5cm]{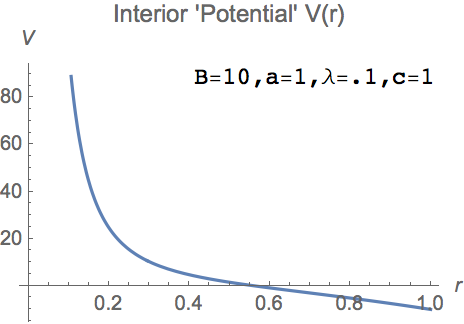}
    \hspace{0.5cm}
    \includegraphics[width=4.5cm]{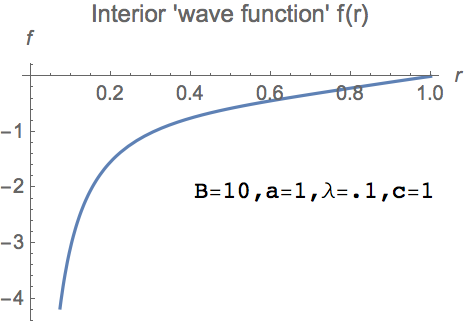}
    \hspace{0.5cm}
    \includegraphics[width=4.5cm]{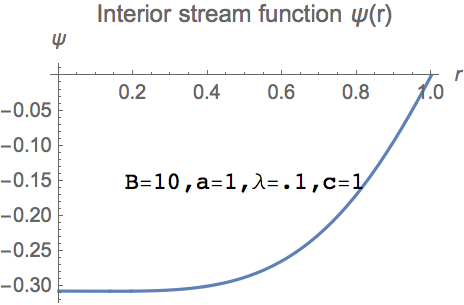}
    \caption{\footnotesize Interior potential $V(r)$ and wave function $f(r)$ for the Schrodinger-like equation for an infinite propagating axisymmetric cylindrical vortex of radius $a$ in a uniform external flow $-c \hat z$. The interior stream function $\psi(r) = r f(r)$ agrees with the exterior $\psi = (c/2) (r^2 - a^2)$ and its gradient at $r = a$.}
    \label{f:cyl-vortex-potn-wfn-stream-fn}
    \end{center}
\end{figure}
\begin{figure}[h]       
    \begin{center}
  \includegraphics[width=4.5cm]{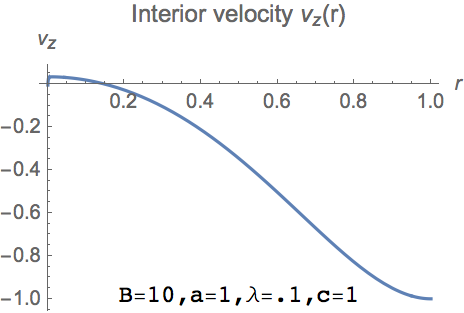}
    \hspace{0.5cm}
  	\includegraphics[width=4.5cm]{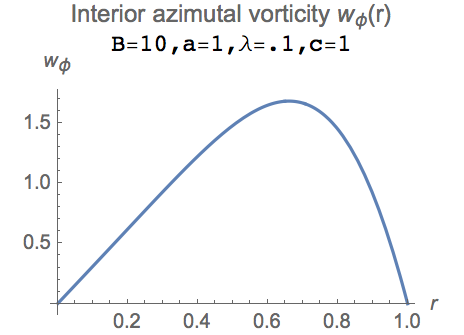}
    \caption{\footnotesize Velocity and vorticity profiles for cylindrical vortex of radius $a$. Velocity is longitudinal and increases in magnitude with increasing distance from the axis of the vortex and reaches the exterior flow value $-c \hat z$ at $r = a$. Voriticity is azimuthal for $r<a$ and matches the irrotational exterior flow at $r=a$. Radial derivative of vorticity is discontinuous across $r=a$ in our simple model.}
	\label{f:cyl-vortex-vel-vorticity}
    \end{center}
\end{figure}
This is a homogeneous second order ODE that may be put in the standard form
	\beq
	f'' + p(r) f' + q(r) f = 0, \quad \text{with} \quad p(r) = \ov{r} \quad \text{and} \quad q(r) = -V(r) = -\ov{r^2} + B r^2 + \la^2 B^2 r^4 + \ldots.
	\eeq
$p$ and $q$ have simple and double poles at $r=0$ and $q$ has simple poles at $r = \pm 1/\sqrt{\la^2 B}$, so the equation has $3$ regular singular points and could be transformed into the Hypergeometric equation. For sufficiently small $\la$, $r = 0$ is the only singular point in the physical region $0 \leq r \leq a$, around which the Frobenius method yields solutions. Making the ansatz $f(r) = r^\al \left[c_0 + c_1 \al + c_2 \al^2 + \ldots \right]$ with $c_0 \ne 0$ and comparing coefficients of $r^{\al - 2}$, we get the indicial equation $c_0 \al (\al -1) + \al c_0 - c_0 = 0$. Its roots $\al_1 = 1$ and $\al_2 = -1$ differ by an integer. Thus we have $2$ linearly independent solutions around $r=0$:
	\beqs
	f_1(r) &=& r \left[ c^{(1)}_0 + c^{(1)}_1 r + c^{(1)}_2 r^2 + \ldots \right] \quad \text{with} \quad c_0^{(1)} \ne 0  \quad \text{and} \cr
	f_2(r) &=& \ov{r} \left[ c_0^{(2)} + c^{(2)}_1 r + c^{(2)}_2 r^2 + \ldots \right] + c^{(2)} f_1(r) \, \log r \quad \text{with} \quad c_0^{(2)}, c^{(2)} \ne 0.
	\eeqs
Comparing coefficients of higher powers $r^{\al -2 + n}$ leads to recursion relations for $c_j^{(1)}$ and $c_k^{(2)}$. $f_1$ has a simple zero at $r=0$ while $f_2$ has a simple pole at $r=0$ in addition to a logarithmic branch cut ending at $r=0$. The solution of our boundary value problem with prescribed BC at $r=a$ is a linear combination of $f_1$ and $f_2$. A generic linear combination $f$ will diverge at $r=0$ like $c^{(2)}_0/r$. Thus we should expect the stream function $\psi = r f(r)$ to linearly approach a non-zero limit $c_0^{(2)}$ as $r \to 0$.

We have solved this ODE with the given BCs numerically. The results are illustrated Fig.\ref{f:cyl-vortex-potn-wfn-stream-fn},\ref{f:cyl-vortex-vel-vorticity} for cylinder radius $a = 1$, regularization length $\la = 1/10$, constant $B = 10$ and exterior flow speed $c = 1$. It is clear that $\psi(r), v_z(r)$ and $w(r)$ are all continuous at $r=a$. However the radial derivative of vorticity $w_r$ is discontinuous at $r =a$. A more careful treatment of a layer of thickness $\sim \la$ around $r=a$ should render $\pdr w/\pdr r$ continuous. Despite this discontinuity, the twirl acceleration $\bfT = \bfw \times (\grad \times \bfw) = r^{-1}w (r w)_r \hat r$ is continuous across the cylindrical surface since $w(a) = 0$. On the axis of the vortex, $\psi$ is divergent, though $v_z$ has a finite value, while $w$ vanishes there. In addition, we see that the radial derivative of $v_z$ is zero at $r=0$, as one expects from axial symmetry and smoothness of $v_z(r)$.

\chapter{Conservative regularization of two-fluid plasmas}
\label{s:two-fluid}

In this Chapter, we extend our local conservative regularization of vortical singularities in compressible ideal MHD to non-relativistic two fluid (ion-electron) plasmas \cite{govind-sonakshi-thyagaraja-two-fluid}. The extension to multi-fluid or electron-positron plasmas is relatively straightforward. As in R-MHD, the continuity equations are unchanged while we introduce regularization terms in the velocity equations for each species ($l = i, e$ with charges $q_l$ and masses $m_l$). In addition to the vortical twirl term $\bfw_l \times (\grad \times \bfw_l)$ analogous to the one in R-MHD, we add a magnetic twirl term $(q_l/m_l)\bfB \times (\grad \times \bfw_l)$ with a {\it common} coupling strength $\la_l^2$. This is similar to the universal coupling of charged particles to both electric and magnetic fields through the electric charge. Here $\la_l$ are (possibly different) regularizing lengths for the two species. The two twirl terms are obtained by a judicious replacement of $\bfw_l$ by $\bfw_l + q_l \bfB/m_l$ in R-MHD. The combination $\bfw + q \bfB/m$ also appears elsewhere, notably in the study of plasmas in non-inertial frames \cite{thyagaraja-mc-clements-2}. The number densities $n_l$ and $\la_l$ must satisfy the constitutive relations $\la_l^2 n_l = C_l$ where $C_l$ must be constant for a conserved energy to exist. These relations are automatic if $\la_{i,e}$ are chosen to be the Debye lengths or skin depths for ions and electrons, where the ideal equations are known to break down. Gauss ($\eps_0 \grad \cdot \bfE = \varrho$), Faraday ($\pdr\bfB/\pdr t = - \grad \times \bfE$) and Amp\`ere ($\mu_0 \eps_0 (\pdr \bfE /\pdr t) = \grad \times \bfB - \mu_0 \bfj_*$) laws take their usual forms with charge density given by $\varrho = \sum_l q_l n_l$. However, the `swirl' current $\bfj_* = \bfj_{\rm flow} + \bfj_{\rm twirl}$ differs from the flow current $\bfj_{\rm flow} = \sum_l q_l n_l \bfv_l$  by an additional regularization term $\bfj_{\rm twirl} = \sum_l q_l n_l \la_l^2 \grad \times \bfw_l$. The constitutive relations ensure that $\bfj_{\rm twirl} = \sum_l \grad \times (\grad \times \la_l^2 \bfj_{{\rm flow},l})$ is solenoidal, thus guaranteeing charge conservation: $\pdr_t \varrho + \grad \cdot \bfj_* = 0$. The constitutive relations and modification of current $\bfj_{\rm flow} \mapsto \bfj_*$ are crucial for obtaining a conserved `swirl' energy including a vortical contribution for {\it compressible} barotropic flow:
	\beq
	E^* = \int \left[ \sum_{l=i,e} \left( \half n_l m_l ({\bfv_l}^2  + \la_l^2 \bfw_l^2) + U_l(n_l m_l) \right) +  \frac{\bfB^2}{2\mu_0} + \frac{\eps_0  \bfE^2}{2} \right] d\bfr 
	\quad \text{where} \quad \grad U_l' = \frac{\grad p_l}{m_l n_l}.
	\label{e:swirl-energy}
	\eeq
Here $p_l$ are the partial pressures. The positive definiteness of $E^*$ along with the constitutive relations ensure that the kinetic and compressional energies as well as the enstrophy of each species is bounded, thus helping to regularize vortical singularities. We also derive local conservation laws for swirl energy, linear and angular momenta in our regularized two-fluid model. Unlike in the single-fluid case, we do not have analogues of conserved magnetic and cross helicities. When the number densities $n_{i,e}$ and $\la_i = \la_e = \la$  are {\it constants} and the compressional and electric energies are omitted, the above equations reduce to a conservative regularization of {\it incompressible} quasineutral two-fluid plasmas. Interestingly, in the incompressible case {\it alone}, if the current in Amp\`ere's law is taken to be $\bfj_{\rm flow}$, we obtain a {\it different} conserved energy that includes terms with both velocity and magnetic field curls: 
	\beqs
	E^*_{\rm inc} &=& \int \left[\sum_{l} \left(\half n m_l \left({\bfv_l}^2  + \la^2 (\grad \times \bfv_l)^2 \right)\right) + \frac{\bfB^2}{2\mu_0}  + \frac{\la^2}{2 \mu_0} (\grad \times \bfB)^2  \right] d\bfr.
	\eeqs
In Section \ref{s:heirarchy-of-models} a hierarchy of regularized plasma models is considered. In many physically interesting situations [eg. tokamak or many astrophysical plasmas \cite{Wesson,Kulsrud}] it is reasonable to sacrifice the generality of the full two-fluid model and assume quasineutrality $(n_i \approx n_e)$ on scales larger than the Debye length $\la_D$ and frequencies less than the plasma frequency $\om_p$. Additionally, in systems such as accretion disks and planetary magnetospheres \cite{Michel}, one may even ignore electron inertia effects (Hall MHD). The passage from our full regularized two fluid model to the corresponding quasineutral, Hall and one-fluid MHD models is achieved via the successive limits $\eps_0 \to 0$ (non-relativistic limit where the displacement current may be ignored), $m_e \to 0$ ($m_e/m_i \ll 1$) and finally electric charge $e \to \infty$ with $\la_e/\la_i \to 1$ ($L \gg \la_D$ and $\om \ll \om_p$). In each case we have a conserved swirl energy guaranteeing boundedness of enstrophy. In the quasineutral limit where $c \to \infty$, $\bfE$ is non-dynamical. It is determined from the electron velocity equation rather than from Gauss' law: 
	\beq
	\bfE = - \bfv_{*e} \times \bfB - \frac{\grad p_e}{en} - \frac{m_e}{e} \left( \pdr_t \bfv_e + \bfw_e \times \bfv_{*e} + \half \grad \bfv_e^2 \right)
	\eeq
where $\bfv_{*e} = \bfv_e + \la_e^2 \grad \times \bfw_e$ is the electron swirl velocity. The situation is analogous to the determination of pressure from the divergence of the Euler equation upon passing to incompressible flow by taking the sound speed $c_s \to \infty$. In the regularized Hall model where electron inertia terms are ignored, magnetic helicity $\int \bfA \cdot \bfB \:d\bfr$ is conserved and in the barotropic case, $\bfB$ is frozen into $\bfv_{*e}$. Finally, when $e \to \infty$ ($L \gg \la_D$) we recover the one-fluid R-MHD model ($\bfv \approx \bfv_i \approx \bfv_e$ and $\la_i =\la_e = \la$) with the magnetic field frozen into the swirl velocity $\bfv_*$.

In Section \ref{s:PB-two-fluid} the Poisson bracket (PB) formalism for regularized compressible two-fluid models is discussed. Interestingly our two-fluid equations follow from the PBs introduced by Spencer-Kaufman \cite{spencer-kaufman} and Holm-Kuperschmidt \cite{holm-kuperschmidt} with the swirl energy $E^*$ taken as the Hamiltonian. Whilst R-MHD admits a Hamiltonian formulation with the Landau-Morrison-Greene PBs \cite{landau,morrison-greene}, we have not identified PBs for the quasineutral two-fluid or Hall MHD models. Moreover, unlike the Hamiltonian and equations of motion (EOM), the two-fluid PBs do not all reduce to the one-fluid PBs under the above limiting processes.

In Section \ref{s:reg-field-curl-PBs-Hamiltonian} we exploit the above PB formulation to propose a way of regularizing magnetic field gradients in compressible one- and two-fluid plasma models. In standard tearing mode theory, (see \cite{Hazeltine-Meiss,Wesson,chandra-thyagaraja}) the magnetic field can have tangential discontinuities associated with current sheets and reconnection. These current density singularities are usually resolved by resistivity; we propose a {\it conservative regularization.} By analogy with the vortical energy densities $(1/2) \la_l^2 \rho_l (\grad \times \bfv_l)^2$ which regularizes velocities we add $(\la_B^2/2\mu_0)(\grad \times \bfB)^2$ to the swirl energy $E^*$ of (\ref{e:swirl-energy}), to prevent $\bfB$ from developing a large curl. Here $\la_B$ is a {\it constant} cut-off length. The equations of motion obtained from this Hamiltonian using the two-fluid PBs can be put in the same form as before by replacing $\mu_0 \bfj_*$ in Amp\`ere's law with $\mu_0 \bfj_* - \la_B^2 \:\grad \times (\grad \times (\grad \times \bfB))
$. On the other hand, the introduction of such a magnetic curl energy in the one-fluid Hamiltonian adds $-(\la_B^2/\rho \mu_0) \bfB \times (\grad \times (\grad \times (\grad \times \bfB)))$ on the RHS of the velocity equation upon use of the one-fluid PBs. In other words, we have a modified Lorentz force term $\bfj_{**} \times \bfB$ where $\mu_0 \bfj_{**} = \grad \times \bfB + \la_B^2 (\grad \times (\grad \times (\grad \times \bfB)))$.
These third derivatives of $\bfB$ could smooth large gradients in current and field across current sheets just as the $u_{xxx}$ term in KdV does across a shock \cite{whitham}. Interestingly, XMHD \cite{kimura-morrison,abdelhamid-kawazura-yoshida} provides an alternate way of regularizing magnetic though not vortical singularities within a one-fluid setup. Indeed, the XMHD Hamiltonian includes $(\grad \times \bfB)^2$ but not $(\grad \times \bfv)^2$. Moreover, the resulting regularization terms in the velocity and Faraday equations are quite different from ours due to the use of different PBs (see Section \ref{s:reg-field-curl-PBs-Hamiltonian}). Another essential difference is that the XMHD cut-off lengths $d_{i,e}$ (normalized collisionless skin-depths) are assumed constant unlike our local cut-offs $\la_{i,e}$.

\section{Regularized compressible two-fluid plasma equations}
\label{s:reg-eqns-two-fluid-compress}

The dynamical variables of a two-fluid plasma are: $\bfE$, $\bfB$, ion and electron velocities $\bfv_{i,e}$, number densities $n_{i,e}$ and partial pressures $p_{i,e}$. The number densities satisfy the continuity equations:
	\beq
	\pdr_t n_{l} + \grad \cdot (n_{l} \bfv_{l}) = 0 \quad \text{where} \quad l = i \;\; \text{or} \;\; e.
	\label{e:reg-cont-eqn-two-fluid}
	\eeq
If $q_{i,e}$ denote the ion and electron charges, then the regularized velocity equations are:
	\beq
	\pdr_t \bfv_l +  \bfv_l \cdot \grad \bfv_l = - \ov{n_l m_l} \grad p_l + \frac{q_l}{m_l} (\bfE + \bfv_l \times \bfB) - \la_l^2 \bfw_l \times (\grad \times \bfw_l) - \frac{\la_l^2 q_l}{m_l} \bfB \times (\grad \times \bfw_l).
	\eeq
The mass densities and vorticities are $\rho_l = m_l n_l$ and $\bfw_l = \grad \times \bfv_l$ while $\la_{i,e}$ are the short distance cut-offs. For barotropic flow, $(\grad p_l)/\rho_l  = \grad h_l$ where $h_l (\rho_l)$ are the specific enthalpies. In this case, the velocity equations may be written as,
	\beq
	\pdr_t \bfv_l + \bfw_l \times \bfv_l = -\grad \sig_l + \frac{q_l}{m_l}  (\bfE + \bfv_l \times \bfB) - \la_l^2 \left[ \bfT^w_l + \frac{q_l}{m_l} \bfT^B_l \right]. 
	\label{e:reg-mom-eq-barotropic-Tw-TB}
	\eeq
Here $\sigma_l = h_l + \half \bfv_l^2$ are the specific stagnation enthalpies. The vortical and magnetic `twirl' regularization terms for each species are denoted $\bfT^w_l = \bfw_l \times (\grad \times \bfw_l)$ and $\bfT^B_l = \bfB \times (\grad \times \bfw_l)$.  As we will see in Section \ref{s:energy-cons-two-fluid}, conservation of energy requires that the strengths $\la_l^2$ of the vortical $\bfT^w_l$ and magnetic $(q_l/m_l) \bfT^B_l$ twirl forces must be the same for a given species. This resembles the universality of the electric charge $q_l$ through which a particle couples to both electric and magnetic fields. The short-distance regulators $\la_{i,e}$ are assumed to satisfy the constitutive relations $\la_l^2 n_l = C_l$ where $C_l$ are constants. We will see that these constitutive relations help to ensure that the EOM admit a conserved energy. Here $\la_{i,e}$ need not be equal (they could, for example, be the ion and electron collisionless skin depths). Yet another way to express the velocity equations is by introducing the swirl velocities $\bfv_{*l} = \bfv_{l} + \la_{l}^2 \grad \times \bfw_{l}$ which allow us to absorb the regularization terms into the vorticity and magnetic Lorentz force terms,
	\beq
	\pdr_t \bfv_l = - \grad \sig_l + \frac{q_l}{m_l} \bfE  + \bfv_{*l} \times \left(\bfw_l + \frac{q_l}{m_l} \bfB \right).
	\label{e:reg-elec-ion-mom-eqn}
	\eeq
We will see that $\bfw_l$ and $\bfB$ often appear in the combination $\bfw_l + q_l \bfB/m_l$ (see, \cite{Ferraro} and also \cite{thyagaraja-mc-clements-2}). In the latter work, it is shown how the vorticity and magnetic fields are intimately linked in non-inertial frames co-moving with a fluid. The evolution equations for vorticities are 	
	\beq
	\pdr_t \bfw_l + \grad \times (\bfw_l \times \bfv_l) = \frac{q_l}{m_l} \grad \times (\bfE + \bfv_l \times \bfB) - \grad \times \left[\la_l^2 \left(\bfT^w_l + \frac{q_l}{m_l} \bfT^B_l \right) \right]
	\label{e:vorticity-eqn-two-fluid}
	\eeq
while the Faraday and Amp\`ere evolution equations are
	\beqs
	\dd{\bfB}{t} &=& - \grad \times \bfE \quad \text{and} \quad
	\mu_0 \eps_0 \dd{\bfE}{t} = \grad \times \bfB - \mu_0 \bfj_*
	\label{e:maxwell-evol-eqns-two-fluid}
	\eeqs
with $c = 1/{\sqrt{\mu_0 \eps_0}}$. Here the total `swirl' current density $\bfj_*$ is related to the velocities and densities of the two species via the constitutive law
	\beq
	\bfj_* = \bfj_{*i} + \bfj_{*e} \quad \text{where} \quad \bfj_{*i,e} = q_{i,e} n_{i,e} \bfv_{* i,e}.
	\eeq
The regularized ion and electron swirl currents are a sum of flow and twirl currents for each species
	\beq
	\bfj_{*l} = \bfj_{{\rm flow},l} + \bfj_{{\rm twirl},l} \equiv q_l n_l \bfv_l + q_l n_l \la_l^2 \grad \times \bfw_l.
	\label{e:swirl-current-two-fluid}
	\eeq 
The constitutive laws $\la_l^2 n_l = C_l$ allow us to write the twirl currents in manifestly solenoidal form:
	\beq
	\bfj_{{\rm twirl},l} = \grad \times (\grad \times \la_l^2 \bfj_{{\rm flow},l}).
	\label{e:j-twirl-as-curlcurl}
	\eeq
Postulating that the current appearing in Amp\`ere's law is $\bfj_*$ rather than the unregularized $\bfj_{\rm flow}$ allows us to derive a conserved energy (\ref{e:energy-density-two-fluid}) in Section \ref{s:energy-cons-two-fluid}. In addition, the electric and magnetic fields must satisfy
	\beq
	\grad \cdot \bfB = 0 \;\; \text{and} \;\; \eps_0 \grad \cdot \bfE = \varrho \;\; \text{where} \;\;\varrho =  n_i q_i + n_e q_e
	\eeq
is the charge density. The consistency of the inhomogeneous Maxwell equations require that $\bfj_*$ and $\varrho$ satisfy the local conservation law $\pdr_t \varrho + \grad \cdot \bfj_* = 0$. Our regularized current does indeed satisfy this condition since $\grad \cdot \bfj_{\rm twirl} = 0$ and by the continuity equations,
	\beq
	\grad \cdot \bfj_{\rm flow} = \grad \cdot \sum_l q_l n_l \bfv_l = - \pdr_t \sum_l q_l n_l = - \pdr_t \varrho.
	\eeq

\subsection{Local conservation laws}

In this section, we show that the compressible regularized two-fluid equations of Section \ref{s:reg-eqns-two-fluid-compress} possess locally conserved energy,  linear and angular momenta and identify the corresponding currents. The conservation of energy depends crucially on the constitutive relations and the modification of Amp\`ere's law to include a regularized `twirl' current in addition to the flow current (\ref{e:swirl-current-two-fluid}). In the limit of constant densities $n_{i,e}$ we obtain a locally conserved energy for incompressible two-fluid plasmas provided the regularization lengths $\la_{i,e}$ are equal. Interestingly, we discover {\it another} way of regularizing the incompressible equations, the difference being that it is $\bfj_{\rm flow}$ and  not $\bfj_*$ that appears in Amp\`ere's law. The resulting conserved energy shows that velocity as well as field curls are regularized. However, this approach does not generalize to the compressible case. Unlike in ideal and twirl regularized one-fluid MHD, magnetic helicity $\int \bfA \cdot \bfB \; d\bfr$ is {\it not} conserved in the general two-fluid model. However, it {\it is} conserved in the Hall two-fluid limit where electron inertia terms are ignored (Section \ref{s:R-Hall-2fluid}). On the other hand, we do not have a two-fluid analogue of the conserved cross helicity of the (regularized) one-fluid MHD equations.

\subsubsection{Local conservation of energy}
\label{s:energy-cons-two-fluid}

The regularized equations (\ref{e:reg-cont-eqn-two-fluid}), (\ref{e:reg-mom-eq-barotropic-Tw-TB}) and (\ref{e:maxwell-evol-eqns-two-fluid}) for barotropic two-fluid plasmas obeying the constitutive laws $\la_l^2 n_l = C_l$ possess a positive definite swirl energy density
	\beq
	{\cal E}^* = \sum_{l=i,e} \left[\half \rho_l ({\bfv_l}^2  + \la_l^2 \bfw_l^2) + U(\rho_l)\right] + \frac{\bfB^2}{2\mu_0} + \frac{\eps_0}{2} \bfE^2
	\label{e:energy-density-two-fluid}
	\eeq
satisfying a local conservation law $\pdr_t {\cal E}^* + \grad \cdot \bff = 0$ where
	\beqs
	\bff =  \sum_{l} \left[\sig_l \rho_l \bfv_l 
	   + \la_l^2 \rho_l \bfw_l \times \left[ \bfv_l \times \bfw_l + \frac{q_l(\bfE + \bfv_l \times \bfB)}{m_l} - \la_l^2 \left[\bfT^w_l + \frac{q_l}{m_l} \bfT^B_l \right] \right] \right] + \frac{\bfE \times \bfB}{\mu_0}.
	\label{e:energy-cons-two-fluid}
	\eeqs
With appropriate BCs (E.g. decaying or periodic) the total swirl energy $\int {\cal E}^* d\bfr$ is a constant of motion. Thus in addition to the kinetic and potential energies of each species, their enstrophies $\int \bfw_l^2 d\bfr$ (or vortical energies) are bounded above. The corresponding kinetic, vortical and potential energy densities in ${\cal E}^*$ will be denoted ${\cal KE}, {\cal VE}$ and ${\cal PE}$. The energy flux may be compactly written in terms of the swirl velocities $\bfv_{l*}$:
	\beqs
	\bff = \sum_{l} \left[\sig_l \rho_l \bfv_l + \bfE \times \left(\frac{\bfB}{\mu_0} - \grad \times \la_l^2 \bfj_{{\rm flow},l} \right) + \la_l^2 \rho_l \bfw_l \times \left( \bfv_{l*} \times \left(\bfw_l + \frac{q_l}{m_l} \bfB \right) \right) \right].
	\label{e:energy-current-2flu-compress}
	\eeqs
The first term comes from ideal flow while the second is the Poynting flux, which is augmented by a regularizing term. It may be noted that the combination $\bfB - \mu_0 \grad \times \la_l^2 \bfj_{{\rm flow},l}$ also appears in Amp\`ere's law (\ref{e:maxwell-evol-eqns-two-fluid}).

Let us sketch the proof of (\ref{e:energy-cons-two-fluid}), which involves some remarkable cancellations. To begin we take the dot product of the velocity equations (\ref{e:reg-elec-ion-mom-eqn}) for each species with $\rho_l \bfv_l$. Since the vorticity and magnetic forces do no work,
	\beq
	\half \rho_l \pdr_t \bfv_l^2 = -\rho_l \bfv_l \cdot \grad \left( h_l + \half \bfv_l^2 \right) + n_l q_l \bfv_l \cdot \bfE - \la_l^2 \rho_l \bfv_l \cdot \bfT^w_l - \la_l^2 n_l q_l \bfv_l \cdot \bfT^B_l
	\eeq
for each $l = i, e$. Using (\ref{e:reg-cont-eqn-two-fluid}) we get
	\beq
	\pdr_t({\cal KE}_l) + \half \bfv_l^2 \grad \cdot (\rho_l \bfv_l) + \rho_l \bfv_l \cdot \grad \left( h_l + \half \bfv_l^2 \right) = n_l q_l \bfv_l \cdot \bfE - \la_l^2 \rho_l \bfv_l \cdot \left[ \bfT^w_l + \frac{q_l}{m_l} \bfT^B_l \right].
	\eeq
Again by the continuity equation, 
	\beq
	\rho_l \bfv_l \cdot \grad h_l = \grad \cdot (\rho_l h_l \bfv_l) - U_l'(\rho_l) \grad \cdot (\rho_l \bfv_l) = \grad \cdot (\rho_l h_l \bfv_l) + \pdr_t U_l.
	\eeq
Thus time derivatives of the sum of kinetic and potential energy densities of each species is
	\beq
	\pdr_t({\cal KE}_l + {\cal PE}_l) = - \grad \cdot (\sig_l  \rho_l \bfv_l ) + n_l q_l \bfv_l \cdot \bfE - \la_l^2 \rho_l \bfv_l \cdot \left( \bfT^w_l + \frac{q_l}{m_l} \bfT^B_l \right).
	\label{e:time-der-of-KE_l+PE_l-two-fluid}
	\eeq
The second term on the RHS is the work done by $\bfE$. To write the work done by the twirl regularization forces in conservation form and introduce the  vortical energy density, we dot the vorticity evolution equation (\ref{e:vorticity-eqn-two-fluid}) for each species with $\la_l^2 \rho_l \bfw_l$:
	\beq
	\pdr_t \left( {\cal VE}_l \right) 
	= \la_l^2 \rho_l \bfw_l \cdot \grad \times \left[ (\bfv_l \times \bfw_l)  
	+ \frac{q_l}{m_l} (\bfE + \bfv_l \times \bfB) -  \la_l^2  \left(\bfT^w_l + \frac{q_l}{m_l} \bfT^B_l \right) \right].
	\label{e:enstrophic-energy-time-der-two-fluid}
	\eeq
The vector identity for the divergence of a cross product allows us to write (\ref{e:enstrophic-energy-time-der-two-fluid}) as
	\beqs
	\pdr_t({\cal VE}_l) 
	&=& \la_l^2 \rho_l \left[ (\bfv_l \times \bfw_l) 
	+ \frac{q_l}{m_l} (\bfE + \bfv_l \times \bfB ) - \la_l^2 \left(\bfT^w_l + \frac{q_l}{m_l} \bfT^B_l \right) \right] \cdot \grad \times \bfw_l \cr
		&& + \la_l^2 \,\rho_l \grad \cdot \left[ \left( \bfv_l \times \bfw_l + \frac{q_l}{m_l} (\bfE + \bfv_l \times \bfB -  \la_l^2 \left(\bfT^w_l + \frac{q_l}{m_l} \bfT^B_l \right) \right)\times \bfw_l \right].
	\eeqs
Using the properties of the scalar triple product and rearranging, the rate of change of vortical energy density of each species is
	\beqs
	( {\cal VE}_l )_t 
	&=& \la_l^2 \rho_l \bfv_l \cdot \left[ \left(\bfw_l + \frac{q_l}{m_l} \bfB \right) \times \grad \times \bfw_l \right] + \bfE \cdot \grad \times (\la_l^2 n_l q_l \bfw_l)  \cr
	&& + \la_l^2 \rho_l \grad \cdot \left[ \left[ \bfv_l \times \bfw_l + \frac{q_l}{m_l} (\bfE + \bfv_l \times \bfB) -  \la_l^2 \left[\bfT^w_l + \frac{q_l}{m_l} \bfT^B_l \right] \right] \times \bfw_l \right] .
	\label{e:time-der-of-EE_l-two-fluid}
	\eeqs
We add (\ref{e:time-der-of-KE_l+PE_l-two-fluid}) and (\ref{e:time-der-of-EE_l-two-fluid}), sum over species and identify the swirl current $\bfj_*$ from (\ref{e:swirl-current-two-fluid}). The work done by the twirl forces $\la_l^2 \rho_l \bfv_l \cdot (\bfT^w_l + (q_l/m_l) \bfT^B_l)$ cancels out giving:
	\beqs
	\pdr_t {\cal E}^* 
	&+& \sum_l \grad \cdot \left[ \sig_l \rho_l \bfv_l - \la_l^2  \rho_l \left[ \bfv_l \times \bfw_l  - \frac{q_l}{m_l} (\bfE + \bfv_l \times \bfB ) + \la_l^2 \left[\bfT^w_l + \frac{q_l}{m_l} \bfT^B_l \right] \right] \times \bfw_l \right]  \cr
	 &=& \bfE \cdot \bfj_*.
	\eeqs
Now we use the regularized Maxwell equations (\ref{e:maxwell-evol-eqns-two-fluid}) to calculate the total work done by the electric field
	\beqs
	\bfE \cdot \bfj_* &=& \frac{\bfE  \cdot (\grad \times \bfB)}{\mu_0} - \eps_0 \bfE \cdot \pdr_t \bfE = \frac{\bfB \cdot \grad \times \bfE}{\mu_0} + \grad \cdot \left( \frac{\bfB \times \bfE}{\mu_0} \right) - \pdr_t \left( \frac{\eps_0 \bfE^2}{2} \right) \cr 
	&=& -\pdr_t \left( \frac{\eps_0 \bfE^2}{2} + \frac{\bfB^2}{2 \mu_0} \right) + \grad \cdot \left( \frac{\bfB \times \bfE}{\mu_0} \right).
	\eeqs
Evidently it is crucial that the current in Amp\`ere's law is taken as the swirl current $\bfj_*$ instead of $\bfj_{\rm flow}$ to obtain the local conservation law for swirl energy ${\cal E}^*$ (\ref{e:energy-density-two-fluid}).

\subsubsection{Conservation of energy in incompressible flow and regularization of $\bfB$}
\label{s:energy-cons-two-fluid-incompress}

For low acoustic Mach numbers $(M_l = |\bfv_l/c^s_{l}|) \ll 1$, the number densities $n_l$ are spatially and temporally constant to leading order. In this limit, the plasma motions while producing changes in $\bfE$ and $\bfB$ do not produce propagating EM waves. This is equivalent to dropping the displacement current in Maxwell's equations ($c \gg c^s_l$). For physical consistency we must take $\eps_0 \to 0$.

By taking $n_{i,e}$ and the regularizing lengths $\la_{i,e}$ to be constants and $\eps_0 \to 0$ we arrive at an incompressible two-fluid model. The continuity equations become $\grad \cdot  \bfv_{i,e} = 0$ and $\eps_0 \to 0$ in Gauss' law implies quasineutrality $(n_i \approx n_e \equiv n$, assuming $q_i = - q_e)$. The velocity equations are 		
	\beq
	\pdr_t \bfv_l + \bfw_l \times \bfv_l = - \grad \sigma_l + \frac{q_l}{m_l} (\bfE + \bfv_l \times \bfB) - \la_l^2 \left(\bfw_l + \frac{q_l}{m_l} \bfB \right) \times (\grad \times \bfw_l)
	\label{e:reg-mom-eqn-incompress-2flu}
	\eeq
where $\sigma_l = p_l/\rho_l + \half \bfv_l^2$ for $l = i, e$. In this limit Amp\`ere's law (\ref{e:maxwell-evol-eqns-two-fluid}) becomes $\grad \times \bfB = \mu_0 \bfj_*$. It follows from Section \ref{s:energy-cons-two-fluid} that upon dropping compressional and electric energies, the energy density, 
	\beq
	{\cal E}^*_{\rm inc} = \sum_{l} \left[\half \rho_l ({\bfv_l}^2  + \la_l^2 \bfw_l^2) \right] + \ov{2 \mu_0} \bfB^2
	\label{e:energy-density-two-fluid-incompressible}
	\eeq
satisfies a local conservation law with the energy current of (\ref{e:energy-current-2flu-compress}). As a consequence, the enstrophy of each species is bounded and velocity curls cannot become too large though there is no a priori bound on field curls.

Remarkably, (as indicated in \cite{thyagaraja-2}) there is another way  of defining the regularized {\it incompressible} two-fluid model (with $\la_i = \la_e = \la$) where the field gradient $\grad \times \bfB$ is also regularized along with $\grad \times \bfv$. This is achieved by keeping the velocity (\ref{e:reg-mom-eqn-incompress-2flu}) and Faraday equations unchanged but postulating that the current in Amp\`ere's law is the flow current $\bfj_{\rm flow} = n \sum_l q_l \bfv_l$ rather than the swirl current $\bfj_*$ (\ref{e:swirl-current-two-fluid}),
	\beq
	\grad \times \bfB = \mu_0 \, \bfj_{\rm flow} 
	\label{e:Ampere-Faraday-2fluid-jflow}.
	\eeq
Under these circumstances, we find a new conserved energy density 
	\beqs
	\tl {\cal E}^*_{\rm inc} = \sum_{l}\left[ \frac{\rho_l}{2} ({\bfv_l}^2  + \la^2 \bfw_l^2) \right] + \frac{\bfB^2}{2\mu_0} 
	+ \frac{\la^2(\grad \times \bfB)^2}{2 \mu_0} \quad
	\label{e:energy-density-two-fluid-incompress-jflow}
	\eeqs
and associated flux 
	\beqs
	\tl \bff &=& \sum_{l} \left[\sig_l \rho_l \bfv_l + \la^2 \rho_l \bfw_l \times \left( \bfv_{l*} \times \left(\bfw_l + \frac{q_l}{m_l} \bfB \right) \right)\right] \cr
	&& + \frac{\bfE \times \bfB}{\mu_0} + \la^2 \left[ \bfE \times (\grad \times \bfj_{\rm flow}) - \bfj_{\rm flow} \times (\grad \times \bfE) \right]
	\label{e:energy-current-incomp-curlB-curlE}
	\eeqs
satisfying a local conservation law $\pdr_t \tl {\cal E}^*_{\rm inc} + \grad \cdot \tl \bff = 0$. This regularization of incompressible flow is remarkable in that the $L^2$ norms of $\bfv, \bfB, \grad \times \bfv$ and $\grad \times \bfB$  are all bounded (say with decaying/periodic BCs). Since in addition, $\grad \cdot \bfv_{i,e} = \grad \cdot \bfB = 0$, we expect vortical singularities as well as singularities in magnetic field gradients to be regularized in this model. The $L^2$-norm of $\bfj_{\rm flow}$ is also bounded as a consequence of Amp\`ere's law (\ref{e:Ampere-Faraday-2fluid-jflow}).

To derive Eqs.~(\ref{e:energy-density-two-fluid-incompress-jflow}) and (\ref{e:energy-current-incomp-curlB-curlE}) we dot the velocity equations (\ref{e:reg-mom-eqn-incompress-2flu}) for each species with $\rho_l \bfv_l$ to get,
	\beq
	\frac{\rho_l}{2} \pdr_t \bfv_l^2 = -\rho_l \bfv_l \cdot \grad \sig_l + n q_l \bfv_l \cdot \bfE - \la_l^2 \rho_l \bfv_l \cdot \left[ \bfT^w_l + \frac{q_l}{m_l} \bfT^B_l \right].
	\eeq
As $\rho_l$ are constants and $\grad \cdot \bfv_l = 0$,
	\beq
	( {\cal KE}_l )_t + \grad \cdot (\sig_l  \rho_l \bfv_l ) = \bfj_{{\rm flow},l} \cdot \bfE - \la_l^2 \rho_l \bfv_l \cdot \left[ \bfT^w_l + \frac{q_l}{m_l} \bfT^B_l \right].
	\label{e:time-der-of-KE_l+PE_l-two-fluid-incompress}
	\eeq
To introduce the vortical energy density, we dot the curl of (\ref{e:reg-mom-eqn-incompress-2flu}) for each species with $\la_l^2 \rho_l \bfw_l$ to get
	\beq
	({\cal VE}_l)_t = \la_l^2 \rho_l \bfw_l \cdot \grad \times \left[ (\bfv_l \times \bfw_l)  
	+ \frac{q_l}{m_l} (\bfE + \bfv_l \times \bfB)  -  \la_l^2  \left(\bfT^w_l + \frac{q_l}{m_l} \bfT^B_l \right) \right].
	\eeq
Vector identities allow us to write
	\beqs
	 ({\cal VE}_l)_t
	&=& \la_l^2 \rho_l \bfv_l \cdot \left\{ \left(\bfw_l + \frac{q_l}{m_l} \bfB \right) \times \grad \times \bfw_l \right\}  + \bfE \cdot \grad \times (\la_l^2 n q_l \bfw_l) \cr
	&& + \la_l^2 \rho_l \grad \cdot \left[ \left( \bfv_l \times \bfw_l + \frac{q_l}{m_l} \left(\bfE + \bfv_l \times \bfB  \right)   - \la_l^2  \left(\bfT^w_l + \frac{q_l}{m_l} \bfT^B_l \right) \right) \times \bfw_l \right].
	\label{e:time-der-of-EE_l-two-fluid-incompress}
	\eeqs
Adding (\ref{e:time-der-of-KE_l+PE_l-two-fluid-incompress}) and (\ref{e:time-der-of-EE_l-two-fluid-incompress}) and summing over species we get 
	\beqs
	 \pdr_t ({\cal KE} + {\cal VE}) 
	 &+& \grad \cdot  \sum_l \left[ \sig_l \rho_l \bfv_l + \la_l^2 \rho_l \bfw_l \times \left( \bfv_l \times \bfw_l + \frac{q_l}{m_l} (\bfE + \bfv_l \times \bfB ) 
	- \la_l^2 \left(\bfT^w_l + \frac{q_l}{m_l} \bfT^B_l \right) \right) \right] \cr
	&=& \: \bfE \cdot \left[ \bfj_{\rm flow} + \bfj_{\rm twirl} \right].
	\label{e:incompr-dt-of-ke+ee}
	\eeqs
where $\bfj_{\rm twirl} = \sum_l \grad \times \grad \times \la_l^2 \bfj_{{\rm flow},l}$ (\ref{e:j-twirl-as-curlcurl}). The work done by $\bfE$ is got from (\ref{e:Ampere-Faraday-2fluid-jflow}) (abbreviating flow and twirl):
	\beqs
	\bfE \cdot \bfj_{\rm fl} 
	&=& -\pdr_t \left( \frac{\bfB^2}{2 \mu_0} \right) + \grad \cdot \left( \frac{\bfB \times \bfE}{\mu_0} \right) \quad \text{and} \cr
	\bfE \cdot \bfj_{\rm tw} &=&
	\sum_l \left[\grad \times \la_l^2 \bfj_{{\rm fl}, l} \cdot \grad \times \bfE  - \grad \cdot ( \bfE \times \grad \times \la_l^2\bfj_{{\rm fl}, l} ) \right] \cr
	&=& \sum_l \left[\la_l^2 \, \bfj_{{\rm fl},l} \cdot \grad \times (\grad \times \bfE) \grad \cdot \left(\la_l^2 \bfj_{{\rm fl},l} \times (\grad \times \bfE) - \bfE \times \grad \times \la_l^2\bfj_{{\rm fl}, l} \right) \right]. \qquad
	\label{e:electric-fld-dot-j-twirl}
	\eeqs
If we assume $\la_i = \la_e = \la$ (constant) then $\sum_l \la_l^2 \bfj_{{\rm fl},l} = \la^2 \bfj_{\rm fl} = (\la^2/\mu_0) \grad \times \bfB$, so that $\bfE \cdot \bfj_{\rm twirl}$ becomes 
	\beq
	- \left( \frac{\la^2 (\grad \times \bfB)^2 }{2 \mu_0} \right)_t + \grad \cdot (\la^2 \bfj_{\rm fl} \times (\grad \times \bfE) - \bfE \times \grad \times \la^2 \bfj_{\rm fl}).
	\label{e:time-der-constant-la-two-fluid-incompress}
	\eeq
Putting this in (\ref{e:incompr-dt-of-ke+ee}) we get the conservation of energy $\tl {\cal E}^*_{\rm inc}$ (\ref{e:energy-density-two-fluid-incompress-jflow}). Notably this trick of replacing $\bfj_*$ by $\bfj_{\rm flow}$ in Amp\`ere's law does {\it not} lead to a conserved energy for compressible flow: $\la_{i,e}$ are not constants and cannot be taken inside the derivatives in (\ref{e:time-der-constant-la-two-fluid-incompress}) to obtain a conserved energy including $(\grad \times \bfB)^2$. As mentioned in Section \ref{s:energy-cons-two-fluid}, for compressible flow, we must include the twirl current in Amp\`ere's law  to obtain the conserved swirl energy (\ref{e:energy-density-two-fluid}).

\subsubsection{Local conservation of linear and angular momenta}

Returning to the compressible two-fluid equations, we obtain a local conservation law $\pdr_t {\cal P}^\al + \pdr_\beta \Pi^{\al \beta} = 0$ for the total momentum density $\vec {\cal P} = \vec {\cal P}_{\rm mech} + \vec {\cal P}_{\rm field} = \sum_{l} \rho_l \bfv_l + \eps_0 (\bfE \times \bfB)$ and symmetric stress tensor,
	\beq
	\Pi^{\al \beta} = p \del^{\al \beta} + \sum_{l} \left[\rho_l v_l^\al v_l^\beta + \la_l^2 \rho_l \left[ \frac{\bfw_l^2 \del^{\al \beta}}{2}  - w_l^\al w_l^\beta \right] \right] + \left[ \frac{\bfB^2 \del^{\al \beta}}{2 \mu_0}  - \frac{B^\al B^\beta}{\mu_0} \right] + \left[ \frac{\eps_0 \bfE^2}{2} \del^{\al \beta} - \eps_0 E^\al E^\beta \right].
	\label{e:mom-flux}
	\eeq
Here $p = p_i + p_e$. The first and last pairs of terms, $\Pi^{\al \beta}_{\rm Euler}$ and $\Pi^{\al \beta}_{\rm field}$ in the flux are familiar from ideal flow and the Poynting flux of electrodynamics. The vortical regularization term in between is similar to the latter with the constants $\la_l^2 \rho_l$ playing the role of $\ov{{\mu_0}}$ and $\eps_0$.

To obtain (\ref{e:mom-flux}), we first multiply the continuity equation (\ref{e:reg-cont-eqn-two-fluid}) by $m_l \bfv_l$ and velocity equation (\ref{e:reg-elec-ion-mom-eqn}) by $\rho_l = n_{l} m_{l}$, add them and sum over species to get
	\beqs
	&& \sum_l \left[ (\rho_l \bfv_l)_t + \rho_l (\bfv_l \cdot \grad \bfv_l) + m_l \bfv_l  \grad \cdot (n_l \bfv_l) \right]	= - \grad p  \cr
	&& + \sum_l \left[ n_l q_l (\bfE + \bfv_l \times \bfB) - \la_l^2 \rho_l \bfw_l \times (\grad \times \bfw_l) - \la_l^2 n_l q_l \bfB \times (\grad \times \bfw_l) \right]. 
	\eeqs
Using Gauss' law, $\eps_0 \grad \cdot \bfE = \sum_l n_l q_l $ and the formulae for flow and twirl currents (\ref{e:swirl-current-two-fluid}) we get
	\beq
	\pdr_t {\cal P}_{\rm mech}^\al + \pdr_\beta \sum_l (\rho_l v_l^\al v_l^\beta) = - \grad^\al p  + \eps_0 E^\al (\grad \cdot \bfE) + (\bfj_* \times \bfB)^\al - \sum_l \la_l^2 \rho_l (\bfw_l \times (\grad \times \bfw_l))^\al.
	\eeq
From Amp\`ere's law $\mu_0 \bfj_* \times \bfB = (\grad \times \bfB) \times \bfB -  \mu_0 \eps_0(\pdr_t \bfE) \times \bfB$ and Faraday's law we get
	\beqs
	\pdr_t {\cal P}_{\rm mech}^\al + \pdr_\beta \Pi^{\al \beta}_{\rm Euler} &=& \eps_0 E^\al \: \grad \cdot \bfE -\ov{\mu_0} (\bfB \times (\grad \times \bfB))^\al - \eps_0 (\pdr_t ( \bfE \times \bfB) + \bfE \times (\grad \times \bfE))^\al 
	\cr
	&& - \sum_l \la_l^2 \rho_l (\bfw_l \times (\grad \times \bfw_l))^\al.
	\eeqs
Using the identity $(\bfS \times (\grad \times \bfS))^\al = \half \pdr^\al \bfS^2 - S^\beta \pdr^\beta S^\al$ and solenoidal nature of $\bfB$ and $\bfw$ we get 
	\beqs
	\pdr_t {\cal P}^\al &+& \pdr_\beta \left[ \Pi^{\al \beta}_{\rm Euler} + \ov{\mu_0}\left(\frac{\bfB^2}{2} \del^{\al \beta} - B^\al B^\beta \right) + \sum_l  \la_l^2 \rho_l \left(\frac{\bfw_l^2}{2} \del^{\al \beta} - w_l^\al w_l^\beta \right) \right] 
	\cr &&
	= \eps_0 \left[ E^\al (\grad \cdot \bfE) - \half \pdr^\al \bfE^2  + E^\beta \pdr^\beta E^\al \right]
	\eeqs
which implies the local conservation law (\ref{e:mom-flux}).

The time derivative of angular momentum density $\vec {\cal L} =  \bfr \times \vec {\cal P} =  {\bf r} \times  \left( \sum_l \rho_l \bfv_l + \eps_0 \bfE \times \bfB \right)$ is calculated using the local conservation law for momentum density and the symmetry of $\Pi_{\al \beta} (\ref{e:mom-flux})$:
	\beq
	\dd{{\cal L}_\al}{t} = \eps_{\al \beta \gamma} r_\beta \pdr_t {\cal P}^\gamma
	= - \eps_{\al \beta \gamma} r_\beta \pdr_\eta \Pi_{\gamma \eta} = - \pdr_\eta \Lambda_{\al \eta}.
	\eeq
Thus $\pdr {\cal L}_\al/ \pdr t + \pdr_\beta \Lambda_{\al \beta} = 0$ where $\Lambda_{\al \beta} = \eps_{\al \gamma \del} r_\gamma \Pi_{\del \beta} $ is the angular momentum flux tensor.

\section{Hierarchy of regularized models}
\label{s:heirarchy-of-models}

The regularized compressible $2$-fluid plasma equations have several free parameters $\eps_0, m_e/m_i$, electric charge $e$ and $\la_i/\la_e$. By successively taking (i) $\eps_0 \to 0$, (ii) $m_e/m_i \to 0$ and (iii) $e \to \infty$ together with $\la_i / \la_e \to 1$ we get the regularized quasineutral two-fluid, Hall and one-fluid MHD models.

\subsection{Regularized quasineutral two-fluid plasma}
\label{s:quasineutral-two-fluid}

For quasineutral plasmas with $q_i = -q_e = e$, the number densities of ions and electrons are approximately equal, $n_i \approx n_e = n$. The equations of such a plasma may be formally obtained from the compressible two-fluid model (Section \ref{s:reg-eqns-two-fluid-compress}) by taking $\eps_0 \to 0$. Indeed, if $n_i, n_e \to n$, Gauss' law $\grad \cdot \bfE = e (n_i - n_e)/\eps_0$ seems to suggest that $\grad \cdot \bfE = 0$. But in fact, the electric field is not divergence free (especially on length scales comparable to the Debye length). We must also let $\eps_0 \to 0$ in such a way that $e (n_i - n_e)/\eps_0$ has a finite limit. The limit $\eps_0 \to  0$ is a convenient way of taking the non-relativistic limit $c = 1/\sqrt{\eps_0 \mu_0} \to \infty$ ($\mu_0$ is a constant) in which $v_{i,e}/c \ll 1$ in the lab frame. In this limit $\bfE$ is not a propagating degree of freedom and we may ignore the displacement current term in Amp\`ere's law (as stated in \ref{s:energy-cons-two-fluid-incompress}). Furthermore, $\bfE$ is no longer determined by Gauss' law but obtained from the electron velocity equation as discussed below.

In the non-relativistic quasineutral limit $\eps_0 \to 0$, the Faraday and Amp\`ere-Maxwell equations become
	\beq
	\grad \cdot \bfB = 0, \quad \dd{\bfB}{t} = - \grad \times \bfE \quad \text{and} \quad
	\grad \times \bfB = \mu_0 \bfj_*.
	\label{e:Maxwell-eqn-quasineutral}
	\eeq
For consistency, $\grad \cdot \bfj_*$ must vanish as we will verify using the continuity equations 
	\beq
	\pdr_t n + \grad \cdot (n \bfv_{i,e}) = 0.
	\label{e:cont-eqn-quasineutral}
	\eeq
The difference between the continuity equations gives
	\beq
	\grad \cdot n( \bfv_i - \bfv_e) = 0.
	\eeq
Multiplying by $e$, we see that the flow current $\bfj_{{\rm flow}} = e n (\bfv_i -\bfv_e)$ is solenoidal. On the other hand, the twirl current $\bfj_{\rm twirl} = \sum_l \grad \times (\grad \times \la_l^2 \bfj_{{\rm flow},l})$ is always divergence free, so the total current $\bfj_* = \bfj_{\rm flow} + \bfj_{\rm twirl}$ for quasineutral plasmas is solenoidal. This also follows from the Amp\`ere-Maxwell equation when $\eps_0 \to 0$.

The ion and electron velocity equations $(l = i, e)$ for quasineutral plasmas are
	\beq
	\pdr_t \bfv_l + \bfw_l \times \bfv_{*l} = - \frac{\grad p_l}{m_l n} - \frac{\grad \bfv_l^2}{2} \pm \frac{e}{m_l} (\bfE + \bfv_{*l} \times \bfB).
	\label{e:reg-ion-elec-mom-quasineutral}
	\eeq
$\bfE$ is determined from the electron velocity equation:
	\beq
	\bfE_{\rm qn} = -  \bfv_{*e} \times \bfB - \frac{\grad p_e}{en} - \frac{m_e}{e}\left [ \pdr_t \bfv_e + \bfw_e \times \bfv_{*e} + \frac{ \grad \bfv_e^2}{2} \right].
	\label{e:E-from-e-mom-eqn}
	\eeq
The relation between general and quasineutral two-fluid plasmas bears a resemblance to that between compressible and incompressible  barotropic neutral flows. In compressible flow, pressure $p$ is obtained from density $\rho$ using the barotropic relation. Similarly, in general two-fluid plasmas $\bfE$ is determined in terms of the charge density from Gauss' law. On the other hand, in the incompressible ($\grad \cdot \bfv = 0$) constant density $(\rho = \rho_0)$ limit, $p$ is no longer determined by the barotropic relation but from the Poisson equation $ [\grad^2 p = -\rho_0 \grad \cdot (\bfv \cdot \grad \bfv)]$ obtained by taking the divergence of the velocity equation. Similarly, in quasineutral plasmas, $\bfE$ is determined from the electron velocity equation rather than from Gauss' law. Moreover, $\eps_0 \to 0$ ($c \to \infty$) is like taking the Mach number to zero (sound speed $c_s \to \infty$).

In this limit, the electric term drops out of the conserved swirl energy for barotropic flow generalizing (\ref{e:energy-density-two-fluid-incompressible}):
	\beq
	{\cal E}_{\rm qn}^* = \sum_{l = i, e} \left(\frac{\rho_l{\bfv_l}^2}{2} + U_l(\rho_l) + \frac{\la_l^2 \rho_l \bfw_l^2}{2} \right) + \frac{\bfB^2}{2\mu_0}.
	\eeq
Here $\rho_l = m_l n$ and $\grad U_{l}' = \grad h_l = \grad p_l/\rho_l$ for $l = i,e$.

\subsection{Regularized Hall MHD without electron inertia}
\label{s:R-Hall-2fluid}

In the limit $m_e/m_i \ll 1$ we drop electron inertia terms to get the regularized Hall model. The Maxwell equations, continuity equations and ion velocity equation are as in the quasineutral theory (Section \ref{s:quasineutral-two-fluid}). In (\ref{e:E-from-e-mom-eqn}) we drop electron inertia terms to get
	\beq
	\bfE_{\rm Hall} = -  \bfv_{*e} \times \bfB - \frac{\grad p_e}{en}
	\label{e:hall-electric-field}
	\eeq
For barotropic flow, where $\grad p_e/n$ is a gradient, Faraday's law becomes $\pdr_t \bfB = \grad \times (\bfv_{*e} \times \bfB)$. Thus unlike in the full two-fluid model, in the R-Hall model the magnetic field is frozen into the electron swirl velocity.

We have an additional conserved quantity: magnetic helicity satisfies the local conservation law
	\beq
	\pdr_t(\bfA \cdot \bfB) + \grad \cdot \left( \phi \bfB + \bfE_{\rm Hall} \times \bfA - \frac{2 \tl h_e \bfB}{e} \right) = 0.
	\label{e:mag-hel-loc-cons-Hall-2fluid}
	\eeq
Here $\phi$ is the scalar potential and we assume the barotropic condition $(\grad p_e)/n = \grad \tl h_e$. To obtain (\ref{e:mag-hel-loc-cons-Hall-2fluid}), we use the homogeneous Maxwell equations and $\bfE = -\grad \phi - \pdr_t \bfA$ to compute
	\beq
	(\bfA \cdot \bfB)_t =  - \bfB \cdot \grad \phi - \bfB \cdot \bfE - \bfA \cdot \grad \times \bfE 
	= - \grad \cdot ( \phi \bfB + \bfE \times \bfA) - 2 \bfE \cdot \bfB.
	\eeq
Using the quasineutral electric field (\ref{e:E-from-e-mom-eqn}) we get
	\beqs
	(\bfA \cdot \bfB)_t &=& - \grad \cdot (\phi \bfB + \bfE_{\rm qn} \times \bfA) + 2 \left[(\bfv_{* e} \times \bfB) +  \frac{\grad p_e}{en} + \frac{m_e}{e} \left[ \pdr_t \bfv_e + \bfw_e \times \bfv_{*e} + \frac{\grad \bfv_e^2}{2} \right] \right] \cdot \bfB\cr
	&=& - \grad \cdot \left[ \phi \bfB + \bfE_{\rm qn} \times \bfA - \frac{2 \tl h_e \bfB}{e} \right] 
	+ \frac{2 m_e}{e} \left[ \pdr_t \bfv_e + \bfw_e \times \bfv_{*e} + \frac{\grad \bfv_e^2}{2} \right] \cdot \bfB.
	\eeqs
When electron inertia terms are ignored, we see that $\bfE_{\rm qn} \to \bfE_{\rm Hall}$ and magnetic helicity satisfies the local conservation law (\ref{e:mag-hel-loc-cons-Hall-2fluid}). The regularization enters through the electron `swirl' velocity $\bfv_{* e}$ in (\ref{e:hall-electric-field}).

However, even in the Hall ($m_e \to 0$) limit, we do not have an analogue of a conserved cross helicity $\bfv \cdot \bfB$ of R-MHD. For instance, using the electron velocity equation (\ref{e:reg-ion-elec-mom-quasineutral}) and the homogeneous Maxwell equations we find
	\beq
	\pdr_t (\bfv_e \cdot \bfB) = - \bfv_e \cdot \grad \times \bfE - \grad \cdot (\sig_e \bfB) + \bfB \cdot \bfv_{*e} \times (\grad \times \bfv_e) - \frac{e}{m_e} \bfE \cdot \bfB.
	\eeq
Substituting for $\bfE_{\rm qn}$ (\ref{e:E-from-e-mom-eqn}), combining terms and taking $m_e \to 0$, we find that unlike for magnetic helicity, the final offending term is not suppressed by $m_e$.
	\beq
	(\bfv_e \cdot \bfB)_t + \grad \cdot (\bfv_{*e} (\bfv_e \cdot \bfB)) = \bfB \cdot \left ( \pdr_t \bfv_e + \bfw_e \times \bfv_{*e} + \grad(\bfv_e \cdot\bfv_{*e}) \right).
	\eeq

\subsection{From R-Hall to one-fluid R-MHD when $e \to \infty$}

To get the regularized one-fluid MHD model of \cite{govind-sonakshi-thyagaraja-pop} from the above R-Hall two-fluid model we let $e \to \infty$, holding $\la_i$ and $\la_e$ fixed. The limit $e \to \infty$ is a convenient way of restricting attention to frequencies small compared to the cyclotron $\om_{c,l} = e B/ m_l$ and plasma $\om_{p,l} = \sqrt{{n_l e^2}/{m_l \eps_0}}$ frequencies and to length scales large compared to the Debye lengths $\la_{D,l} = \sqrt{{ k_B T_l \eps_0}/{n_l e^2}}$,  gyroradii $r_l = v_{th,l}/\om_{c,l} = \sqrt{k_B T_l m_l}/eB$ and collisionless skin depths $\del_l = c/\om_{p,l} = \sqrt{{m_l}/{\mu_0 n_l e^2}}$.

To switch to one-fluid variables we express $\bfv_i$ and $\bfv_e$ in terms of center of mass velocity $\bfv = (m_i \bfv_i + m_e \bfv_e)/m$ and $\bfj_{\rm flow} = e n (\bfv_i - \bfv_e)$
	\beq
	\bfv_{i,e} = \bfv \pm \frac{m_{e,i}}{m} \frac{\bfj_{\rm flow}}{e n}.
	\label{e:v_i-v_e-to_v-j}
	\eeq
Here $m = m_i + m_e$. The continuity equation $\pdr_t \rho = -\grad \cdot (\rho \bfv)$ for the total mass density $\rho = nm$ is obtained by taking a mass-weighted average of the continuity equations in (\ref{e:cont-eqn-quasineutral})
	\beq
	\pdr_t ((m_i + m_e )n) =  -\grad \cdot (n m_i v_i + n m_e v_e)
	\eeq
The evolution equation for the center of mass velocity $\bfv$ is similarly  obtained from (\ref{e:reg-ion-elec-mom-quasineutral}),
	\beq
	\bfv_t + \frac{m_i}{m}\bfw_i \times \bfv_{*i} + \frac{m_e}{m}\bfw_e \times \bfv_{*e} = - \ov{n m} \grad (p_i + p_e) -\frac{1}{2m} \grad (m_i \bfv_{i}^2 + m_e \bfv_{e}^2)  + \frac{e}{m}(\bfv_{*i} - \bfv_{*e}) \times \bfB.
	\eeq
Neglecting terms of order $m_e/m \ll 1$ and introducing $\bfj_* = e n (\bfv_{*i} - \bfv_{*e})$ and $p = p_i + p_e$ we get 
	\beq
	\pdr_t \bfv + \bfw_i \times \bfv_{*i} = -  \ov{\rho} \grad p - \half \grad \bfv_{i}^2 + \frac{1}{\rho}(\bfj_* \times \bfB).
	\eeq
Next we take the limit $e \to \infty$ in (\ref{e:v_i-v_e-to_v-j}) keeping $\bfj_{\rm flow}$ finite so that $\bfv, \bfv_i$ and $\bfv_e$ are all equal, as are $\bfw, \bfw_i$ and $\bfw_e$. Defining $\la = \la_i $, $\bfv_{*i}$ = $\bfv_* = \bfv + \la^2 \grad \times \bfw$. Thus, we arrive at the velocity equation for one-fluid R-MHD,
	\beq
	\pdr_t \bfv + \bfw \times \bfv_{*} = -  \ov{\rho} \grad p - \half \grad \bfv^2 + \frac{1}{\rho}(\bfj_* \times \bfB).
	\label{e:one-fluid-reg}
	\eeq
However unlike in the two-fluid model $\bfj_*$ is no longer given by $e n (\bfv_{*i} - \bfv_{*e})$. Instead, it is obtained from Amp\`ere's law $\mu_0 \bfj_* = \grad \times \bfB$. On the other hand, taking the limit $e \to \infty$ in the Hall electric field (\ref{e:hall-electric-field}) the pressure gradient term drops out and we get
	\beq
	\bfE_{\rm one-fluid} = - \bfv_{*e} \times \bfB = - \bfv_{*} \times \bfB.
	\eeq
This identification of $\bfv_{*e}$ with the one-fluid swirl velocity $\bfv_*$ requires that $\la_e = \la$. Thus, to get the one-fluid R-MHD model we need to take $\la_i = \la_e = \la$. Finally, Faraday's law (\ref{e:Maxwell-eqn-quasineutral}) becomes $\pdr_t \bfB = \grad \times (\bfv_* \times \bfB)$ implying that the solenoidal $\bfB$ is frozen into $\bfv_* $.

\section{Poisson brackets for regularized compressible two-fluid plasmas}
\label{s:PB-two-fluid}

Poisson brackets for (unregularized) two-fluid plasmas were proposed by Spencer and Kaufman \cite{spencer-kaufman} and Holm and Kuperschmidt \cite{holm-kuperschmidt}. The non-trivial PBs are given by 
	\beqs
	&& \{v_l^{\al} (x), v_l^{\beta} (y) \} = \frac{\eps^{\al \beta \gamma}}{m_l n_l} \left( w_l^{\gamma} + \frac{q_l B^{\gamma}}{m_l}\right) \del( x - y), \;\{\bfv_l(x), n_l(y) \} \
	= \frac{\grad_y}{m_l} \del( x - y),
	\cr 
	&& \{E^{\al} (x), B^{\beta} (y) \} = \frac{\eps^{\al \beta \gamma}} {\eps_0} \pdr_{y^{\gamma}}\del( x - y)
	 \quad \text{and} \quad \{v_l^{\al} (x), E^{\beta} (y) \} = \frac{q_l}{m_l \eps_0} \del^{\al \beta}\del(x - y).
	\eeqs
Here, $l = i,e$ labels species while $\al, \beta ,\gamma$ label Cartesian components. The velocity PBs for a given species are obtained from the Landau PBs $\{v^{\al}, v^{\beta} \} = \eps^{\al \beta \gamma} w^{\gamma} \del(x -y)/\rho$ of fluid mechanics by replacing $\bfw$ by $\bfw + q \bfB/m$ and $\rho$ by $m n$ for each species. This is reminiscent of the results established in \cite{thyagaraja-mc-clements-2}, already mentioned. Similarly, $\{ \bfv_l, n_l \}$ is obtained from Landau's PB $\{ \bfv(x) , \rho(y) \} = \grad_y \del(x-y)$. The rest of the PBs vanish $\{\bfB(x) , \bfB(y) \} = \{v_l, \bfB \} = \{\bfB , n_l \} = \{\bfE , n_l \} = \{\bfE , \bfE \} = \{n_l , n_{l'} \}  = \{\bfv_i , \bfv_e \} = \{\bfv_e, n_i \} = \{\bfv_i , n_e  \}  = 0$. In particular, unlike in one-fluid MHD \cite{landau,morrison-greene,govind-sonakshi-thyagaraja-pop}, velocities and $\bfB$ commute. Vorticity behaves in a manner similar to $\bfB$: $\{ \bfw_l, n_{l'} \} = \{ \bfw_l, \bfB \} = 0$; $\{ \bfE, \bfw_l \}$ is similar to $\{ \bfE, \bfB \}$:
	\beq 
	\{ E^\al(x) , w_l^\beta(y) \} = \frac{\eps^{\al \beta \gamma} q_l}{\eps_0 m_l} \pdr_{y^{\gamma}}\del(x - y).
	\eeq
Our twirl regularization is natural in the sense that the regularized equations follow from these PBs with the swirl energy (\ref{e:energy-density-two-fluid}) as Hamiltonian. We sketch how this happens. It follows from the PBs that only the kinetic energies contribute to the continuity equations,
	\beqs
	\pdr_t n_l(x) &=& \{ n_l, KE_l \} = \int m_l n_l \bfv_l \cdot \{ n_l(x) , \bfv_l(y) \} dy
	\cr
	&=& \int n_l \bfv_l \cdot \grad_y \del(x-y) = - \grad \cdot (n_l \bfv_l).
	\eeqs
To obtain the velocity equations we note that the following relations hold for the electric (EE), kinetic (KE$_l$), compressional (PE$_l$) and vortical (VE$_l$) energies:
	\beqs
	\{ \bfv_l(x) , {\rm EE} \} &=& \eps_0 \int E^\beta (y) \{ \bfv_l(x) , E^\beta(y) \} dy = \frac{q_l}{m_l} \bfE, \cr
	 \{ \bfv_l(x) , {\rm PE}_l \} &=& \int U'_l \{ \bfv_l(x) , \rho_l (y) \} dy = - \grad U'_l = -\grad h_l,
	\{ \bfv_l(x) , {\rm KE}_l \} \cr
	&=& \int \left( \rho_l v_l^\beta(y) \{ \bfv_l (x), v_l^\beta(y) \}  + \frac{\bfv_l^2}{2} \{ \bfv_l (x), \rho_l(y) \} \right) dy 
	\cr
	&=& \bfv_l \times  \left( \bfw_l + \frac{q_l \bfB}{m_l} \right) - \half \grad \bfv_l^2 
	 \cr
	\{ v_l^\al(x) , {\rm VE}_l \} &=& \la_l^2 \rho_l  \int w_l^\beta(y) \eps_{\beta \g \del } \pdr_{y^\g} \{ v_l^\al (x), v_l^\del (y) \} dy = - \eps_{\al \eta \del} \la_l^2 \left( w_l^{\eta} + \frac{q_l B^{\eta}}{m_l}\right) \eps_{\del \g \beta} \pdr_\g  w_l^\beta 
	\cr
	&=& - \la_l^2 \left[ \left( \bfw_l + \frac{q_l \bfB}{m_l}\right) \times  (\grad \times  \bfw_l) \right]^\al.
	\eeqs
Thus, using $\sig_l = h_l + \half \bfv_l^2$, we get the velocity equations (\ref{e:reg-mom-eq-barotropic-Tw-TB}) for $l=i,e$. If $\{ \bfv_i, n_e \} \ne 0$, the electron pressure would contribute to the ion velocity equation. Faraday's law receives a contribution only from the electric energy:
	\beq
	\pdr_t \bfB(x) 
	=  \eps_0 \int \bfE(y) \cdot \{ \bfB(x), \bfE(y) \} \: dy = - \grad \times \bfE.
	\eeq
Only KE, VE and magnetic energy (ME) contribute to Amp\`ere's law:
	\beqs
	\{ \bfE(x), {\rm KE}_l \} &=& m_l \int n_l v_l^\al \{ \bfE(x), v^\al_l(y) \} dy
	= -\frac{\bfj_{{\rm flow},l}}{\eps_0}, \cr
	\{ \bfE(x), {\rm VE}_l \} &=& \la_l^2 n_l m_l \int w_l^\al \{ \bfE(x), w^\al_l(y) \} dy 
	= - \frac{\la_l^2 n_l q_l}{\eps_0} (\grad \times \bfw_l)
	= -\frac{\bfj_{{\rm twirl},l}}{\eps_0} \;\; \text{and}
	\cr
	\{ \bfE(x), {\rm ME} \} &=& \int \frac{B^\al}{\mu_0} \{ \bfE(x), B^\al(y) \} dy = \frac{\grad \times \bfB}{\mu_0 \eps_0}.
	\label{e:E-fld-PBs-two-fluid}
	\eeqs
Combining, we see that the swirl current $\bfj_*$ in Amp\`ere's law is the sum of flow and twirl currents:
	\beq
	\pdr_t \bfE = - \ov{\eps_0} \sum_{l} \left( \bfj_{{\rm flow},l} + \bfj_{{\rm twirl},l} \right) + \ov{\mu_0 \eps_0} \grad \times \bfB.
	\label{e:ampere-law-from-pb-reg-two-fluid}
	\eeq

\section{Regularization of $\grad \times \bfB$ in single and two-fluid models}
\label{s:reg-field-curl-PBs-Hamiltonian}

The twirl terms $\bfw_l \times (\grad \times \bfw_l)$ and $\bfB \times (\grad \times \bfw_l)$ in the EOM and the corresponding vortical energies $\half \la_l^2 n_l m_l \bfw_l^2$ can smooth out large velocity gradients and regularize vortical singularities. Similarly, we would like to identify appropriate terms in the EOM to regularize magnetic field gradients and current sheets. Recall from Section \ref{s:energy-cons-two-fluid-incompress} that in the quasineutral incompressible case the term $(\la^2/2\mu_0) (\grad \times \bfB)^2$ automatically arose in the conserved energy if the current in Amp\`ere's law is chosen to be the flow current $\bfj_{\rm flow}$ and $\la_i = \la_e = \la$. This approach however does not generalize to compressible flow. In the compressible case, the current in Amp\`ere's law must be the swirl current $\bfj_*$ to guarantee energy conservation. On the other hand, the Poisson bracket formulation gives us a natural way of introducing field gradient energies in compressible flow. Adding the simplest possible positive definite magnetic gradient energy (MGE) term $\int \la_B^2 (\grad \times \bfB)^2/{2 \mu_0} \, d\bfr$ to the Hamiltonian of the single and two-fluid models and using the relevant PBs to obtain the EOM, we ensure the $L^2$ boundedness of $\grad \times \bfB$.

\subsection{Regularization of $\grad \times \bfB$ in R-MHD}
\label{s:one-fluid-curl-B-reg}

We augment the R-MHD Hamiltonian with a magnetic gradient energy taking $\la_B$ to be a constant cut-off length
	\beq
	H = \int \left[ \frac{\rho \bfv^2}{2}  + U + \frac{\la^2 \rho \bfw^2}{2}  + \frac{\bfB^2}{2 \mu_0} + \frac{\la_B^2}{2 \mu_0} (\grad \times \bfB)^2 \right] \: d\bfr.
	\eeq
Using the non-trivial one-fluid PBs \cite{landau,morrison-greene}, \; $\{ \rho(x), \bfv(y) \} = \grad_\bfy \del(x-y)$,
	\beq
	\{ v_\al(x), v_\beta(y) \} =\frac{\eps_{\al \beta \g} w_\g}{\rho} \del(x-y) \quad \text{and}  \quad \{ v_\al(x) , B_\beta(y) \} = \frac{\eps_{\al \g \sig} \eps_{\beta \eta \sig}}{\rho(x)} B_\g(x) \pdr_{x^\eta} \del(x-y),
	\eeq
the continuity and Faraday equations are unchanged
	\beq
 	\pdr_t \rho + \grad \cdot (\rho \bfv) = 0 \quad \text{and} \quad 
	\pdr_t \bfB = \grad \times (\bfv_* \times \bfB).
	\eeq
On the other hand, the velocity equation is modified by
	\beqs
	 \{v_{\al} (x), {\rm MGE} \} &=& \frac{\la_B^2}{2 \mu_0}  \int \{ v_{\al} (x), (\grad \times \bfB)^2  \} dy \cr
	 &=& \frac{\la_B^2}{\mu_0 \rho} \: \eps_{jkl} \eps_{\al mn}\eps_{lpn} B_m \pdr_{x^p} \int \left[(\grad \times \bfB)_j \pdr_{y^k} \del(x - y) \right] dy \cr
	 &=& - \frac{\la_B^2}{\rho \mu_0} \left[\bfB \times \left( \grad \times \left( \grad \times (\grad \times \bfB) \right) \right) \right]_\al.
	\eeqs
Combining this with contributions from  kinetic, potential, vortical and magnetic energies, the velocity equation takes the same form as (\ref{e:one-fluid-reg}) with $\bfj_*$ replaced by the regularized `magnetic swirl' current 
	\beq
	\mu_0 \bfj_{**} = \grad \times \bfB + \la_B^2 \grad \times \left( \grad \times \left(\grad \times \bfB \right) \right) = (1 - \la_B^2 \grad^2) (\grad \times \bfB).
	\label{e:magnetic-swirl-current}
	\eeq
Evidently, $\mu_0 \bfj_{**}$ is the magnetic analogue of $\bfv_* = \bfv + \la^2 \grad \times (\grad \times \bfv)$. Furthermore, $\grad \times \bfB$ is a smoothed version of the regularized current obtained through the application of the integral operator $(1 - \la_B^2 \grad^2)^{-1}$:
	\beq
	\grad \times \bfB = \mu_0 (1 - \la_B^2 \grad^2)^{-1} \, \bfj_{**}.
	\eeq
A similar smoothing operator appears in the {\it non-local} Euler-$\alpha$ equations \cite{holm-marsden-ratiu}. As noted in the introduction, these additional terms in the velocity and Faraday equations are quite different from those that appear in XMHD \cite{kimura-morrison,abdelhamid-kawazura-yoshida}. The latter involves the introduction of a $\bfB^* = \bfB + d_e^2 \grad \times ((\grad \times \bfB)/\rho)$  where $d_e$ is a constant normalized electron skin depth, rather than a swirl current $\bfj_{**}$. For instance, this leads to a new term $\bfj \times \bfB^*$ in both the velocity equation and in the electric field in XMHD.

\subsection{Regularization of field curl in the two-fluid model}

As for the single fluid, we augment the two-fluid Hamiltonian (\ref{e:energy-density-two-fluid}) with a magnetic gradient energy:
	\beq
	H = \int \left[ \sum_l \left( \half m_l n_l \left( \bfv_l^2  + \la_l^2 \bfw_l^2 \right) + U_l(\rho_l) \right) + \frac{\bfB^2}{2 \mu_0} + \frac{\eps_0 \bfE^2}{2} + \ov{2 \mu_0} \la_B^2 (\grad \times \bfB)^2 \right] \: d\bfr.
	\eeq
Like before, $\la_B$ is a constant cut-off length. Using the two-fluid PBs of Section \ref{s:PB-two-fluid}, we see that the momentum, continuity and Faraday equations remain unchanged since $\bfv_i, \bfv_e, n_i, n_e$ and $\bfB$ commute with the magnetic field. We do not introduce a $(\grad \times \bfE)^2$ term in $H$ as it would modify Faraday's law. The evolution equation for the electric field is modified by the term:
	\beqs
	\left\{ \bfE(x), {\rm MGE} \right\}
	= \frac{\la_B^2}{\mu_0 \eps_0}\grad \times (\grad \times (\grad \times \bfB))
	= - \frac{\bfj_B}{\eps_0}. \quad \;
	\label{e:j_B-defn-two-fluid}
	\eeqs
Combining with (\ref{e:E-fld-PBs-two-fluid}), Amp\`ere's law (\ref{e:ampere-law-from-pb-reg-two-fluid}) becomes
	\beq
	\mu_0 \eps_0 \pdr_t \bfE = \grad \times \bfB - \mu_0 \bfj_* - \mu_0 \bfj_B.
	\eeq
Here, $\bfj_* = \bfj_{\rm flow} + \bfj_{\rm twirl}$. Now, we can define a new current density $\bfj_{**} = \bfj_* + \bfj_B$. Note that (\ref{e:j_B-defn-two-fluid}) implies $\grad \cdot \bfj_B = 0$. Thus $\bfj_B$ and $\bfj_{\rm twirl}$ are like magnetization currents in material media/plasmas. We notice that the introduction of the MGE in the Hamiltonian has apparently very different effects in the single and two-fluid models. In the former, the velocity equation is modified while it is the Amp\`ere equation that is modified in the latter. However, the two are closely related. In fact, upon taking the limits $\eps_0 \to 0,m_e \to 0 $ and $e \to \infty$, the two-fluid current density $\bfj_{**}$ exactly matches the magnetic swirl current (\ref{e:magnetic-swirl-current}) appearing in the Lorentz force term of the single fluid velocity equation.

\chapter{Dispersive regularization of inviscid gas dynamics}
\label{s:r-gas-dynamics}

\begin{figure}
	\begin{center}
 	\includegraphics[width=10cm]{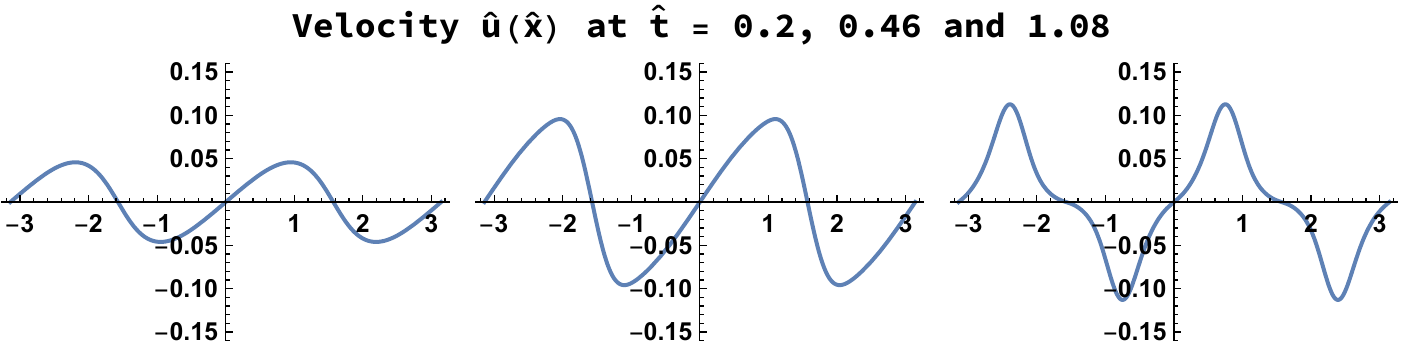}
 	\caption{\footnotesize Numerical evolution of velocity showing how the gradient catastrophe is averted through the formation of a pair of solitary waves in 1D with periodic boundary conditions and initial condition $\hat \rho = 1 + 0.1 \cos 2\hat x$ and $\hat u = 0$ in dimensionless (hatted) variables \cite{govind-sachin-sonakshi-thyagaraja-r-gas-dynamics}.}
	\label{f:gradient-catastrophe-r-gas-soliton}
 	\end{center}
	\end{figure}

In this Chapter we give a very brief summary of our work in \cite{govind-sachin-sonakshi-thyagaraja-r-gas-dynamics} where we have extended the idea of a conservative regularization from vortical singularities to shock-like singularities with discontinuous density/velocity in ideal gas dynamics. This subject is treated in detail in \cite{sachin-thesis}. Viscosity typically regularizes such singularities and leads to a shock structure. On the other hand, in 1D, singularities in the Hopf equation $u_t + u u_x = 0$ can be non-dissipatively smoothed via a Korteweg-de Vries (KdV) dispersion term $u_{xxx}$ \cite{dauxois-peyrard}. We have extended this idea to a minimal conservative regularization of 3D ideal adiabatic flow of a gas with polytropic exponent $\gamma$. It may be regarded as a way of extending the single-field KdV equation to include the dynamics of density, velocity and pressure and also to dimensions higher than one. It is achieved by augmenting the ideal gas dynamics Hamiltonian by a capillarity energy $\beta(\rho) (\nabla \rho)^2$:
	\beq
    H = \int {\cal E} \; d\bfr \equiv  \int \left[ \half \rho \bfv^2 + \frac{p}{\g-1} + \beta(\rho) \frac{(\grad \rho)^2}{2} \right] d\bfr.
    \label{e:3d-hamiltonian}
	\eeq
Such a term arose in the work of van der Waals and Korteweg \cite{vdW,korteweg,dunn-serrin,gorban-karlin,huang-wang-wang-yang} in the context of capillarity, but can be important even away from interfaces in any region of rapid density variation, especially when dissipative effects are small, such as in weak shocks, cold atomic gases, superfluids and collisionless plasmas. The regularized gas dynamics equations follow from the above Hamiltonian and the standard Poisson brackets \cite{morrison-greene}:
	\beq
	\{ \bfv(\bfx), \rho(\bfy) \} = \grad_y \del(\bfx - \bfy), \;\;\;
	\{ \bfv (\bfx), s(\bfy) \} = \frac{\grad s}{\rho} \del ( \bfx - \bfy) \quad \text{and} \quad \{ v_i(\bfx), v_j(\bfy) \} = \frac{\eps_{ijk} w_k}{\rho} \del (\bfx - \bfy).
	\eeq
We find that the simplest capillarity coefficient leading to local conservation laws for mass, momentum, energy and entropy is $\beta(\rho) = \beta_*/\rho$ for constant $\beta_*$ with dimensions $L^4 T^{-2}$. It can be taken as $\la^2 c^2$ where $\la$ is a short-distance cut-off (say shock-width) and $c$ a typical speed (say sound speed). The resulting continuity, velocity and energy equations are
	\beqs
	&&\rho_t + \grad \cdot (\rho \bfv) = 0, \quad \bfv_t + \bfv \cdot \grad \bfv + \frac{\grad p}{\rho} 
	= \beta_* \grad \left[ \half \frac{(\grad \rho)^2}{\rho^2} + \grad \cdot \left( \frac{\grad \rho}{\rho} \right) \right] = 2 \beta_* \grad \left[ \frac{\grad^2 \sqrt{\rho}}{\sqrt{\rho}} \right]\cr
	&& \text{and} \quad {\cal E}_t + \grad \cdot \left[ \frac{\rho \bfv^2}{2} \bfv + \frac{\g p\bfv}{\g - 1} \right] + \beta_* \grad \cdot\left[\frac{\grad \rho}{\rho} \grad \cdot (\rho \bfv) - \rho \bfv \grad \cdot \left( \frac{\grad \rho}{\rho} \right)  - \frac{\rho \bfv}{2}\frac{ (\grad \rho)^2}{\rho^2}  \right] = 0. \quad 
	\label{e:3d-vel-eqn}
	\eeqs
We see that the velocity equation now involves a new nonlinear body-force term  which is related to the Gross quantum pressure \cite{gross} and is given by the gradient of the Bohm potential $\left( \frac{\grad^2 \sqrt{\rho}}{\sqrt{\rho}} \right)$ \cite{Bohm-1}. The momentum equation is also in conservation form:
	\beq
	\pdr_t (\rho v_i) + \pdr_j \left( \rho v_i v_j + \sig_{ij} \right) = 0 \quad \text{where} \quad \sig_{ij} = p \, \del_{ij} + \beta_* \left( \frac{(\pdr_i \rho) (\pdr_j \rho)}{\rho}  - \pdr_i \pdr_j \rho \right).
	 \label{e:stress-R-gas-dyn-3d}
	\eeq
A consequence of the constitutive law $\beta = \beta_*/\rho$ is that the ideal momentum equation now involves a term with three derivatives of $\rho$ corresponding to a Kortweg-type {\it grade 3} \footnote{`` In order to model more complex spatial interaction effects in an elastic material of grade $N$, the constitutive quantities are permitted to depend not only on the first gradient of deformation, the strain, but also on all gradients of the deformation less than or equal to $N$'', see \cite{dunn-serrin}.}  \cite{dunn-serrin,gorban-karlin}. Just like KdV, our equations admit sound waves with a leading cubic dispersion relation, solitary waves and periodic traveling waves. As with KdV, there are no steady continuous shock-like solutions satisfying the Rankine-Hugoniot conditions. Nevertheless, in 1D, for $\gamma = 2$, numerical solutions show that the gradient catastrophe is averted through the formation of pairs of solitary waves which can display approximate phase-shift scattering (see Fig. \ref{f:gradient-catastrophe-r-gas-soliton}).
Numerics also indicate recurrent behavior in periodic domains. These observations are related to an equivalence between our regularized equations for potential flow $(\bfv = \grad \phi)$ in the isentropic case (globally constant entropy and $p = K (\g - 1) \rho^\gamma$) and the defocussing nonlinear Schr\"odinger equation. The equivalence is achieved via the Madelung transformation  \cite{madelung} $\psi = \sqrt{\rho} \exp\left( i\phi/2\sqrt{\beta_*} \right)$ with $\beta_*$ playing the role of $\hbar^2$:
	\beq
	i \sqrt{\beta_*} \frac{\pdr \psi}{\pdr t} = - {\beta_*} \grad^2 \psi + \frac{\g K}{2} |\psi|^{2(\g - 1)} \psi.		\eeq 
This transformation may be regarded as a conservative analog of the Cole-Hopf transformation for Burgers, applies in any dimension, and results in a defocusing NLSE with $|\psi|^{2(\g - 1)} \psi$ nonlinearity, so that one obtains the celebrated cubic NLSE for $\gamma = 2$. The latter is known to admit an infinite number of conservation laws and display recurrence. Thus, our regularization of gas dynamics may be viewed as a generalization of both the single field KdV \cite{dauxois-peyrard} and nonlinear Schr\"odinger equations \cite{faddeev-takhtajan} to include the adiabatic dynamics of density, velocity, pressure and entropy in two, three or more dimensions.

\chapter{Conclusions and discussion}
\label{s:discussion}

The motivation for regularizing conservative, continuum systems like Eulerian ideal fluid mechanics
and ideal MHD was explained in the Introduction (see also \cite{thyagaraja}). A brief discussion of two famous examples should suffice here as a recapitulation of the arguments provided earlier: the well-known example of Dirac-Pauli-Heisenberg Quantum Electrodynamics with its divergences beyond the first order was regularized and renormalized by Feynman, Schwinger and Tomonaga and shown to work at all orders of the covariant perturbation theory by Dyson. The result of this profound set of ideas was a powerful tool which provided agreement between theory and experiment to remarkable accuracy. The modern recognition that non-abelian gauge theories share this remarkable renormalizability has rightly focused researchers into constructing such theories. 

A simpler but still deep example is provided by the KdV equation $u_t + u u_x = u_{xxx}$ which is a conservatively  regularized version of the one dimensional Hopf or ``kinematic wave'' equation (KWE) $u_t + u u_x = 0$.  The KWE is limited in its utility as a research tool due to its well-known failure to have single-valued solutions with finite gradients for all time. It might be argued that it should be the underlying physics which should provide the regularizing terms for macro dynamical fluid systems: indeed, such systems can be regarded as suitable limits of fundamentally kinetic/particle systems, and as such must have dissipation terms like viscosity and thermal conduction which provide regularization. In fact, the simplest regularization of KWE is provided by Burgers' equation $u_t + u u_x = \nu u_{xx}$, which indeed is dissipative and even exactly soluble by the Cole-Hopf transformation into the heat equation. 

In spite of the general validity of dissipative, i.e., entropy producing regularizations of ideal fluid dynamics and MHD arising from underlying kinetic theories, experience has shown that the studies of purely conservative physical models often provide indispensable physical insight. For instance, planetary motion involves many dissipative processes which render singular phenomena such as simultaneous three-body collisions perfectly regular. Yet, most of Newtonian mechanics of point particles and rigid bodies profits enormously from using tools like Hamiltonian mechanics and action principles which are the hallmark of conservative dynamics \cite{arnold}. 

We have therefore adopted the principle that singularities such as unbounded enstrophies and density gradients in Eulerian fluid/gas dynamics and finite time failure of the models should be removed, if possible, by suitable {\it local} regularizing terms in the governing equations, in the spirit of Landau's effective field theory. These terms are required to satisfy certain strict physical criteria: i) The discrete and continuous symmetries (usually global) obeyed by the original  unregularized system must be obeyed by the regularized system.  ii) The added terms must be ``minimal'' and ``small'' in some sense and should not alter the macro or meso-scale behaviour of the original system, although short-wavelength or ultraviolet catastrophes will have been significantly modified. iii) One should be able to derive appropriate conservation laws for the regularized equations for suitable boundary conditions. These should be extended versions of the same laws for the original singular system. iv) The system dynamics must admit Lagrange stability \cite{Nem} (but not necessarily integrability or Lyapunov stability), ergodicity and a valid statistical mechanics as in the case of 2D vortex systems considered by Onsager, London and Feynman in fluids [starting with the work of Kirchoff and Lamb] and successfully applied to 2D MHD by Edwards and Taylor \cite{edwards-taylor} and many others. It has been shown by several researchers (cf. \cite{Miura,Drazin,arnold}) that KdV has all of the above and many other interesting properties, like a Hamiltonian formulation, infinite number of conserved quantities  in involution and corresponding local conservation laws, soliton scattering and exact solubility via the inverse scattering transform. Apart from the incompressible systems considered in \cite{thyagaraja} we are not aware of any 3D continuum dynamical models with the characteristics we have demonstrated for our regularized systems (R-Euler and one and two-fluid R-MHD).

We have, in loose analogy with Dyson's concept of renormalizable field theories, introduced the idea of ``regularizable conservative continuum field theories''. Such theories must satisfy the criteria enumerated above. In the case of the Navier-Stokes equations or the visco-resistive MHD equations, and the Fokker-Planck kinetic equation of plasma theory (a regularized form of Vlasov collisionless kinetics), we have dissipative regularizations. It is not yet fully clear to us if the Navier-Stokes (NS) equations admit continuous, unique solutions to initial-boundary value problems for reasonable data. We note that it has been shown in \cite{ladyzhenskaya-1} that NS can be regularized by adding a ``hyper-viscosity''. We conjecture that it may be possible that R-NS systems incorporating the twirl acceleration terms (\ref{e:R-Euler-eqn}) will, by definition, lead to a dynamical system with bounded enstrophy and one could demonstrate unique, classical solutions to such systems for the initial-boundary value problem for small, but non-zero values of $\lambda,c_{n}; n\geq 0$ (see Section \ref{s:other-const-laws-and-regs}). However, we do not attempt any proof of the existence of classical continuous solutions to the initial-boundary value problems in this work. A 1D analogue of the twirl regularized viscous fluid and visco-resistive MHD models is the KdV-Burgers equation investigated by Grad and Hu \cite{Grad-Hu,Hu-KdV-Burgers} in the context of weak plasma shocks propagating perpendicular to a magnetic field. Mathematical examples of divergent series being ``summed'' to give perfectly well-defined and finite answers in Fourier analysis using summability methods of Abel and C\'{e}saro exemplify our approach to regularizability and its utility. Unlike the above dissipative regularizations, we focus here on the complementary question of ``conservative regularizability'' of continuum fluid models.  

In numerical simulations, using the R-Euler system would enable one to avoid finite-time singularities in the enstrophy distribution and control the number of effective modes used depending on the initial data. A careful evaluation of our conserved swirl energy and other integral invariants should help to monitor the quality of simulations (see Section \ref{s:integral-inv-v-star}). Furthermore, we believe that the numerical study of plasma and fluid turbulence at very low collisionality (i.e. very high experimentally relevant Reynolds, Mach and Lundquist numbers) will be greatly facilitated by the use of our regularization.

\noindent {\bf Summary of regularization of Euler, MHD and two-fluid plasmas}

Dissipative systems like NS are only associated with semi-groups and the system motion does not take place on a fixed manifold reversibly in time and can  involve ``strange attractors'' with complicated fractal properties. The regularization in the R-Euler dynamics is provided by the ``twirl acceleration'' $- \la^2 \bfT = - \la^2 \bfw \times (\grad \times \bfw)$ while in R-MHD we also have a magnetic twirl term $-\la^2 \bfB \times (\grad \times \bfw)$. The size of this is determined by a parameter $\la$ with dimensions of length. The twirl term is expected to be important in high speed flows with vorticity or flows with large vorticity and its curl. At any given Reynolds number, it should dominate the viscous term for sufficiently high vorticity. The parameter $\la$ is a constant micro-length scale in incompressible R-Euler and R-MHD. For compressible flow, we have found that it satisfies a physically meaningful constitutive relation: $\lambda^{2}\rho$ is a constant. In plasma physics there are natural length-scales which are inversely proportional to the square-root of the number density. For example, the electron collisionless skin-depth $\del_e = c/\om_{pe} \propto 1/\sqrt{n_e}$ where $\om_{pe} = \sqrt{\frac{n_e e^2}{m \eps_0}}$ is the electron plasma frequency. In any event, it is well-known that ideal MHD is not valid at length scales of order $\del_e$. Another example is provided by the electron Debye length $\la_{D,e} = \sqrt{k_B T_e \eps_0/n_e e^2}$ in an isothermal plasma with electron temperature $T_e$. Thus, having a cut-off of this kind will provide a finite upper bound to the enstrophy of the system and a valid statistical mechanics. More generally, we have shown that a much wider class of constitutive relations is possible, some of which lead to bounded higher moments of the square of vorticity  (see Section \ref{s:other-const-laws-and-regs}).  

In Section \ref{s:formulation-r-euler-r-mhd}, we obtained the R-Euler and R-MHD equations which constitute the regularized equations of compressible flow and one-fluid plasmas. These equations have a positive-definite `swirl' energy that includes contributions from kinetic energy, compressional potential energy, magnetic energy and the square of vorticity. We have shown for the above constitutive law, this nonlinear energy functional is a constant of the motion for suitable boundary conditions and thus prevents the unboundedness of enstrophy.  
 
The system motion takes place in the function space of $\rho(\bfx),\bfv(\bfx)$ and $\bfB(\bfx)$  which is ``foliated'' by the closed, nested surfaces formed by the constant energy functional. The regularized systems are shown to be time reversible and to satisfy the symmetries of the Euler and ideal MHD equations and have conservation laws corresponding with and generalizing those of the ideal systems. There are even generalized Kelvin-Helmholtz and Alfv\'en freezing-in theorems and associated integral invariants. Furthermore, we have employed the elegant non-canonical Poisson Brackets (PBs) developed by Landau \cite{landau} (in quantum hydrodynamics), Morrison, Greene \cite{morrison-greene} and others to show that the R-Euler and R-MHD equations can be derived from the energy functional using  these PBs. This fact is remarkable in that we have demonstrated the existence and properties of a regularization which preserves the Poisson structure, conservation laws and global symmetries of the Euler and ideal MHD equations while guaranteeing the boundedness of enstrophy. Our formalism implies that the system evolves on the intersection of the level hypersurfaces of energy and any other prevalent constants of motion, through a Hamiltonian, PB-mediated, infinitesimal 1-parameter group of time translations. 

In Section \ref{s:reg-eqns-two-fluid-compress}, we have extended the conservative twirl regularization of Euler and one fluid MHD described in Section \ref{s:formulation-r-euler-r-mhd} to dissipationless compressible two-fluid plasmas. This involves vortical and magnetic twirl terms $\lambda_l^2 (\bfw_l + \frac{q_l}{m_l} \bfB) \times (\grad \times \bfw_l)$ in the velocity equations for ions and electrons $( l = i, e )$. We find that $\la_l^2 n_l$ must be constant for energy conservation, so that $\la_l$ behaves likes $\la_{D}$ or $c/\om_{p,l}$. The key difference between the regularized and unregularized two-fluid models is that the flow current $\bfj_{\rm flow} = \sum_l q_l n_l \bfv_l$ in Amp\`ere's law is augmented by a solenoidal `twirl' current $\sum_l \grad \times (\grad \times \la_l^2 \bfj_{{\rm flow},l})$ analogous to magnetization currents in material media. This leads to locally conserved momenta and a positive definite swirl energy $E^*$. In addition to kinetic, compressional and electromagnetic contributions, $E^*$ includes a vortical energy $\int \sum_l \la_l^2 n_l m_l \bfw_l^2 \: d\bfr$, thus placing an a priori upper bound on the enstrophy of each species.  It is noteworthy that our twirl-regularized two-fluid equations follow from the Hamiltonian $E^*$ using unchanged the Poisson brackets of \cite{spencer-kaufman,holm-kuperschmidt}. This PB formalism shows that among regularizations preserving the continuity equations and symmetries of the ideal system, our twirl regularization terms are unique and minimal in nonlinearity and space derivatives of velocities. It is also employed to regularize magnetic field curls in the compressible models by adding $(\la_B^2/2\mu_0) \int (\grad \times \bfB)^2 \: d\bfr$ to $E^*$ so that field and velocity curls are $L^2$-bounded. By taking suitable successive limits we get a hierarchy of compressible and incompressible regularized plasma models (quasineutral two-fluid, Hall and one-fluid MHD). Interestingly, in the incompressible two-fluid case alone, it is also possible to choose the current as $\bfj_{\rm flow}$, which leads to a conserved swirl energy that automatically includes a $(\la^2/2\mu_0) \int (\grad \times \bfB)^2 \: d\bfr$ term in $E^*$. Furthermore, the assumption of local short-distance cut-offs $\la_l$ limits the number of effective degrees of freedom, thus considerably extending results on the Charney-Hasegawa-Mima (CHM) model \cite{lashmore-mccarthy-thyagaraja} to the full 3-D two-fluid equations. This feature is crucial to numerical modeling of conservative plasma dynamics and consequently provides a viable  framework to investigate statistical theories of turbulence in these systems.

\noindent{\bf Applications to specific problems}

A natural question concerns the effect of our twirl regularization in specific fluid and plasma systems of interest. We have examined this in a few representative steady flows (see Section \ref{s:examples}): a rotating columnar (Rankine) vortex and its extension to MHD, a vortex sheet, compressible plane flow, channel flow and variants of Hill's vortex. In all these steady flows, the nonlinear regularized equations are under-determined as in ideal Euler or ideal MHD. For instance, in our rotating columnar vortex model for a tornado with core radius $a$, the equations determine the density if the vorticity distribution is prescribed. In a layer whose width can be of order the regularization length $\lambda \ll a$, the vorticity smoothly drops from its value in the core to that in the periphery. We find that the regularization relates this decrease in vorticity to a rise in density (\ref{e:density-vorticity-balance-across-layer}). On the other hand, vorticity is allowed to have an unrestricted jump across the layer in the unregularized model while $\rho$ is continuous and its increase is unrelated to the drop in vorticity. Similarly, the regularization can smooth the vorticity in a magnetized columnar vortex. Given vorticity and current profiles, the density profile is determined. While the Lorentz force tends to pinch the column, the twirl force points outwards for radially decreasing vorticity. An analogue of Hill's vortex, a cylindrical vortex in pipe-like flow was also considered in Section \ref{s:examples}. The flow is irrotational outside an infinite circular cylinder of radius $a$ with vorticity purely azimuthal inside the cylinder. The regularized equations with appropriate boundary conditions were solved numerically and unlike in the unregularized case, the vorticity was found to be continuous across $r = a$. In modeling a vortex sheet, we found steady solutions to the regularized equations that smooth discontinuous changes in vorticity over a layer of thickness $\approx \lambda$. A regularized analogue of a Bernoulli-like equation implies a reduction in density on the sheet compared to its asymptotic values: depending on the relative flow Mach number, the decrease can be significant when the thickness of the sheet is comparable to the regulator $\lambda$. 

These examples show that twirl-regularized steady flows can be more regular than the corresponding ideal ones. They also serve as a starting point for numerical simulations of time-dependent flows. An interesting example that is currently under investigation concerns the effect of our regularizations on the growth of perturbations to vortex/current sheets and their nonlinear saturation.

{\flushleft \bf Some directions for future research}

\begin{enumerate}

\item A problem of fundamental importance is the initial value problem in 3D, say with periodic BCs. We would like to numerically simulate the regularized equations of Chapters \ref{s:r-euler-r-mhd} and \ref{s:two-fluid}. In particular, it would be interesting to determine the long-time behaviour of spectral distributions of energy and enstrophy. 

\item We would like to study linear instabilities in a conservatively regularized vortex sheet/rotating vortex, follow their growth and nonlinear saturation due to the bound on enstrophy. The a priori bound on enstrophy and kinetic energy demands a purely conservative nonlinear saturation of any linearly growing mode. This would involve a conservative compressible analogue of the Orr-Sommerfeld equation. The behaviour of such nonlinear dynamics could provide insight into the statistics and kinematics of turbulent motions in the inertial range.

\item We would also like to model oblique shocks and the Sedov-Taylor spherical blast wave problem using our regularized gas dynamic equations. 

\item We note that kinetic approaches such as the Chapman-Enskog method based on, for example the Fokker-Planck equation of plasma theory, typically lead in higher orders in the mean-free-path asymptotic expansion to both ``entropy conserving reactive'' terms and to dissipative (entropy producing) terms in the stress tensor and the heat-flux vector (see  \cite{Braginskii,Lifshitz-Pitaevski} and the more recent work \cite{karlin-gorban,struchtrup-torrilhon,gorban-karlin,huang-wang-wang-yang}). It is possible that the ``twirl-acceleration'' and ``density gradient'' terms (introduced here essentially as formal conservative regularizing effects) could arise in higher order asymptotics (like the Burnett expansion) of kinetic equations. Somewhat analogous formal regularizers are commonly  encountered in effective field theory (E.g. the short-range repulsive Skyrme term with $4$ derivatives is believed to stabilize the singularity in the soliton solution of the QCD effective chiral Lagrangian \cite{bal-book}). String theory attempts to provide a relativistically acceptable short-range cut-off to the divergences encountered in the Einstein-Hilbert formulation of classical General Relativity in a manner which resembles in spirit the regularizers we have advocated for compressible fluid flow. Thus, it would be interesting to investigate if our conservative regularization terms can arise from kinetic theory using a Chapman-Enskog-like expansion in Knudsen number.

\item It is useful to note that a possible approach to the statistical mechanics of the R-Euler system is through the approach pioneered by E Hopf (see the extensive discussion by Stanisic, \cite{stan}). This was originally conceived as a method of investigating the statistical theory of hydrodynamic turbulence governed by the NS equations. However, it would seem that the ideas relating to the Hopf functional can certainly be of value in R-Euler statistical mechanics. Our PBs allow us to formulate Hopf's equation (analogue of the Liouville equation) $F_t + \{F, H \} = 0$ for the functional $F[\rho, \bfv, t]$. The Hamiltonian structure of the flow on the constant energy hyper-surface leads to micro-canonical statistical mechanics, and more generally to a canonical distribution (Boltzmann-Gibbs or Fermi-Dirac, \cite{LanLif}). A statistical mechanics of entangled 3D regularized vortex tubes with bounded enstrophy and energy in dissipationless motion would be a significant extension of the 2D Onsager theory of line vortices, quantized or otherwise.

\item The ideas due to Koopman and von Neumann (see the account given in \cite{RieszNagy}) in ergodic theory are also directly relevant provided a suitable measure can be developed for the constant energy surface on which the system motion takes place. The possibility of mapping the nonlinear evolution of the R-Euler flow on to unitary transformations in a function space of effectively a finite number of degrees of freedom could have many practical applications.

\end{enumerate}

\appendix
\chapter{Some properties of the Poisson brackets}
\label{a:PB-properties}

\section{Poisson Brackets in terms of scalar and vector potentials} 

We express the PBs among $\rho$ and $\bfv$ in terms of scalar and vector potentials. For irrotational flows these non canonical PBs may be expressed in terms of canonical Bose fields. To begin with, the Helmholtz theorem allows us to write $\bfv$ as a sum of curl-free and divergence-free fields $\bfv^{\rm irrot}$  and $\bfv^{\rm sol}$. The irrotational and incompressible fields admit scalar and vector potentials:
	\beq
	\bfv = \bfv^{\rm irrot} + \bfv^{\rm sol} =  - \grad C + \grad \times \bfQ.
	\eeq
Note that $C$ and $\bfQ$ are non-local in $\bfv$. If the flow domain is $\mathbb{R}^3$ and $\bfv$ falls off faster than $1/r$, then
	\beq
	C(\bfr) = \ov{4\pi} \int \frac{\grad_s \cdot \bfv(s)}{|\bfr- \bfs|} d\bfs \quad \text{and} \quad
	\bfQ(\bfr) = \ov{4\pi} \int \frac{\bfw(s)}{|\bfr- \bfs|} d\bfs \quad \text{with} \quad \grad \cdot \bfQ = 0.
	\label{e:scalar-and-vector-potentials-for-v}
	\eeq
We may treat $\rho, C$ and $\bfQ$ as dynamical variables in place of $\rho$ and $\bfv$. It is interesting to identify their PBs. Now, $\bfQ$ commutes with $\rho$ since $\bfw$ does. On the other hand $\{ C(x), \rho(y) \} = \del(x-y)$ since $\{ \bfv(s) , \rho(y) \} = - \grad_s \del(y-s)$ and $\grad^2_s (1/|\bfr - \bfs|) = -4\pi \del(\bfr -\bfs)$. The PBs of $\bfQ$ and $C$ are more involved:
\small
	\beqs
	16 \pi^2 \{ C(\bfx) , Q_j(\bfy)\}
	&=& \int  \left [\frac{(\bfx - \bfr) \cdot (\bfy - \bfr) w_j - (\bfx - \bfr) \cdot \bfw (y_j - r_j)}{\rho(\bfr) \: |\bfx - \bfr|^3 |\bfy - \bfr|^3} \right] \; d\bfr  \cr
	\text{and} \quad 16\pi^2 \{ Q_i(\bfx), Q_j(\bfy) \} &=& \int \left[ \frac{\eps_{ijk} (y_k - r_k) (\bfx-\bfr) \cdot \bfw - (x_i - r_i) ((\bfy - \bfr) \times \bfw)_j}{\rho(\bfr) \: |\bfx - \bfr|^3 |\bfy - \bfr|^3} \right] d \bfr.
	\label{e:AA-Aphi-PB}
	\eeqs \normalsize	
Since $\{ \rho(x), \rho(y) \} = 0$ it is natural to ask whether $\{ C, C \} = 0$ so that $C$ and $\rho$ would be canonically conjugate. We find \small
	\beq
	16 \pi^2 \{ C(\bfa) , C(\bfb)\} = \int \frac{\pdr_{r^i}\pdr_{s^j} \{v_i (r), v_j(s)\}}{|\bfa - \bfr||\bfb - \bfs|} d\bfr \: d\bfs
	= \int \frac{(a_i - r_i)(b_j - r_j)}{|\bfa - \bfr|^3 |\bfb - \bfr|^3} \frac{\om_{ij}}{\rho(\bfr)} d\bfr 
	= \int \frac{(\bfa - \bfr) \times (\bfb - \bfr) \cdot \bfw}{ \rho(\bfr) |\bfa - \bfr|^3 |\bfb - \bfr|^3} d\bfr.
	\label{e:phiphi-PB}
	\eeq \normalsize
The integrals in (\ref{e:scalar-and-vector-potentials-for-v}, \ref{e:AA-Aphi-PB}, \ref{e:phiphi-PB}) are finite as may be seen in spherical coordinates centered at $\bfr = \bfa$. Defining $\tl \bfr = \bfr - \bfa$, the double pole at $\tl \bfr = 0$ is cancelled by the double zero in the volume element $\tl r^2 d\tl r d\Om$. The same applies to a neighborhood of $\bfr = \bfb$.

Note that, $\{ C(\bfa) , C(\bfa)\} = 0$, consistent with anti-symmetry. But $C$ at distinct locations don't generally commute. It suffices to show this in a special case. We take $\bfa = (0,0,0)$, $\bfb = (0,1,0)$, asymptotically constant $\rho = z/(z^2 + 1)$ and rapidly decaying $\bfv = x^2 (y - 1)^4 e^{-r^2}\: \hat z$. This ensures (\ref{e:phiphi-PB}) is manifestly convergent, $\bfw$ has zeros at $\bfa$ and $\bfb$ to cancel the apparent triple poles: \small
	\beqs
	\bfw &=& e^{-r^2}\left[ 2x^2 (y-1)^3 (2 -y(y-1)) \; \hat x + 2 x (y-1)^4 (x^2 - 1) \: \hat y\right] \quad \text{and} \cr
	 \{C (\bfa) , C(\bfb)\} &=&  \ov{16\pi^2} \int^{\infty}_{-\infty} \frac{\left(z^2 + 1\right)e^{-r^2}\left( 2 x^2 (y-1)^3 [2 - y(y-1)] \right)}{r^3 \: \left({x^2 + (y - 1)^2 + z^2}\right)^{3/2}} \, dx\, dy\,dz \approx -0.026 \ne 0.
	 \eeqs \normalsize
Thus $\rho$ and $C$ are {\it not} canonically conjugate in general. But in irrotational flow, $\bfw = \bfQ = 0$ so $\{ C(a), C(b) \} \equiv 0$ and $\rho, C$ are canonically conjugate. This is reminiscent of the number density-phase PB and suggests the introduction of the complex field $\psi = \sqrt{\rho} e^{i C/\kappa}$ where $\kappa$ is a constant with dimensions of diffusivity\footnote{A natural choice is $\kappa = c_s L$ where $c_s$ is a sound speed and $L$ a macroscopic length associated with the flow. In quantum theory $\kappa = \hbar/m$.}. The $C$-$\rho$ PB (for $\bfw = 0$) then imply that $\psi$ and $\psi^*$ satisfy canonical Bose PB: $\{ \psi, \psi \} = \{ \psi^*, \psi^* \} = 0$ and $\{ \psi(x), \psi^*(y) \} = (i/\kappa) \del(x-y)$. The evolution equation for $\psi$ in the irrotational case is reminiscent of the 3D Gross-Pitaevskii or nonlinear Schr\"odinger equation (especially for $\gamma = 2$ where $U'(\rho) \propto \rho = |\psi|^2$)
\beq
i \kappa \{ \psi , H\} = i \kappa \frac{\pdr \psi}{\pdr t}  = - \frac{\bfv^2}{2} \psi - U'(\rho) \: \psi - \frac{i \kappa}{2 \rho} \grad \cdot (\rho \bfv) \psi \;\; \text{where} \;\; \bfv = \frac{\kappa}{2i}\left(\frac{\psi \grad \psi^* - \psi^* \grad \psi }{|\psi|^2}\right) \;\; \text{and} \;\; \rho = |\psi|^2.
\eeq
However, the above calculation implies that $\psi$ and $\psi^*$ are not canonical Bose fields for flows with vorticity. For flows with vorticity, Clebsch potentials give a way of identifying canonically conjugate variables (see Section \ref{a:Lagrangian}).

\section{Poisson brackets of mass current and swirl velocity}
\label{a:pb-mass-curr-and-v*}

The PB of mass current ${\bf M} = \rho \bfv$ are of particular interest. Suppose $\bf a$ and $\bf b$ are a pair of constant vectors, then using (\ref{e:PB-among-basic-var}),
	\beqs
	(a) && \{ \bfM(\bfx) , \rho(\bfy) \} = - \rho(\bfx) \grad_\bfx \del(\bfx-\bfy), 
	\cr
	(b) && \{ {\bf a} \cdot {\bf M}(\bfx) , {\bf b} \cdot \bfv(\bfy) \} = \left[ ({\bf a} \times {\bf b}) \cdot \bfw(\bfx) - ({\bf a} \cdot \bfv)(\bfx) \, {\bf b} \cdot \grad_\bfx \right] \del(\bfx-\bfy),
	\cr
	(c) &&  \{ \bfa \cdot \bfM(\bfx) , \bfb \cdot \bfM(\bfy) \} = \left[\rho(\bfy) (\bfa \cdot \bfv(\bfx)) (\bfb \cdot \grad_\bfy) - (\bfa,\bfx \leftrightarrow \bfb, \bfy) + \rho \, (\bfa \times \bfb) \cdot \bfw  \right] \del(\bfx-\bfy),
	\cr
	(d) && \{ \bfa \cdot \bfM(\bfx) , \bfb \cdot \bfw(\bfy) \} = \rho(\bfx) \, \sum_i \left( \bfb \times \grad_\bfy \left( \rho^{-1} (\bfw \times \bfa)_i \del(\bfx-\bfy) \right) \right)_i.
	\eeqs
Given the important dynamical role that the swirl velocity $\bfv_* = \bfv + \la^2 \grad \times \bfw$ plays, we mention some of its PBs. For e.g. the PB of $\bfv_*$ with $\rho$ is the same as that of $\bfv$ with $\rho$ (as $\la$ and $\bfw$ commute with $\rho$):
	\beq
	\{ \bfv_*(\bfx), \rho(\bfy) \} = \{ \bfv(\bfx) + \la(\bfx)^2 \grad \times \bfw, \rho(\bfy) \} = \{ \bfv(\bfx) , \rho(\bfy) \}.
	\eeq
Using $\grad \times (\grad \times \bfv) = \grad (\grad \cdot \bfv) - \grad^2 \bfv$, the swirl velocity may be got from $\bfv$ by the action of the tensor operator $T_{ik}(\bfx)$:
	\beq
	v_{*i} = \left[ \del_{ik} + \la^2 \left(\pdr_i \pdr_k - \del_{ik} \grad^2 \right) \right] v_k \equiv T_{ik} v_k.
	\eeq
The PB of $\bfv_*$ with other quantities can be conveniently expressed in terms $T_{ik}$\footnote{$T_{ik}$ commutes with $\bfw$ and $\rho$, but not $\bfv$. Moreover $\grad \times \bfw = \la^{-2} (T - I) \bfv$. The non-dynamical $\la^{-2} (T_{ik} - \del_{ik}) = (\pdr_i \pdr_k - \del_{ik} \grad^2)$ commutes with everything.}:
	\begin{enumerate}
	\item $\left\{ v_{*i}(\bfx) , w_j(\bfy) \right\} = T_{ik}(\bfx) \left\{ v_k(\bfx), w_j(\bfy) \right\}$, 
	\item $\left\{ v_{*i}(\bfx) , \frac{w_j(\bfy)}{\rho(\bfy)} \right\} = - \{ v_i(\bfx), \rho(\bfy) \} \frac{w_j(\bfy)}{\rho^2(\bfy)} + \ov{\rho(\bfy)} T_{ik}(\bfx) \left\{ v_k(\bfx), w_j(\bfy) \right\}$, 
	 \item $\{ v_{*i}(\bfx), v_j(\bfy) \} = T_{ik}(\bfx) \{ v_k(\bfx) , v_j(\bfy) \} + (\grad \times \bfw)_i(\bfx) \left\{ \la^2(\bfx), v_j(\bfy) \right\}$, 
	 \item $\!\begin{aligned}[t]
                      \{ v_{*i}(\bfx), v_{*j}(\bfy) \} &= (\grad \times \bfw)_i(\bfx) \{ \la^2(\bfx) , v_j(\bfy) \}  + (\grad \times \bfw)_j(\bfy) \{ v_i(\bfx), \la^2(\bfy) \} \\
                    &\quad + T_{ik}(\bfx) T_{jl}(\bfy) \{ v_k(\bfx), v_l(\bfy) \}
           \end{aligned}$
	\item $\{ \rho(\bfx) v_{*i}(\bfx) , \rho(\bfy) \} = \rho(\bfx) \{ v_i(\bfx), \rho(\bfy) \}$, 
	\item $\{ \rho v_{*i}(\bfx), v_j(\bfy) \} = \rho(\bfx) T_{ik}(\bfx) \left\{ v_k(\bfx), v_j(\bfy) \right\} + v_i(\bfx) \{ \rho(\bfx) , v_j(\bfy) \}$, 
	\item $\{\rho v_{*i}(\bfx), w_j(\bfy) \} = \rho(\bfx) T_{ik}(\bfx) \{ v_k(\bfx), w_j(\bfy) \}$.
	\end{enumerate}

\section{PBs of solenoidal and irrotational linear functionals}
\label{a:solenoidal-irrot-pb}

The PB (\ref{e:pb-between-functionals-of-rho-v}) of (especially linear) functionals of $\bfv$ and $\rho$ have interesting properties. Suppose $F[\bfv] = \int \bff \cdot \bfv \, d\bfr$ and $G[\bfv] = \int \bfg \cdot \bfv \, d\bfr$ are two {\it linear} functionals of $\bfv$, with $\bff$ and $\bfg$ a pair of test vector fields vanishing sufficiently fast at infinity. Then
	\beq
	\{ F[\bfv] , G[\bfv] \} = \int (\bfw/\rho) \cdot (\bff \times \bfg) \: d\bfr.
	\eeq
An interesting sub-class of such linear functionals are the `solenoidal' ones $F_s[\bfv] = \int \bff \cdot \bfv \, d\bfr$ where $\bff$ is solenoidal $\grad \cdot \bff = 0$. Writing $\bff = \grad \times \bfA$\footnote{In terms of Clebsch potentials $\alpha, \beta$ for the solenoidal field $\bff = \grad \alpha \times \grad \beta = \grad \times (\alpha \grad \beta)$, we may take $\bfA = \alpha \grad \beta$.} and assuming $\bfA$ vanishes at infinity, $F$ can be written as a linear functional of vorticity:
	\beq
	F_s[\bfv] = \int (\grad \times \bfA) \cdot \bfv \, d\bfr 
	= \int \bfA \cdot (\grad \times \bfv) \, d\bfr + \int \grad \cdot (\bfA \times \bfv) \, d\bfr = \int \bfA \cdot \bfw \, d\bfr.
	\eeq
Since $\{\bfw, \rho \} = 0$, it follows that a solenoidal $F_s$ commutes with any functional of $\rho$: $\{ F_s[\bfv], H[\rho] \} = 0$. Associated to a solenoidal $F_s$, we may define the functional $F^\rho_s = \int \bff \cdot \rho \bfv \, d\bfr$. Then one checks that $\{ F_s, F_s^\rho \} = 0$:
	\beq
	\{ F_s, F_s^\rho \} = \int \left[ \frac{\bfw}{\rho} \cdot (\bff \times \rho \bff) - \bff \cdot \grad (\bff \cdot \bfv) \right] d\bfr = \int (\bff \cdot \bfv) (\grad \cdot \bff) \, d\bfr = 0.
	\eeq
Similarly, if $\phi$ is any function (independent of $\rho$ and $\bfv$) then it follows that $F^\phi_s = \int \bff \cdot \phi \bfv d\bfr$ commutes with $F_s$ (but not with $F^\rho_s$ in general) if $\bff$ is solenoidal.

Similar to solenoidal linear functionals we may define irrotational linear functions $F_i[\bfv] = \int \bff \cdot \bfv \, d\bfr$ where $\bff = \grad \alpha$ is irrotational. Then $F_i[\bfv] = - \int \alpha(\bfr) (\grad \cdot \bfv) \, d\bfr$. The PB of two irrotational functionals is in general non-zero:
         \beq
         \{ F_i,G_i \} = \int ({\bf w}/{\rho}) \cdot [\nabla \alpha \times \nabla \beta] d\bfr \quad
         = \; -\int {\bf v}\cdot \grad \times \frac{ [\nabla \alpha \times \nabla \beta]}{\rho} d\bfr. 
         \eeq
They commute if the potentials $\alpha$ and $\beta$ are functionally dependent. The PB of an irrotational functional with a linear functional of density $H[\rho] = \int h \rho \, d\bfr$ is also non-zero in general
	\beq
	\{ F_i[\bfv], H[\rho] \} = - \int \bff \cdot \grad h \, d\bfr = - \int \grad \alpha \cdot \grad h \, d\bfr,
	\eeq
but vanishes if $\bff = \grad \alpha$ and $\grad h$ are orthogonal. 

\section{Proof of Jacobi identity for 3 linear functionals of $\bfv$ and $\rho$}
\label{a:jacobi}

Suppose $F, G$ and $H$ are three {\em linear} functionals of velocity and density \small
	\beq
	F = \int \left[ \bff(\bfr) \cdot \bfv(\bfr) + \tl f(\bfr) \rho(\bfr) \right] \: d\bfr, \;\;
	G = \int \left[ \bfg(\bfr) \cdot \bfv(\bfr) + \tl g(\bfr) \rho(\bfr) \right] \: d\bfr, \;\;
	H = \int \left[ \bfh(\bfr) \cdot \bfv(\bfr) + \tl h(\bfr) \rho(\bfr) \right] \: d\bfr,
	\eeq \normalsize
where $\bff, \bfg, \bfh$ are three smooth test vector fields and $\tl f, \tl g, \tl h$ are three test functions all vanishing sufficiently fast at infinity. We prove that the Jacobi expression $\{ \{ F, G \} , H \} + {\rm cyclic} = 0$. This is a non-trivial special case of the Jacobi identity. As a corollary, the Jacobi identity for three linear functionals of vorticity is also satisfied. For, we can write any linear functional of vorticity $F[\bfw] = \int \bfA \cdot \bfw \: d\bfr = \int (\grad \times \bfA) \cdot \bfv \, d\bfr$ as a solenoidal functional of velocity and use the previous result.

We will first obtain an interesting formula (\ref{e:jacobi-expr-3-lin-fnals-of-v}) for the Jacobi expression. Recall from (\ref{e:pb-between-functionals-of-rho-v}) that the PB of two linear functionals is $\{ F, G \}	= \int \left[ \rho^{-1} \bfw \cdot (\bff \times \bfg) - \bff \cdot \grad \tl g + \bfg \cdot \grad \tl f \right] \: d\bfr$. To find $\{ \{ F, G \} , H \}$ we need the functional derivatives
	\beqs
	\deldel{\{ F, G \}}{\rho} &=& - \frac{\bfw}{\rho^2} \cdot (\bff \times \bfg) \quad \text{and} \quad
	\deldel{\{ F, G \}}{\bfv} = \grad \times \left( \frac{\bff \times \bfg}{\rho} \right). \quad {\rm Thus} ,
	\cr
	J_1 &=& \{ \{ F, G \} , H \} = \int \left[ \deldel{\{ F, G \}}{\rho} \grad \cdot \deldel{H}{\bfv} + \left[ \grad \cdot \deldel{\{ F,G \}}{\bfv} \right] \deldel{H}{\rho} - \deldel{\{ F, G \}}{\bfv} \cdot \left[ \frac{\bfw}{\rho} \times \deldel{H}{\bfv} \right]  \right] \: d\bfr \cr
	&=& \int \left[ \frac{\bfw \cdot (\bff \times \bfg)}{\rho^2} \grad \cdot \bfh - \grad \times \left( \frac{\bff \times \bfg}{\rho} \right) \cdot \frac{\bfw \times \bfh}{\rho} \right] \: d\bfr.
	\eeqs
Notice that $J_1$ is independent of the test functions $\tl f, \tl g$ and $\tl h$ so that the dependence of $F,G$ and $H$ on $\rho$ does not play any role in the Jacobi condition. We would like to separate the dependence on dynamical variables $\bfw$ and $\rho$ from the dependence on $\bff,\bfg$ and $\bfh$. Using the curl of a cross product we arrive at
	\beqs
	J_1 &=& \{ \{ F,G \}, H \} = \int \frac{\grad \rho}{\rho^3} \cdot \left[ \bfw \cdot (\bfh \times \bff) \: \bfg + \bfw \cdot (\bfg \times \bfh) \: \bff \right] \: d\bfr
	\cr && + \int \frac{\bfw}{\rho^2} \cdot \left[ \bfh \times [\bff, \bfg] + (\bff \times \bfg) (\grad \cdot \bfh) - (\bfg \times \bfh) (\grad \cdot \bff) - (\bfh \times \bff) (\grad \cdot \bfg) \right] d\bfr.
	\eeqs
Here $[\bff , \bfg] = (\bff \cdot \grad) \bfg - (\bfg \cdot \grad) \bff$ is the commutator of vector fields. Notice that the $1^{\rm st}$ term involves the gradient of $\rho$ while the $2^{\rm nd}$ does not. $J_2$ and $J_3$ are obtained by cyclic permutations of $\bff,\bfg,\bfh$. Adding $J_1 + J_2 + J_3 = J$, several terms cancel leaving
\small
	\beqs
	J &=& J^{\pdr \rho} + J^\rho  = \int \Bigg( \Bigg. \frac{\bfw}{\rho^2} \cdot \Bigg( \Bigg. \left( \bff \times [\bfg, \bfh] + \bfg \times [\bfh, \bff] + \bfh \times [\bff, \bfg]  \right)  \cr
	&+& \left\{ (\bfh \times \bfg) (\grad \cdot \bff) + (\bff \times \bfh) (\grad \cdot \bfg) + (\bfg \times \bff) (\grad \cdot \bfh)  \right\} \Bigg. \Bigg) \cr
	&-& \grad\left(\rho^{-2} \right) \cdot [(\bfw \cdot (\bff \times \bfg)) \bfh + (\bfw \cdot (\bfg \times \bfh)) \bff + (\bfw \cdot (\bfh \times \bff)) \bfg ]\Bigg. \Bigg) d\bfr.
	\label{e:jacobi-expr-3-lin-fnals-of-v}
	\eeqs \normalsize
For the Jacobi identity to be satisfied, this must vanish for arbitrary test vector fields $\bff,\bfg,\bfh$ and any fixed $\rho$ (asymptotically constant) and $\bfw$ (vanishing at infinity). The $1^{\rm st}$ term involves $\grad \rho$, so we call it $J^{\pdr \rho}$ while the second term is called $J^\rho$. In the integrand of $J^\rho$ the dependence on $\bfw,\rho$ is factorized from the dependence on $\bff,\bfg,\bfh$. This is not quite the case with $J^{\pdr \rho}$.

{\noindent \bf Proof that $J=0$:} We expand the test vector fields as a linear combination of fields along the coordinate directions $\hat x, \hat y, \hat z$ and write the linear functionals\footnote{As remarked above, the dependence on $\rho$ of the {\it linear} functionals $F,G$ and $H$ does not enter the Jacobi expression.} as a sum $F[\bfv] = \sum_i \int \bff_i \cdot \bfv = F_1[\bfv] + F_2[\bfv] + F_3[\bfv]$. Thus the Jacobi expression becomes
	\beq
	J = \{ \{ F, G \} , H \} + \text{cyclic} = \sum_{i,j,k=1}^3 \{ \{ F_i, G_j \} , H_k \} + \text{cyclic}.
	\eeq
There are 27 terms of the form $\{ \{ F_i, G_j \} , H_k \}$ plus their cyclic permutations. Consider any one of the $27$ terms. There are three possibilities: (1) $i,j,k$ all distinct  (mutually orthogonal test fields); (2) $i=j=k$, (collinear test fields) (3) two indices the same and one distinct. We show below that the Jacobi identity is satisfied for three linear functionals of velocity $F,G,H$ falling into any one of the above categories. Consequently $J=0$ for any three linear functionals of velocity and density.

\subsection{Jacobi identity for 3 orthogonal test fields}
Let the three linear functionals $F,G,H$ in the Jacobi expression (\ref{e:jacobi-expr-3-lin-fnals-of-v}) point along $\hat x, \hat y \:$and$\: \hat z$, i.e. $\bff = \al(\bfr) \hat x$, $\bfg = \beta(\bfr) \hat y$, $\bfh = \gamma(\bfr) \hat z$ where $\al, \beta, \gamma$ are three test functions. Beginning with $(\bfw \cdot (\bff \times \bfg)) \bfh = \al \beta \gamma \, (\bfw \cdot \hat z) \hat z$ and their cyclic permutations we get (subscripts on $\al, \beta, \gamma$ denote partial derivatives)
	\beqs
	(\bfw \cdot (\bff \times \bfg)) \bfh + \text{cyclic} = (\al \beta \gamma) \bfw \quad \imply \quad 
	J^{\pdr \rho} = - \int (\alpha \beta \gamma) \: \bfw \cdot\grad(\rho^{-2}) \; d\bfr.
	\eeqs
The quantity $\al \beta \gamma = (\bff \times \bfg) \cdot \bfh$ is the volume of the parallelepiped spanned by the test vector fields. On the other hand $J^\rho$ is evaluated using $\bff \times [\bfg , \bfh] + (\bfh \times \bfg) (\grad \cdot \bff) + \text{cyclic} = - \grad(\al \beta \gamma)$. By the divergence theorem and $\grad \cdot \bfw = 0$ we get
	\beq
	J^\rho = - \int \frac{\bfw}{\rho^2} \cdot \grad (\al \beta \gamma) \, d \bfr = \int (\al \beta \gamma) \grad \cdot (\bfw/\rho^2) \, d\bfr 
	= \int (\al \beta \gamma) \bfw \cdot \grad (\rho^{-2}) \, d\bfr.
	\eeq
So, $J = J^\rho + J^{\pdr \rho} = 0$ and the Jacobi identity (\ref{e:jacobi-expr-3-lin-fnals-of-v}) for three orthogonal test fields is proved.

\subsection{Jacobi identity for three test fields in the same direction}

Let all three vector fields be collinear, say:
	$\bff = \alpha \hat x, \: \bfg = \beta \hat x \; \text{and} \; \bfh = \gamma \hat x.$ Since their cross product vanishes, $J^{\pdr \rho} = 0$ and the Jacobi expression (\ref{e:jacobi-expr-3-lin-fnals-of-v}) reduces to
	\beq
	J = \int \frac{\bfw}{\rho^2} \cdot \left[ \bff \times [\bfg, \bfh] + \bfg \times [\bfh, \bff] + \bfh \times [\bff, \bfg]  \right] \, d\bfr.
	\eeq
All the commutators point along $\hat x$, e.g. $[\bfg,\bfh] = (\beta \gamma_x - \gamma \beta_x) \hat x.$
It follows that the cross product of the vector fields and the commutators is zero, so $J = 0$.

\subsubsection{Jacobi identity for two collinear test fields and one orthogonal to them}

Let $2$ of the test fields be collinear and the $3^{\rm rd}$ point orthogonally. Without loss of generality we take $\bff = \alpha \hat x, \; \bfg = \beta \hat x, \; \bfh = \gamma \hat y$. Now, $\bfw \cdot(\bff \times \bfg) \bfh = 0, \;\bfw \cdot(\bfg \times \bfh) \bff = \al \beta \gamma w_z \hat x \; \text{and} \; \bfw \cdot(\bfh \times \bff) \bfg = - \al \beta \gamma w_z \hat x$. Therefore $J^{\pdr \rho} = - \int \grad\left(\rho^{-2} \right) \cdot [(\bfw \cdot (\bff \times \bfg)) \bfh + (\bfw \cdot (\bfg \times \bfh)) \bff + (\bfw \cdot (\bfh \times \bff)) \bfg] \: d\bfr = 0$. To compute $J^{\rho}$ (\ref{e:jacobi-expr-3-lin-fnals-of-v}) we need $\bff \times [\bfg, \bfh] + (\bfh \times \bfg) (\grad \cdot \bff) + \text{cyclic}$. Now, 
	\beq
	(\bfh \times \bfg) (\grad \cdot \bff) = - \al_x \beta \gamma \hat z, \quad
	(\bff \times \bfh) (\grad \cdot \bfg) = \al \beta_x \gamma \hat z, \quad
	\text{and} \quad
	(\bfg \times \bff) (\grad \cdot \bfh) = 0
	\eeq
So $(\bfh \times \bfg) (\grad \cdot \bff) + \text{cyclic} = \left(\al \beta_x \gamma - \al_x \beta \gamma \right) \hat z$. On the other hand, $\bff \times [\bfg , \bfh] =\al \beta \gamma_x \hat z, \;\bfg \times [\bfh , \bff] = - \al \beta \gamma_x \hat z\;\text{and} \; \bfh \times [\bff , \bfg] = (- \al \beta_x \gamma + \al_x \beta \gamma) \hat z.$
Adding these 
        \beq
	\bff \times [\bfg, \bfh] + \text{cyclic} = (\al_x \beta \gamma - \al \beta_x \gamma) \hat z.
	\eeq
It follows that $J = 0$, so the Jacobi identity is satisfied if two of the test fields are collinear and the third points orthogonally.

\subsection{Proof of Jacobi identity for nonlinear functionals}
\label{s:Jacobi-general-proof}

Consider exponentials of three linear functionals of $\rho$ and $\bfv$:
	\beq
	{\cal F}[\rho, \bfv] = \exp i F[\rho,\bfv] \quad \text{where} \quad F[\rho, \bfv] = \int \left( \bff \cdot \bfv + \tl f \rho \right) \: d\bfr.
	\eeq
Here $\bff, \tl f$ are test field and test function as in \ref{a:jacobi}. ${\cal G}$ and $\cal H$ are defined similarly. Then
	\beq
	\deldel{\cal F}{\rho} = i \, \tl f \: {\cal F} \quad \text{and} \quad \deldel{\cal F}{\bfv} = i \, \bff \: {\cal F}, \quad {\rm e.t.c.}
	\eeq
Using (\ref{e:pb-between-functionals-of-rho-v}) the PB between the exponential functional $\cal F$ and an arbitrary functional $K$ is
	\beq
	\{{\cal F},K\}
     = i{\cal F}\int \left[ \frac{\bfw}{\rho} \cdot \left( \deldel{F}{\bfv} \times \deldel{K}{\bfv} \right) - \deldel{F}{\bfv} \cdot \grad \deldel{K}{\rho} + \deldel{K}{\bfv} \cdot \grad F_{\rho} \right] \: d\bfr = i \: {\cal F} \: \{F,K\}.
	\eeq
Taking $K= {\cal G}$, we have $\{{\cal F}, {\cal G}\}=-{\cal F}{\cal G} \{F,G\}$. Thus, the first term in the Jacobi expression becomes
	\beq
	{\cal J}_1 = \{ {\cal F}, \{ {\cal G}, {\cal H} \} \} = {\cal F} {\cal G} {\cal H} \left[ -i \{ F, \{ G, H \} \} + \{ G, H \} \: \{ F, G + H \} \right].
	\eeq
The product of PBs cancels out upon adding cyclic permutations, resulting in the Jacobi expression
	\beq
	{\cal J} \equiv {\cal J}_1 + \text{cyclic} = -i {\cal F} {\cal G} {\cal H} \left[  \{ F, \{ G, H \} \} + \{ G, \{ H, F \} \} + \{ H, \{ F, G \} \} \right] = 0.
	\eeq
So remarkably, the Jacobi expression for three exponential functionals is proportional to the corresponding expression for three linear functionals, which was shown to vanish in Appendix \ref{a:jacobi}. Thus we have proved the Jacobi identity for nonlinear functionals that are exponentials of linear functionals!

The Jacobi identity for finite linear combinations of exponential functionals follows from linearity of the PBs. Now we propose that an arbitrary nonlinear functional $P[\rho,\bfv]$ can be formed by the following {\it functional} Fourier transform:
	\beq
	P[\rho,\bfv]=\int D[\tl p,{\bf p}] \; \hat{P}[\tl p,{\bf p}] \exp\left[ i \int [\tl p\rho + {\bf p} \cdot \bfv] d\bfx \right]
	\eeq
where $\int D[\tl p,{\bf p}]$ denotes functional integration over the test fields and test functions. Now, suppose $\hat{P}[\tl p,{\bf p}],\hat{Q}[\tl q,{\bf q}]$ and $\hat{R}[\tl r,{\bf r}]$ are suitable [functionally integrable] functionals of the test functions and test fields. $P,Q,R$ are clearly linear combinations of the exponential functionals considered above. It is then clear, by the linearity of PBs in each argument, that the nonlinear functionals $P,Q,R$ must satisfy Jacobi's identity since any three exponential functionals do, as shown above. A rigorous treatment of the above functional Fourier transform is beyond the scope of this work. We observe that this type of functional calculus is freely used in the Hopf functional theory and in modern quantum field theories based on the Wiener measure and Feynman's path integrals. This approach may also be applied to proving the Jacobi identity for other PBs including the canonical $\{x,p \}$ Poisson brackets of particle mechanics.

\chapter{Lagrangian and PBs for R-Euler in Clebsch variables}
\label{a:Lagrangian}
Following the treatment of \cite{ecg-nm}, consider a Lagrangian for a system with $n$ degrees of freedom, $q_1, \cdots, q_n$ that is linear in velocities: 
	\beq
	L(q_s, \dot{q}_s)  =  \sum_{s=1}^n A_s(q)\dot{q}_s - V(q).
	\label{e:cm Lagrangian}
	\eeq
The Euler-Lagrange equations are first order in time and do not contain any `acceleration' terms
	\beq
	\sum_{r=1}^n \eta_{sr}(q)\dot{q}_r = \dd{V}{q_s}, \quad \text{for} \quad s = 1, 2, \cdots, n, \quad \text{where} \quad \eta_{sr}(q) = \dd{A_r}{q_s} - \dd{A_s}{q_r}.
	\eeq
Assuming the matrix $\eta_{sr}$ is non-singular, denote its inverse by $\eta^{sr}$ (which should be anti-symmetric). Then the equations of motion can be written as,
	\beq
	\dot{q}_s = \sum_{r=1}^n \eta^{sr}(q) \dd{V}{q_r}.
	\eeq
These equations can be written in Hamiltonian form by defining $H = V$ and Poisson bracket of functions of $q$ by
	\beq
	\{ f(q), g(q) \} = \sum_{r,s} \eta^{rs} \dd{f}{q_r} \dd{g}{q_s} 
	\eeq
One checks that the equations of motion follow:
	\beq
	\dot{q}_a = \{ q_a,  V(q) \} = \sum_{r,s} \eta^{rs}\delta_{ar} \dd{V}{q_s} =  \sum_s \eta^{as}\dd{V}{q_s}.
	\eeq
The Euler equation admits a Lagrangian formulation when the velocity and vorticity fields are expressed in terms of three scalar Clebsch potentials $\phi$, $\al$ and $\beta$:
	\beq
	\bfv = - \grad \phi - \al \grad \beta \quad \text{and} \quad
	\bfw = - \grad \al \times \grad \beta.
	\eeq
As explained in \cite{ecg-nm}, convenient variables for a Lagrangian formulation are $\rho, \phi, \beta$ and $\tl \al = \rho \al$. It turns out that the first order Lagrangian density ${\cal L} = \rho \phi_t + \tl \al \beta_t - {\cal H}$ where ${\cal H} = \rho \bfv^2/2 + U(\rho)$ is the conserved Hamiltonian density leads to the Euler equations. 
We show here that the Clebsch variables permit a Lagrangian formulation of the R-Euler equations as well with ${\cal L} = \rho \phi_t + \tl \al \beta_t - {\cal H}$ where $\cal H$ is now the conserved swirl energy density 
	\beq
	H = \int \left[ \half \rho \bfv^2 + U(\rho) + \half \la^2 \rho \bfw^2  \right]d\bfr.
	\eeq
Thus consider
	\beq
	{\cal L} = \rho \phi_t + \tl \al \beta_t 
	- \half \rho \left[ \grad \phi + \left( \frac{\tl \al}{\rho} \right) \grad \beta \right]^2 - U(\rho) - \half \la^2 \rho \left[ \grad \left(\frac{\tl \al}{\rho} \right) \times \grad \beta \right]^2.
	\label{e:lagrangian-density-R-Euler}
	\eeq
If $L = \int {\cal L} d\bfx$ is the Lagrangian, then the EL EOM are $\deldel{L}{\psi} = \DD{}{t} \deldel{L}{\psi_t}$. There are 4 EOM. (a) The evolution of $\phi$ is got by varying $L$ in $\rho$:
	\beqs
	\deldel{L}{\rho}  &=& \deldel{}{\rho} \int \left[\rho \phi_t - \frac{\rho}{2} \left(\grad \phi + \left( \frac{\tl \al}{\rho} \right) \grad \beta \right)^2 - U(\rho) \right]dy - \deldel{EE}{\rho}  \cr
	&=& \phi_t - \frac{(\grad \phi)^2}{2} + \frac{{\al}^2 (\grad \beta)^2}{2}  - U'(\rho) - \deldel{EE}{\rho}=0.
	\eeqs
The variation of the enstrophic energy $EE$ with density adds a new term to the evolution equation for $\phi$, which contributes to the twirl term in the evolution equation for $\bfv$:
	\beqs
	\deldel{EE}{\rho(x)} &=& \deldel{}{\rho(x)} \int \half \la^2 \rho \left[\grad \left(\frac{\tl \al}{\rho}\right) \times \grad \beta \right]^2 dy = \la^2 \rho \int \bfw \cdot \left[\grad \left(\frac{\tl \al \del(x-y)}{\rho^2}\right) \times \grad \beta \right]dy \cr
	&=& -\la^2 \rho \int \bfw\times \grad \beta \cdot \grad \left[\frac{\tl \al \del(x-y)}{\rho^2}\right] dy = \la^2 \grad \cdot (\bfw \times \grad \beta) \frac{\tl \al}{\rho} = \la^2 \al\grad \beta \cdot \grad \times \bfw.
	\eeqs
Thus the evolution equation for $\phi$ is
	\beq
	\phi_t = \frac{(\grad \phi)^2}{2} - \frac{{\al}^2 (\grad \beta)^2}{2} + h(\rho) + \la^2 \al \grad \times \bfw \cdot \grad \beta \quad \text{where} \quad h(\rho) = U'(\rho).
	\eeq
(b) The continuity equation $\rho_t + \grad \cdot (\rho \bfv) = 0$ is obtained from $\DD{}{t}\deldel{L}{\phi_t} = \rho_t = \deldel{L}{\phi}$ since
	\beq
	\deldel{L}{\phi} = - \int \rho \left[ \left(\grad \phi + \left(\frac{\tl \al}{\rho}\right) \grad \beta \right) \cdot \grad \del(x-y)\right]dy = \int \rho \bfv \cdot \grad \del(x-y) dy = -\grad \cdot (\rho \bfv).
	\eeq
(c) Similarly, the evolution of $\beta$ is obtained by varying in $\tl \al$:
	\beq
	\deldel{L}{\tl \al} = \beta_t  + \int \rho \bfv \cdot \left( \frac{\del(x -y)}{\rho} \grad \beta \right) dy - \deldel{EE}{\tl \al} = \beta_t + \bfv \cdot \grad \beta -\deldel{EE}{\tl \al} = 0.
	\eeq
The contribution to $\beta_t$ from the enstrophic energy is
	\beq
	\deldel{EE}{\tl \al} = - \la^2 \rho \int \bfw \cdot \left[\grad \left(\frac{\del(x-y)}{\rho}\right) \times \grad \beta \right]dy = \la^2 \rho \int \grad \left(\frac{\del(x-y)}{\rho}\right) \cdot (\bfw \times \grad \beta)\;dy = -\la^2 \grad \times \bfw  \cdot \grad \beta.
	\eeq
Thus, evolution equation for $\beta$ is $\beta_t + \bfv_* \cdot \grad \beta = 0$ where $\bfv_* = \bfv + \la^2 \grad \times \bfw$. (d) The evolution equation for $\tl \al$ is $\DD{}{t}\deldel{L}{\beta_t} = \tl \al_t = \deldel{L}{\beta}$ where
	\beqs
	\deldel{L}{\beta(x)} &=& \int \rho  \bfv \cdot \left( \frac{\tl \al}{\rho} \grad \del(x - y) \right) dy + \la^2 \rho \int \bfw \cdot \left[ \grad \left(\frac{\tl \al}{\rho}\right) \times \grad \del(x-y)\right] dy \cr
	&=& \grad \cdot(\tl \al \bfv) - \la^2 \rho \int \left( \bfw \times\grad \al\right) \cdot \grad \del(x-y)dy 
	= \grad \cdot(\tl \al \bfv) + \la^2 \rho \grad \cdot \left( \bfw \times \grad \al\right) \cr
	&=& \grad \cdot(\tl \al \bfv) +\la^2 \rho \grad \al \cdot \grad \times \bfw 
	=  \grad \cdot(\tl \al \bfv)+ \grad \left(\la^2\tl \al\right)\cdot \grad \times \bfw 
	= \grad \cdot(\tl \al \bfv)+\grad \cdot \left(\tl \al \la^2 \grad \times \bfw\right) \cr
	&=& \grad (\tl \al \bfv_*).
	\eeqs
Thus we get $\tl \al_t + \grad \cdot (\tl \al \bfv_*) = 0$. The above evolution equations for $\tl \al = \rho \al$ and $\rho$ together imply that $\al_t + \bfv_* \cdot \grad \al = 0$. It remains to show that the above evolution equations for $\phi, \al$ and $\beta$ derived from the Lagrangian (\ref{e:lagrangian-density-R-Euler}) imply the correct R-Euler equation. To this end, we calculate $\bfv_t$:
	\beqs
	\bfv_t  &=& - \grad \phi_t - \al_t \grad \beta - \al \grad \beta_t 
	= - \grad \phi_t + (\bfv_* \cdot \grad \al)\grad \beta + \al \grad (\bfv_* \cdot \grad \beta) \cr
	&=& - \grad \left(\frac{(\grad \phi)^2}{2} - \frac{\al^2 (\grad \beta)^2}{2} + h(\rho) + \la^2 \al (\grad \times \bfw) \cdot \grad \beta\right)- \left(\grad \phi + \al \grad \beta - \la^2 (\grad \times \bfw) \cdot \grad \al \right)\grad \beta \cr 
	&& - \al \grad \left(\grad \phi + \al \grad \beta - \la^2 \grad \times \bfw \right) \cdot \grad \beta.
	\eeqs
We split the above expression into terms of ${\cal O}(\la^0)$ and ${\cal O}(\la^2)$:
	\beqs
	\bfv_t &=& - \grad \left(\half (\grad \phi)^2 - \half \al^2 (\grad \beta)^2 + h(\rho)  \right)- \left((\grad \phi + \al \grad \beta) \cdot \grad \al\right)\grad \beta - \al \grad(\grad \phi + \al \grad \beta)\cdot \grad \beta\cr 
	&& - \grad \left(\la^2 \al (\grad \times \bfw) \cdot \grad \beta\right) + \la^2 \grad \beta (\grad \times \bfw) \cdot \grad \al  + \al\grad(\la^2 \grad \times \bfw \cdot \grad \beta)\cr 
	&=& -  \frac{\grad (\grad \phi)^2}{2} + \grad \left(\frac{\al^2}{2} (\grad \beta)^2\right) - \grad h - (\grad \phi \cdot \grad \al)\grad \beta - (\al \grad \beta \cdot \grad \al) \grad \beta - \grad( \al \grad \phi \cdot \grad \beta)\cr
	&&  + (\grad \phi \cdot \grad \beta) \grad \al - \al (\grad \al)(\grad \beta)^2 -  \al^2 \grad (\grad \beta)^2 - \al\grad(\la^2\grad \times \bfw \cdot \grad \beta)-  \grad \al (\la^2\grad \times \bfw \cdot \grad \beta) \cr 
	&&+ \la^2 \grad \beta (\grad \times \bfw) \cdot \grad \al  + \al\grad(\la^2 \grad \times \bfw \cdot \grad \beta).
	\eeqs
Now we use the Leibnitz rule to write $ - \al^2 \grad (\grad \beta)^2 =  - \grad \left( \al^2 (\grad \beta)^2 \right) + 2\al (\grad \al) (\grad \beta)^2 $ and combine it with $-\al (\grad \al)(\grad \beta)^2$ to get
	\beqs
	\bfv_t &=&  -\grad h - (\grad \phi \cdot \grad \al)\grad \beta + (\grad \phi \cdot \grad \beta) \grad \al + \al (\grad \al \cdot \grad \beta) \grad \beta + \al \grad \al (\grad \beta)^2 - \frac{\grad (\grad \phi)^2}{2} \cr
	&& - \frac{ \grad \left(\al^2 (\grad \beta)^2\right)}{2} - \grad (\al \grad \phi \cdot \grad \beta)- \la^2 \left[ \grad \al (\grad \times \bfw \cdot \grad \beta)- \grad \beta (\grad \times \bfw \cdot \grad \al) \right] \cr
	\bfv_t &=&  - \grad h -\bfw \times \bfv  - \half \grad \bfv^2 - \la^2 \bfT = - \grad h -(\bfv \cdot \grad)\bfv - \la^2 \bfw \times (\grad \times \bfw).
	\eeqs 
In the last step we used the following formulae for $\grad \bfv^2$, vorticity and twirl accelerations\footnote{\scriptsize Note that the curl of vorticity and advection terms take the following forms 
	\beqs \nonumber
	\grad \times \bfw &=& \grad^2 \beta \: \grad \al - \grad^2 \al \: \grad \beta + (\grad \beta \cdot \grad) (\grad \al) - (\grad \al \cdot \grad) (\grad \beta),	\cr
	\bfv \cdot \grad \bfv  &=& (\grad \phi \cdot \grad \al) \grad \beta - (\grad \phi \cdot \grad \beta) \grad \al + \al (\grad \al \cdot \grad \beta)\grad \beta - \al (\grad \al) (\grad \beta)^2 + \frac{\grad}{2} \left((\grad \phi)^2 + \al^2 (\grad \beta)^2\right) + \grad(\al \grad \phi \cdot \grad \beta).
	\eeqs}
\normalsize
	\beqs
	\grad \bfv^2 &=& \grad \left((\grad \phi)^2 + \al^2 (\grad \beta)^2 + 2 \al \grad \phi \cdot \grad \beta)\right), \cr
	\bfw \times \bfv &=& (\grad \phi \cdot \grad \al) \grad \beta - (\grad \phi \cdot \grad \beta) \grad \al + \al (\grad \al \cdot \grad \beta)\grad \beta - \al (\grad \al) (\grad \beta)^2 \quad \text{and}\cr 
	\bfT &=& \bfw \times (\grad \times \bfw)= \grad \al\: (\grad \times \bfw \cdot \grad \beta)- \grad \beta \:(\grad \times \bfw \cdot \grad \al).
	\label{e:gradv^2-vortcity-and-twirl-acc-clebsch}
	\eeqs
Thus, we recover the R-Euler momentum equation from the proposed Lagrangian.

\section{Poisson brackets in terms of Clebsch variables}

As discussed in \cite{ecg-nm} (page 129, 423), the first-order Lagrangian ${\cal L} = \rho \phi_t + \tl \al \beta_t - {\cal H}$ leads to a Hamiltonian formulation with Hamiltonian $H = \int {\cal H} d\bfr$ and canonical Poisson brackets
	\beq
	\{\phi(x), \rho(y)\} = \del(x - y) = \{ \beta(x), \tl \al (y)\}.
	\label{e:clebsch-pb-canonical}
	\eeq
In other words, the equations of motion take the form $\rho_t = \{ \rho, H \}, \beta_t = \{ \beta, H \}$ etc. Thus Clebsch potentials furnish canonical (i.e. Darboux) coordinates unlike the non-canonical Landau PBs among $\rho$ and $\bfv$. Indeed we may use (\ref{e:clebsch-pb-canonical}) to recover the Landau PBs among $\rho$ and $\bfv$. To see this, it is convenient to write the PB in terms of $\al = \tl \al/\rho$ instead of $\tl \al$:
	\beq
	\{ \phi(x), \al (y) \} = - \frac{\del(x - y) \al(y)}{\rho(y)} \quad \text{and} \quad \{\beta(x), \al(y) \} = \frac{\del(x -y)}{\rho(y)}.
	\eeq
Writing $\bfv = - \grad \phi - \al \grad \beta$ we get,
	\beq
	\{\rho(x), v_i(y) \} = \{\rho(x), - \pdr_{y^i} \phi - \al(y)\pdr_{y^i} \beta(y) \} = - \pdr_{y^i}\{ \rho(x), \phi(y)\} - \al(y) \pdr_{y^i} \{\rho(x), \beta(y)\} = \pdr_{y^i} \del(x -y).
	\eeq
Similarly we get $\{v_i(x), v_j(y)\} = \ov{\rho}\left((\pdr_i\beta) (\pdr_j \al) - (\pdr_j \beta) (\pdr_i \al)\right) \del( x - y) = \ov{\rho}\eps_{ijk} w_k \del(x -y)$.

\chapter{Minimality of twirl regularization}
\label{a:minimality}

Here we address the question of minimality/uniqueness of the twirl regularization, firstly in the context of neutral flows. We show that the twirl term $\la^2 \bfw \times (\grad \times \bfw)$ is the minimal symmetry-preserving conservative regularization term that can be added to the Euler equation while retaining the usual continuity equation and standard Hamiltonian formulation. The Euler equation is invariant under space-time translations, rotations, time reversal $T$ and parity $P$. We seek regularization term(s) involving $\rho$, $\bfv$ and derivatives of $\bfv$ that may be added to the Euler equation while preserving these symmetries. Any such term must be even under $T$, odd under $P$, not involve either $\bfr$ or $t$ explicitly, and transform as a vector under rotations. Furthermore, we seek terms with as few spatial derivatives, no time derivatives and as low a nonlinearity in $\bfv$ as possible. The term must preferably involve a (possibly dynamical) length $\la$ that can play the role of a short-distance cut-off. However, there are very many such terms even if we restrict to those quadratic in $\bfv$ with at most three derivatives [E.g. $\la^2 \bfw \times (\grad \times \bfw), \la^2 (\bfw \cdot \grad) \bfw \;\text{or}\; \la^2 \eps^{ijk} \pdr_j w_l \pdr_l v_k$] and it is an arduous task to identify all of them. We may simplify our task by requiring that the regularized equations follow from a Hamiltonian and the standard Landau PBs. Thus we seek a positive definite regularization term ${\cal H}_R$ involving $\bfv$ and its derivatives (dependence on $\rho$ is then fixed by dimensional arguments) that may be added to the ideal Hamiltonian density ${\cal H}_I = (1/2) \rho v^2 + U(\rho)$. The possibility of including derivatives of $\rho$ in ${\cal H}_R$ will be considered elsewhere (see Chapter \ref{s:r-gas-dynamics}). The advantage of working with the Hamiltonian is that we need only consider scalars rather than the more numerous vectors [regularizations that do not admit a Hamiltonian-PB formulation would however not be identified by this approach]. Due to the PB structure $(\{\bfv , \bfv \} \propto \pdr \bfv)$, the number of spatial derivatives in the velocity equation $\bfv_t = \{ \bfv, H \}$ is one more than that in $H$ and the degree of nonlinearity in $\bfv$ is the same as in $H$. Thus, ${\cal H}_R(v_i, \pdr_j v_i, \ldots)$ must be a $P$ and $T$-invariant scalar with a minimal number of derivatives and minimal nonlinearity in $\bfv$. It would be natural to ask that ${\cal H}_R$ be non-trivial in the incompressible limit, so that it may regularize vortical singularities in such flows. However, we find that such a restriction is not necessary. On the other hand, we do require that the regularization leave the continuity equation $\rho_t = \{\rho, H\}= -\grad \cdot (\rho \bfv)$ unaltered i.e., $\{ \rho, H_R\} = 0$, assuming decaying or periodic boundary conditions (BCs) in a box. Now, for ${\cal H}_R$ to be $P$-even, the sum of the number of spatial derivatives and degree of nonlinearity in $\bfv$ must be even. $T$-invariance as well as positive definiteness require that the degree of ${\cal H}_R$ in $\bfv$ be even. Thus we begin by listing all scalars at most quadratic in $\bfv$ with at most two derivatives. They are obtained by picking coefficient tensors $C^{ijk\ldots}$ below as linear combinations of products of the rotation-invariant tensors $\del^{ij}$ and $\eps^{ijk}$:
	\beqs
	1v,1\pdr: \: &&  C^{ij} \pdr_i v_j = \del^{ij} \pdr_i v_j  = \grad \cdot \bfv, \cr
	1v,2\pdr: \: &&  C^{ijk} \pdr_i \pdr_j v_k = \eps^{ijk}\pdr_i \pdr_j v_k = 0, \cr
	2v,0\pdr: \: && C^{ij} v_i v_j = \del^{ij}v_i v_j =\bfv^2, \cr
	2v,1\pdr: \: && C^{ijk} v_i \pdr_j v_k = \bfv \cdot \bfw; C^{ijk} \pdr_i (v_j v_k) = 0.
		\eeqs
$T$-invariance eliminates $\grad \cdot \bfv$, $P$-invariance eliminates $\bfv \cdot \bfw$ while $\bfv^2$ is already present in ${\cal H}_I$. Thus we are left with quadratic scalars with two derivatives:
	\beqs
	 C^{ijkl} v_i \pdr_j \pdr_k v_l &=& (c_1 + c_3) \bfv \cdot \grad (\grad \cdot \bfv) + c_2 \bfv \cdot \grad^2 \bfv \cr
	 C^{ijkl} \pdr_i v_j \: \pdr_k v_l &=& c_4 (\pdr_i v_j)^2 + c_5 \pdr_i v_j \: \pdr_j v_i + c_6 (\grad \cdot \bfv)^2 
	\cr
	C^{ijkl} \pdr_i \pdr_j (v_k v_l) &=& c_7 \grad^2 \bfv^2 + (c_8 + c_9) (2 \bfv \cdot \grad (\grad \cdot \bfv)  \cr
	&& + (c_8 + c_9)((\grad \cdot \bfv)^2 + \pdr_i v_j \: \pdr_j v_i).
	\eeqs
Here, $C^{ijkl}$ has been written as a linear combination of the products $\del^{ij} \del^{kl}$, $\del^{il} \del^{jk}$ and $ \del^{ik} \del^{jl}$. Note that the order of indices in $C^{ijk\cdots}$ does not matter: E.g., the space of scalars spanned by $C^{ijkl} \pdr_i \pdr_j (v_k v_l)$ and $C^{ljki} \pdr_i \pdr_j (v_k v_l)$ are the same. The coefficients in the linear combination must be functions of $\rho$ alone and on dimensional grounds must be constants $c_n = \la_n^2 \rho$ where $\la_n$ are position-dependent short-distance cutoffs. The identity $\grad^2 \bfv^2 = 2 \bfv \cdot \grad^2 \bfv + 2 (\pdr_i v_j)^2$ implies there are only five such linearly independent scalars. Since enstrophy density $\bfw^2 = (\pdr_i v_j)^2 - (\pdr_i v_j)(\pdr_j v_i)$ is a physically interesting linear combination, it is convenient to choose the basis for such scalars as $S_1 = \bfw^2$, $S_2 = \bfv \cdot \grad^2 \bfv$, $S_3 = (\pdr_i v_j)(\pdr_j v_i)$, $S_4 = (\grad \cdot \bfv)^2$ and $S_5 = \bfv \cdot \grad (\grad \cdot \bfv)$. We will now argue that $\bfw^2$ is the only independent regularizing term. Consider first the incompressible case where $S_4 = S_5 = 0$. Integrating by parts, $\int S_3 d\bfr = 0$ for decaying/periodic BCs. Furthermore, $\int S_2 \: d\bfr = \int \bfv \cdot \left[\grad (\grad \cdot \bfv) - \grad \times \bfw \right] d\bfr = \int \bfw^2 d\bfr$. Thus for incompressible flow we have shown that $\la^2 \rho \bfw^2$ is the only independent, positive definite $(\la^2 \rho > 0)$, Galilean-invariant regularization term. For compressible flow, we will not consider regularizations that alter the continuity equation, leaving that possibility for the future. Thus we require $\{ \rho, H_R \} = 0$. Since $\{\rho, \bfw\} = 0$, the term  $\bfw^2$ will not affect the continuity equation. On the other hand, the four other possibilities do modify it:
	\beq
	\{ \rho, \int (S_3 \;,\; S_4 \;, \; -S_2 \;, \; -S_5 ) \: d\bfr \} 
	= 2 \grad^2 (\grad \cdot \bfv).
	\eeq
To preserve the continuity equation, we may consider sums or differences of the above terms. Thus we replace the $S_{1, \cdots, 5}$ basis with the new basis $\tl S_1 = \bfw^2$, $\tl S_2 = \bfv \cdot \grad^2 \bfv + (\pdr_i v_j)(\pdr_j v_i)$, $\tl S_3 = \bfv \cdot \grad^2 \bfv + (\grad \cdot \bfv)^2$, $\tl S_4 = \bfv \cdot \grad (\grad \cdot \bfv) + (\pdr_i v_j)(\pdr_j v_i)$ and $\tl S_5 = \bfv \cdot \grad (\grad \cdot \bfv) + (\grad \cdot \bfv)^2$. As before, $\int \tl S_2 \: d\bfr =  \int \tl S_3 \: d\bfr = -\int \bfw^2 \: d\bfr$ and $\int \tl S_4 \: d\bfr = \int \tl S_5 \: d\bfr = 0$. Subject to these BCs, we have shown that $H_R = \int \la^2 \rho \bfw^2 \: d\bfr$ is the only positive-definite velocity-dependent regularizing term in the Hamiltonian that (a) preserves parity, time-reversal, translation, rotation and boost symmetries of the system, (b) does not alter the continuity equation and (c) involves at most two spatial derivatives and is at most quadratic in $\bfv$. We conclude that with the standard PBs, the twirl term $- \la^2 \bfw \times (\grad \times \bfw)$ with the constitutive relation $\la^2 \rho = $ const., is the only possible regularizing term in the Euler equation that is at most quadratic in $\bfv$ with at most 3 derivatives while possessing properties (a) and (b).

Extending these arguments to two-fluid plasmas, we may add a linear combination of $\bfw_i^2$, $\bfw_e^2$ and $\bfw_i \cdot \bfw_e$ to the Hamiltonian density. The cross term $\bfw_i \cdot \bfw_e$ leads to {\it direct} interspecies interaction in the velocity equations which we wish to avoid, preferring the ions and electrons to interact via the electromagnetic field. Thus we are left with $\bfw_i^2$ and $\bfw_e^2$ which lead to the vortical energies of ions and electrons considered in Section \ref{s:reg-eqns-two-fluid-compress}.

\chapter{A time averaged inequality}
\label{a:time-averaged-inequality}

The evolution of $\bfv$ in R-MHD is given by 
	\beq
	\dd{\bfv}{t} + \bfw \times \bfv_* = - \grad \sigma + \frac{\bfj \times \bfB}{\rho}.
	\eeq
Taking an inner product with $\rho \bfv$, we get an expression for the time derivative of the kinetic energy (which is known to be bounded)
	\beqs
	\rho \bfv \cdot \dd{\bfv}{t} + \rho \bfv \cdot(\bfw \times \bfv_*) &=& - \rho \bfv \cdot	\grad \sigma + \bfv \cdot (\bfj \times \bfB) \cr
	\imply	\qquad \dd{}{t}\left(\frac{\rho \bfv^2}{2}\right) - \frac{\bfv^2}{2}\dd{\rho}{t} &=& - \rho \bfv \cdot	\left( \grad \sigma + \la^2 \bfT - \frac{\bfj \times \bfB}{\rho} \right) \cr
	\imply	\qquad 	\dd{}{t}\left(\frac{\rho \bfv^2}{2}\right)  &=& - \rho \bfv 	\cdot \left( \grad \sigma + \la^2 \bfT - \frac{\bfj \times \bfB}{\rho}\right) -\frac{\bfv^2}{2}\grad \cdot (\rho \bfv) .
	\eeqs
where we have used the continuity equation for $\rho$. Now integrating over the flow domain and over time from $t = 0$ to $t = T$,
	\beqs
	\int_{0}^{T} \dd{(KE)}{t} \: dt &=& - \int_0^T \left( \rho \bfv \cdot \left(\grad \sigma + \la^2 \bfT -\frac{\bfj \times \bfB}{\rho}\right)+ \frac{\bfv^2}{2}\grad \cdot (\rho \bfv)\right) \: d\bfr \: dt \cr
	KE(T) - KE(0) &=& - \int_0^T \left( \rho \bfv \cdot \left(\grad \sigma + \la^2 \bfT -\frac{\bfj \times \bfB}{\rho}\right)+ \frac{\bfv^2}{2}\grad \cdot (\rho \bfv)\right) \: d\bfr \:dt.
	\eeqs 
Since $0 \leq KE \leq E^*$ at all times we have the inequality,
	\beq
	\left|\int_0^T \left( \rho \bfv \cdot \left(\grad \sigma + \la^2 (\bfw \times (\grad \times \bfw)) -\frac{\bfj \times \bfB}{\rho}\right)+ \frac{\bfv^2}{2}\grad \cdot (\rho \bfv)\right) \: d\bfr \:dt \right| \leq E^*.
	\eeq
We note that this inequality involves the twirl force which involves derivatives of vorticity, unlike our a priori bounds on kinetic energy and enstrophy (\ref{e:enstrophy-bound}).


\end{document}